\documentclass[10pt,twoside,openright]{book}

\usepackage[utf8]{inputenc}
\usepackage[danish,english]{babel}
\usepackage[sc]{mathpazo}
\usepackage[T1]{fontenc}
\usepackage{xcolor}
\definecolor{aaublue}{RGB}{0,0,0}% black
\usepackage{graphicx}
\usepackage{caption}
\usepackage{subcaption}
\usepackage{array,booktabs}
\usepackage{tabularx}
\usepackage{multirow}
\usepackage{framed}
\usepackage{tikz}
\usepackage[boxed,linesnumbered,algosection]{algorithm2e}
\usepackage{amsmath}
\usepackage{amssymb}
\usepackage{stmaryrd}
\usepackage{bbm}
\usepackage{amsthm}
\usepackage{thmtools}
\usepackage[framemethod=TikZ]{mdframed}
\usepackage{csquotes}
\usepackage{enumitem}
\usepackage{ebproof}
\newcommand\hypo{\Hypo}
\newcommand\infer{\Infer}

\usepackage[
  paperwidth=17cm, % width of a page
  paperheight=24cm, % height of a page
  outer=2.5cm, % right margin on an odd page
  inner=2.5cm, % left margin on an odd page
  top=2.5cm, % top margin
  bottom=2.5cm % bottom margin
  ]{geometry}
\usepackage{titlesec}
\titleformat*{\section}{\normalfont\Large\bfseries\color{aaublue}}
\titleformat*{\subsection}{\normalfont\large\bfseries\color{aaublue}}
\titleformat*{\subsubsection}{\normalfont\normalsize\bfseries\color{aaublue}}
\addto\captionsenglish{%this line is required when using the babel package
}

\usepackage{fancyhdr}
\usepackage{calc}
\usepackage{mparhack}

\usepackage[sectionbib]{chapterbib}
\usepackage{cite}

\usepackage{etoolbox}
\patchcmd{\thebibliography}{\section*{\refname}}{}{}{}

\usepackage[nottoc,section,numbib]{tocbibind}
\usepackage{appendix}
\usepackage{lastpage}

\usepackage{hyperref}

\newcommand{\papertitlepage}[5]{
  \chapter{#1}\label{#2}
  \chaptermark{}
  \vspace{2cm}
  \begin{center}
    \large #3
  \end{center}
  \vspace{3cm}
  \begin{center}
  \normalsize #4
  \end{center}
  \vspace*{\fill}
  \newpage\thispagestyle{empty}
  \vspace*{\fill}
    #5\par
  \vspace*{\fill}
  \cleardoublepage
}

\newenvironment{abstract}{\section*{Abstract}\it}{}

\newcommand{\defaultbib}{%
  \bibliographystyle{IEEEtranS}
  \bibliography{mybib}
}

\declaretheorem[numberwithin=section,style=plain]{theorem}
\declaretheorem[numberlike=theorem,style=definition,qed=$\blacktriangle$]{definition} 
\declaretheorem[numberlike=theorem,style=definition,qed=$\blacklozenge$]{example}
\declaretheorem[numberlike=theorem,style=definition,qed=$\blacklozenge$]{remark}
\declaretheorem[numberlike=theorem,style=plain]{lemma}
\declaretheorem[numberlike=theorem,style=plain]{proposition}
\declaretheorem[numberlike=theorem,style=plain]{corollary}
\declaretheorem[numberlike=theorem,style=plain]{conjecture}
\numberwithin{section}{chapter}
\numberwithin{figure}{section}
\numberwithin{table}{section}

\theoremstyle{definition}
\newmdtheoremenv[%
  outerlinewidth=1,%
  roundcorner=5pt,%
  outerlinecolor=gray,%
  leftmargin=60,%
  rightmargin=40,%
  backgroundcolor=gray!20,%
  innertopmargin=0pt,%
  ntheorem=false,%
]{contribution}{Contribution}

\usetikzlibrary{arrows,automata,positioning,calc}
\tikzset{
    every node/.style={
        font={\fontsize{8pt}{12}\selectfont}
    },
    >=latex
}
\tikzstyle{WTS}=[every node/.style={font=\fontsize{20}{40}\selectfont}, every state/.style={circle, draw=black, minimum size=15mm}, node distance=1cm, thick, ->, scale=0.5, transform shape]

\newcommand{\transm}[2]{\theta_{\mathcal{M}}\left(#1\right)\left(#2\right)}
\newcommand{\trans}[2]{\theta \left(#1\right)\left(#2\right)}
\newcommand{\transl}[2]{\theta^{-} \left(#1\right)\left(#2\right)}
\newcommand{\transr}[2]{\theta^{+} \left(#1\right)\left(#2\right)}
\newcommand{\quot}{h}

\renewcommand{\P}{\mathbb{P}}

\newcommand{\NP}{\mathbf{NP}}
\newcommand{\coNP}{\mathbf{coNP}}
\newcommand{\PTIME}{\mathbf{PTIME}}
\newcommand{\PSPACE}{\mathbf{PSPACE}}
\newcommand{\EXPTIME}{\mathbf{EXPTIME}}
\newcommand{\EXPSPACE}{\mathbf{EXPSPACE}}

\newcommand{\ain}{\mathtt{In}}
\newcommand{\aout}{\mathtt{Out}}

\newcommand{\ft}{\preceq}

\newcommand{\R}{\mathbb{R}}

\newcommand{\N}{\mathbb{N}}

\newcommand{\set}[1]{\left\{ #1 \right\}}

\newcommand{\A}{\mathcal{A}}
\newcommand{\B}{\mathcal{B}}
\newcommand{\C}{\mathcal{C}}
\newcommand{\Convex}{\mathrm{Convex}}
\newcommand{\Conv}{\mathrm{Conv}}

\newcommand{\M}{\mathcal{M}}

\newcommand{\D}{\mathcal{D}}

\newcommand{\simul}{\precsim}
\newcommand{\diam}[1]{\langle #1 \rangle}
\newcommand{\sat}[1]{\llbracket #1 \rrbracket}

\newcommand{\xra}[1]{\xrightarrow{\smash{#1}}}

\newcommand{\mlp}{\mathrm{TML}}
\newcommand{\mlpleq}{\mlp^{\leq}}
\newcommand{\mlpgeq}{\mlp^{\geq}}
\newcommand{\pert}[2]{(#1)_{#2}}
\newcommand{\upw}[1]{{\uparrow}_{#1}}
\newcommand{\cylinder}{\mathfrak{C}}
\newcommand{\prob}{\mathbb{P}}
\newcommand{\tas}[1]{\llparenthesis #1 \rrparenthesis}
\newcommand{\fft}{\sqsubseteq}

\newcommand{\simdist}{d}
\newcommand{\lsimdist}{\simdist^{\log}}

\newcommand{\supp}[1]{\mathrm{supp}(#1)}
\newcommand{\comp}[1]{\mathbin{\|_{#1}}}

\newcommand{\sproj}[1]{[ #1 ]}
\newcommand{\tproj}[1]{\langle #1 \rangle}
\newcommand{\aproj}[1]{\llbracket #1 \rrbracket}
\newcommand{\Exp}[1]{\mathit{Exp}\!\left[#1\right]}
\newcommand{\unif}[2]{\mathit{Unif}\!\left[#1,#2\right]}
\newcommand{\depth}{dpt}
\newcommand{\dirac}[1]{\mathit{Dirac}\!\left[#1\right]}
\newcommand{\dist}{\mathcal{D}}
\newcommand{\distone}{\dist_{=1}}

\newcommand{\bigo}{\mathcal{O}}
\newcommand{\due}{\mathcal{C}_\Lambda}
\newcommand{\wrt}[1]{\mathop{\mathrm{d}#1}}

\newcommand{\eventually}[1]{\Diamond^{#1}}
\newcommand{\paths}{\Pi}
\newcommand{\spaths}[2]{\paths_{#1}[#2]}
\newcommand{\ap}{\mathcal{AP}}
\newcommand{\subdist}{\dist_{\leq}}
\newcommand{\eqft}{\equiv}
\newcommand{\mon}[2]{\mathbin{\lessapprox^{#1}_{#2}}}
\newcommand{\smon}[2]{\mathbin{\leqq^{#1}_{#2}}}

\newcommand{\borel}{\mathbb{B}}
\newcommand{\cl}{\mathrm{cl}}
\newcommand{\len}{\mathrm{len}}

\DeclareSymbolFont{Symbols}{OMS}{cmsy}{m}{n}
\DeclareMathSymbol{\Emptyset}{\mathord}{Symbols}{"3B}

\begin{document}
%frontmatter
\frontmatter
\pagestyle{empty} %disable headers and footers
\pagenumbering{roman} %use roman page numbering in the frontmatter
%% Front page
\pdfbookmark[0]{Front page}{label:frontpage}%
\begin{titlepage}
  \addtolength{\hoffset}{0.5\evensidemargin-0.5\oddsidemargin} %set equal margins on the frontpage - remove this line if you want default margins
  \noindent%
  \begin{tabular}{@{}p{\textwidth}@{}}
    \toprule[2pt]
    \midrule
    \vspace{0.2cm}
    \begin{center}
    \Huge{\textbf{
      Behavioural Preorders on Stochastic Systems - Logical, Topological, and Computational Aspects\\% insert your title here
    }}
    \end{center}
    \vspace{0.2cm}\\
    \midrule
    \toprule[2pt]
  \end{tabular}
  \vspace{4 cm}
  \begin{center}
    {\large
      Ph.D. Dissertation%Insert document type (e.g., Project Report)
    }\\
    \vspace{0.2cm}
    {\Large
      Mathias Ruggaard Pedersen%Insert your group name or real names here
    }
  \end{center}
  \vfill
  \begin{center}
  Dissertation submitted October, 2018
  \end{center}
\end{titlepage}
\clearpage

%% Colophon
\thispagestyle{empty}
\noindent
\begin{tabularx}{\textwidth}{@{}lX}
    Thesis submitted: & October, 2018\\
    & \\
    PhD Supervisor: & Prof.\ Dr.\ Radu Mardare\\
                    & Aalborg University\\
    & \\
    Assistant PhD Supervisor: & Prof.\ Kim Guldstrand Larsen\\
                    & Aalborg University\\
    & \\
    PhD Committee:  & Assoc. Prof.\ Lisbeth Fajstrup\\ & Aalborg University\\
                    & \\
                    & Assoc. Prof.\ Dr.\ Ana Sokolova\\  & University of Salzburg\\
                    & \\
                    & Assoc. Prof.\ Marco Carbone\\ & IT University of Copenhagen\\
    & \\
    PhD Series:     & The Technical Faculty of IT and Design\\
                    & Aalborg University\\
    & \\                
    Department:     & Department of Computer Science\\
\end{tabularx}
\strut\vfill
\noindent
\begin{tabularx}{\textwidth}{@{}lX}
    ISSN: & xxxx-xxxx\\
    ISBN: & xxx-xx-xxxx-xxx-x\\
\end{tabularx}
\strut\vfill
\noindent Published by:\newline
Aalborg University Press\newline
Langagervej 2\newline
DK – 9220 Aalborg Ø\newline
Phone: +45 99407140\newline
aauf@forlag.aau.dk\newline
forlag.aau.dk
\strut\vfill
\noindent \copyright{} Copyright by Mathias Ruggaard Pedersen\newline
\strut\vfill
\noindent Printed in Denmark by Rosendahls, 2018
%\strut\vfill\vfill\vfill
%\noindent Normalsider: XXX sider (á 2.400 anslag inkl. mellemrum).\\
%Standard pages: XXX pages (2,400 characters incl. spaces).
\clearpage

%% Abstract
\chapter*{Abstract\markboth{Abstract}{Abstract}}\label{ch:Abstract}
\addcontentsline{toc}{chapter}{Abstract}
Computer systems can be found everywhere:
in space, in our homes, in our cars, in our pockets, and sometimes even in our own bodies.
For concerns of safety, economy, and convenience,
it is important that such systems work correctly.
However, it is a notoriously difficult task to ensure that
the software running on computers behaves correctly and does not contain any bugs.

One approach to ease this task is that of model checking,
where a model of the system is made using some mathematical formalism.
Requirements expressed in a formal language can then be verified against the model
in order to give guarantees that the model satisfies the requirements.
If the model is faithful to the system being modelled,
then the system itself will also satisfy the requirements.

For many computer systems such as satellites,
airbags, and traffic lights, time is an important factor.
As such, we need our formalisms and requirement languages
to be able to incorporate real time.

In this thesis,
we therefore develop formalisms and algorithms
that allow us to compare and express properties about real-time systems.
We first introduce a logical formalism for reasoning about upper and lower bounds on time,
and study the properties of this formalism,
including axiomatisation and algorithms for checking when a formula is satisfied.

We then consider the question of when a system is faster than another system.
We show that this is a difficult question which can not be answered in general,
but we identify some special cases where this question can be answered.
We also show that under this notion of faster-than,
a local increase in speed may lead to a global decrease in speed.
This is known as a timing anomaly,
and we take steps toward avoiding such timing anomalies.

Finally, we consider how to compare the real-time behaviour of systems
not just qualitatively, but also quantitatively.
Thus, we are not just interested in knowing whether a system is faster or slower than another system,
but also how much faster or slower it is.
This is done by introducing a distance between systems.
We show how to compute this distance
and that it behaves well with respect to properties expressed in a certain logical formalism.

\chapter*{Resumé\markboth{Resumé}{Resumé}}\label{ch:Resume}
\addcontentsline{toc}{chapter}{Resumé}
Computersystemer kan findes overalt: i rummet, i vores hjem, i vores biler, i vores lommer og endda nogle gange i vores egne kroppe.
På grund af økonomiske, sikkerheds- og bekvemmelighedsmæssige problemstillinger
er det vigtigt at sådanne systemer virker korrekt.
Det er dog et notorisk svært problem at sørge for at softwaren,
der kører på computere, opfører sig korrekt og ikke indeholder fejl.

En tilgang til at gøre dette problem nemmere er modeltestning,
hvor man laver en model af systemet i en matematisk formalisme.
Derefter kan krav, som er udtrykt i et formelt sprog,
blive verificeret op mod modellen
for at give en garanti for, at modellen opfylder kravene.

For mange computersystemer såsom satellitter, airbags og trafiklys,
er tid en vigtig faktor.
På grund af dette har vi behov for formelle sprog til at udtrykke krav og for formalismer,
som kan inkorporere realtid.

I denne afhandling udvikler vi derfor formalismer og algoritmer,
som lader os sammenligne og udtrykke egenskaber vedrørende realtidssystemer.
Først introducerer vi en logisk formalisme til at udtrykke egenskaber om øvre og nedre grænser på tid,
og vi studerer denne formalismes egenskaber,
herunder aksiomatisering og algoritmer til at fastslå om en formel er opfyldt.

Derefter betragter vi spørgsmålet om, hvornår et system er hurtigere end et andet.
Vi viser at dette er et svært spørgsmål, som generelt ikke kan besvares,
men vi identificerer nogle specialtilfælde hvor spørgsmålet kan besvares.
Vi viser også at med denne opfattelse af hurtigere-end,
kan en lokal forøgelse af hastighed føre til et globalt fald i hastighed.
Dette kaldes en tidsanomali,
og vi påbegynder en undersøgelse af,
hvordan sådanne tidsanomalier kan undgås.

Til slut beskæftiger vi os med hvordan vi kan sammenligne systemers realtidsadfærd
ikke kun kvalitativt, men også kvantitativt.
Dette betyder at vi ikke kun er interesserede i at vide, om et system er hurtigere eller langsommere end et andet system,
men også hvor meget hurtigere eller langsommere det er.
Dette gør vi ved at introducere en afstand mellem systemer.
Vi viser hvordan man kan beregne denne afstand,
og vi viser at den kan beskrives ved hjælp af en bestemt logisk formalisme.

\cleardoublepage
\pdfbookmark[0]{Contents}{label:contents}
\pagestyle{fancy} %enable headers and footers again
\tableofcontents

%% Preface
\chapter*{Preface\markboth{Preface}{Preface}}\label{ch:preface}
\pagestyle{fancy}
\addcontentsline{toc}{chapter}{Preface}
The research described in this thesis was carried out at Aalborg University from September 2015 to October 2018
as part of the research project Approximate Reasoning for Stochastic Markovian Systems
(project number 4181-00360) funded by The Danish Council for Independent Research (DFF-FNU).
The aim of this project is to develop an approximation theory for stochastic Markovian systems
from a logical, topological, and computational point of view.

The content of this thesis contributes to that aim by developing a notion of a faster-than relation
in the context of semi-Markov decision processes,
which subsume the popular formalism of continuous-time Markov chains.
This allows us to approximate a system by another system which operates slower or faster than the former system.
Furthermore, we extend this relation to a distance which can give quantitative information about how closely
one system approximates another.

From the logical point of view, we give a logical characterisation
of both the faster-than relation and the distance which extends it.
Furthermore, we consider common aspects in a logical analysis,
such as axiomatisation, satisfiability, and model checking.

From the topological point of view,
we consider the topology induced by the distance,
and how properties given by a logical specification behave in this topology.
In particular,
we show that approximate reasoning in the limit is sound,
meaning that when approximating closer and closer to the real system,
properties enjoyed by the approximations are preserved by the real system.

Finally, from the computational point of view,
we develop efficient algorithms for deciding the faster-than relation
and for computing the distance.

%% Acknowledgements
\chapter*{Acknowledgements\markboth{Acknowledgements}{Acknowledgements}}\label{ch:ack}
\pagestyle{fancy}
\addcontentsline{toc}{chapter}{Acknowledgements}

The making of this thesis would not have been possible without the involvement and support of a number of people.

First of all, I would like to thank the Danish Council for Independent Research
for funding the DFF-FNU project Approximate Reasoning for Stochastic Markovian Systems
which has paid for my PhD studies.

Also thanks to my supervisors Radu Mardare and Kim Guldstrand Larsen
for giving me the opportunity to pursue a PhD degree,
and for giving me both freedom and support in the process.

Thanks also go to my co-authors Mikkel Hansen, with whom I shared an office and bounced ideas off,
Nathanaël Fijalkow, from whom I have learnt much and who kindly hosted me for a week in London,
and in particular to Giorgio Bacci, who has been a great help through much of the process.

I thank Christel Baier and Daniel Gburek as well as the rest of the group on
algebraic and logical foundations of computer science at TU Dresden for hosting me for three months during my studies.

Special thanks go to Daniel Hillerström, my sparring partner throughout my undergrad studies,
with whom I shared stories and worries about the PhD life.

Lastly, but most importantly, I wish to thank my friends and family for their love and support,
in particular my parents who I can never thank enough for being the kind of people that I strive to be like,
and for always believing in me.

\vfill
\hfill Mathias Ruggaard Pedersen

\hfill Aalborg University, October 31, 2018

\cleardoublepage

%mainmatter
\mainmatter
\part{Overview}\label{chap:overview}

  \setcounter{figure}{0}
  \setcounter{subfigure}{0}
  \setcounter{section}{0}
  \setcounter{subsection}{0}
  \setcounter{table}{0}
  \setcounter{equation}{0}
  \newcommand{\introtitle}{Introduction}
\chapter{\introtitle}\label{chap:intro}

Computer systems today are ubiquitous,
from the tiny chips in watches and pacemakers
to the massive server farms that power the search engines
and other web services that we use every day.
For economic, safety, and costumer satisfaction reasons,
it is important that such systems function correctly:
If a fault is found in a system that has already been mass produced and sold,
the manufacturer may have to repair or replace a large number of systems,
resulting in additional expenses for the manufacturer.
Furthermore, we place high importance on the correctness of safety-critical systems
where lives may be at stake,
since even small and rare errors may result in serious injury or death,
such as the case of the Therac-25 radiation therapy machine,
where at least one person died from radiation poisoning
due to a software error~\cite{LT93}.
Lastly, because computer systems are so common in our everyday lives,
we, as customers and users of such systems, have an interest in them working correctly,
to save us confusion and frustration.

One of the important aspects of many computer systems is time,
especially for \emph{real-time systems} which operate under time constraints.
This means that when analysing the correctness of such systems,
we want to be able to understand how the system reacts to and evolves over time.
Consider the following examples.
\begin{description}
  \item[Solar-powered satellites:]
    Most satellites orbiting the earth rely on electricity from on-board solar panels.
    Since such satellites spend periods in the earth's shadow where it can not collect solar power,
    it is important to correctly plan and schedule the power consumption of the satellite
    in order to maintain the condition of the battery. %\cite{HKN17}.
  \item[Airbags:]
    The effectiveness of airbags is extremely sensitive to time.
    Airbags fully inflate in a matter of milliseconds,
    and if they inflate too soon, they may already have deflated at the moment of impact,
    whereas if they inflate too late, they may cause more harm than good. %\cite{Wallis490,AFGKLL09}.
  \item[Intelligent traffic lights:]
    Many traffic lights today are equipped with sensors to detect cars arriving at the traffic light.
    Based on this information, the traffic light can decide how to direct the traffic.
    Ideally, this could reduce the waiting time for road users in traffic lights.
\end{description}
In all of the above examples, we see that time is an important factor,
and numerous such examples exist.

Many of the properties of interest in real-time systems
are \emph{non-functional requirements}.
Non-functional requirements put constraints on the way in which a system can be realised,
in order to make the end user experience more pleasant \cite{CL09}.
Some of the key non-functional requirements that interact with time are reliability, throughput, and response time.
\begin{description}
  \item[Reliability:] A system should function correctly over long periods of time,
    even as components start to deteriorate.
  \item[Throughput:] A system should be able to produce or accept output at a consistent and high rate.
  \item[Response time:] A system should react quickly to inputs given to it.
\end{description}

We therefore need techniques and methods that will allow us
to verify properties such as non-functional requirements in real-time systems.
One successful approach to verifying the correctness of computer systems
is that of \emph{model checking} \cite{EC80,QS82,Clarke08}.
The aim of model checking is to build an abstract model of the system in question
using some precise mathematical formalism,
and then verifying that this model satisfies some specification.
Such specifications are often expressed by formulas of some logical formalism,
but they can also be represented by another model.

The act of translating a real system into a mathematical formalism,
also known as \emph{modeling}, is therefore central to model checking.
However, modeling has many difficulties attached to it.
One such difficulty is what we will call the \emph{approximate modeling problem},
which is the problem of accurately representing quantitative information such as time in the formalism.
All measurements in the real world are made within some error margin,
even for highly advanced measuring equipment.
The modelling formalism used should therefore be able to accommodate this inaccuracy.
Furthermore, many systems have some uncertainty attached to them,
such as robots that operate in physical environments
or environments where the robot interacts with other agents.
Finally, the act of modeling itself requires abstracting away some parts of the real system
in order to arrive at a formal model of the essential parts of the system.
This abstraction process also introduces an element of error,
since elements may be modeled wrongly, or key elements may be left out.
It is up to the skill and experience of the person doing the modeling to prevent this from happening.

All of these issues together makes the process of modeling difficult,
and we therefore need to develop formalisms, specification languages, and techniques
that allow us to account for this uncertainty and approximation.
That is the aim of this thesis.

\section{Structure of Thesis}
This thesis is split into two parts. Part~\ref{chap:overview} gives an overview of the current state of the art
and the papers that are part of this thesis, including the contributions that the papers make to further the state of the art
as well as the mathematical preliminaries necessary for the results of the papers.

Part~\ref{chap:papers} include the full versions of the papers outlined in Part~\ref{chap:overview}.
The papers presented in this thesis are extended versions,
including detailed proofs and additional material.

Each chapter has its own separate bibliography.

\section{State of the Art}
We first survey the formalisms, specification languages, and techniques
that have already been developed for reasoning about real-time systems.

\subsection{Models}
The two most common formalisms for modeling real-time systems are timed automata and Markov chains~\cite{AD94, BK08}.

\textbf{Timed automata.} Timed automata were introduced by Alur and Dill~\cite{AD90, AD94} in the early '90s
as a way of modeling time in automata theory.
The key concept in timed automata is that of \emph{clocks},
each of which keep track of time and can be independently reset.
Transitions can then be constrained such that a transition is only allowed
when the current value of each clock satisfies the constraints.
Many extensions of timed automata have been considered,
many of which include probabilistic behaviour.
The most significant, non-probabilistic extension of timed automata
is that of timed I/O automata~\cite{KLSV10,DLLNW10},
in which an important distinction is made between input and output actions.
Probabilistic timed automata~\cite{KNSS00,KNSS02} are timed automata
in which the transitions are given probabilistically,
and furthermore, the value to which the clocks reset are also given by a probability distribution.
Stochastic timed automata~\cite{BBBMBGJ14} modify the semantics of timed automata,
i.e. how the behaviour of a timed automaton is interpreted,
rather than modifying the timed automaton itself,
which then gives rise to a probabilistic process.
There is also another kind of stochastic timed automata,
which give different semantics to timed automata.
In this semantics, the focus is on many timed automata
operating in a network~\cite{DLLMPVW11,JLLMPS16}.
Here the semantics is a race between the different components,
in the sense that the component that has the smallest delay gets to choose the output.

\textbf{Markov chains.} Markov chains were developed in the early 20th century by Markov
and extended to continuous time by Kolmogorov.
Nowadays, Markov chains are used in almost all fields of engineering and science.
The use of Markov chains in model checking began in the second half of the '80s~\cite{vardi85,CY88,LS91},
but probabilistic automata, a generalisation of Markov chains,
have been studied in automata theory since their introduction in 1963 by Rabin~\cite{rabin63}.
Markov chains operate by choosing its transition to the next state according to a probability distribution.
In order to model real-time systems,
continuous-time Markov chains are often used instead.
In continuous-time Markov chains, the transitions are probabilistic as in Markov chains,
but in addition the waiting time in each state is governed by an exponential distribution.
Various aspects of model checking and specification have been extensively studied for continuous-time Markov chains~\cite{ASSB00,BHHK03,BKHW05,CHKM11},
even for the case of infinite-state chains~\cite{HHWZ09}.
However, many phenomena that occur in practice are not exponentially distributed,
so it is useful to extend the model of continuous-time Markov chains with distributions other than the exponential.
Semi-Markov chains therefore allow the waiting time in a state to be governed by an arbitrary distribution.
Aspects of model checking have also been investigated for semi-Markov chains~\cite{LHK01},
although they have received much less attention in the literature.
One can also generalise even further to obtain generalised semi-Markov processes~\cite{AB06,GJP06}
that allow different distributions for different actions.
Generalised semi-Markov processes are in fact close in spirit to probabilistic timed automata.
They operate by having a set of clocks, one for each possible action.
The clocks run down, and when a clock reaches $0$,
the action associated with that clock is fired.
This affects a probabilistic transition to a new state,
as well as a probabilistic reset of the clocks according to arbitrary distributions.
Another important variant of Markov chains used in model checking is that of Markov decision processes,
in which the actions of the process are determined by an outside controller~\cite{EKVY08,HMZPC12}.
Finally we mention the model of interactive Markov chains,
which combine continuous-time Markov chains with non-deterministic behaviour~\cite{hermanns2002}.

\textbf{Reactive and generative.}
When discussing probabilistic models,
one important distinction is that between reactive and generative models~\cite{GSS95}.
Reactive systems are those that react to input by taking a transition depending on the input given.
On the other hand, generative systems are those that generate output as it executes transitions.
As such, reactive systems take inputs, whereas generative systems create outputs.
Examples of reactive systems are probabilistic automata,
that takes words as input and either accepts or rejects the word,
and Markov decision processes, whose behaviour depends on the input given by the controller.
An example of a generative system is that of Segala automata~\cite{segala1995},
where each state can non-deterministically choose
between a number of different generative transitions.
Of course, the generative and reactive models can be combined
to have both inputs and outputs, as in the model of timed I/O automata.

\subsection{Logical Specification Languages}
Many different logical specification languages have been introduced in the literature for expressing properties related to time.

\textbf{Weighted logics.}
For weighted logics, weighted monadic second order logic was introduced to capture the behaviour of weighted automata~\cite{droste2005}.
This was later extended to many different, closely related models~\cite{babari2016,droste2006a,droste2006b,meinecke2006,fichtner2011}.
There have also been attempts to understand the connection between weighted monadic second order logic and probabilistic logics~\cite{bollig2009}.
Weighted modal logic~\cite{larsen1} was introduced to reason about the consumption of resources in weighted transition systems.
This formalism was later extended to handle recursion~\cite{larsen2014b,larsen2014a} and concurrency~\cite{LMX15}.
A weighted extension of the expressive $\mu$-calculus has also been developed~\cite{larsen2015}.

\textbf{Timed logics.}
The two most influential timed logics are linear temporal logic (LTL)~\cite{pnueli77} and computation tree logic (CTL)~\cite{CE81},
both of which are subsumed by CTL$^*$~\cite{EH86},
which in turn is subsumed by the $\mu$-calculus~\cite{kozen83}.
In LTL, time is linear, meaning that at each moment in time, there is only one possible future,
whereas in CTL, time is branching, meaning that at each moment in time,
we may simultaneously branch out into different future paths.
Both LTL and CTL have operators meaning ``in the next step, a property will hold'',
``one property will hold until another property holds'', and ``a property will eventually hold'',
as well as many other derived operators.
Both LTL and CTL in their original form consider time to be discrete.
Therefore real-time extensions have been developed,
such as timed CTL~\cite{ACD93} for CTL as well as metric interval temporal logic~\cite{AFH96} and timed LTL~\cite{dsouza03,BLS06} for LTL.

\textbf{Probabilistic logics.}
Lastly we discuss logical specification languages for reasoning about probabilistic behaviour.
CTL has been extended with probabilistic operators in PCTL~\cite{HJ94},
where one can express properties such as ``with probability at least $p$, a property will eventually hold''.
However, like CTL, time is discrete in PCTL.
To extend CTL with both probability and real time,
continuous stochastic logic (CSL) has been introduced~\cite{ASSB96,BHHK00,DP03}.
In another direction, Markovian logic~\cite{MCL12,KMP13} has been introduced,
based on earlier work on knowledge and probability~\cite{Fagin,aumann99a,aumann99b}.
Markovian logic has two kinds of operators,
one which states ``with at least probability $p$, we can go to a state satisfying some property'',
and another which states ``with at most probability $p$, we can go to a state satisfying some property''.

\subsection{Relations Among Computational Processes}
There are many ways in which one can compare and relate processes.

\textbf{Bisimulation.}
The most popular way to compare processes is that of bisimulation,
which was introduced by van Benthem~\cite{benthem76} under the name of zigzag connection,
and independently by Milner and Park~\cite{park1981,milner89} who introduced the name bisimulation.
In the context of Markov chains, bisimulation was defined by Larsen and Skou~\cite{LS91}
and has close connections to the notion of lumpability~\cite{buchholz94}.
Bisimulation is a notion of behavioural equivalence,
in which each process must be able to mimic the behaviour of the other process.

Since the concept of bisimulation has been so successful,
notions of bisimulation have been studied for most of the systems that are studied in the literature,
including timed automata~\cite{cerans92,ACH94}, probabilistic timed automata~\cite{ST10}, timed I/O automata~\cite{KLSV10},
Markov chains~\cite{LS91,jou1990,DEP02}, continuous-time Markov chains~\cite{hillston2005,BKHW05},
continuous-time Markov decision process~\cite{neuhausser2007}, and generalised semi-Markov processes~\cite{GJP06}.

Many of these notions of bisimulation follow from a more general theory of processes as coalgebras~\cite{rutten00,DEP02}.

\textbf{Simulation.}
Instead of asking when processes are behaviourally equivalent,
we can ask when one process can simulate or mimic the behaviour of another process.
This is the idea behind simulation relations.

Although simulation relations are not as ubiquitous in the literature as bisimulation relations,
they have still been studied for many types of systems, including timed automata~\cite{TAKB96},
probabilistic timed automata~\cite{ST10}, timed I/O automata~\cite{KLSV10},
Markov chains~\cite{JL91,DGJP03,BMOW05}, and continuous-time Markov chains~\cite{BKHW05}.

Simulation has also been studied from the coalgebraic point of view~\cite{HJ04}.

\textbf{Trace equivalence and inclusion.}
Another notion of behavioural equivalence,
which comes from automata theory,
is that of trace equivalence,
in which two processes are said to be equivalent
if they have the same possible executions, known as traces.
Generally speaking, it has proven more difficult
to reason about trace equivalences than to reason about bisimulation~\cite{HS96}.

Trace equivalences have been studied for timed automata~\cite{ACH94},
Markov chains~\cite{bernardo06}, and continuous-time Markov chains~\cite{WBM06}

A related, but non-symmetric notion of trace inclusion
has also been studied for Markov decision processes~\cite{FKS16}.

\textbf{Bisimulation distances.}
Although the concept of bisimulation has been influential,
the concept is not satisfactory for quantitative systems,
due to the approximate modeling problem discussed in the introduction.
This was originally emphasised by Jou and Smolka~\cite{jou1990,GJS90},
who instead of the qualitative bisimulation relations
advocated a quantitative bisimulation distance,
which not only says when processes behave differently,
but also by how much their behaviour differs.

Such a distance has been developed for timed automata~\cite{HMP05}
by measuring the difference between the time points along traces of the automata.
Distances have also been developed for weighted transition systems~\cite{AFS09,LFT11,FTL11},
including not only bisimulation distances, but also simulation distances that generalise
simulation rather than bisimulation.

Much work has been dedicated to studying bisimulation distances for probabilistic systems.
One way of defining such a distance is by allowing an approximation factor $\varepsilon$
on the probabilities involved in the definition of bisimulation.
This leads to the notion of $\varepsilon$-bisimulation,
and the distance is then given by the smallest $\varepsilon$
that allows for a $\varepsilon$-bisimulation~\cite{GJS90,DLT08}.
Another approach is to make use of the Kantorovich\footnote{This distance has many other names, the most notable alternatives being the Wasserstein, Hutchinson, and earth-mover distance.} distance~\cite{DD09}
between probability distributions.
This approach has been successfully applied to define bisimulation distances for
Markov chains~\cite{DGJP04,BW05,DCPP06}, Markov decision processes~\cite{FPP11},
continuous-time Markov chains~\cite{BBLM17}, and generalised semi-Markov processes~\cite{GJP06}.

Just as bisimulation distances generalise bisimulation to a quantitative setting,
so one could also generalise trace equivalence to a quantitative setting.
Together with bisimulation distances, such distances are often called behavioural distances,
since they quantify the dissimilarity between the behaviour of systems.
Using the total variation distance, such generalisations of trace equivalence has been studied for
Markov chains~\cite{kiefer18} and semi-Markov chains~\cite{BBLM15},
and various other ways to generalise trace equivalence have been considered for non-deterministic Markov chains~\cite{castiglioni18}.

\textbf{Faster-than relations.}
Another way to relate the behaviour of two processes
is to ask when one process is faster than another.
For non-probabilistic and discrete-time systems,
i.e. systems with no probabilities and where each transition or step takes one time unit,
such faster-than relations have been well-studied~\cite{corradini1995,LV01,luttgen2006,MT91,satoh1994}.
In particular, they have been studied for timed automata~\cite{GNA12}
and for Petri nets~\cite{vogler1995a,vogler1995b},
which are systems that model the production and consumption
of resources as transitions are taken.
However, for continuous-time systems, very little work has been done.
To the best of our knowledge, the only work that has been done
on faster-than relations for continuous-time probabilistic systems
is on continuous-time Markov chains~\cite{BKHW05},
where the simulation relation for continuous-time Markov chains
contains a condition which informally says that the process
which is simulating another process is allowed to fire faster than the process it is simulating.

\subsection{Algorithms}
In order to make use of the concepts we have surveyed so far,
one needs algorithms that allow us to determine if e.g. a formula in a logical language can be satisfied
or two models are in a certain relation.
Preferably, we also want the algorithms to be efficient,
so that we can actually run them on computers and get an answer within a reasonable time frame.

\textbf{Model checking.}
The model checking problem asks,
given a model and a formula, whether the model satisfies that formula.
For most logical specification languages, this is a simple problem to verify,
and efficient algorithms therefore exist.
The model checking problem is in $\PTIME$ for CTL~\cite{CE81,CES86},
PCTL~\cite{HJ94}, and CSL~\cite{BHHK00}.
For LTL~\cite{SC85} and timed CTL~\cite{ACD93},
the problem is $\PSPACE$-complete,
whereas it is $\EXPSPACE$-complete for metric interval temporal logic~\cite{AFH96}.

\textbf{Satisfiability.}
Another natural problem for a logical language is the satisfiability problem,
which asks whether a given formula can be satisfied at all,
i.e. whether we can find some model which satisfies the formula.
This can be seen as a sanity check for a formula:
If a formula is not satisfiable,
then it is unreasonable to ask for a system
which has the property expressed by the formula.
The satisfiability problem is often harder than the model checking problem,
since in the model checking problem, we only have to consider a single model,
whereas in the satisfiability problem, we have to consider all models.
For LTL, the satisfiability problem is $\PSPACE$-complete~\cite{SC85,VW94},
for CTL it is $\EXPTIME$~\cite{emerson90}, for metric interval temporal logic
it is $\EXPSPACE$-complete~\cite{AFH96}, and for timed CTL it is undecidable~\cite{ACD93}.
For PCTL, the decidability (and hence also complexity) of the satisfiability problem
is a longstanding open problem, but various fragments
have been successfully studied~\cite{HS86,BFKK08,BFS12,MM15,katoen:sat,KR18}.
For a certain weighted logic called recursive weighted logic,
the satisfiability problem has been shown to be decidable~\cite{LMX18}.

\textbf{Bisimulation and simulation.}
Since bisimulation, and to some extent simulation,
are core concepts when reasoning about the behavior of systems,
algorithms have been developed to decide whether two systems
are in a simulation or bisimulation relation.
Most algorithms for timed automata use the notion of regions,
which is a way to partition the state space into finitely many classes.
For timed automata, both simulation and bisimulation are $\EXPTIME$-complete~\cite{LW97,cerans92,LS00},
and the two are in $\EXPTIME$ as well for probabilistic timed automata~\cite{ST10}.
Algorithms for deciding simulation and bisimulation
between Markov chains make use of the maximum flow problem for networks~\cite{CHM90}.
These algorithms are in $\PTIME$ for Markov chains,
continuous-time Markov chains, and Markov decision processes~\cite{baier2000,zhang09}.

\textbf{Trace equivalence and inclusion.}
Generally speaking, reasoning about traces is much harder than reasoning about bisimulation.
For timed automata, both trace equivalence~\cite{AD94} and trace inclusion~\cite{OW04,AD94}
are undecidable. See~\cite{AM04} for a survey on these and related results.
For probabilistic automata, the trace inclusion problem is also undecidable~\cite{CL89,blondel2003}.
However, somewhat surprisingly, the trace equivalence problem for probabilistic automata
is in fact decidable and in $\PTIME$~\cite{schutzenberger61,FKS16}, using techniques from linear algebra.
See~\cite{Fijalkow17} for an overview of undecidability results for probabilistic automata.

\textbf{Bisimulation distances.}
Algorithms to compute bisimulation distances are different in nature
from the algorithms we have surveyed so far,
since we do not just need a yes/no answer, but must output a number.
Therefore it also makes sense to sometimes approximate this number up to an error margin,
meaning that instead of computing the number,
we compute a number that is sufficiently close to the actual number.
For probabilistic systems, computing the bisimulation distance
makes use of linear programming to solve the transportation problem
(see e.g.~\cite[pp. 221-223]{schrijver86}).
This results in a $\PTIME$-complete algorithm for Markov chains~\cite{CBW12,BW06}.
An algorithm also exists for continuous-time Markov chains~\cite{BBLM17},
however, technically speaking, this is an approximation algorithm,
since the actual value may be irrational.
If we instead consider the total variation distance,
then the threshold problem, which asks whether the distance is greater than a given threshold,
is undecidable~\cite{kiefer18}.
However, the distance can be approximated in $\PSPACE$~\cite{BBLM15,kiefer18}.

\section{Contributions}
This section summarises the most significant contributions
that this thesis adds to the state of the art.
The content of this thesis is based on the following papers.

\begin{itemize}
  \item Paper A: \textbf{Reasoning About Bounds in Weighted Transition Systems} is under submission for Logical Methods in Computer Science~\cite{HLMP17} and is an extended version of \textbf{A Complete Approximation Theory for Weighted Transition Systems}, which was published in the proceedings of the Second International Symposium on Dependable Software Engineering: Theories, Tools, and Applications (2016)~\cite{hansen2016}.\\
        {\small\textbf{Co-authors:} Mikkel Hansen, Kim Guldstrand Larsen, and Radu Mardare.}
  \item Paper B: \textbf{Timed Comparisons of Semi-Markov Processes} was published in
    the proceedings of the 12th International Conference on Language and Automata Theory and Applications (2018)~\cite{PFBLM18}. \\
        {\small\textbf{Co-authors:} Nathana\"el Fijalkow, Giorgio Bacci, Kim Guldstrand Larsen, and Radu Mardare.}
  \item Paper C: \textbf{A Faster-Than Relation for Semi-Markov Decision Processes} is based on an unpublished manuscript~\cite{PBL18}. \\
        {\small\textbf{Co-authors:} Giorgio Bacci and Kim Guldstrand Larsen.}
  \item Paper D: \textbf{A Hemimetric Extension of Simulation for Semi-Markov Decision Processes}
    was published in the proceedings of the 15th International Conference on Quantitative Evaluation of Systems (2018)~\cite{PBLM18}. \\
        {\small\textbf{Co-authors:} Giorgio Bacci, Kim Guldstrand Larsen, and Radu Mardare.}
\end{itemize}
  
Paper A takes a simple view of time,
where the time that something takes is given explicitly
as the weight of taking a transition in a graph.
Such systems are called \emph{weighted transition systems}.

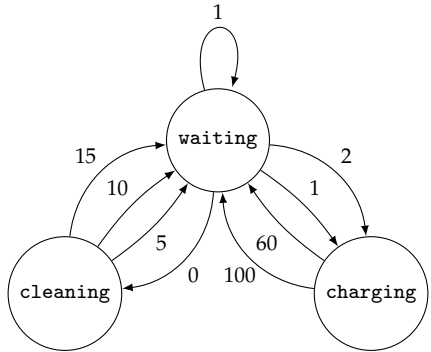
\begin{figure}
  \centering
  \begin{tikzpicture}
    \node[state] (0) {\tt waiting};
    \node[state] (1) [below left = of 0] {\tt cleaning};
    \node[state] (2) [below right = of 0] {\tt charging};
    
    \path[->] (0) edge [loop above] node {$1$} (0);
    
    \path[->] (0) edge [bend left = 10, above] node [above, xshift=1.5mm] {$1$} (2);
    \path[->] (0) edge [bend left = 40, above] node [above, xshift=1.5mm] {$2$} (2);
    \path[->] (2) edge [bend left = 10, below] node [below, xshift=-2mm] {$60$} (0);
    \path[->] (2) edge [bend left = 40, below] node [below left, xshift=2mm] {$100$} (0);
    
    \path[->] (0) edge [bend left = 40, below] node [below, xshift=1mm] {$0$} (1);
    \path[->] (1) edge [bend right = 10, below] node [below, xshift=1mm] {$5$} (0);
    \path[->] (1) edge [bend left = 10, above] node [above, xshift=-2mm] {$10$} (0);
    \path[->] (1) edge [bend left = 40, above] node [above, xshift=-2mm] {$15$} (0);
  \end{tikzpicture}
  \caption{A simple model of a robot vacuum cleaner.}
  \label{fig:wts-example-intro}
\end{figure}

\begin{example}\label{ex:wts-intro}
  Figure \ref{fig:wts-example-intro} shows an example of a weighted transition system.
  It is a model of a robot vacuum cleaner
  which has three states: a {\tt waiting} state, a {\tt cleaning state}, and a {\tt charging} state.
  In the {\tt waiting} state, the system can choose to simply keep waiting for another minute.
  However, it can also choose to immediately start cleaning by going to the {\tt cleaning} state.
  Depending on how dirty the floor is, this cleaning could take 5, 10, or 15 minutes,
  after which the system returns to the {\tt waiting} state.
  In the {\tt waiting state}, the system can furthermore decide that it needs to recharge its batteries.
  This is done by going to the {\tt charging} state,
  which may take 1 or 2 minutes depending on how far away the robot is from the charging station.
  The charging itself takes either 60 or 100 minutes,
  depending on how depleted the batteries are,
  after which the system returns to the {\tt waiting} state.
\end{example}

However, because of the approximate modeling problem,
putting an exact value for the amount of time that something takes
is not feasible.
Paper A therefore suggests that we instead reason about
lower and upper bounds on the time taken,
which is a more manageable engineering task.

This is done by introducing a logical specification language
which lets us express properties such as
\begin{displayquote}
  ``it takes at most 0.1 seconds to go to a state where the airbag is deployed''
\end{displayquote}
and
\begin{displayquote}
  ``it takes at least 0.01 seconds to go to a state where the airbag is deployed''.
\end{displayquote}
The main contributions of Paper A are then to study the
properties of this language.

We show first of all that the language describes
exactly the behaviour of systems.
This means that if two systems have the same behaviour,
then they must also satisfy the same properties of our language,
and vice versa.
This kind of property is known by various names in the literature,
including bisimulation invariance, logical characterisation, adequacy,
Hennessy-Milner property, and full abstraction.

\begin{contribution}
  We present a language for reasoning about lower and upper bounds
  in weighted transition systems
  and we show that this language characterises exactly those
  systems that have the same kind of behaviour.
\end{contribution}

We then present a proof system given by a set of axioms
which fully describe our logical language,
in the sense that anything that can be proved from the axioms must be true,
and anything which is true can be proved from the axioms.
This means that the axioms are both sound and complete.

Finally, we present two decision algorithms.
The first algorithm decides the \emph{model checking problem},
which asks whether a given system satisfies some formula.
The second algorithm decides whether,
for a given formula in a our language,
there exists some weighted transition system in which that formula is true.
If such a system exists,
the formula is said to be \emph{satisfiable}.

\begin{contribution}
  We provide a complete axiomatisation of the logical specification language,
  and give an algorithm for deciding the model checking problem
  and an algorithm for deciding satisfiability of a formula.
\end{contribution}

Papers B, C, and D take a different view of time.
Here, the approximate modeling problem
is tackled by introducing probability into the modeling formalism.
We thus consider \emph{semi-Markov processes},
where both the time spent in a given state,
and the transition step to a next state
are governed by probability distributions.

\begin{figure}
  \centering
  \begin{tikzpicture}
    \node[state, circle split] (1) {$0.1$ \nodepart{lower} $g_1, r_2$};
    \node[state, circle split] (2) [right = 5cm of 1] {$2$ \nodepart{lower} $g_1, r_2$};
    \node[state, circle split] (3) [below = 3cm of 1] {$0.1$ \nodepart{lower} $r_1, g_2$};
    \node[state, circle split] (4) [right = 5cm of 3] {$2$ \nodepart{lower} $r_1, g_2$};
    
    \path[->, thick] (1) edge [bend left] node [above] {$\texttt{car}_1?, \texttt{stay}! : 1$} (2);
    \path[->, thick] (1) edge [bend right = 10] node [left] {$\overline{\texttt{car}_1}?, \texttt{change}! : 1$} (3);
    
    \path[->, thick] (2) edge [loop above] node [above] {$\texttt{car}_1?, \texttt{stay}! : 0.9$} (2);
    \path[->, thick] (2) edge [bend left = 10] node [left, very near start, xshift=6mm, yshift=6mm] {$\texttt{car}_1?, \texttt{change}! : 0.1$} (3);
    \path[->, thick] (2) edge [bend left = 25] node [right, very near start] {$\overline{\texttt{car}_1}?, \texttt{change}! : 1$} (3);
    
    \path[->, thick] (3) edge [bend right] node [below] {$\texttt{car}_2?, \texttt{stay}! : 1$} (4);
    \path[->, thick] (3) edge [bend right = 10] node [right] {$\overline{\texttt{car}_2}?, \texttt{change}! : 1$} (1);
    
    \path[->, thick] (4) edge [loop below] node [below] {$\texttt{car}_2?, \texttt{stay}! : 0.9$} (4);
    \path[->, thick] (4) edge [bend right = 10] node [left, very near start, xshift=6mm, yshift=-6mm] {$\texttt{car}_2?, \texttt{change}! : 0.1$} (1);
    \path[->, thick] (4) edge [bend right = 25] node [right, very near start] {$\overline{\texttt{car}_2}?, \texttt{change!} :1$} (1); 
  \end{tikzpicture}
  \caption{A simple model of an intelligent traffic light.}
  \label{fig:smp-example-intro}
\end{figure}
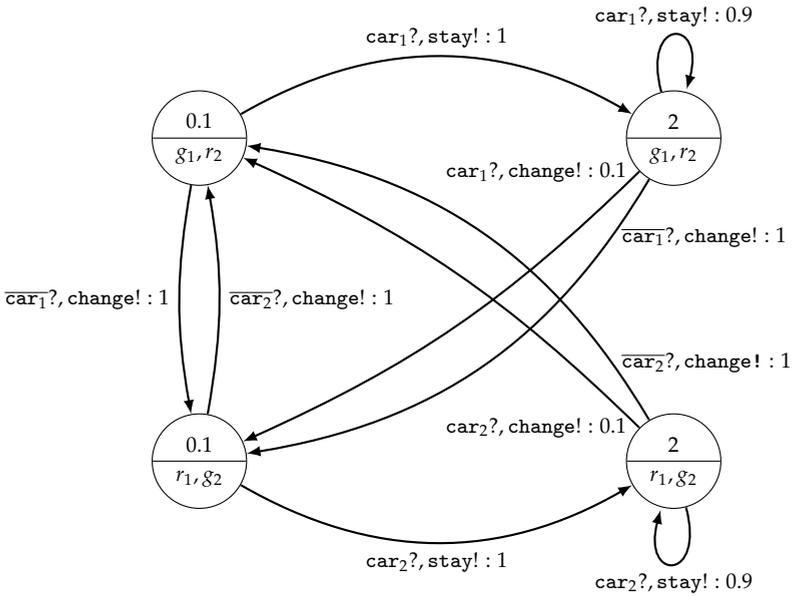

\begin{example}\label{ex:smp-intro}
  Figure \ref{fig:smp-example-intro} shows an example of a semi-Markov process,
  modeling a simple traffic light at an intersection with two roads.
  In this example, there are four states,
  where the top two states show green light for the first road and red light for the second,
  and the bottom two states show red light for the first road and green light for the second.
  This is indicated by the labels $g_1$ and $r_2$ in the top two states,
  and $r_1$ and $g_2$ in the bottom two states.
  The number in each state denotes the rate of an exponential distribution,
  meaning that the higher the number,
  the faster that state will take a transition.
  The labels on the edges between states denote what action is taken,
  whether it is an input or an output, and the probability of taking that action.
  Inputs are denoted by $?$ and outputs by $!$.
  For example, $\texttt{car}_2?, \texttt{stay}! : 0.9$
  means that when receiving the input $\texttt{car}_2$,
  the system has a $90\%$ chance to take this transition,
  which also outputs \texttt{stay}.
  
  Imagine that the system starts in the top left state,
  where the light is green for the first road and red for the second.
  If now the system sees a car approaching on the first road,
  indicated by the action $\texttt{car}_1?$,
  then the system changes to the top right state,
  in which the system has a high chance of keeping the light green for the first road.
  Hence, seeing a car approaching on the first road will tend to cause the light to remain green for that road.
  On the other hand, if a car is observed on the second road, both roads, or none of the roads,
  all of which is indicated by the action $\overline{\texttt{car}_1}?$,
  then the system changes the red and green lights by going to the bottom left state,
  in which the light is red for the first road and green for the second, and outputting $\texttt{change}!$.
  The top right state has a higher rate than the top left state,
  since in this state the system needs to be able to quickly react to a car approaching on the second road,
  so that the second road is not stuck with red light for too long.
  In this state, when another car is observed on the first road, again indicated by $\texttt{car}_1?$,
  then the system has probability $0.9$ of staying in this state and outputting $\texttt{stay}!$,
  thus keeping the light green for the first road a little longer.
  However, even if another car is observed on the first road,
  the system may still change the lights with probability $0.1$,
  out of fairness to pedestrians who are not covered by the model.
  If a car is observed on the second road, both roads, or none of the roads while in this state,
  the system again changes to the bottom left state, thus changing the light to green for the second road.
  
  The bottom two states function symmetrically,
  with the light being green for the first road and red for the second road.
\end{example}

Papers B and C consider how to compare semi-Markov processes
with respect to the amount of time it takes to execute
sequences of actions.
Here, the idea is that one process should be faster than another
if whatever sequence of actions the slow process can do,
the fast process can do faster and with a higher probability.
This is the so-called trace-based semantics of semi-Markov processes,
hence we will refer to this as the trace-based faster-than relation.

We show that the trace-based faster-than relation is undecidable,
meaning that there is no algorithm which can determine
whether a given process is faster than another.
This undecidability result is quite robust,
since the relation remains undecidable
even if we consider approximating the relation
up to a multiplicative error term.

\begin{contribution}
  We show that deciding the trace-based faster-than relation
  is a difficult problem.
  In particular, the relation is undecidable
  and approximating it up to a multiplicative constant is impossible.
\end{contribution}

However, we still obtain some positive results.
If we consider approximation up to an additive constant
rather than a multiplicative constant,
then we can recover decidability under the following two assumptions.
The first assumption is that we only consider
the behaviour of the processes up to some finite point in time.
Thus we allow the fast process to become slower than the other process,
as long as this happens only sufficiently far into the future.
The second assumption is that a process must spend some non-zero
amount of time in each state that it visits.
In other words, instantaneous change of state is not allowed,
which is a reasonable assumption from a practical point of view.
Such processes are called slow.

\begin{contribution}
  We give an algorithm for approximating
  a time-bounded version of the trace-based faster-than relation
  up to an additive constant for slow processes.
\end{contribution}

Another way to recover decidability is to
restrict ourselves to so-called unambiguous processes.
An unambiguous process is one in which
the next state is determined uniquely by the output,
so if you know the current state of the process
and you know what is output,
then you know exactly what state the process ends up in next.

\begin{contribution}
  We give an algorithm for unambiguous processes
  which can decide whether one process is trace-based faster than another.
\end{contribution}

We also study the trace-based faster-than relation from the logical point of view.
We describe a simple logical language which is expressive enough to characterise exactly those states
that are related by the trace-based faster-than relation.
We study the properties of the language and show in particular that both the satisfiability
and the model checking problem are decidable.

\begin{contribution}
  We introduce a logical language which characterises the trace-based faster-than relation
  and we show that both the satisfiability problem and the model checking problem
  for this language are decidable.
\end{contribution}

Lastly we consider the compositional aspects of the trace-based faster-than relation.
When considering a number of components operating in parallel,
we would like to replace one or more of these components
with another component which is faster.
However, we show that doing so may lead to parallel timing anomalies,
meaning that although the replaced component is faster than the previous component,
this may result in the overall system becoming slower.
In an attempt to better understand such parallel timing anomalies
and how to avoid them,
we identify a set of conditions which will ensure that parallel timing anomalies do not occur.
Furthermore, we give an algorithm to check whether these conditions are satisfied.

\begin{contribution}
  We give examples of parallel timing anomalies occuring
  for the trace-based faster-than relation.
  However, we also describe some conditions under which
  parallel timing anomalies can not occur,
  and we develop an algorithm for checking whether these conditions are met.
\end{contribution}

Paper D also considers semi-Markov processes.
However, here we furthermore address the approximate modeling problem
by not only comparing processes qualitatively, but also quantitatively.
Thus we are not only able to say whether a process is faster than another,
but are also able to quantify how much slower or faster it is.

For this we consider what we will call the simulation-based faster-than relation,
so called because it is based on the idea of one process simulating another.
Roughly speaking, a process is simulation-based faster than another process
if every step in the slow process can be simulated by the faster process,
except the faster process is allowed to do the step faster.
To turn this into a quantitative measure,
we introduce an acceleration factor through which we obtain a distance between processes.
The acceleration factor allows us to increase the speed of a process,
so that by accelerating a process by this factor,
it may become faster than another process which it was originally slower than.
We then define the distance between from one process to another
as the smallest acceleration factor necessary to make second process faster than the first.
Our first result is an algorithm for computing this distance.

\begin{contribution}
  We describe an algorithm for computing the distance from one process to another.
  This algorithm runs in polynomial time using known techniques,
  making it relevant for use and implementation in practice.
\end{contribution}

We then consider the compositional aspects of the distance.
Here the relevant notion is that of non-expansiveness:
Composition should not expand the distance between processes.
If composition is non-expansive,
then it also follows that we are guaranteed that parallel timing anomalies do not occur.
We show that under some mild conditions,
which are satisfied by many types of composition found in the literature,
composition is indeed non-expansive with respect to our distance.

\begin{contribution}
  We show that, under mild assumptions,
  composition is non-expansive with respect
  to the distance between semi-Markov processes.
\end{contribution}

Finally we consider a logical language which we call timed Markovian logic.
This language can express properties such as
\begin{displayquote}
  ``with probability at least 0.99 we leave the state where the traffic light is red before 10 seconds have passed''
\end{displayquote}
and
\begin{displayquote}
  ``with probability at most 0.5 we end up in a state where the traffic light is green.''
\end{displayquote}
We show that this language gives a logical characterisation of the simulation-based faster-than relation.
Furthermore, we extend this to a quantitative generalisation of logical characterisation for our distance.

\begin{contribution}
  We introduce a logical specification language called timed Markovian logic
  and show that this language characterises both
  the simulation-based faster-than relation
  and the distance between semi-Markov processes.
\end{contribution}

The rest of Part I is structured as follows.
In Chapter \ref{chap:prelim} we introduce some mathematical concepts and notation
that we will use throughout the thesis,
including the definition of weighted transition systems
and semi-Markov processes that we have mentioned here.
We describe in more detail the results of Paper A in Chapter~\ref{chap:logic-intro},
the results of Paper B and Paper C in Chapter~\ref{chap:tft},
and the results of Paper D in Chapter~\ref{chap:sft}.
For the complete details, see the full papers in Part~\ref{chap:papers}.

{\small\bibliographystyle{IEEEtranS}\bibliography{mybib}}

  \setcounter{figure}{0}
  \setcounter{subfigure}{0}
  \setcounter{section}{0}
  \setcounter{subsection}{0}
  \setcounter{table}{0}
  \setcounter{equation}{0}
  \newcommand{\prelimtitle}{Preliminaries}
\chapter{\prelimtitle}\label{chap:prelim}

In this chapter we introduce some of the key concepts
and standard results that we will use throughout the paper.
The material in this chapter is not novel,
and can be found in any standard textbook on each of the subjects discussed.

We will use $\mathbb{N}$, $\mathbb{Q}$, and $\mathbb{R}$
to denote the natural, rational, and real numbers, respectively.
Furthermore, we will use $\mathbb{Q}_{\geq 0}$ and $\mathbb{R}_{\geq 0}$
to denote the non-negative rational and real numbers, respectively,
and $\mathbb{R}_{> 0}$ denotes the strictly positive real numbers. 
For expressions involving $\infty$, we will adopt the convention that
\[\infty + x = x + \infty = \infty\]
whenever $x \in \mathbb{R}$ and
\[\infty \cdot x = x \cdot \infty = \infty\]
whenever $x \in \mathbb{R}_{> 0}$.

\section{Set Theory}
We assume the reader is familiar with the basic concepts of set theory.
Given two sets $A$ and $B$,
$A \cup B$, $A \cap B$, $A \times B$ is the union, intersection, and Cartesian product, respectively, of $A$ and $B$.
We denote by $2^A$ the power set of $A$ and by $A^c$ the complement of $A$.
For a function $f : A \rightarrow B$, the preimage of a set $Y \subseteq B$ under $f$ is given by
\[f^{-1}(Y) = \{x \in A \mid f(x) \in Y\}.\]

A \emph{relation} on a set $A$ is simply a subset $R \subseteq A \times A$.
We will sometimes write $a R b$ to mean $(a,b) \in R$.
An \emph{equivalence relation} on a set $A$ is a relation $R \subseteq A \times A$ that has the following properties.
\begin{description}
  \item[Reflexivity:] $(a,a) \in R$ for any $a \in A$.
  \item[Symmetry:] If $(a,b) \in R$, then $(b,a) \in R$.
  \item[Transitivity:] If $(a,b) \in R$ and $(b,c) \in R$, then $(a,c) \in R$.
\end{description}
An equivalence relation partitions a set into \emph{equivalence classes} such that
every element of the set is in exactly one equivalence class.
A \emph{preorder} is a relation which satisfies reflexivity and transitivity.

A set which can be put into bijection with
a subset of the natural numbers is said to be \emph{countable},
otherwise it is said to be \emph{uncountable}.

\section{Boolean Algebra}
Boolean algebra is a formalisation and generalisation of the rules of logic as initially introduced by George Boole.
For a comprehensive introduction to Boolean algebras,
see the excellent textbook by Givant and Halmos \cite{halmos2009}.

\begin{definition}
  A \emph{Boolean algebra} is a set $A$ together with two binary operations $\wedge$ and $\vee$,
  a unary operation $\neg$, and two distinguished elements $\top$ and $\bot$.
  These must satisfy the following conditions, for any elements $p,q,r \in A$.
  
  \[\begin{array}{c c l}
    p \wedge \top = p & p \vee \bot = p & \text{(identity laws)} \\
    & & \\
    p \wedge \neg p = \bot & p \vee \neg p = \top & \text{(complement laws)} \\
    & & \\
    p \wedge q = q \wedge p & p \vee q = q \vee p & \text{(commutative laws)} \\
    & & \\
    \multicolumn{2}{c}{\begin{aligned}
      p \wedge (q \vee r) &= (p \wedge q) \vee (p \wedge r) \\
      p \vee (q \wedge r) &= (p \vee q) \wedge (p \vee r)
    \end{aligned}} & \text{(associative laws)}
  \end{array} \]  
\end{definition}

The two operations $\wedge$ and $\vee$ are known as \emph{meet} and \emph{join}, respectively,
whereas the operation $\neg$ is known as \emph{complement}.
The elements $\top$ and $\bot$ are known as the \emph{top} and \emph{bottom} elements, respectively.

\begin{example}
  The two-element Boolean algebra has only the elements $\top$ and $\bot$.
  In this algebra, we interpret $\top$ as true and $\bot$ as false.
  We then recover the usual logical connectives, as
  $\wedge$ becomes conjunction, $\vee$ becomes disjunction, and $\neg$ becomes negation.
\end{example}

\begin{example}
  The power set $2^X$ of a set $X$ is a Boolean algebra
  where meet is intersection, join is union,
  and complement is set complement.
\end{example}

From the definition of a Boolean algebra follows naturally a notion of \emph{order}
between the elements of a Boolean algebra.
%By introducing the notion of \emph{order} into a Boolean algebra,
%we can arrive at a definition of filter that is perhaps more intuitive.

\begin{definition}
  We will write $p \rightarrow q$ and say that $p$ is \emph{below} $q$
  if $p \wedge q = p$.
\end{definition}

This order also explains the names top and bottom for $\top$ and $\bot$,
since we now have $\bot \rightarrow p \rightarrow \top$ for any $p \in A$,
meaning that $\bot$ is below every element and every element is below $\top$.

A map from one Boolean algebra to another that preserves the structure of the Boolean algebra is called a homomorphism.

\begin{definition}
  Let $A$ and $B$ be two Boolean algebras.
  A \emph{homomorphism} from $A$ to $B$
  is a map $f : A \rightarrow B$ such that
  \begin{align*}
    f(p \wedge q) &= f(p) \wedge f(q) \\
    f(p \vee q)   &= f(p) \vee f(q) \\
    f(\neg p)     &= \neg f(p) \qedhere
  \end{align*}
\end{definition}

It is simple to show that if $f$ is a homomorphism, then we also get
\[f(\top) = \top \quad \text{ and } \quad f(\bot) = \bot.\]

%\begin{definition}
%  Given a Boolean logic $\mathcal{L}$, i.e. a logic with disjunction, conjunction, and negation,
%  we can construct the \emph{Lindenbaum algebra} of $\mathcal{L}$ as follows.
%  Let $\varphi \in \mathcal{L}$ be a formula and denote by $[\varphi]$
%  the set of all formulas in $\mathcal{L}$ that are logically equivalent to $\varphi$,
%  also known as the equivalence class of $\varphi$.
%  Let $A$ be the collection of all such equivalence classes.
%  Then $A$ becomes a Boolean algebra by letting
%  \[[\varphi] \wedge [\psi] = [\varphi \land \psi] \quad [\varphi] \vee [\psi] = [\varphi \lor \psi] \quad \neg[\varphi] = [\neg \varphi] \quad \top = [\ffalse] \quad \bot = [\ttrue]\]
%\end{definition}

An important example of homomorphism comes from the quotient constructed from a congruence.

\begin{definition}
  A \emph{congruence} on a Boolean algebra $A$ is an equivalence relation $\mathcal{R}$ such that whenever $p \mathcal{R} r$ and $q \mathcal{R} s$ we also have
  \begin{align*}
    (p \land q) &\mathcal{R} (r \land s) \\
    (p \lor q) &\mathcal{R} (r \lor s) \\
    (\neg p) &\mathcal{R} (\neg r) \qedhere
  \end{align*}
\end{definition}

Given a congruence $\mathcal{R}$ on a Boolean algebra $A$,
we can now construct the \emph{quotient} of $A$ under $\mathcal{R}$,
which is denoted by $A / \mathcal{R}$.
The quotient of $A$ under $\mathcal{R}$ consists of all equivalence classes of $\mathcal{R}$,
and is in fact a Boolean algebra by defining meet, join, and complement as
\begin{align*}
  [p] \land [q] &= [p \land q] \\
  [p] \lor [q] &= [p \lor q] \\
  \neg [p] &= [\neg p],
\end{align*}
where $[p]$ denotes the equivalence class of $p$.

\begin{proposition}[{\hspace{1sp}\cite[Chapter 17]{halmos2009}}]
  Let $A$ be a Boolean algebra and $\mathcal{R}$ a congruence on $A$.
  Define the function $f : A \rightarrow A / \mathcal{R}$ by $f(p) = [p]$.
  Then $f$ is a homomorphism, known as the \emph{projection} from $A$ unto $A / \mathcal{R}$.
\end{proposition}

\begin{example}
  Consider a Boolean algebra $A$.
  Define $p \leftrightarrow q$ if and only if $p \rightarrow q$ and $q \rightarrow p$.
  For example, it is easy to see that $(p \land \top) \leftrightarrow p$.
  Then the relation
  \[\mathcal{R}_\leftrightarrow = \{(p,q) \in A \times A \mid p \leftrightarrow q\},\]
  is an equivalence relation and can be shown to be a congruence on $A$.
  The quotient of $A$ under $\mathcal{R}_\leftrightarrow$
  is known as the \emph{Lindenbaum} algebra.
\end{example}

An important concept for Boolean algebras is that of a filter.

\begin{definition}\label{def:filter}
  Given a Boolean algebra $A$, a \emph{filter} is a subset $F \subseteq A$ such that
  \begin{enumerate}%[label=(F\arabic*), series=f]
    \item $\top \in F$, %\label{enum:f1}
    \item if $p \in F$ and $q \in F$, then $p \wedge q \in F$, and %\label{enum:f2}
    \item If $p \in F$ and $p \rightarrow q$, then $q \in F$. \qedhere
  \end{enumerate}
\end{definition}

%With the notion of order, we can replace condition \ref{enum:f3} in Definition \ref{def:filter} by the following condition.
%\begin{enumerate}[label=(F\arabic*), resume*=f]
%  \item If $p \in F$ and $p \leq q$, then $q \in F$.
%\end{enumerate}

In other words, a filter is a subset which contains $\top$,
is closed under meet, and is upward-closed.

\begin{definition}
  Given a Boolean algebra $A$, $U$ is an \emph{ultrafilter} if
  \begin{itemize}
    \item $U$ is a filter,
    \item $U \neq A$, and
    \item if $F$ is another filter such that $U \subseteq F$,
      then either $F = U$ or $F = A$. \qedhere
  \end{itemize}
\end{definition}

An ultrafilter is therefore a filter which is maximal,
in the sense that we can not add anything to the filter
while having it remain a filter.
Ultrafilters have the following nice property.

\begin{lemma}[{\hspace{1sp}\cite[{Chapter 20, Lemma 1}]{halmos2009}}]
  Let $A$ be a Boolean algebra.
  $U$ is an ultrafilter if and only if for any $p \in A$
  we have either $p \in U$ or $\neg p \in U$, but not both.
\end{lemma}

\section{Metric Spaces}
The theory of metric spaces is concerned with notions of distance.
The most basic distance is that of Euclidean distance,
which is simply the length of the straight line between two points in a Euclidean space.
However, this notion quickly becomes too simplistic.
For example, the distance one has to travel to go from Denmark to England
depends on whether one is going by airplane, by ferry, by train, or something else.
Furthermore, the Euclidean distance is symmetric: If the distance from $A$ to $B$ is $x$,
then the distance from $B$ to $A$ is also $x$.
However, some natural notions of distance are not symmetric.
For example, the distance when traveling by car in a city from point $A$ to point $B$
may not be the same as the distance from $B$ to $A$,
due to the existence of one-way streets.

\begin{definition}
  Let $X$ be a set, and $d : X \times X \rightarrow [0,\infty]$ a function.
  Consider the following conditions on $d$.
  \begin{enumerate}[label=(D\arabic*)]
    \item $d(x,y) = 0$ implies $x = y$. \label{enum:d1}
    \item $x = y$ implies $d(x,y) = 0$. \label{enum:d2}
    \item $d(x,y) = d(y,x)$. \label{enum:d3}
    \item $d(x,z) \leq d(x,y) + d(y,z)$. \label{enum:d4}
  \end{enumerate}
  
  The function $d$ is
  \begin{itemize}
    \item a \emph{metric} if it satisfies \ref{enum:d1}-\ref{enum:d4},
    \item a \emph{pseudometric} if it satisfies \ref{enum:d2}-\ref{enum:d4}, and
    \item a \emph{hemimetric} if it satisfies \ref{enum:d2} and \ref{enum:d4}. \qedhere
  \end{itemize}
\end{definition}

Condition \ref{enum:d3} is known as symmetry and condition \ref{enum:d4} is known as the triangle inequality.
We will use the term \emph{distance} to mean any of the above three.

\begin{example}
  For any set $X$, define the function
  \[d(x,y) = \begin{cases} 0 & \text{if } x = y \\ 1 & \text{if } x \neq y.\end{cases}\]
  This is known as the \emph{discrete} metric.
\end{example}

\begin{example}
  Consider the real numbers $\mathbb{R}$
  and define the function
  \[d(x,y) = |y - x| = \sqrt{(y - x)^2}.\]
  This is known as the \emph{Euclidean distance},
  and simply measures the length of the straight line between the two points $x$ and $y$.
  This can easily be generalised to $\mathbb{R}^n$ for any $n$.
\end{example}

A metric space is then a set $X$ together with a metric on $X$,
and similarly for a pseudometric and hemimetric space.
Metric spaces are the most common and well-studied of the three types of spaces in the literature.
However, for us, the pseudometric and hemimetric spaces are more natural.
This is because we want systems that have the same behaviour to be at distance $0$ from each other,
but having the same behaviour does not necessarily mean that the systems are equal.
Hence condition \ref{enum:d1} becomes unnatural.

Given a distance $d$ on $X$, we can define the closed and open sets of $X$ as follows.
For a point $x \in X$ and a radius $r > 0$, the open ball of radius $r$ centered in $x$ is
\begin{equation}\label{eq:ball}
  \mathcal{B}_r(x) = \{y \in X \mid d(x,y) < r\}.
\end{equation}

\begin{definition}
  Given a space $X$ with a distance $d$,
  a subset $U \subseteq X$ is said to be \emph{open} if for any $x \in U$
  there exists $r > 0$ such that $\mathcal{B}_r(x) \subseteq U$.
  A subset $V \subseteq X$ is said to be \emph{closed} if its complement
  $V^c$ is open.
\end{definition}

In other words, a set is open if for any point in the set,
we can find an open ball around that point
such that the open ball is contained in the set.
Note that a set can be both open and closed at the same time,
such a set is called \emph{clopen}.

The open sets (or equivalently, the closed sets) of a space $X$ with a distance $d$ form a topology on $X$.

\begin{definition}
  Given a set $X$, a \emph{topology} on $X$ is a collection of subsets $\tau \subseteq 2^X$
  such that
  \begin{itemize}
    \item $\Emptyset \in \tau$ and $X \in \tau$,
    \item any union of elements of $\tau$ is again in $\tau$, and
    \item any finite intersection of elements of $\tau$ is again in $\tau$. \qedhere
  \end{itemize}
\end{definition}

A topological space is then a set $X$ together with a topology $\tau$ on $X$.
Topological spaces are more general than metric, pseudometric, or hemimetric spaces,
since the distance of a metric, pseudometric, or hemimetric space induces a topology,
and thus these are also topological spaces,
whereas a topological space need not have a distance at all.

In Equation \eqref{eq:ball},
we used $d(x,y)$, the distance from $x$ to $y$,
for the definition of an open ball.
For metric and pseudometric spaces,
it would not have made a difference if we instead had written $d(y,x)$,
the distance from $y$ to $x$,
since in these spaces, the distance is symmetric.
However, for hemimetric spaces,
we will distinguish between the \emph{left-centered open balls} $\mathcal{B}_r^L(x)$
and the \emph{right-centered open balls} $\mathcal{B}_r^R(x)$, defined as
\[\mathcal{B}_r^L(x) = \{y \in X \mid d(x,y) < r\} \quad \text{and} \quad \mathcal{B}_r^R(x) = \{y \in X \mid d(y,x) < r\}.\]
The left-centered and the right-centered open balls
both give rise to a topology, but these two topologies will in general be quite different.

The intuition for open sets is that they do not necessarily contain their border,
whereas closed sets must contain their border,
so that whenever the points in the set get infinitely close to some other point,
then that other point must also be in the set.
To make this intuition precise,
we introduce the notion of convergence.

\begin{definition}
  Let $X$ be a space with a distance $d$.
  We say that a sequence of points $(x_n)_{n \in \mathbb{N}}$
  \emph{converges} to $x \in X$ if for every $\varepsilon > 0$
  we can find $N \in \mathbb{N}$ such that for every $n \geq N$
  we have that $d(x,x_n) < \varepsilon$.
\end{definition}

If a sequence $(x_n)_{n \in \mathbb{N}}$ converges to $x$,
we will also say that $x$ is a \emph{limit} of $(x_n)_{n \in \mathbb{N}}$.
In metric spaces limits are unique,
so that we may speak of \emph{the} limit of a sequence.
However, this is not the case for pseudometric and hemimetric spaces.
We can now define what it means to be sequentially closed.

\begin{definition}
  A set $U \subseteq X$ is \emph{sequentially closed}
  if whenever a sequence $(x_n)_{n \in \mathbb{N}}$,
  where $x_n \in U$ for all $n \in \mathbb{N}$,
  converges to some $x$, then also $x \in U$.
\end{definition}

A set is therefore sequentially closed if the limit of any sequence is again in that set.
The two notions of closed set and sequentially closed set coincide.

\begin{lemma}[{\hspace{1sp}\cite[Exercise 4.7.14 and Lemma 6.3.6]{goubault2013}}]
  For metric, pseudometric, and hemimetric spaces,
  a set is closed if and only if it is sequentially closed.
\end{lemma}

\section{Measure Theory}
The aim of measure theory is to generalise the notion of size
by ``measuring'' the size of sets.
This is usually simple for finite and countable sets,
but becomes very subtle for uncountable sets
such as the real numbers.
We will only concern ourselves here with those measures
that assign a probability to sets,
the so-called probability measures.
The starting point of measure theory is the notion of a $\sigma$-algebra.

\begin{definition}
  Let $X$ be a set.
  A \emph{$\sigma$-algebra} on $X$ is a non-empty collection of subsets $\Sigma \subseteq 2^X$ such that
  \begin{enumerate}[label=(A\arabic*)]
    \item $X \in \Sigma$, \label{enum:a1}
    \item $A \in \Sigma$ implies $A^c \in \Sigma$, and \label{enum:a2}
    \item $A_1, A_2, A_3, \dots \in \Sigma$ implies $\bigcup_{n=1}^\infty A_n \in \Sigma$. \label{enum:a3}
  \end{enumerate}
  
  A \emph{measurable space} is a set $X$ together with a $\sigma$-algebra on $X$.
\end{definition}

Condition \ref{enum:a2} is known as closure under complement,
and condition \ref{enum:a3} is known as closure under countable union.
Because we have closure under complement,
condition \ref{enum:a1} could be replaced by $\Emptyset \in \Sigma$.
Furthermore, conditions \ref{enum:a2} and \ref{enum:a3}
together also imply closure under countable intersection.
An element of a $\sigma$-algebra will be called a \emph{measurable set}.
%These will be the sets to which we can assign a probability,
%and any set which is not a measurable set will not have a defined probability.

\begin{example}
  For any set $X$, $\{\Emptyset, X\}$ is a $\sigma$-algebra on $X$.
  This is known as the \emph{trivial} or \emph{indiscrete} $\sigma$-algebra on $X$.
\end{example}

\begin{example}
  For any set $X$, the power set $2^X$ is a $\sigma$-algebra on $X$,
  known as the \emph{discrete} $\sigma$-algebra on $X$. 
\end{example}

\begin{example}
  If we start with a topological space $X$ with topology $\tau$,
  then the \emph{Borel $\sigma$-algebra} is the smallest $\sigma$-algebra
  containing all the elements of $\tau$.
  The elements of the Borel $\sigma$-algebra are called \emph{Borel sets}.
\end{example}

When speaking about the real numbers,
we will always assume that they are equipped with the Borel $\sigma$-algebra,
which we denote by $\borel$.

The structure-preserving maps between measurable spaces are the measurable functions.

\begin{definition}
  Given two measurable spaces $X$ and $Y$ with $\sigma$-algebra $\Sigma_X$ and $\Sigma_Y$, respectively,
  a function $f : X \rightarrow Y$ is \emph{measurable} if $f^{-1}(E) \in \Sigma_X$ for any $E \in \Sigma_Y$.
\end{definition}

We will now introduce the central concept of a measure.
A measure assigns a numerical value to each measurable set,
which we may interpret as the size of that set.

\begin{definition}
  Given a measurable space $X$ with $\sigma$-algebra $\Sigma$,
  a \emph{measure} is a function $\mu : \Sigma \rightarrow \mathbb{R}_{\geq 0}$ such that
  \begin{enumerate}[label=(M\arabic*)]
    \item $\mu(\Emptyset) = 0$ and \label{enum:m1}
    \item for any countable collection $A_1, A_2, A_3, \dots \in \Sigma$ of pairwise disjoint sets it holds that \label{enum:m2}
      \[\mu\left(\bigcup_{i \in \mathbb{N}} A_i\right) = \sum_{i \in \mathbb{N}} \mu(A_i).\]
  \end{enumerate}
  
  A measure $\mu : \Sigma \rightarrow [0,1]$ is called a \emph{subprobability measure} if $\mu(X) \leq 1$
  and a \emph{probability measure} if $\mu(X) = 1$. 
  
  Given a set $X$, we will use the notation
  \begin{itemize}
    \item $\dist(X)$ to denote the set of all subprobability measures on $X$, and
    \item $\distone(X)$ to denote the set of all probability measures on $X$. \qedhere
  \end{itemize}
\end{definition}

Condition \ref{enum:m2} is known as countable additivity.
A probability measure is thus a function that assigns a probability
to the measurable sets, with the condition that
the probability of \emph{something} happening must be $1$.
A measure $\mu \in \dist(X)$ will be said to be \emph{finitely supported} if its support
\[\supp{\mu} = \{x \in X \mid \mu(x) > 0\}\]
is finite.

An important example of measure is the product measure.

\begin{proposition}[Product measure {\cite[Theorem 18.2]{billingsley1995}}]
  Let $X$ and $Y$ be two measurable spaces with $\sigma$-algebra $\Sigma_X$ and $\Sigma_Y$, respectively.
  We will then denote by $X \times Y$ the \emph{product space}, which is a measurable space
  with $\sigma$-algebra $\Sigma_X \otimes \Sigma_Y$,
  defined as the smallest $\sigma$-algebra containing the sets $E \times F$ for $E \in \Sigma_X$ and $F \in \Sigma_Y$.
  
  Given two measures $\mu : \Sigma_X \rightarrow \mathbb{R}_{\geq 0}$ and $\nu : \Sigma_Y \rightarrow \mathbb{R}_{\geq 0}$,
  the \emph{product measure} $\mu \times \nu : \Sigma_X \otimes \Sigma_Y \rightarrow \mathbb{R}_{\geq 0}$,
  is the unique measure such that
  \[(\mu \times \nu)(E \times F) = \mu(E) \cdot \nu(F)\]
  for all $(E,F) \in \Sigma_X \times \Sigma_Y$.
\end{proposition}

One of the important concepts derived from a probability measure
is the cumulative distribution function.

\begin{definition}
  Given a probability measure $\mu \in \dist(\mathbb{R}_{\geq 0})$,
  the \emph{cumulative distribution function (CDF)}, or simply \emph{distribution function},
  of $\mu$ will be denoted by $F_\mu$ and is given by $F_\mu(t) = \mu([0,t])$.
\end{definition}

Consider now (sub)probability measures on the non-negative real numbers,
i.e. $\mu \in \dist(\mathbb{R}_{\geq 0})$,
where we interpret $\mathbb{R}_{\geq 0}$ as time.
If we consider $\mu(X)$ for some measurable $X$ to be the probability that
an event, such as a system taking an action, has happened within the time interval given by $X$.
Then $F_\mu(t)$ gives the probability that an event has occurred before the time point $t$.
CDFs have the following nice properties.

\begin{description}
  \item[Monotonicity:] If $x \leq y$, then $F_\mu(x) \leq F_\mu(y)$.
  \item[Right-continuity:] Let $x \in \mathbb{R}_{\geq 0}$. For every $\varepsilon > 0$ there exists a $\delta > 0$
    such that if $x < y < x + \delta$, then $F_\mu(y) - F_\mu(x) < \varepsilon$.
\end{description}

\begin{figure}
  \centering
  \begin{subfigure}{0.3\textwidth}
    \centering
    \includegraphics[scale=0.25]{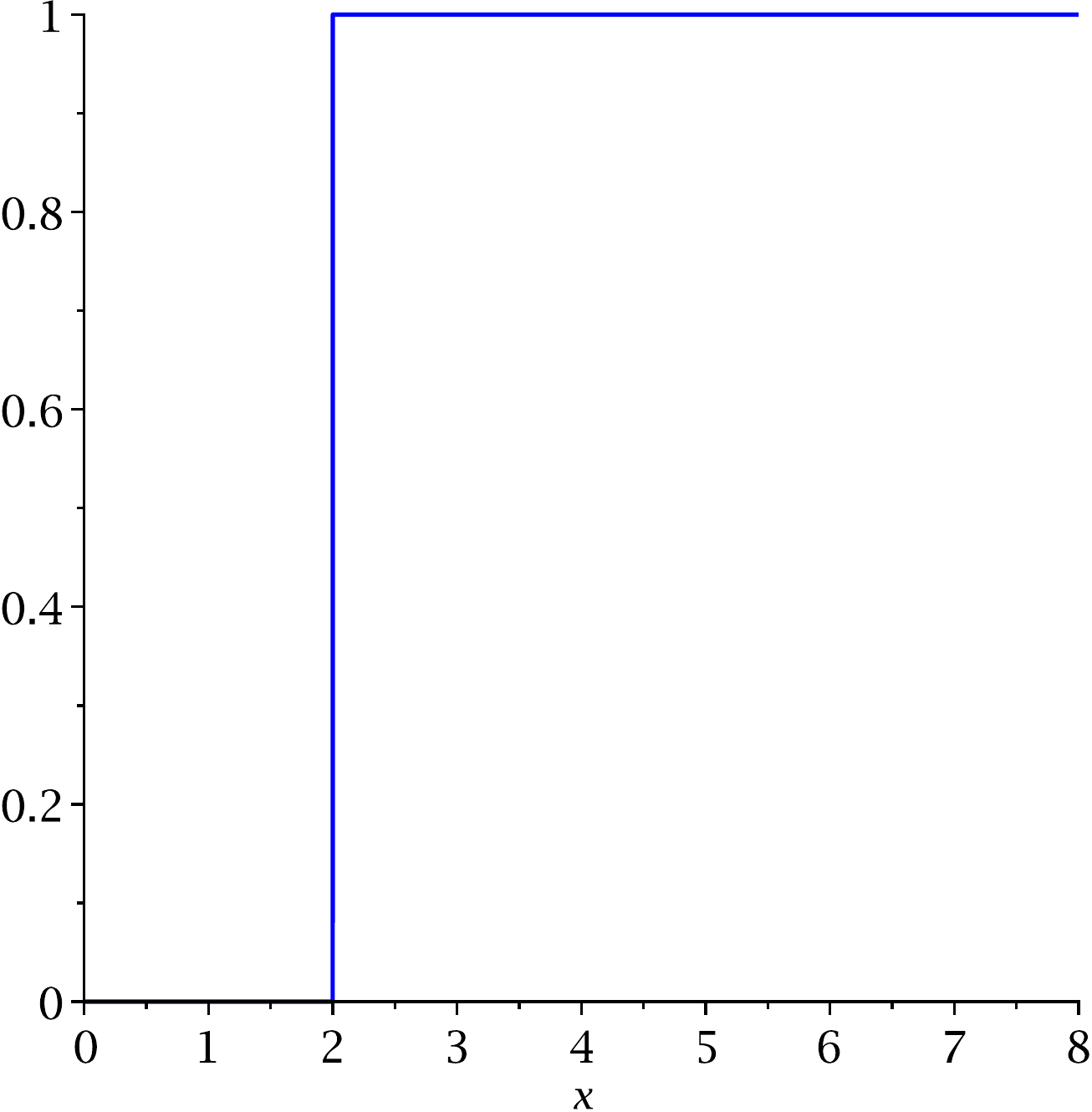}
    \caption{Dirac}
    \label{fig:dirac}
  \end{subfigure}%
  \begin{subfigure}{0.3\textwidth}
    \centering
    \includegraphics[scale=0.25]{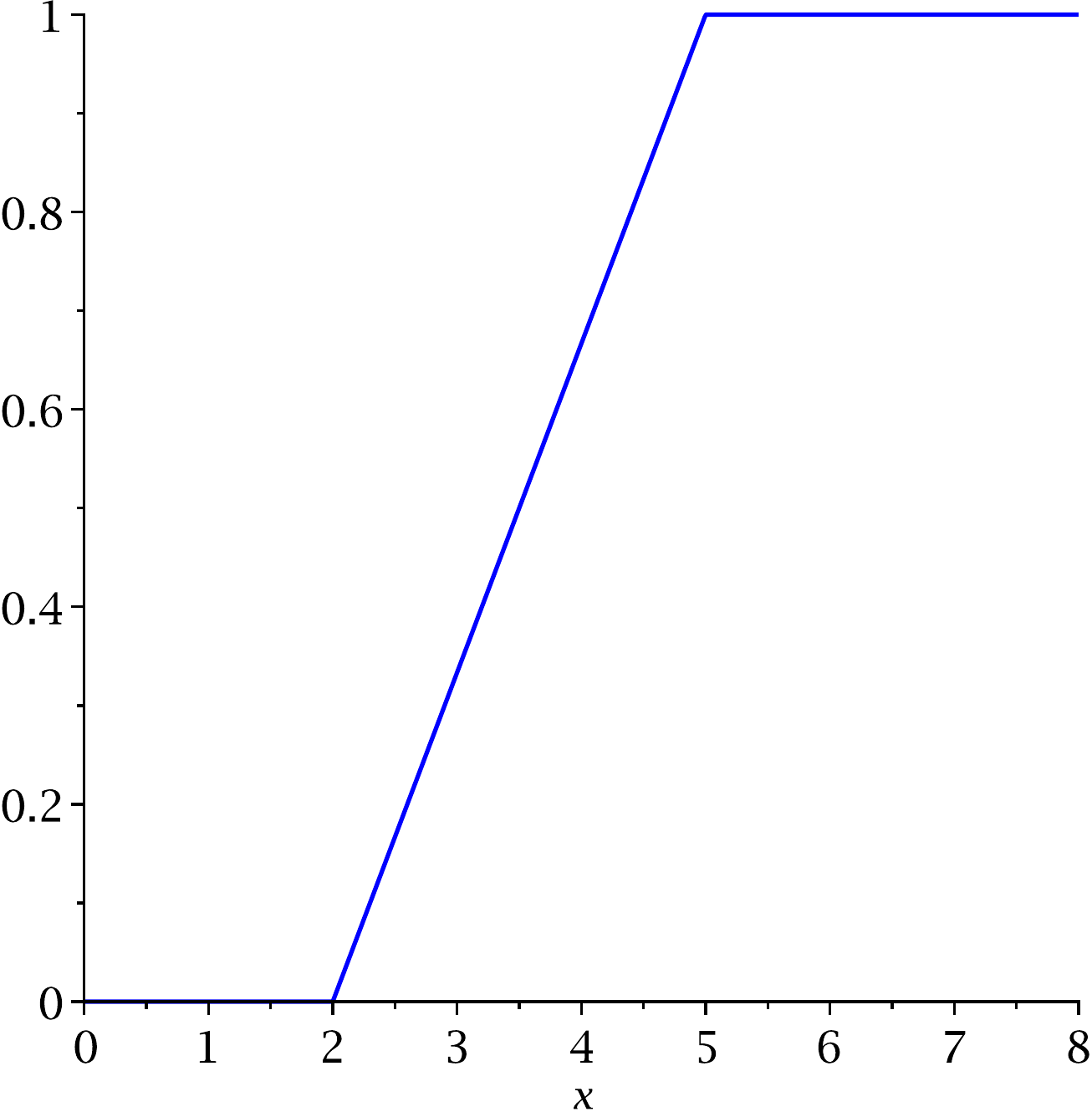}
    \caption{Uniform}
    \label{fig:unif}
  \end{subfigure}%
  \begin{subfigure}{0.3\textwidth}
    \centering
    \includegraphics[scale=0.25]{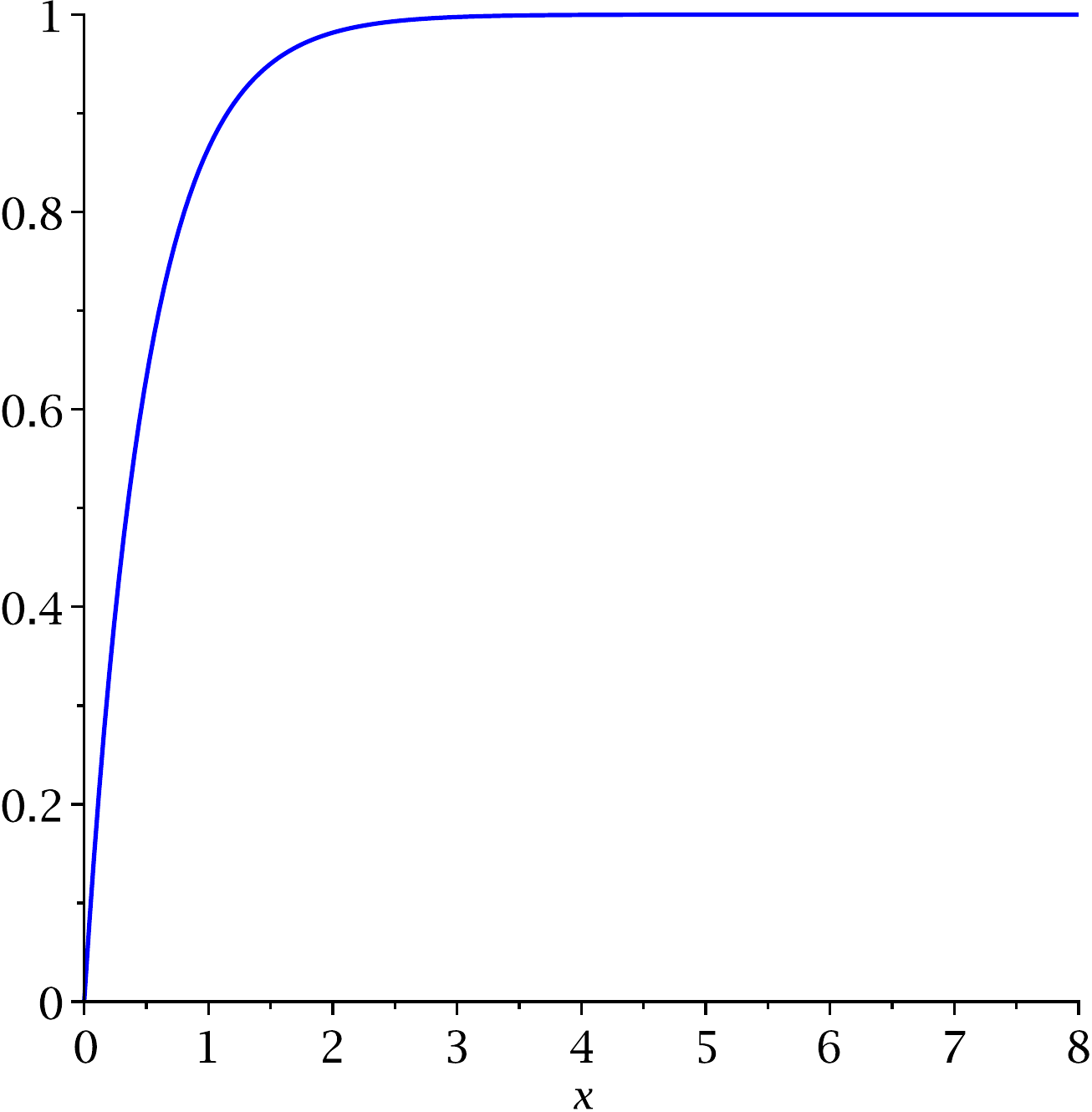}
    \caption{Exponential}
    \label{fig:exp}
  \end{subfigure}
  \caption{Plots of the CDFs of a Dirac distribution at $2$, a uniform distribution with parameters $a = 2$ and $b = 5$,
    and an exponential distribution with rate $2$.}
  \label{fig:distributions}
\end{figure}

\begin{example}
  The \emph{Dirac measure at $x$}, denoted by $\delta_x$,
  is given by
  \[\delta_x(E) = \begin{cases} 1 & \text{if } x \in E \\ 0 & \text{if } x \notin E \end{cases}\]
  for any measurable $E$.
  We will denote the CDF of $\delta_x$ by $\dirac{x}$,
  which has the property that
  \[\dirac{x}(t) = \begin{cases} 1 & \text{if } t \geq x \\ 0 & \text{otherwise.}\end{cases}\]
  The CDF of the Dirac measure at $2$ is plotted in Figure \ref{fig:dirac}.
\end{example}

\begin{example}
  A \emph{uniform} distribution is given by two parameters
  $a,b \in \mathbb{R}_{\geq 0}$ such that $a < b$.
  We will denote its CDF by $\unif{a}{b}$, which is defined by
  \[\unif{a}{b}(t) = \begin{cases} 1 & \text{if } t < a \\ \frac{t-a}{b-a} & \text{if } a \leq t < b \\ 0 & \text{if } x \geq b. \end{cases}\]
  The CDF of a uniform distribution with parameters $a = 2$ and $b = 5$ is plotted in Figure \ref{fig:unif}.
\end{example}

\begin{example}
  An \emph{exponential} distribution is given by a parameter $\theta > 0$,
  often called the \emph{rate}.
  Its CDF will be denoted by $\Exp{\theta}$, and is defined by
  \[\Exp{\theta}(t) = 1 - e^{-\theta t}.\]
  The CDF of an exponential distribution with rate $2$ is plotted in Figure \ref{fig:exp}.
\end{example}

Since it is often difficult to define a measure directly on a $\sigma$-algebra,
some of the most useful results in measure theory are the extension theorems
that allow us to only define something simpler,
after which the extension theorem guarantees us that this definition
can be extended to a measure on the $\sigma$-algebra.
The extension theorem that we are interested in requires the following definition.

\begin{definition}\label{def:premeasure}
  Let $A$ be a Boolean algebra of sets.
  A function $\mu_0 : A \rightarrow \mathbb{R}_{\geq 0}$ is called a \emph{pre-measure} if
  \begin{enumerate}[label=(P\arabic*)]
    \item $\mu_0(\Emptyset) = 0$ and \label{enum:p1}
    \item $\mu_0(\bigcup^{\infty}_{n = 1} E_n) = \sum^{\infty}_{n=1} \mu_0(E_n)$ \label{enum:p2}
      whenever $E_1, E_2, \dots \in A$ are disjoint sets such that $\bigcup^\infty_{n=1} E_n \in A$. \qedhere
  \end{enumerate}
\end{definition}

A pre-measure is therefore much like a measure,
except that it is defined on a Boolean algebra rather than a $\sigma$-algebra.
However, by the following theorem, any pre-measure can be uniquely extended to a measure on a $\sigma$-algebra.

\begin{theorem}[Hahn-Kolmogorov Theorem {\cite[Theorem 1.7.8]{tao2013}}]\label{thm:hahn-kolmogorov}
  Let $A$ be a Boolean algebra of sets.
  Any pre-measure $\mu_0 : A \rightarrow \mathbb{R}_{\geq 0}$
  can be uniquely extended to a measure $\mu : \Sigma \rightarrow \mathbb{R}_{\geq 0}$,
  where $\Sigma$ is the smallest $\sigma$-algebra containing $A$,
  such that $\mu_0(E) = \mu(E)$ whenever $E \in A$.
\end{theorem}

\subsection{Integration}
An important part of measure theory is the theory of integration.
We will not concern ourselves here with the intricacies
of defining integration through measure theory.
Instead, we will state some properties of the Lebesgue integral
that are useful when manipulating integrals.
In the following, we therefore assume that all functions are integrable.
We use the notation
\[\int_E f(x) \; \mu(\wrt{x})\]
to denote that we are integrating the function $f$ over the measurable set $E$
with respect to the measure $\mu$ viewed as a function of $x$.
When $E = [0,t]$, we instead write
\[\int_0^t f(x) \; \mu(\wrt{x}).\]

\begin{lemma}[Linearity of integrals {\cite[Theorem 16.1(ii)]{billingsley1995}}]
  The integral is linear, meaning that
  \[\int_E a f(x) + b g(x) \; \mu(\wrt{x}) = a \int_E f(x) \; \mu(\wrt{x}) + b \int_E g(x) \; \mu(\wrt{x})\]
  when $a,b \in \mathbb{R}_{\geq 0}$ and $E$ is measurable.
\end{lemma}

\begin{lemma}[Fubini's theorem {\cite[Theorem 3.16]{panangaden2009}}]
  Let $X$ and $Y$ be measurable spaces with $\sigma$-algebra $\Sigma_X$ and $\Sigma_Y$, respectively,
  and let $\mu : \Sigma_X \rightarrow \mathbb{R}_{\geq 0}$ and $\nu : \Sigma_Y \rightarrow \mathbb{R}_{\geq 0}$ be measures.
  If $f : X \times Y \rightarrow [0,\infty]$ is a measurable function, then
  \begin{align*}
    \int_{E \times F} f(x,y) \; (\mu \times \nu)(\wrt {(x,y)}) &= \int_E \int_F f(x,y) \; \nu(\wrt{y}) \mu(\wrt{x}) \\
                                                               &= \int_F \int_E f(x,y) \; \mu(\wrt{x}) \nu(\wrt{y})
  \end{align*}
  for any $E \in \Sigma_X$ and $F \in \Sigma_Y$.
\end{lemma}

\begin{lemma}[Change of variable {\cite[Proposition 3.8]{panangaden2009}}]
  Let $X$ and $Y$ be measurable spaces with $\sigma$-algebra $\Sigma_X$ and $\Sigma_Y$, respectively.
  Furthermore, let $T : X \rightarrow Y$ be a measurable function,
  and define the measure $\nu$ by $\nu = \mu \circ T^{-1}$.
  
  If $f : Y \rightarrow \mathbb{R}_{\geq 0}$ is a measurable function, then
  \[\int_{T^{-1}(E)} (f \circ T)(x) \; \mu(\wrt{x}) = \int_E f(y) \; \nu(\wrt{y})\]
  for any $E \in \Sigma_Y$.
\end{lemma}

Using integration, we can now define the important concept of convolution.
Whereas the CDF $F_\mu(t)$ gives the probability that a single event has occurred before time $t$,
we are often interested in the probability that multiple events have all occurred before time $t$.
For this, we need the notion of convolution.

\begin{definition}
  Given two measures $\mu,\nu \in \dist(\mathbb{R}_{\geq 0})$,
  the \emph{convolution} of $\mu$ and $\nu$ is given by
  \[(\mu * \nu)([0,t]) = \int_0^t \nu([0,t-x]) \; \mu(\wrt{x}). \qedhere\]
\end{definition}

$(\mu * \nu)([0,t])$ is then the probability that the event governed by $\mu$
and the event governed by $\nu$ have both occurred, in sequence, before time $t$.
Convolution is both commutative and associative, meaning that
\[\mu * \nu = \nu * \mu \quad \text{and} \quad \mu * (\nu * \eta) = (\mu * \nu) * \eta.\]
Furthermore, the Dirac measure at $0$ is the identity for convolution, so that
\[\mu * \delta_0 = \mu.\]

\section{Complexity Theory}
One of the most important concepts in computer science is that of an algorithm.
Informally, an algorithm is a mechanical procedure for producing a set of outputs.
Many equivalent ways of formalising the notion of algorithm have been proposed,
most notably those of Turing machines, recursive functions, and the $\lambda$-calculus.
We will take our underlying model of computation to be that of Turing machines.
However, as is commonly done, we will describe algorithms informally as pseudocode,
with the understanding that such pseudocode could, if needed, be translated into one of the above-mentioned formalisms.
The textbook by Sipser provides a gentle introduction to the theory of computability and complexity \cite{sipser2006}.

One of the most profound results of computability theory is the fact that,
for some problems, there can not exist an algorithm to solve that problem.

\begin{definition}
  A \emph{decision problem} is the problem of deciding whether a given element is a member of some set.
  A decision problem is said to be \emph{decidable} if there exists an algorithm that solves the decision problem.
  If no such algorithm exists, the decision problem is said to be \emph{undecidable}.
\end{definition}

Sometimes we want our algorithm to not just answer yes or no,
but to compute some value, for example the value of a function.
For this, we have the notion of computability.

\begin{definition}
  We will say that a function is \emph{computable} if there exists an algorithm which
  computes the value of the function for any input in the domain of the function.
\end{definition}

For practical issues, we are not only interested in whether or not an algorithm exists,
but also in how fast that algorithm runs: Does it finish in seconds, or do we have to wait years for the output?
In order to talk about the running time of algorithms,
we need the following notation.

\begin{definition}
  Given two functions $f,g : \mathbb{N} \rightarrow \mathbb{R}_{\geq 0}$,
  we will say that $g$ is an \emph{asymptotic upper bound} for $f$
  and write $f(n) \in \bigo(g(n))$ if we can find $c,N \in \mathbb{N}$
  such that for any $n \geq N$ we have $f(n) \leq c g(n)$.
\end{definition}

Here, the word ``asymptotic'' means ``as we approach infinity'',
meaning that the function $g$ will eventually become an upper bound for the function $f$,
if we let $n$ become large enough.
$\bigo$ therefore suppresses smaller terms and constant factors.

\begin{example}
  When $f$ is a polynomial,
  we can simply pick out the largest term in the polynomial,
  which will then be an asymptotic upper bound for $f$.
  For example, if $f(n) = 7n^4 + 3n^2 + 5$,
  then $7n^4$ is the largest term, and we can drop the constant $7$,
  so we obtain that $f(n) \in \bigo(n^4)$.
\end{example}

We can now use the concept of asymptotic upper bound
to describe the complexity of algorithms.

\begin{definition}
  We will say that an algorithm \emph{uses time} $\bigo(g(n))$
  if $f(n) \in \bigo(g(n))$,
  where $f$ is a function that returns the number of steps
  that the algorithm goes through when given an input of size $n$.
  
  Likewise, we will say that an algorithm \emph{uses space} $\bigo(g(n))$
  if $f(n) \in \bigo(g(n))$,
  where $f$ is a function that returns the amount of space or memory
  that the algorithm uses when given an input of size $n$.
\end{definition}

We can now classify the complexity of different algorithms,
leading to a veritable zoo of complexity classes~\cite{complexityzoo}.
We will describe here only some of the most important complexity classes.
Let $g(n)$ be a polynomial function.

\begin{description}
  \item[$\mathbf{P}$ (or $\PTIME$)] is the class of problems that can be solved by an algorithm that uses $\bigo(g(n))$ time.
  \item[$\NP$] is the class of problems such that whenever the answer is ``yes'',
    there exists a proof or witness of this fact, and furthermore, there exists an algorithm in $\mathbf{P}$
    that can verify whether a given proof is correct.
  \item[$\coNP$] is the class of problems such that whenever the answer is ``no'',
    there exists a proof or witness of this fact, and furthermore, there exists an algorithm in $\mathbf{P}$
    that can verify whether a given proof is correct.
  \item[$\PSPACE$] is the class of problems that can be solved by an algorithm that uses $\bigo(g(n))$ space.
  \item[$\EXPTIME$] is the class of problems that can be solved by an algorithm that uses $\bigo(2^{g(n)})$ time.
  \item[$\EXPSPACE$] is the class of problems that can be solved by an algorithm that uses $\bigo(2^{g(n)})$ space.
\end{description}

\section{Models}
In this thesis we will make use of two different kinds of models:
Weighted transition systems and semi-Markov processes,
both of which are standard in the literature.
Weighted transition systems are similar to labelled transitions systems~\cite{BK08}
except they have weights on transitions instead of labels.
Semi-Markov processes extend continuous-time Markov processes by allowing the sojourn time to follow any distribution,
not just exponential distributions~\cite{puterman94}.

\subsection{Weighted Transition Systems}
Weighted transition systems are systems in which each transition from a state to a new state
is associated with some weight. This weight can be interpreted as the cost of taking that transition,
the time spent taking that transition, the amount of resources spent taking that transition, etc.
Assume that we have a countable set of \emph{atomic propositions},
denoted by $\ap$.

\begin{definition}
  A \emph{weighted transition system (WTS)} is a tuple $M = (S, \rightarrow, \ell)$,
  where
  \begin{enumerate}
    \item $S$ is a set of \emph{states},
    \item $\rightarrow \subseteq S \times \mathbb{R}_{\geq 0} \times S$ is the \emph{transition relation}, and
    \item $\ell: S \rightarrow 2^{\ap}$ is the \emph{labelling function}. \qedhere
  \end{enumerate}
\end{definition}

We will write $s \xra{r} t$ to mean that $(s,r,t) \in \rightarrow$.
A WTS is said to be \emph{image-finite} if for any $s \in S$,
there are only finitely many $t \in S$ such that $s \xra{r} t$ for some $r \in \mathbb{R}_{\geq 0}$.

A WTS operates by non-deterministically choosing, from a state $s$,
a transition $s \xra{r} t$, after which the system ends up in state $t$, and from there can do further transitions.
In a state $s$, $\ell(s)$ gives all the atomic propositions that are true in that state.
One can also think of $\ell(s)$ as the labels that we put on that state.

\begin{example}
  An example of a WTS was given in Example \ref{ex:wts-intro}.
  In this example, we have three states, so $S = \{s_1, s_2, s_3\}$,
  where $\ell(s_1) = \{\texttt{waiting}\}$, $\ell(s_2) = \{\texttt{cleaning}\}$, and $\ell(s_3) = \{\texttt{charging}\}$.
  The transition relation is given by the arrows between the states,
  so that e.g. $s_1 \xra{0} s_2$ and $s_3 \xra{60} s_1$.
\end{example}

The standard way of determining whether two WTSs behave the same is that of bisimulation \cite{blackburn}.

\begin{definition}
  Let $M = (S, \rightarrow, \ell)$ be a WTS.
  A \emph{weighted bisimulation relation} is an equivalence relation $\mathcal{R} \subseteq S \times S$
  such that $(s,t) \in \mathcal{R}$ implies
  \begin{description}
    \item[(Atomic harmony)] $\ell(s) = \ell(t)$,
    \item[(Zig)] if $s \xra{r} s'$, then there exists $t' \in S$ such that $t \xra{r} t'$ and $(s',t') \in \mathcal{R}$, and
    \item[(Zag)] if $t \xra{r} t'$, then there exists $s' \in S$ such that $s \xra{r} s'$ and $(s',t') \in \mathcal{R}$. \qedhere
  \end{description}
\end{definition}

We will say that $s,t \in S$ are \emph{weighted bisimilar} and write $s \sim_W t$ if
there exists a weighted bisimulation relation $\mathcal{R}$ such that $(s,t) \in \mathcal{R}$.
Weighted bisimilarity, denoted by $\sim_W$, is the largest weighted bisimulation.

Atomic harmony says that any states in the relation must have the same labels.
The zig and zag conditions are responsible for ensuring that the transition behaviour of the two states is the same.
Zig says that if $s$ can do a transition, then $t$ can do the same transition,
and the states that we end up in after taking these transitions are also bisimilar.
Zag is symmetric, saying that $s$ can match transitions taken by $t$.

\subsection{Semi-Markov Processes}\label{sec:smps-intro}
A semi-Markov process is a system in which both the time that is spent in each state
and the next state reached after taking a transition is determined probabilistically.
We will take here the view of semi-Markov processes as graphs or transition systems,
rather than as a sequence of random variables.
This also means that we will implicitly assume that all semi-Markov processes are time-homogeneous.
We assume a countable set $\ain$ of input actions and a countable set $\aout$ of output actions.

\begin{definition}
  A \emph{semi-Markov process (SMP)} is a tuple $\mathcal{M} = (S, \tau, \rho, \ell)$,
  where
  \begin{enumerate}
    \item $S$ is a countable set of \emph{states},
    \item $\tau : S \times \ain \rightarrow \dist(S \times \aout)$ is the \emph{transition function},
    \item $\rho : S \rightarrow \dist(\mathbb{R}_{\geq 0})$ is the \emph{time-residence function}, and
    \item $\ell : S \rightarrow 2^{\ap}$ is the \emph{labelling} function. \qedhere
  \end{enumerate}
\end{definition}

The operational behaviour of a SMP is as follows.
Starting in a state $s$, the SMP receives some input $a$ from the environment.
It then probabilistically goes to a new state $s'$ while outputting some $b$ after waiting some time $t$,
the probability of which is given by $\tau(s,a)(s',b) \cdot \rho(s)([0,t])$.
The labelling function tells us which labels each state has.
For a state $s \in S$, we will write $F_s$ for the CDF $\rho(s)$,
i.e. $F_s(t) = \rho(s)([0,t])$.

We will say that an SMP $\mathcal{M} = (S, \tau, \rho, \ell)$ is \emph{finite} if $S$ is a finite set.
If $\mathcal{M}$ is finite, we will denote by $|\mathcal{M}|$ the size of the state space of $\mathcal{M}$.

\begin{example}
  An example of a SMP was given in Example \ref{ex:smp-intro}.
  Here we have four states $S = \{s_1, s_2, s_3, s_4\}$,
  where $s_1$ is the top left state, $s_2$ the top right state,
  $s_3$ the bottom left state, and $s_4$ the bottom left state.
  $\rho$ is then given by
  \[\rho(s_1) = \rho(s_3) = \Exp{0.1} \quad \text{and} \quad \rho(s_2) = \rho(s_4) = \Exp{2},\]
  and $\ell$ is given by
  \[\ell(s_1) = \ell(s_2) = \{g_1, r_2\} \quad \text{and} \quad \ell(s_3) = \ell(s_4) = \{r_1, g_2\}.\]
  The transition function $\tau$ is given by the arrows in Figure \ref{fig:smp-example-intro}
  and their associated probability, so that for example $\tau(s_2, \texttt{car}_1?)(s_3, \texttt{change}!) = 0.1$
  and $\tau(s_4, \texttt{car}_2?)(s_4, \texttt{stay}!) = 0.9$.
\end{example}

From this general definition of semi-Markov process,
we can obtain the following standard models as special cases.

\begin{description}
  \item[Generative:] We get generative semi-Markov processes by letting $\ain$ be a singleton set.
    For a generative process, we will simply write $\tau(s)(s',a)$ for the transition function.
  \item[Reactive:] Reactive semi-Markov processes, also known as semi-Markov decision processes,
    are obtained by letting $\ain = \aout$, and in addition requiring that $\tau(s,a)(s',b) > 0$ only when $a = b$.
    For a reactive process, we write the transition function as $\tau(s,a)(s')$.
  \item[Continuous-time:] If for every $s \in S$,
    $\rho(s)$ is an exponential distribution for some rate $\theta > 0$,
    then we obtain the popular model of continuous-time Markov chains.
  \item[Discrete-time:] If we let $\rho(s) = \rho(s')$ for every $s,s' \in S$,
    then we obtain discrete-time Markov chains.
\end{description}

These can of course be combined to obtain e.g. continuous-time Markov decision processes,
which are reactive continuous-time Markov chains.
For the majority of this thesis, we will focus on the special cases of reactive and generative SMPs.

\begin{remark}
  Continuous-time processes are often described in a different way than presented here.
  In the more common definition, there is no residence-time function,
  and instead there are rates given on the transitions.
  When adding actions on the transitions, this leads to the definition of (reactive) continuous-time Markov decision processes \cite{neuhausser2007}.
  We recall here this alternative definition.
  
  \begin{definition}[Alternative]
    A \emph{continuous-time Markov decision process} is a pair $M = (S, \mathbf{R}, \ell)$,
    where $S$ is a countable set of \emph{states},
    $\mathbf{R} : S \times \ain \times S \rightarrow \mathbb{R}_{\geq 0}$ is the \emph{rate matrix},
    and $\ell : S \rightarrow 2^{\ap}$ is the \emph{labelling function}.
  \end{definition}
  
  We then have the following derived quantities. The \emph{exit rate} of $s \in S$ under $a \in \ain$
  is given by $E(s,a) = \sum_{s' \in S} \mathbf{R}(s,a,s')$,
  and the \emph{probability} of going from $s$ to $s'$ under $a$ is given by
  $\prob(s,a,s') = \frac{\mathbf{R}(s,a,s')}{E(s,a)}$.
  
  Unfortunately, the two definitions are not equivalent.
  However, we can view our definition as a special case of the alternative definition
  by letting $\mathbf{R}(s,a,s') = \theta_s \cdot \tau(s,a)(s')$,
  where $\theta_s$ denotes the rate of the exponential distribution given by $\rho(s)$.
  Then we see that
  \[E(s,a) = \sum_{s' \in S} \mathbf{R}(s,a,s') = \theta_s \cdot \sum_{s' \in S} \tau(s,a)(s') = \theta_s,\]
  so that $E(s,a) = E(s,a')$ for all $a,a' \in \ain$.
  We also get
  \[\prob(s,a,s') = \frac{\mathbf{R}(s,a,s')}{E(s,a)} = \tau(s,a)(s'),\]
  as expected.
  
  The reason that the two definitions are not equivalent is that in our definition,
  we must have $E(s,a) = E(s,a')$ for any $a,a' \in \ain$,
  meaning that every action has the same exit rate,
  whereas in the alternative definition, each action can have a different exit rate.
  One could attempt to remedy this by modifying our definition to allow
  a different residence-time function $\rho_a$ for each $a \in \ain$,
  but we will not explore this idea further in this thesis.
  
  Despite this difference from the more common definition,
  we have chosen the definition given in this thesis since it generalises more easily to semi-Markov processes.
\end{remark}

Often we wish to construct systems by putting together smaller components.
In order to do this, we need to define what it means to compose systems.
Because our systems have real-time behaviour,
in particular we need to describe how the real-time behaviour of the components
influence the combined system.
In order to accommodate different choices for combining real-time behaviour from the literature,
we let this be described by a generic composition function.

\begin{definition}
  A function $\star : \dist(\mathbb{R}_{\geq 0}) \times \dist(\mathbb{R}_{\geq 0}) \rightarrow \dist(\mathbb{R}_{\geq 0})$
  is a \emph{residence-time composition function} if it is commutative, meaning that
  \[\star(\mu,\nu) = \star(\nu, \mu) \quad \text{for all } \mu,\nu \in \dist(\mathbb{R}_{\geq 0}). \qedhere\]
\end{definition}

For technical reasons, we will only consider composition of reactive SMPs.

\begin{definition}
  Let $\star$ be a residence-time composition function.
  Then the $\star$-composition of $\mathcal{M}_1 = (S_1, \tau_1, \rho_1, \ell_1)$
  and $\mathcal{M}_2 = (S_2, \tau_2, \rho_2, \ell_2)$,
  denoted $\mathcal{M}_1 \comp{\star} \mathcal{M}_2 = (S, \tau, \rho, \ell)$,
  is given by
  \begin{enumerate}
    \item $S = S_1 \times S_2$,
    \item $\tau((s_1,s_2),a)((s_1',s_2')) = \tau_1(s_1,a)(s_1') \cdot \tau_2(s_2,a)(s_2')$,
    \item $\rho((s_1,s_2)) = \star(\rho_1(s_1),\rho_2(s_2))$, and
    \item $\ell((s_1,s_2)) = \ell(s_1) \cup \ell(s_2)$.
  \end{enumerate}
  
  We will also write $s_1 \comp{\star} s_2$ to mean that $(s_1, s_2) \in S$.
\end{definition}

Similarly to the case of weighted transition systems,
the standard notion of behavioural equivalence for SMPs is that of bisimulation.

\begin{definition}
  Let $M = (S, \tau, \rho, \ell)$ be a SMP.
  A \emph{bisimulation relation} is a relation $\mathcal{R} \subseteq S \times S$
  such that $s_1 \mathcal{R} s_2$ implies
  \begin{enumerate}[label=(B\arabic*)]
    \item $\ell(s_1) = \ell(s_2)$, \label{enum:b1}
    \item $F_{s_1}(t) = F_{s_2}(t)$ for all $t \in \mathbb{R}_{\geq 0}$, and \label{enum:b2}
    \item for all $a \in \ain$ there exists a \emph{weight function} $\Delta_a \in \dist(S \times S \times \aout)$ such that \label{enum:b3}
      \begin{enumerate}
        \item $\Delta_a(s,s',b) > 0$ implies $s \mathcal{R} s'$,
        \item $\tau(s_1,a)(s,b) = \sum_{s' \in S} \Delta_a(s,s',b)$, and
        \item $\tau(s_2,a)(s',b) = \sum_{s \in S} \Delta_a(s,s',b)$. \qedhere
      \end{enumerate}
  \end{enumerate}
\end{definition}

The idea behind bisimulation for SMPs is the same as that for WTSs.
Conditions \ref{enum:b1} and \ref{enum:b2} ensure that the information in the states is the same,
by requiring that they have the same labels and the same residence-time function.
Instead of the zig and zag conditions, we have the concept of a weight function in condition \ref{enum:b3}.
The weight function matches the probability mass of the transitions available to $s_1$
with the probability mass of the transitions available to $s_2$
in such a way that the bisimulation relation is preserved by the successor states.
This means that when the probability mass of going from $s_1$ to a successor state $s$ and outputting $b$,
is matched with the probability mass of going from $s_2$ to $s'$ and outputting $b$,
then it must also hold that $s$ and $s'$ are in the bisimulation relation.

If there exists a bisimulation relation $\mathcal{R}$ such that $s_1 \mathcal{R} s_2$,
then we will say that $s_1$ and $s_2$ are \emph{bisimilar} and write $s_1 \sim s_2$.
\emph{Bisimilarity} is the largest bisimulation relation and is denoted by $\sim$.

A related notion is that of simulation.
Whereas bisimulation guarantees that the processes have the same behaviour,
simulation guarantees that one process can simulate any behaviour from the other process.
Therefore, if $s_1$ simulates $s_2$, $s_1$ will be able to do anything that $s_2$ can do,
but may also be able to do some things that $s_2$ can not do.

\begin{definition}\label{def:simul-intro}
  Let $\mathcal{M} = (S, \tau, \rho, \ell)$ be an SMP.
  A \emph{simulation relation} is a relation $\mathcal{R} \subseteq S \times S$
  such that $s_1 \mathcal{R} s_2$ implies
  \begin{enumerate}[label=(S\arabic*)]
    \item $\ell(s_1) = \ell(s_2)$, \label{enum:s1}
    \item $F_{s_1}(t) \leq F_{s_2}(t)$ for all $t \in \mathbb{R}_{\geq 0}$, and \label{enum:s2}
    \item for all $a \in \ain$ there exists a \emph{weight function} $\Delta_a \in \dist(S \times S \times \aout)$ such that \label{enum:s3}
      \begin{enumerate}
        \item $\Delta_a(s,s',b) > 0$ implies $s \mathcal{R} s'$,
        \item $\tau(s_1,a)(s,b) = \sum_{s' \in S} \Delta_a(s,s',b)$, and
        \item $\tau(s_2,a)(s',b) = \sum_{s \in S} \Delta_a(s,s',b)$. \qedhere
      \end{enumerate}
  \end{enumerate}
\end{definition}

The only difference between bisimulation and simulation is in conditions \ref{enum:b2} and \ref{enum:s2},
where equality is used for bisimulation, whereas an inequality is used for simulation.
This means that while bisimilarity is an equivalence relation,
similarity is only a preorder.

If there exists a simulation relation $\mathcal{R}$ such that $s_1 \mathcal{R} s_2$,
then we will say that $s_2$ \emph{simulates} $s_1$ and write $s_1 \simul s_2$.
\emph{Similarity} is the largest simulation relation and is denoted by $\simul$.

Both simulation and bisimulation are at their core coinductive definitions,
meaning that they compare elements step by step.
However, sometimes we want to compare the entire history of an execution of a system
with the execution of another system, rather than comparing them stepwise.
We therefore also need to know what the probability of such an execution is.
In order to define this probability, we first introduce the space of timed paths.
The observable behaviour to keep track of in an execution of a SMP is the states that it visited,
the time at which transitions were made, and the output actions that were performed.
An \emph{execution} or \emph{path} is therefore an infinite sequence
\[\pi = (s_1, t_1, a_1), (s_2, t_2, a_2), (s_3, t_3, a_3), \dots \in (S \times \mathbb{R}_{\geq 0} \times \aout)^{\omega}.\]
Given a path $\pi$ and $i \in \mathbb{N}$, we let
\[\pi\sproj{i} = s_i, \quad \pi\tproj{i} = t_i, \quad \pi\aproj{i} = a_i,\]
\[\pi|_i = (s_1, t_1, a_1), \dots, (s_i, t_i, a_i), \quad \text{and} \quad \pi|^i = (s_i, t_i, a_i), (s_{i+1}, t_{i+1}, a_{i+1}), \dots\]
We denote by $\paths(M)$ the set of all paths in $M$
and by
\[\paths_n(M) = \{\pi|_n \mid \pi \in \paths(M)\}\]
the set of all prefixes of length $n$ of paths in $M$.

In order to turn $\paths(M)$ into a measurable space,
we need to construct a suitable $\sigma$-algebra.
We will do this through the standard cylinder set construction.
Given $n \geq 1$ and a set $E \subseteq \paths_n(M)$,
the \emph{cylinder set} of rank $n$ is the set of all paths
whose prefix up to the $n$th position agrees with that of $E$.
The cylinder set of rank $n$ is therefore given by
\[\cylinder(E) = \{\pi \in \paths(M) \mid \pi|_n \in E\}.\]
This means that all paths in $\cylinder(E)$ begin exactly as prescribed by $E$,
but after the $n$th step, they may start to differ.
For notational convenience, given a set
\[E = S_1 \times O_1 \times R_1 \cdots \times S_n \times O_n \times R_n \subseteq \paths_n(M),\]
we will often write $\cylinder(E)$ as
\[\cylinder(S_1 \dots S_n, O_1 \dots O_n, R_1 \dots R_n).\]
A cylinder $\cylinder(S_1 \dots S_n, O_1 \dots O_n, R_1 \dots R_n)$
is said to be \emph{measurable} if $S_i \in 2^S$, $O_i \in 2^{\aout}$, and $R_i \in \borel$
for all $1 \leq i \leq n$.

\begin{lemma}[{\hspace{1sp}\cite[Section 2.7]{ash1999}}]\label{lem:bool}
  The set of measurable cylinders forms a Boolean algebra of sets.
\end{lemma}

\begin{definition}
  Let $M = (S, \tau, \rho, \ell)$ be a SMP.
  The \emph{measurable space of paths} is the set of paths $\paths(M)$
  together with the $\sigma$-algebra $\Sigma$, defined as the smallest $\sigma$-algebra
  containing all measurable cylinders.
\end{definition}

Now that we have a $\sigma$-algebra for $\paths(M)$,
we wish to define a probability measure on paths.
However, in order to do so,
we must somehow resolve the non-determinism that is given by the input actions.
In other words, we must decide on how the environment behaves when choosing inputs.
This will be done by \emph{schedulers},
which are also known in the literature as controllers, policies, or adversaries.

\begin{definition}
  A \emph{scheduler} is a function $\sigma : S^* \rightarrow \dist(\ain)$.
\end{definition}

Intuitively, a scheduler looks at the history of visited states so far,
and based on this information, it probabilistically chooses an input.
One can also consider more complicated schedulers,
such as schedulers that take into account the timed history \cite{WJ06},
but we will not do so in this thesis.

\begin{definition}\label{def:prob-intro}
  Given a sequence of states $w = s_1 \dots s_k$ and a scheduler $\sigma$,
  we define the subprobability $\prob^\sigma(w)$ inductively on measurable cylinders as
  \[\prob^\sigma(w)(\Emptyset) = 0\]
  \[\prob^\sigma(w)(\cylinder(S, O, R)) = \rho(s_k)(R) \cdot \sum_{s' \in S}\sum_{a \in \ain}\sum_{b \in O} \tau(s_k,a)(s',b) \cdot \sigma(w)(a), \text{ and}\]
  \begin{align*}
    &\phantom{{}={}}\prob^\sigma(w)(\cylinder(S_1 S_2 \dots S_n, O_1 O_2 \dots O_n, R_1 R_2 \dots R_n)) \\
    &= \rho(s_k)(R_1) \cdot \sum_{s' \in S_1}\sum_{a \in \ain}\sum_{b \in O_1} \tau(s_k, a)(s',b) \cdot \sigma(w)(a) \\
    &\phantom{{}={}} \cdot \prob^{\sigma}(ws')(\cylinder(S_2 \dots S_n, O_2 \dots O_n, R_2 \dots R_n)) \qedhere
  \end{align*}
\end{definition}

For generative systems, $\ain$ is a singleton,
and therefore there is only one possible scheduler,
namely the one that assigns probability one to the single element of $\ain$ for any $w \in S^*$.
When considering generative systems, we can therefore forget about schedulers.

Notice that we have only defined $\prob^\sigma(w)$ on the Boolean algebra of measurable cylinders.
We will now invoke the Hahn-Kolmogorov theorem to show that we can extend it to the measurable space of paths.

\begin{lemma}\label{lem:unfold-intro}
  \begin{align*}
    &\phantom{{}={}}\prob^\sigma(w)(\cylinder(S_1 \dots S_n, O_1 \dots O_n, R_1 \dots R_n)) \\
    &= \sum_{s_1' \in S_1}\sum_{a_1 \in \ain}\sum_{b_1 \in O_1} \dots \sum_{s_n \in S_n}\sum_{a_n \in \ain}\sum_{b_n \in O_n} \tau(s_k,a_1)(s_1',b_1) \cdots \tau(s_{n-1}',a_n)(s_n',b_n) \\
    &\phantom{{}={}}\cdot \sigma(w)(a_1) \cdots \sigma(ws_1' \cdots s_{n-1}')(a_n) \\
    &\phantom{{}={}}\cdot \rho(s_k) \times \rho(s_1') \times \dots \times \rho(s_{n-1}')(R_1 \times \dots \times R_n)
  \end{align*}
\end{lemma}
\begin{proof}
  The result follows from unfolding the induction in Definition \ref{def:prob-intro}.
\end{proof}

\begin{theorem}
  The subprobability $\prob^\sigma(w)$ can be uniquely extended to a subprobability on the measurable space of paths.
\end{theorem}
\begin{proof}
  We will first argue that $\prob^\sigma(w)$ is a pre-measure.
  Let $\Sigma_0$ denote the set of measurable cylinders.
  By Lemma \ref{lem:bool}, $\Sigma_0$ is a Boolean algebra.
  In order to show that $\prob^\sigma(w)$ is a pre-measure,
  we therefore only need to show that conditions \ref{enum:p1} and \ref{enum:p2}
  from Definition \ref{def:premeasure} are satisfied.
  \ref{enum:p1} is satisfied by definition.
  
  Next we consider condition \ref{enum:p2}.
  For notational convenience, let
  \[\bigcup_{m=1}^\infty \cylinder(S_{m,1} \dots S_{m,n_m}, O_{m,1} \dots O_{m,n_m}, R_{m,1} \dots R_{m,n_m}) = \bigcup_{m=1}^\infty \cylinder_m.\]
  We must then show that
  \[\prob^\sigma(w)\left(\bigcup_{m=1}^\infty \cylinder_m\right) = \sum_{m=1}^\infty \prob^\sigma(w)(\cylinder_m)\]
  whenever $\bigcup_{m=1}^\infty \cylinder_m$
  is a disjoint union of measurable cylinders
  such that $\bigcup_{m=1}^\infty \cylinder_m = \cylinder$
  for some measurable cylinder
  \[\cylinder = \cylinder(S_1 \dots S_n, L_1 \dots L_n, R_1 \dots R_n).\]
  Note first that we can make all cylinders in $\bigcup_{m=1}^\infty \cylinder_m$
  have the same length without affecting disjointness.
  This is because if
  \[\cylinder(S_{i,1} \dots S_{i,n_{i}}, O_{i,1} \dots O_{i,n_i}, R_{i,1} \dots R_{i,n_i})\]
  and
  \[\cylinder(S_{j,1} \dots S_{j,n_j}, O_{j,1} \dots O_{j,n_j}, R_{j,1} \dots R_{j,n_j})\]
  are two disjoint cylinders with $n_i < n_j$, then we can extend the first with
  \begin{align*}
    &\phantom{{}={}}\cylinder(S_{i,1} \dots S_{i,n_{i}}, O_{i,1} \dots O_{i,n_i}, R_{i,1} \dots R_{i,n_i}) \\
    &= \cylinder(S_{i,1} \dots S_{i,n_{i}} \underbrace{S \dots S}_{l \text{ times}}, O_{i,1} \dots O_{i,n_i} \underbrace{\aout \dots \aout}_{l \text{ times}}, R_{i,1} \dots R_{i,n_i} \underbrace{\mathbb{R} \dots \mathbb{R}}_{l \text{ times}})
  \end{align*}
  where $l = n_j - n_i$, and the two cylinders are still disjoint.
  We can therefore assume without loss of generality that all the cylinders have length $n$.
  Let $w = s_1 \dots s_k$. By Lemma \ref{lem:unfold-intro} we then get
  \begin{align*}
    &\phantom{{}={}}\prob^\sigma(w)\left(\bigcup_{m=1}^\infty \cylinder_m \right) \\
    &=\prob^\sigma(w)(\cylinder(S_1 \dots S_n, O_1 \dots O_n, R_1 \dots R_n) \\
    &=\sum_{s_1' \in S_1}\sum_{a_1 \in \ain}\sum_{b_1 \in O_1} \dots \sum_{s_n \in S_n}\sum_{a_n \in \ain}\sum_{b_n \in O_n} \tau(s_k,a_1)(s_1',b_1) \cdots \tau(s_{n-1}',a_n)(s_n',b_n) \\
    &\phantom{{}={}}\cdot \sigma(w)(a_1) \cdots \sigma(ws_1' \cdots s_{n-1}')(a_n) \\
    &\phantom{{}={}}\cdot \rho(s_k) \times \rho(s_1') \times \dots \times \rho(s_{n-1}')(R_1 \times \dots \times R_n) \\
    &=\sum_{s_1' \in \bigcup\limits_{m=1}^\infty S_{m,1}}\sum_{a_1 \in \ain} \sum_{b_1 \in \bigcup\limits_{m=1}^\infty O_{m,1}} \dots \sum_{s_n' \in \bigcup\limits_{m=1}^\infty S_{m,n}}\sum_{a_n \in \ain}\sum_{b_n \in \bigcup\limits_{m=1}^\infty O_{m,n}} \tau(s_k,a_1)(s_1',b_1) \\
    &\phantom{{}={}}\cdot \tau(s_1',a_2)(s_2',b_2) \cdots \tau(s_{n-1}',a_n)(s_n',b_n) \\
    &\phantom{{}={}}\cdot \sigma(w)(a_1) \cdots \sigma(ws_1' \cdots s_{n-1}')(a_n) \\
    &\phantom{{}={}}\cdot \rho(s_k) \times \rho(s_1') \times \dots \times \rho(s_{n-1}')\left(\bigcup\limits_{m=1}^\infty R_{m,1} \times \dots \times \bigcup\limits_{m=1}^\infty R_{m,n}\right) \\
    &=\sum_{m=1}^\infty\sum_{s_1' \in S_{m,1}}\sum_{a_1 \in \ain}\sum_{b_1 \in O_{m,1}} \dots \sum_{s_n' \in S_{m,n}}\sum_{a_n \in \ain}\sum_{b_n \in O_{m,n}} \tau(s_k,a_1)(s_1',b_1) \\
    &\phantom{{}={}}\cdot \tau(s_1',a_2)(s_2',b_2) \cdots \tau(s_{n-1}',a_n)(s_n',b_n) \\
    &\phantom{{}={}}\cdot \sigma(w)(a_1) \cdots \sigma(ws_1' \cdots s_{n-1}')(a_n) \\
    &\phantom{{}={}}\cdot \rho(s_k) \times \rho(s_1') \times \dots \times \rho(s_{n-1}')(R_{m,1} \times \dots \times R_{m,n}) \\
    &= \sum_{m=1}^\infty \prob^\sigma(w)(\cylinder_m),
  \end{align*}
  and we conclude that condition \ref{enum:p2} is also satisfied.
  
  We have thus shown that $\prob^\sigma(w)$ is a pre-measure.
  Since the measurable space of paths is defined as the smallest $\sigma$-algebra containing all measurable cylinders,
  it then follows from the Hahn-Kolmogorov theorem (Theorem \ref{thm:hahn-kolmogorov})
  that $\prob^\sigma(w)$ can be uniquely extended to the measurable space of paths.
\end{proof}

\defaultbib

  \setcounter{figure}{0}
  \setcounter{subfigure}{0}
  \setcounter{section}{0}
  \setcounter{subsection}{0}
  \setcounter{table}{0}
  \setcounter{equation}{0}
  \newcommand{\logictitle}{Logical Specification Language for Reasoning About Bounds}
\chapter{\logictitle}\label{chap:logic-intro}
%\addcontentsline{toc}{chapter}{\logictitle}

This chapter summarises the content of Paper A ``Reasoning About Bounds in Weighted Transition Systems''~\cite{HLMP17}.
For the full paper, see Part~\ref{chap:papers}.

When using weighted transition systems (WTSs) as a modeling formalism,
we encounter the approximate modeling problem:
the quantities of interest may be irrational,
whereas we can only ever measure rational values, and even then only with some uncertainty.
It is therefore not clear which weight we should assign to each transition.
Consider again the WTS in Example \ref{ex:wts-intro},
where cleaning takes either $5$, $10$, or $15$ minutes.
These numbers are rather arbitrary, and we may easily imagine
that cleaning could also take $6$ minutes or $12.5$ minutes.
One way of approaching this problem is to allow \emph{intervals} of transitions,
and then reasoning about these. 
This is the approach taken by for example interval Markov chains \cite{JL91}
and interval weighted modal transition systems \cite{Juhl2012408}.

We will consider here a different, but related approach to the problem.
Instead of reasoning about individual transitions or intervals of transitions,
we reason about upper and lower bounds on the transitions.
In Example \ref{ex:wts-intro}, we may for instance say that cleaning takes at most $15$ time units,
so $15$ is an upper bound on the time it takes to clean.
Likewise, $5$ is a lower bound.
We argue that this is a reasonable point of view from an applications perspective,
partly because bounds on quantities are easier to engineer than very precise measurements,
and partly because many requirements that we are interested in verifying use upper and lower bounds,
such as
\begin{displayquote}
  ``the airbag must inflate within at most 2 milliseconds''
\end{displayquote}
and
\begin{displayquote}
  ``traffic light must be green for at least 8 seconds before changing.''
\end{displayquote}

While the aforementioned approach of using intervals has some similarities with our approach of using bounds,
namely that the endpoints of the intervals are also bounds,
the two approaches are not equivalent.
This is because when we talk about bounds,
transitions with weights that are in between the bounds may or may not exist.
To see this, consider again Example~\ref{ex:wts-intro}.
$5$ is a lower bound on transitions from the {\tt cleaning} state to the {\tt waiting state},
and $15$ is an upper bound.
However, the only transition that is allowed in between these bounds has weight $10$.
In contrast, interval approaches allow transitions with any weight that is within the interval.

Our focus is on logical aspects:
we will introduce a logical specification language, which we call weighted logic with bounds (WLWB),
and develop its metatheory, including a complete axiomatisation.
We also argue that WLWB is the correct language for speaking about bounds in WTSs,
by showing that it characterises a modified version of weighted bisimulation
in which states are behaviourally equivalent when they have the same upper and lower bounds on behaviours.
Furthermore, we give algorithms for solving the satisfiability and model checking problems for WLWB.

\section{Weighted Logic With Bounds}
First we introduce weighted logic with bounds (WLWB).
We will denote formulas of WLWB by $\mathcal{L}$, and they are induced by the abstract syntax
\[\mathcal{L}: \quad \varphi, \psi ::= p \mid \neg \varphi \mid \varphi \land \psi \mid L_r \varphi \mid M_r \varphi\]
where $r \in \mathbb{Q}_{\geq 0}$ is a non-negative rational number and $p \in \mathcal{AP}$ is an atomic proposition.
$\neg$ and $\land$ are the standard Boolean negation and conjunction,
and the rest of the Boolean operators, such as disjunction and implication,
can be derived from these.

The novel formulas are the ones of the form $L_r \varphi$ and $M_r \varphi$.
Intuitively, $L_r \varphi$ says that it is possible to take a transition
with weight \emph{at least} $r$ to a state where $\varphi$ is true.
Similarly, $M_r \varphi$ says that it is possible to take a transition
with weight \emph{at most} $r$ to a state where $\varphi$ is true.

In order to define the semantics of WLWB, we introduce some notation.
Consider a WTS $\mathcal{M} = (S, \rightarrow, \ell)$.
We then define the \emph{image set} for a given state $s \in S$ and subset $T \subseteq S$ as
\[\trans{s}{T} = \{r \in \mathbb{R}_{\geq 0} \mid \text{there exists } t \in T \text{ such that } s \xra{r} t\}.\]
In other words $\theta(s)(T)$ is the set of all weights
with which we can take a transition from $s$ to some state in $T$.
Furthermore, we let
\[\transl{s}{T} = \begin{cases} -\infty & \text{if } T = \Emptyset \\ \inf \trans{s}{T} & \text{otherwise}\end{cases}\]
and
\[\transr{s}{T} = \begin{cases} \infty & \text{if } T = \Emptyset \\ \sup \trans{s}{T} & \text{otherwise,}\end{cases}\]
so $\transl{s}{T}$ is a lower bound on $\trans{s}{T}$ and $\transr{s}{T}$ is an upper bound.

Given a WTS $\mathcal{M} = (S, \rightarrow, \ell)$,
we now define the semantics of WLWB by the satisfaction relation $\models$ as follows.
\[
  \begin{array}{l l l}
    \mathcal{M},s \models p                   & \text{ if and only if } & p \in \ell(s), \\
    \mathcal{M},s \models \neg \varphi        & \text{ if and only if } & \mathcal{M},s \not\models \varphi, \\
    \mathcal{M},s \models \varphi \wedge \psi & \text{ if and only if } & \mathcal{M},s \models \varphi \;\text{and}\; \mathcal{M},s \models \psi, \\
    \mathcal{M},s \models L_r \varphi         & \text{ if and only if } & \transl{s}{\sat{\varphi}_{\mathcal{M}}} \geq r,\\
    \mathcal{M},s \models M_r \varphi         & \text{ if and only if } & \transr{s}{\sat{\varphi}_{\mathcal{M}}} \leq r,\\
  \end{array}
\]
where $\sat{\varphi}_{\mathcal{M}} = \{s \in S \mid \mathcal{M},s \models \varphi\}$
is the set of all states of $\mathcal{M}$ that satisfies the formula $\varphi$.
We will often omit the subscript $\mathcal{M}$ when it is clear which WTS is referred to.

The semantics of $L_r$ and $M_r$ is illustrated in Figure \ref{fig:semantics-intro}.
The horizontal arrows represent the real number line,
and the arches represent the part of the real number line in which elements of $\trans{s}{\sat{\varphi}}$ may lie.
The endpoints of the arches therefore correspond to $\transl{s}{\sat{\varphi}}$ and $\transr{s}{\sat{\varphi}}$.
This means that if a state satisfies $L_r \varphi$, then $r$ must be to the left of the arch,
and if a state satisfies $M_r \varphi$, then $r$ must be to the right of the arch.

\begin{figure}
  \centering
  \begin{subfigure}{0.45\textwidth}
    \centering
    \begin{tikzpicture}[scale=.5]
      %% X-axis
      \draw [->,thick] (0,0) -- (8,0);

      %% Arc
      \draw [-,semithick] (6,0) arc (0:180:2cm) node[above, xshift=10mm, yshift=10mm] {$\trans{s}{\sat{\varphi}}$};

      %% r
      \draw[shift={(1,0)},-] (0pt,5pt) -- (0pt,-5pt) node[below] {$r$};

      %% Lower bound
      \draw[->] (2,0) ++ (0,-.5) -- (2,0) node[below,at start] {$\theta^{-}$};
      %% Upper bound
      \draw[->] (6,0) ++ (0,-.5) -- (6,0) node[below,at start] {$\theta^{+}$};
    \end{tikzpicture}
    \caption{$\mathcal{M},s \models L_r \varphi$}
    \label{fig:Lsemantics-intro}
  \end{subfigure}
  \begin{subfigure}{0.45\textwidth}
    \centering
    \begin{tikzpicture}[scale=.5]
      %% X-axis
      \draw [->,thick] (0,0) -- (8,0);

      %% Arc
      \draw [-,semithick] (6,0) arc (0:180:2cm) node[above, xshift=10mm, yshift=10mm] {$\trans{s}{\sat{\varphi}}$};

      %% r
      \draw[shift={(7,0)},-] (0pt,5pt) -- (0pt,-5pt) node[below] {$r$};

      %% Lower bound
      \draw[->] (2,0) ++ (0,-.5) -- (2,0) node[below,at start] {$\theta^{-}$};
      %% Upper bound
      \draw[->] (6,0) ++ (0,-.5) -- (6,0) node[below,at start] {$\theta^{+}$};
    \end{tikzpicture}
    \caption{$\mathcal{M},s \models M_r \varphi$}
    \label{fig:Msemantics-intro}
  \end{subfigure}
  \caption{The semantics of $L_r$ and $M_r$. If $\mathcal{M},s \models L_r \varphi$,
           then $r$ is to the left of $\transl{s}{\sat{\varphi}}$, and if $\mathcal{M},s \models M_r \varphi$,
           then $r$ is to the right of $\transr{s}{\sat{\varphi}}$.}
  \label{fig:semantics-intro}
\end{figure}

The operators $L_r$ and $M_r$ are inspired by similar operators in Markovian logic \cite{cardelli2011a,KMP13},
which in turn were inspired by logics for Harsanyi type spaces~\cite{aumann99a,aumann99b}.
However, although the intuition behind the operators are the same in both cases,
they behave quite differently, since Markovian logic considers probabilities,
whereas we consider weights, which have less structure between them.
We will see some of these differences when we consider axiomatisation in Section \ref{sec:axiom}.

WLWB can express the usual ``necessarily'' and ``possibly'' operators from modal logic, written $\Box$ and $\Diamond$, respectively.
To see this, note that $\Diamond \varphi$ means that it is possible to take a transition to where $\varphi$ holds.
Since all our weights are non-negative, $L_0 \varphi$ is true if and only if it is possible to take a transition to where $\varphi$ holds.
The two are therefore equivalent, so we get
\[\Diamond \varphi = L_0 \varphi \quad \text{and} \quad \Box \varphi = \neg L_0 \neg \varphi.\]

\section{Bisimulation Using Bounds}
We now argue that WLWB is the right language to reason about bounds in weighted transition systems.
In order to do this, we define a new notion of bisimulation,
which we call generalised weighted bisimulation.
This notion of bisimulation only compares the upper and lower bounds
of the possible behaviour of the systems.
We then show that WLWB characterises exactly those states
that are in a generalised weighted bisimulation relation with each other.

\begin{definition}
  Let $\mathcal{M} = (S, \rightarrow, \ell)$ be a WTS.
  An equivalence relation $\mathcal{R} \subseteq S \times S$ is a \emph{generalised weighted bisimulation relation}
  if for any states $s,t \in S$ we have that $s \mathcal{R} t$ implies
  \begin{description}
    \item[(Atomic harmony)] $\ell(s) = \ell(t)$,
    \item[(Lower bound)] $\transl{s}{T} = \transl{t}{T}$, and
    \item[(Upper bound)] $\transr{s}{T} = \transl{t}{T}$
  \end{description}
  for any $\mathcal{R}$-equivalence class $T \subseteq S$.
\end{definition}

We will say that $s$ and $t$ are generalised weighted bisimilar and write $s \sim t$
if there exists a generalised weighted bisimulation relation $\mathcal{R}$
such that $s \mathcal{R} t$.
Generalised weighted bisimilarity, denoted $\sim$,
is the largest generalised weighted bisimulation relation.

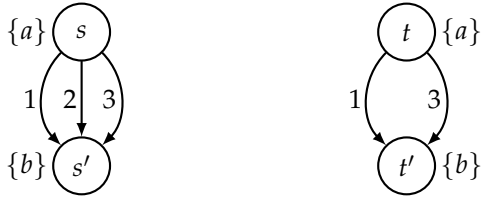
\begin{figure}
  \centering
  \hfill
  \begin{tikzpicture}[WTS, node distance=2cm]
    \node[state, label=left:{$\{a\}$}]  (s0)               {$s$};
    \node[state, label=left:{$\{b\}$}]  (s1) [below=of s0] {$s'$};

    \path (s0) edge[bend right=45] node[left] {$1$} (s1);
    \path (s0) edge                node[left] {$2$} (s1);
    \path (s0) edge[bend left=45]  node[left] {$3$} (s1);
  \end{tikzpicture}
  \hfill
  \begin{tikzpicture}[WTS, node distance=2cm]
    \node[state, label=right:{$\{a\}$}] (t0)  {$t$};
    \node[state, label=right:{$\{b\}$}] (t1) [below=of t0] {$t'$};
    
    \path (t0) edge[bend right=45] node[left] {$1$} (t1);
    \path (t0) edge[bend left=45]  node[left] {$3$} (t1);
  \end{tikzpicture}
  \hfill \
  \caption{$s \sim t$ but $s \not \sim_W t$.}
  \label{fig:bisim-intro}
\end{figure}

\begin{example}\label{ex:bisim-intro}
  Consider the states $s$ and $t$ of the WTS in Figure \ref{fig:bisim-intro}.
  We will show that the relation
  \[\mathcal{R} = \{(s,s), (t,t), (s',s'), (t',t'), (s,t), (t,s), (s',t'), (t',s')\}\]
  is a generalised weighted bisimulation relation.
  It is clearly an equivalence relation.
  
  For $(s',t') \in \mathcal{R}$, we have $\ell(s') = \{b\} = \ell(t')$,
  and since neither of the two states have any outgoing transitions,
  their lower and upper bounds also match.
  For $(s,t) \in \mathcal{R}$, we have $\ell(s) = \{a\} = \ell(t)$,
  so atomic harmony is satisfied.
  For the bounds, note that $T = \{s',t'\}$ is the only equivalence class
  that can be reached from $s$ and $t$,
  so this is the only equivalence class we need to consider.
  We have
  \[\transl{s}{T} = \min\{1,2,3\} = 1 = \min\{1,3\} = \transl{t}{T}\]
  and
  \[\transr{s}{T} = \max\{1,2,3\} = 3 = \max\{1,3\} = \transr{t}{T},\]
  so the lower and upper bounds also match.
  The remaining elements in the relation can be verified in a similar and symmetric manner.
  We therefore conclude that $s \sim t$.
  
  On the other hand, it is not the case that $s \sim_W t$.
  To see this, simply note that $s \xra{2} s'$,
  which can not be matched by $t$,
  i.e. there is no state $t''$ such that $t \xra{2} t''$.
\end{example}

Example \ref{ex:bisim-intro} also shows the essential difference between
weighted bisimilarity and generalised weighted bisimilarity:
weighted bisimilarity looks at all the individual transitions,
whereas generalised weighted bisimilarity ignores the transitions in between
the upper and lower bounds.

It is easy to see that if a relation is a weighted bisimulation,
then it must also be a generalised weighted bisimulation,
since if all transition weights match, then their lower and upper bounds must also match.
Hence we get the following result, relating the two notions of bisimulation.

\begin{theorem}
  \[\sim_W \mathbin{\subseteq} \sim \quad \text{and} \quad \sim_W \mathbin{\neq} \sim.\]
\end{theorem}

In order to prove the claim that WLWB characterises exactly those states that are generalised weighted bisimilar,
we must restrict ourselves to a certain class of WTSs,
namely those that are image-finite.
The notion of image-finiteness is well-known in the literature,
and is also necessary for other modal logics \cite{HM80,blackburn}.

\begin{definition}
  A WTS $\mathcal{M} = (S, \rightarrow, \ell)$ is said to be \emph{image-finite}
  if for any state $s \in S$ there are only finitely many states $t \in S$
  such that $s \xra{r} t$ for some $r \in \mathbb{R}_{\geq 0}$.
\end{definition}

In other words, a WTS is image-finite if any state can only reach finitely many other states in one step.

\begin{theorem}
  For image-finite WTSs, we have
  \[s \sim t \quad \text{if and only if} \quad \text{for all } \varphi, s \models \varphi \text{ if and only if } t \models \varphi.\]
\end{theorem}

\section{Complete Axiomatisation}\label{sec:axiom}
Having argued for why WLWB is an interesting logical specification language to consider,
we now explore the properties of this language.
As a first result, we will give a sound and complete axiomatisation of WLWB.
Along the way, we will also obtain the finite model property.

The axiomatic system that we will consider is given by the axioms of propositional logic
in addition to the axioms given in Table \ref{tab:axioms}.
The axiom (A$1$) says that it is not possible to take a transition to where $\bot$ holds.
Axioms (A$2$) and (A$2'$) give some monotonicity properties of $L_r$ and $M_r$,
whereas axioms (A$3$), (A$3'$), and (A$4$) show how $L_r$ and $M_r$ distribute over $\land$ and $\lor$.
There is no axiom (A$4'$), which should be the obvious variant of (A$4$) with $M_r$ instead of $L_r$.
This is not because such an axiom would not be sound,
but rather because it can be proved from the remaining axioms.
Axioms (A$5$) and (A$5'$) say that if there is no transition to where $\psi$ holds,
then the upper and lower bounds for going to $\varphi$
coincide with the upper and lower bounds for going to $\varphi \lor \psi$.
Axioms (A$6$) and (A$7$) show how the $L_r$ and $M_r$ operators interact.

The rules (R$1$) and (R$1'$) say that if $\varphi$ implies $\psi$,
and we know that there is some transition to where $\varphi$ holds,
then an upper or lower bound for $\psi$ is also an upper or lower bound for $\varphi$.
Lastly, the rule (R$2$) says that if $\varphi$ implies $\psi$,
then if there is a transition to where $\varphi$ holds,
there must also be a transition to where $\psi$ holds.

\begin{table}
  \centering
  \begin{tabular}{l l l}
    \hline
    (A$1$):   & $\vdash \neg L_0 \bot$ & \\
    (A$2$):   & $\vdash L_{r + q}\varphi \rightarrow L_r \varphi$ & if $q > 0$ \\
    (A$2'$):  & $\vdash M_r\varphi \rightarrow M_{r + q} \varphi$ & if $q > 0$ \\
    (A$3$):   & $\vdash L_r \varphi \land L_q \psi \rightarrow L_{\min\{r,q\}}(\varphi \lor \psi)$ & \\
    (A$3'$):  & $\vdash M_r \varphi \land M_q \psi \rightarrow M_{\max\{r,q\}}(\varphi \lor \psi)$ & \\
    (A$4$):   & $\vdash L_r(\varphi \lor \psi) \rightarrow L_r \varphi \lor L_r \psi$ & \\
    (A$5$):   & $\vdash \neg L_0 \psi \rightarrow (L_r \varphi \rightarrow L_r(\varphi \lor \psi))$ & \\
    (A$5'$):  & $\vdash \neg L_0 \psi \rightarrow (M_r \varphi \rightarrow M_r(\varphi \lor \psi))$ & \\
    (A$6$):   & $\vdash L_{r + q}\varphi \rightarrow \neg M_r\varphi$ & if $q > 0$ \\
    (A$7$):   & $\vdash M_r \varphi \rightarrow L_0 \varphi$ & \\
    (R$1$):   & $\vdash \varphi \rightarrow \psi \implies \vdash (L_r \psi \land L_0 \varphi) \rightarrow L_r \varphi$ & \\
    (R$1'$):  & $\vdash \varphi \rightarrow \psi \implies \vdash (M_r \psi \land L_0 \varphi) \rightarrow M_r \varphi$ & \\
    (R$2$):   & $\vdash \varphi \rightarrow \psi \implies \vdash L_0 \varphi \rightarrow L_0 \psi$ & \\
    \hline
  \end{tabular}
  \caption{The axioms for our axiomatic system, where $\varphi, \psi \in \mathcal{L}$ and $q,r \in \mathbb{Q}_{\geq 0}$.}
  \label{tab:axioms-intro}
\end{table}

The axioms of Table \ref{tab:axioms-intro} are sound,
meaning that anything derived from the axioms must also be true semantically.

\begin{theorem}[Soundness]
    \[\vdash \varphi \quad\text{implies}\quad \models \varphi.\]
\end{theorem}

We will say that a formula $\varphi$ is \emph{consistent} if we can not derive $\bot$ from $\varphi$ using the axioms.
For a given consistent formula $\varphi$, we can construct a finite model $\mathcal{M}_\varphi$
with ultrafilters as states such that $\mathcal{M}_\varphi, s \models \varphi$ for some state $s$.

\begin{theorem}[Finite model property]
  For any consistent formula $\varphi \in \mathcal{L}$,
  there exists a finite WTS $\mathcal{M} = (S, \rightarrow, \ell)$
  and a state $s \in S$ such that $\mathcal{M},s \models \varphi$.
\end{theorem}

As an immediate consequence of the finite model property,
we get that our axiomatisation is complete.

\begin{theorem}[Completeness]
  \[\models \varphi \quad\text{implies}\quad \vdash \varphi.\]
\end{theorem}

\section{Satisfiability and Model Checking}
Lastly we will consider some decision problems for WLWB.
First we consider the problem of deciding whether a given formula is satisfiable,
i.e. whether it has a model or not.
In order to solve this problem,
we will construct a tableau for a given formula $\varphi$.
If the tableau is successful, meaning that it contains no inconsistencies,
then we can construct a model for $\varphi$ from the tableau.
Otherwise, if the tableau is not successful,
then we will know that there is no model for $\varphi$.

Given a formula $\varphi$,
we start with the tuple $\langle \{\varphi\}, [0,0], [0,0] \rangle$
and then successively apply the rules of Table \ref{tab:rules-intro} until no further rules can be used.
The (mod) rule may only be applied when no other rules can be applied.

The intuition behind a tuple $\langle \Gamma, \mathcal{I}^L, \mathcal{I}^M \rangle$,
where $\Gamma$ is a set of formulas and $\mathcal{I}^L$ and $\mathcal{I}^M$ are intervals,
is that the current state must satisfy all the formulas in $\Gamma$,
and any transitions to the current state must have a lower bound within the interval $\mathcal{I}^L$
and an upper bound within the interval $\mathcal{I}^M$.
The (mod) rule signifies a state change,
at which point all formulas in $\Gamma$ have been broken down into either literals,
i.e. formulas of the form $p$ or $\neg p$ where $p \in \mathcal{AP}$, or modal formulas.
The modal formulas then determine what transitions, if any,
there must be from the current state to the next states.

\begin{table}
  \centering
  \begin{tabular}{c}
    \hline
    
    \\
    
    {\begin{prooftree}
      \hypo{\langle \Gamma \cup \{\varphi \land \psi\}, \mathcal{I}^L, \mathcal{I}^M \rangle}
      \infer[left label = {($\land$)}]1{\langle \Gamma \cup \{\varphi, \psi\}, \mathcal{I}^L, \mathcal{I}^M \rangle}
    \end{prooftree}} \\
    
    \\
    
    {\begin{prooftree}
      \hypo{\langle \Gamma \cup \{\neg (\varphi \land \psi)\}, \mathcal{I}^L, \mathcal{I}^M \rangle}
      \infer[left label = {($\neg \land$)}]1{\langle \Gamma \cup \{\neg \varphi\}, \mathcal{I}^L, \mathcal{I}^M \rangle \quad \langle \Gamma \cup \{\neg \psi\}, \mathcal{I}^L, \mathcal{I}^M \rangle}
    \end{prooftree}} \\
    
    \\
  
    {\begin{prooftree}
      \hypo{\langle \Gamma \cup \{\neg \neg \varphi\}, \mathcal{I}^L, \mathcal{I}^M \rangle}
      \infer[left label = {($\neg\neg$)}]1{\langle \Gamma \cup \{\varphi\}, \mathcal{I}^L, \mathcal{I}^M \rangle}
    \end{prooftree}} \\
    
    \\
    
    {\begin{prooftree}
      \hypo{\langle \Gamma \cup \{N^1_{r_1} \varphi_1, \dots, N^n_{r_n} \varphi_n\} \cup \{\neg O^1_{r_1'} \varphi_1', \dots, \neg O^{n'}_{r_{n'}'} \varphi_{n'}'\}, \mathcal{I}^L, \mathcal{I}^M \rangle}
      \infer[left label = {(mod)}]1{\langle \{\psi_1\}, \mathcal{I}^L_1, \mathcal{I}^M_1 \rangle \quad \cdots \quad \langle \{\psi_k\}, \mathcal{I}^L_k, \mathcal{I}^M_k \rangle}
    \end{prooftree}} \\
    
    \\
    
    \parbox{10cm}{
      if $N^i \in \{L,M\}$ for all $1 \leq i \leq n$, $O^j \in \{L,M\}$ for all $1 \leq j \leq n'$,
      and no formula in $\Gamma$ is of the form $N_r \varphi$ or $\neg N_r \varphi$ where $N \in \{L,M\}$.
    } \\
    
    \\
    
    \hline
  \end{tabular}
  \caption{Tableau rules.}
  \label{tab:rules-intro}
\end{table}

\begin{example}
  We illustrate how the conclusions in the (mod) rule are constructed.
  Consider the tuple
  \[\langle \{p_1, p_2, L_2 p_1, L_4(p_1 \land p_2), L_0 p_3, \neg L_5 p_2, \neg M_6 p_3\}, \mathcal{I}^L, \mathcal{I}^M \rangle.\]
  By letting
  \[\Gamma = \{p_1, p_2\}, \Gamma' = \{L_2 p_1, L_4 (p_1 \land p_2), L_0 p_3\}, \text{ and } \Gamma'' = \{\neg L_5 p_2, \neg M_6 p_3\},\]
  we get that $\langle \Gamma \cup \Gamma' \cup \Gamma'', \mathcal{I}^L, \mathcal{I}^M\rangle$ has the correct form for the hypothesis of the (mod) rule.

  Now, the modal formulas in $\Gamma'$ put requirements on the transitions to the next states.
  Consider the formulas $L_2 p_1$ and $L_4 (p_1 \land p_2)$.
  Any successor state where $p_1 \land p_2$ holds must also satisfy $p_1$.
  Hence we will not create two next states for these two formulas,
  but only one, which will be the most restrictive of the formulas,
  in this case $p_1 \land p_2$.

  So we need two successor states, one that satisfies $p_1 \land p_2$ and one that satisfies $p_3$.
  It only remains to determine the weights on the transitions to these new states.
  For the state satisfying $p_1 \land p_2$, we see that the lower bounds must be at least $4$,
  and the formula $\neg L_5 p_2 \in \Gamma''$ tells us that the lower bound must be strictly less than $5$.
  However, since no $M_r$ formulas speak about $p_1$ or $p_2$, the upper bound has no restrictions.
  Hence the interval for the lower bound is $\mathcal{I}^L = [4,5)$
  and the interval for the upper bound is $\mathcal{I}^M = [0, \infty)$.
  Likewise we see that for the state satisfying $p_3$,
  the lower bound must be in the interval $[0,\infty)$,
  and the upper bound must be in the interval $[6,\infty)$.

  Applying the mod rule to the tuple therefore gives the following result.
  \[
    \begin{prooftree}
      \hypo{\langle \{p_1, p_2, L_2 p_1, L_4(p_1 \land p_2), L_0 p_3, \neg L_5 p_2, \neg M_6 p_3\}, \mathcal{I}^L, \mathcal{I}^M \rangle}
      \infer[left label = {(mod)}]1{\langle \{p_1 \land p_2\}, [4,5), [0,\infty) \rangle \quad \langle \{p_3\}, [0,\infty) (6,\infty) \rangle}
    \end{prooftree} \qedhere
  \]
\end{example}

A tableau constructed from the tableau rules will be said to be \emph{successful}
if we can find a suitable subtree of the tableau such that for each tuple $\langle \Gamma, \mathcal{I}^L, \mathcal{I}^M \rangle$
in the subtree, the set of formulas $\Gamma$ is consistent,
and the intervals $\mathcal{I}^L$ and $\mathcal{I}^M$ are well-formed intervals.

The significance of the tableau construction follows from the next two lemmas.

\begin{lemma}
  $\varphi$ is satisfiable if and only if there exists a successful tableau for $\varphi$.
\end{lemma}

\begin{lemma}
  Given a successful tableau for $\varphi$,
  we can construct a model $\mathcal{M} = (S, \rightarrow, \ell)$ and a state $s \in S$
  such that $\mathcal{M},s \models \varphi$.
\end{lemma}

This gives a decision procedure for the satisfiability problem:
To determine whether $\varphi$ is satisfiable,
simply construct a tableau from the tableau rules of Table \ref{tab:rules-intro} starting with the tuple $\langle \{\varphi\}, [0,0], [0,0] \rangle$.
If the tableau is successful, then $\varphi$ is satisfiable, otherwise it is not satisfiable.
Furthermore, if the tableau is successful, we can actually construct a model for $\varphi$ from the successful tableau.

\begin{theorem}
  The satisfiability problem for WLWB is decidable.
\end{theorem}

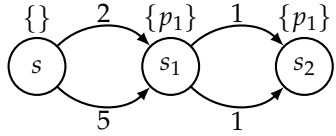
\begin{figure}
  \centering
  \begin{tikzpicture}[WTS, node distance=2cm]
    \node[state, label=above:{$\{\}$}]    (st)               {$s$};
    \node[state, label=above:{$\{p_1\}$}] (s1) [right=of st] {$s_1$};
    \node[state, label=above:{$\{p_1\}$}] (s2) [right=of s1] {$s_2$};

    \path (s0) edge[bend right=45] node[below] {$5$} (s1);
    \path (s0) edge[bend left=45]  node[above] {$2$} (s1);

    \path (s1) edge[bend right=45] node[below] {$1$} (s2);
    \path (s1) edge[bend left=45]  node[above] {$1$} (s2);
  \end{tikzpicture}
  \captionof{figure}{The model for the successful tableau in Example \ref{ex:sat-intro}.}
  \label{fig:sat-ex-intro}
\end{figure}

\begin{example}\label{ex:sat-intro}
  Consider the formula $\varphi = \neg(\neg (L_2 p_1 \land M_5L_1 p_1) \land \neg M_2 p_2 ))$.
  Using the tableau rules, we get the following tableau for $\varphi$.
  \[
    \begin{prooftree}[proof style=downwards]
      \hypo{\langle \{p_1\}, [1,\infty), [0,\infty) \rangle}
      \infer[left label = {(mod)}]1{\langle \{p_1, L_1 p_1\}, [2, \infty), [5, \infty) \rangle}
      \infer[left label = {(mod)}]1{\langle \{L_2 p_1, M_5L_1 p_1\}, [0,0], [0,0] \rangle}
      \infer[left label = {($\land$)}]1{\langle \{L_2 p_1 \land M_5L_1 p_1\}, [0,0], [0,0] \rangle}
      \infer[left label = {($\neg \neg$)}]1{\langle \{\neg\neg (L_2 p_1 \land M_5L_1 p_1)\}, [0,0], [0,0] \rangle}
      \hypo{\langle \{p_2\}, [0,\infty), [0, 2] \rangle}
      \infer[left label = {(mod)}]1{\langle \{M_2 p_2\}, [0,0], [0,0] \rangle}
      \infer[left label = {($\neg\neg$)}]1{\langle \{\neg\neg M_2 p_2\}, [0,0], [0,0] \rangle}
      \infer[left label = {($\neg \land$)}]2{\langle \{\neg (\neg (L_2 p_1 \land M_5L_1 p_1) \land M_2 p_2)\}, [0,0], [0,0] \rangle}
    \end{prooftree}
  \]
  In this case the tableau is successful,
  every leaf and every node before a (mod) rule is consistent.
  From this tableau we can construct a model that satisfies $\varphi$.
  This model is shown in Figure \ref{fig:sat-ex-intro}
\end{example}

Lastly we consider the model checking problem for WLWB.
This problem asks us to decide for a given model $\mathcal{M} = (S, \rightarrow, \ell)$,
state $s \in S$, and formula $\varphi$ whether $\mathcal{M}, s \models \varphi$.

We can solve this problem in polynomial time by adapting the classical model checking algorithm by Clarke et al. \cite{CES86} to our setting.
This algorithm constructs a function $F_\varphi$,
which assigns to each state the set of subformulas of $\varphi$ that are true in that state.
$F_\varphi$ is built iteratively by first considering the smallest subformulas of $\varphi$ (i.e. atomic propositions),
then the second smallest subformulas (i.e. subformulas of the form $\neg p_1$, $p_1 \land p_2$, $L_r p_1$, or $M_r p_1$,
where $p_1,p_2 \in \ap$), and so on until we get to $\varphi$ itself.
At each step we can use information about the smaller subformulas that have already been assigned by $F_\varphi$
to determine which formulas must be assigned in the current step.

\begin{lemma}\label{lem:model-checking-intro}
  Given a model $\mathcal{M} = (S, \rightarrow, \ell)$, state $s \in S$, and formula $\varphi$,
  it holds that $\mathcal{M},s \models \varphi'$ if and only if $\varphi' \in F_\varphi(s)$
  for any subformula $\varphi'$ of $\varphi$.
\end{lemma}

By Lemma \ref{lem:model-checking-intro}, we can therefore decide the model checking problem
by checking whether $\varphi \in F_\varphi(s)$.

\begin{theorem}
  The model checking problem for WLWB is decidable in polynomial time.
\end{theorem}

\defaultbib

  \setcounter{figure}{0}
  \setcounter{subfigure}{0}
  \setcounter{section}{0}
  \setcounter{subsection}{0}
  \setcounter{table}{0}
  \setcounter{equation}{0}
  \newcommand{\tfttitle}{Trace-Based Faster-Than Relation}
\chapter{\tfttitle}\label{chap:tft}
%\addcontentsline{toc}{chapter}{\tfttitle}

%\section{Trace-Based Faster-Than Relation}\label{sec:introtrace}

This chapter summarises Paper B ``Timed Comparisons of Semi-Markov Processes''~\cite{PFBLM18}
and paper C ``A Faster-Than Relation for Semi-Markov Decision Processes''~\cite{PBL18}.
For the full papers, see Part~\ref{chap:papers}.

For real-time systems,
non-functional requirements such as reliability, throughput, and response time are important to consider.
It is therefore of interest to improve the worst-case timing guarantees on such systems,
which leads us to investigate how to compare the timing behaviour of systems.
In particular, we want to be able to describe when a system is faster than another,
which will allow incremental timing improvements in a system.

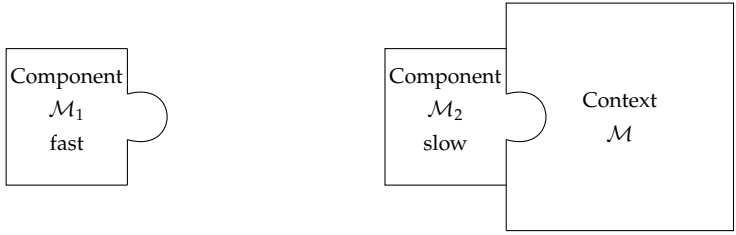
\begin{figure}
  \center
  \begin{tikzpicture}
    % Context
    \draw (0,0) -- (3,0) -- (3,3) -- (0,3) -- (0,1.8);
    \draw (0,1.2) -- (0,0);
    \draw (0,1.8) .. controls  (0.7,2.0) and (0.7,1.0) .. (0,1.2);
    \node[align=center] at (1.5,1.5) {Context \\ $\mathcal{M}$};
    
    % Component 1
    \draw (0,0.6) -- (-1.6,0.6) -- (-1.6,2.4) -- (0,2.4);
    \node[align=center] at (-0.8,1.6) {Component \\ $\mathcal{M}_2$ \\ slow};
    
    % Component 2
    \draw (-5,0.6) -- (-6.6,0.6) -- (-6.6,2.4) -- (-5,2.4);
    \draw (-5,2.4) -- (-5,1.8);
    \draw (-5,1.2) -- (-5,0.6);
    \draw (-5,1.8) .. controls (-4.3,2.0) and (-4.3,1.0) .. (-5,1.2);
    \node[align=center] at (-5.8,1.6) {Component \\ $\mathcal{M}_1$ \\ fast};
    
  \end{tikzpicture}
  \caption{The context $\mathcal{M}$ operates in parallel with the component $\mathcal{M}_2$.
           If the component $\mathcal{M}_1$ is faster than $\mathcal{M}_2$, then if we replace $\mathcal{M}_2$ with $\mathcal{M}_1$,
           we would expect the overall behaviour to also be faster.}
  \label{fig:ft-intro}
\end{figure}

Another important aspect of real-time systems is compositionality,
which allows us to describe complex systems in terms of smaller components that together make up the whole system~\cite{CLM89}.
This leads to the picture in Figure~\ref{fig:ft-intro},
where we have a complex system consisting of $\mathcal{M}$ and the component $\mathcal{M}_2$,
and we have a new component $\mathcal{M}_1$ which is faster than $\mathcal{M}_2$.
The idea is then to replace $\mathcal{M}_2$ by $\mathcal{M}_1$ to obtain a faster system.

However, it is not always the case that replacing a slower component with a faster one
leads to an overall system that is also faster.
In other words, a local increase in timing behaviour may lead to a global decrease in timing behaviour.
This is known as a \emph{(parallel) timing anomaly}~\cite{lundqvist1999,kirner2009}.
We therefore also wish to investigate how to avoid such timing anomalies.

\section{Faster-Than Relation}
We first consider the question of what it means for one system to be faster than another.
In this chapter, the systems we will consider are semi-Markov processes (SMPs),
and we will take a trace-based view of SMPs.
Intuitively, we will say that a state $s_1$ is faster than another state $s_2$
if any sequence of actions that $s_2$ can do within some time $t'$,
$s_1$ can do within some time $t \leq t'$ and with at least the same probability.

In order to do this, we must define the probability of completing a sequence of actions
within a given time bound.
Consider therefore an SMP $\mathcal{M} = (S, \tau, \rho, \ell)$,
and recall from Section \ref{sec:smps-intro} that given $n$ sets of states
\[S_1, \dots S_n \in 2^S,\]
$n$ sets of time points
\[R_1, \dots, R_n \in \borel,\]
and $n$ sets of output actions
\[O_1, \dots, O_n \in 2^\aout,\]
as well as a scheduler $\sigma : S^* \rightarrow \dist(\ain)$,
then
\[\prob^\sigma(s)(\cylinder(S_1 \dots S_n, O_1 \dots O_n, R_1 \dots R_n))\]
is the probability of starting in $s$,
then going to a state $s_1 \in S_1$ within time $t_1 \in R_1$ while outputting $o_1 \in O_1$,
then going from $s_1$ to a state in $s_2 \in S_2$ within time $t_2 \in R_2$ while outputting $o_2 \in O_2$,
and so on.

\begin{definition}
  Given a finite sequence of actions $o_1, \dots, o_n \in \aout$ and a time bound $t \in \mathbb{R}_{\geq 0}$,
  we will say that
  \[\cylinder(a_1 \dots a_n, t) = \{\pi \in \paths(\mathcal{M}) \mid \forall 1 \leq i \leq n, \pi\aproj{i} = a_i \text{ and } \sum_{j = 1}^n \pi\tproj{j} \leq t\}\]
  is a \emph{time-bounded cylinder}.
\end{definition}

A time-bounded cylinder $\cylinder(a_1 \dots a_n, t)$ therefore denotes all paths
where the first $n$ steps output the sequence $a_1 \dots a_n$ and are done within time $t$.
For example, the time-bounded cylinder $\cylinder(aa,2)$ denotes that the first two output labels must be $a$'s,
and that the first two steps must be completed within $2$ time units.

We will consider \emph{pointed} SMPs, meaning that each SMP has a designated initial state.
Given an SMP $\mathcal{M}$, we will write $(\mathcal{M}, s)$ to mean that $s$ is the initial state of $\mathcal{M}$.

\begin{definition}
  We will say that $s_1$ is \emph{faster than} $s_2$ and write $s_1 \ft s_2$
  if for all schedulers $\sigma$ there exists a scheduler $\sigma'$ such that
  \[\prob^{\sigma'}(s_1)(C) \geq \prob^{\sigma}(s_2)(C)\]
  for all time-bounded cylinders $C$.
  
  Given two pointed SMPs $(\mathcal{M}_1, s^*_1)$ and $(\mathcal{M}_2, s^*_2)$,
  we will say that $\mathcal{M}_1$ is \emph{faster than} $\mathcal{M}_2$ and write $\mathcal{M}_1 \ft \mathcal{M}_2$
  if $s^*_1 \ft s^*_2$.
\end{definition}

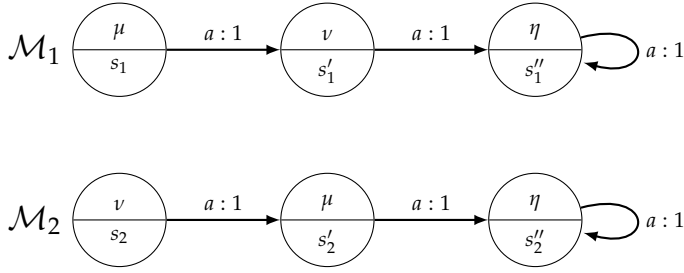
\begin{figure}
  \centering
  \hfill
  \begin{tikzpicture}\tikzset{every state/.style={minimum size=35pt}}
    %Nodes
    \node[state, circle split] (3) [below = of 0]{$\mu$ \nodepart{lower} $s_1$};
    \node[state, circle split] (4) [right = 1.5cm of 3]{$\nu$ \nodepart{lower} $s_1'$};
    \node[state, circle split] (5) [right = 1.5cm of 4]{$\eta$ \nodepart{lower} $s_1''$};
    
    \node[state, circle split] (6) [below = of 3]{$\nu$ \nodepart{lower} $s_2$};
    \node[state, circle split] (7) [right = 1.5cm of 6]{$\mu$ \nodepart{lower} $s_2'$};
    \node[state, circle split] (8) [right = 1.5cm of 7]{$\eta$ \nodepart{lower} $s_2''$};
    
    \node[xshift=-0.5cm] at (3.west) {\large{$\mathcal{M}_1$}};
    \node[xshift=-0.5cm] at (6.west) {\large{$\mathcal{M}_2$}};
    
    % Edges
    \path[thick, ->] (3) edge [above] node {$a:1$} (4);
    \path[thick, ->] (4) edge [above] node {$a:1$} (5);
    \path[thick, loop right, ->] (5) edge [right] node {$a:1$} (5);
    
    \path[thick, ->] (6) edge [above] node {$a:1$} (7);
    \path[thick, ->] (7) edge [above] node {$a:1$} (8);
    \path[thick, loop right, ->] (8) edge [right] node {$a:1$} (8);
  \end{tikzpicture}
  \hfill \
  \caption{$\mathcal{M}_1$ is faster than $\mathcal{M}_2$.}
  \label{fig:ft-example-intro}
\end{figure}

\begin{example}\label{ex:tft1}
  Consider the two SMPs $\mathcal{M}_1$ and $\mathcal{M}_2$ in Figure~\ref{fig:ft-example-intro}
  and assume that $F_\mu(t) \geq F_\nu(t)$ for all $t \in \mathbb{R}_{\geq 0}$.
  
  \textbf{(Case $n = 1$)} In this case we get
  \[\prob(s_1)(\cylinder(a,t)) = F_\mu(t) \quad \text{and} \quad \prob(s_2)(\cylinder(a,t)) = F_\nu(t).\]
  Since we have assumed $F_\mu(t) \geq F_\nu(t)$, it follows that
  \[\prob(s_1)(\cylinder(a,t)) \geq \prob(s_2)(\cylinder(a,t)).\]
  
  \textbf{(Case $n > 1$)} In this case we get
  \[\prob(s_1)(\cylinder(a^n,t)) = (\mu * \nu * \eta^{*(n-2)})([0,t])\]
  and
  \[\prob(s_2)(\cylinder(a^n,t)) = (\nu * \mu * \eta^{*(n-2)})([0,t]),\]
  where $\eta^{*n}$ is the $n$-fold convolution of $\eta$.
  However, since convolution is commutative,
  it follows that
  \[\prob(s_1)(\cylinder(a^n,t)) = \prob(s_2)(\cylinder(a^n,t)).\]
  
  Thus we have $\prob(s_1)(C) \geq \prob(s_2)(C)$ for all time-bounded cylinders $C$,
  and therefore $s_1 \ft s_2$.
\end{example}

\subsection{Comparison With Simulation and Bisimulation}
In this section we compare our notion of a faster-than relation to the standard notions of simulation and bisimulation.
For this, we also introduce the notion of two processes being equally fast.

\begin{definition}
  $\mathcal{M}_1$ and $\mathcal{M}_2$ are \emph{equally fast}, written $\mathcal{M}_1 \eqft \mathcal{M}_2$,
  if $\mathcal{M}_1 \ft \mathcal{M}_2$ and $\mathcal{M}_2 \ft \mathcal{M}_1$.
\end{definition}

\begin{figure}
  \centering
  % First picture
  \hfill
  \begin{tikzpicture}[->, >=latex, baseline={(current bounding box.center)}]
    % Nodes
    \node[state, circle split] (0) {$\mu$ \nodepart{lower} $s$};
    %
    %\node[xshift=-1cm] at (0.west) {\large{$U$}};
    %
    % Edges
    \path[thick] (0) edge [out=310,in=280,looseness=6] node[right] {$a:1$} (0);
    \path[thick] (0) edge [out=230,in=260,looseness=6] node[left] {$b:1$} (0);
  \end{tikzpicture}
  \hfill
  % Second picture
  \begin{tikzpicture}[->, >=latex, baseline={(current bounding box.center)}]
    % Nodes
    \node[state, circle split] (2) [below = 2 cm of 0] {$\mu$ \nodepart{lower} $s_0$};
    \node[state, circle split] (3) [below left = 0.7cm of 2] {$\mu$ \nodepart{lower} $s_1$};
    \node[state, circle split] (4) [below right = 0.7cm of 2] {$\mu$ \nodepart{lower} $s_2$};
    %
%      \node[xshift=-2.5cm,yshift=-0.6cm] at (2.west) {\large{$V$}};
    %
    % Edges
    \path[thick] (2) edge [bend right, left] node {$a:1$} (3);
    \path[thick] (2) edge [bend left, right] node {$b:1$} (4);
    \path[thick] (3) edge [out=310,in=280,looseness=6] node[right] {$a:1$} (3);
    \path[thick] (3) edge [out=230,in=260,looseness=6] node[left] {$b:1$} (3);
    \path[thick] (4) edge [out=310,in=280,looseness=6] node[right] {$a:1$} (4);
    \path[thick] (4) edge [out=230,in=260,looseness=6] node[left] {$b:1$} (4);
  \end{tikzpicture}
  \hfill \ 
  
  \caption{Example showing that the faster-than relation and the simulation relation are incomparable.}
  \label{fig:comparison-intro}
\end{figure}
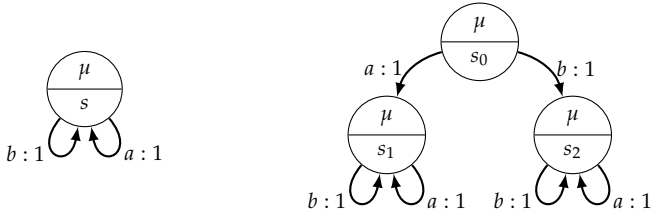

\begin{example}\label{ex:tft2}
  Consider the SMP in Figure \ref{fig:comparison-intro} with the same probability measure $\mu$ in all states.
  It is easy to see that $s$ is bisimilar to $s_0$, and hence $s_0$ also simulates $s$.
  However, we can show that $s \not \ft s_0$ in the following way.
  Construct the (memoryless) scheduler $\sigma$ by letting
  \[\sigma(s_0)(a) = 0.5, \; \sigma(s_0)(b) = 0.5, \; \sigma(s_1)(a) = 1, \, \text{ and } \, \sigma(s_2)(b) = 1.\]
  Now, for any scheduler $\sigma'$, we must have either
  $\sigma'(s)(a) < 1$ or $\sigma'(s)(b) < 1$.
  If $\sigma'(s)(a) < 1$, then $\sigma'(s)(a) > (\sigma'(s)(a))^2 > \dots > (\sigma'(s)(a))^n$.
  Furthermore, we see that
  \[\prob^{\sigma}(s_0)(\cylinder(a^n,t)) = 0.5 \cdot \mu^{*n}(t) \text{ and } \prob^{\sigma'}(s)(\cylinder(a^n, t)) = (\sigma'(s)(a))^n \cdot \mu^{*n}(t)\]
  for $n > 1$.
  Take some $n$ such that $(\sigma'(s)(a))^n < 0.5$.
  In that case we get $\prob^{\sigma'}(s)(\cylinder(a^n, t)) < \prob^{\sigma}(s_0)(\cylinder(a^n, t))$.
  The same procedure can be used in case $\sigma'(s)(b) < 1$.
  
  For schedulers with memory, notice that, starting from $s$, in each step either the probability of a trace consisting
  only of $a$'s or the probability of a trace consisting only of $b$'s must decrease.
  After some number of steps, the probability of one of these two must therefore decrease below $0.5$,
  and then the rest of the argument is as before. 
  Hence we conclude that $s \not \ft s_0$, and therefore also that $s \not \eqft s_0$.
\end{example}

\begin{example}\label{ex:tft3}
  Consider again Figure~\ref{fig:ft-example-intro}
  and let $F_\mu = \Exp{\theta_1}$ and $F_\nu = \Exp{\theta_2}$ with $\theta_1 > \theta_2 > 0$.
  Then, as shown in Example~\ref{ex:tft1} we have $s_1 \ft s_2$.
  However, we have both $s_1 \not\simul s_2$ and $s_1 \not\sim s_2$.
\end{example}

From Examples \ref{ex:tft2} and \ref{ex:tft3}, we obtain the following theorem.

\begin{theorem}
  $\simul$ and $\ft$ are incomparable, $\sim$ and $\ft$ are incomparable, and $\sim \mathbin{\subsetneq} \eqft$.
\end{theorem}

\subsection{Algorithmic Considerations}
When discussing algorithms for SMPs,
we have to consider how residence-time distributions are handled by the algorithms.
They must be described by some finite number of rational parameters
that can be given as input to the algorithm,
and since we are interested in the faster-than relation,
we also want to be able to make comparisons between the distributions.

\begin{definition}
  We say that a class of distributions $\mathcal{C}$ is \emph{effective} if for any $\varepsilon > 0$,
  $b \in \mathbb{R}_{\geq 0}$, and $\mu,\nu \in \Conv(\mathcal{C})$,
  \[\{t \in \mathbb{R}_{\geq 0} \mid \mu([0,t]) \geq \nu([0,t]) - \varepsilon \text{ and } t \leq b\}\]
  is a semialgebraic set,
  where $\Conv(\mathcal{C})$ is the closure of $\mathcal{C}$ under convex combinations and convolutions.
\end{definition}

Semialgebraic sets are essentially those sets that can be described in the first-order theory of the reals,
and since this theory is decidable~\cite{Tarski51},
this allows us to compare distributions.

\section{Hardness Results}
In this section we consider the \emph{faster-than problem}:
\[\text{Given } s_1 \text{ and } s_2, \text{ is it the case that } s_1 \ft s_2?\]
Unfortunately, this problem turns out to be a difficult one.
In particular, the faster-than problem is undecidable, and even approximating it is impossible.

In order to show these hardness results we rely on a connection to probabilistic automata~\cite{rabin63}.
A probabilistic automaton is a tuple
\[\mathcal{A} = (Q, A, q_0, \Delta : Q \times A \rightarrow \distone(Q), F),\]
where $Q$ is a set of states, $A$ is the alphabet, $q_0$ is the initial state,
$\Delta$ is the transition function, and $F$ is a set of accepting states.
Many important problems for probabilistic automata are undecidable~\cite{Fijalkow17}.
The problem that we will make use of is the \emph{universality problem} for probabilistic automata
which asks whether a given automaton $\mathcal{A}$ satisfies $\prob_{\mathcal{A}}(w) \geq \frac{1}{2}$
for all words $w$.
Here, $\prob_{\mathcal{A}}(w)$ is the probability for $\mathcal{A}$ to accept the word $w$.
In other words, the universality problem asks whether all words are accepted with at least probability $\frac{1}{2}$.
The universality problem is known to be undecidable~\cite{GO10,Paz71}.

Given a probabilistic automaton $\mathcal{A}$,
we construct the \emph{derived} (generative) discrete-time Markov chain with output alphabet $A$
\[\mathcal{M}(\mathcal{A}) = (S, \tau, \ell),\]
where
\begin{itemize}
  \item $S = (Q \times \{\ell, r\}) \cup \{\top\}$ for some new state $\top$,
  \item $\ell(s) = \Emptyset$ for all $s \in S$, and
  \item $\tau$ is given by
\end{itemize}
\begin{align*}
  \tau((p,\ell))((q,\ell),a) &= \frac{1}{2|A|} \Delta(p,a)(q)  &\quad
  \tau((p,\ell))(\top,a)     &= \frac{1}{2|A|} \text{ if } p \in F \\
  \tau((p,r))((q,r),a)       &= \frac{1}{2|A|} \Delta(p,a)(q)  &\quad 
  \tau((p,r))(\top,a)        &= \frac{1}{4|A|} .
\end{align*}

Now let $s_1 = (q_0, \ell)$ and $s_2 = (q_0,r)$
We then get
\[\prob(s_1)(\cylinder(wa)) = \frac{1}{(2|A|)^{|w| + 1}}(1 + \prob_{\mathcal{A}}(w))\]
and
\[\prob(s_2)(\cylinder(wa)) = \frac{1}{(2|A|)^{|w| + 1}}\left(1 + \frac{1}{2} \right).\]

From this it follows that the faster-than problem for generative SMPs is undecidable,
and a small extension of the argument shows that the same is true for reactive SMPs.

\begin{theorem}
  The faster-than problem is undecidable for both reactive and generative SMPs,
  and hence also for general SMPs.
\end{theorem}

Note that the undecidability result does not depend on the real-time behaviour of the systems,
since the derived Markov chain is discrete-time.
This means that the difficulty with the faster-than problem is not actually the real-time behaviour of the systems,
but rather the probabilistic branching structure.

We discuss three approaches to recover decidability:

\begin{itemize}
  \item Imposing \emph{structural restrictions} on the underlying graph,
  \item restricting the \emph{observations} (i.e. input and output alphabet), and
  \item using \emph{approximations}.
\end{itemize}

\subsection{Structural Restrictions}
The undecidability result for probabilistic automata already applies in the case of acyclic graphs,
meaning that the only loops allowed are self-loops.
Hence restricting to acyclic graphs will not help.
However, there is another kind of structural restriction which has proved fruitful for probabilistic automata,
which is that of \emph{unambiguous} automata~\cite{FRW17}.
We will show in Section~\ref{sec:unambiguous-intro} that this notion also allows us to recover decidability for the faster-than problem
in the case of generative systems.

\subsection{Observations}
The undecidability of the universality problem for probabilistic automata holds even when the alphabet only has two elements.
Interestingly, the decidability of the universality problem is still an open problem when considering unary probabilistic automata,
i.e. probabilistic automata where the alphabet only has a single symbol~\cite{AGKV16}.
However, in this case the problem also has connections to the positivity problem for linear recurrence sequences~\cite{AAOW15},
which has been a major open problem for decades.
The positivity problem asks whether all terms of a given linear recurrence sequence are positive.
It has been shown~\cite{AAOW15} that the universality problem is at least as hard as the positivity problem.
Hence we get the following.

\begin{theorem}
  For generative processes with one output label, the faster-than problem is at least as hard as the positivity problem.
\end{theorem}

Note that we have only been able to show the above for generative processes,
since in that case the reduction from the universality problem works for only one symbol.
However, in the reactive case the reduction we give requires us to introduce a new symbol,
meaning that we need at least two symbols.

\subsection{Approximations}
By exploiting once again the connection to probabilistic automata,
we can show that approximating the faster-than problem up to a multiplicative constant is impossible.
This result relies on the following impossibility theorem for probabilistic automata.

\begin{theorem}[\hspace{1sp}\cite{CL89,Fijalkow17}]
  Let $0 < \alpha < \beta < 1$ be two constants.
  There is no algorithm which, given a probabilistic automaton $\A$,
    \begin{itemize}
      \item returns \textbf{YES} if for all $w$ we have $\P_\A(w) \geq \beta$ and
      \item returns \textbf{NO} if there exists $w$ such that $\P_\A(w) \leq \alpha$.
    \end{itemize}
\end{theorem}

This in turn gives us the following impossibility result for approximating the faster-than problem.

\begin{theorem}
  Let $0 < \varepsilon < \frac{1}{3}$ be a constant.
  There is no algorithm which, given a discrete-time Markov chain $\mathcal{M}$ and two states $s$ and $s'$,
  \begin{itemize}
    \item returns \textbf{YES} if for all $w$ we have $\prob(s)(\cylinder(w)) \geq \prob(s')(\cylinder(w))$ and
    \item returns \textbf{NO} if there exists $w$ such that $\prob(s)(\cylinder(w)) \leq \prob(s')(\cylinder(w)) \cdot (1 - \varepsilon)$.
  \end{itemize}
\end{theorem}

However, in Section~\ref{sec:approx-intro} we will show that approximation up to an additive constant can be done
for a special kind of residence-time distributions,
if we only consider what happens up to some given point in time.

\section{Time-Bounded Additive Approximation}\label{sec:approx-intro}
Although multiplicative approximation is impossible,
we will now show that time-bounded additive approximation is possible.
More precisely, we will show that the \emph{time-bounded additive approximation problem} is decidable
for a suitable class of residence-time distributions.
This problem asks, given $\varepsilon > 0$, a time bound $b \in \mathbb{R}_{\geq 0}$, and two states $s_1$ and $s_2$,
whether for all schedulers $\sigma$ there exists a scheduler $\sigma'$ such that
\begin{equation}\label{eq:add-approx-intro}
  \prob^\sigma(s_1)(C) \geq \prob^{\sigma'}(s_2)(C) - \varepsilon
\end{equation}
for all time-bounded cylinders $C = \cylinder(a_1 \dots a_n, t)$ where $t \leq b$.

Our decidability result holds for residence-time distributions that are \emph{slow}.
The formal definition of slow residence-time distributions is somewhat technical,
but the idea is that slow residence-time distributions must use some non-zero amount of time to take a transition.
This ensures that the process can not do infinitely many transitions within a given time bound,
thus ruling out so-called Zeno behaviour.
Furthermore, this means that the probability of time-bounded cylinders above some specific length
must be less than $\varepsilon > 0$ for that given time bound,
and hence the inequality in \eqref{eq:add-approx-intro} is trivially satisfied.
Therefore we only need to consider finitely many time-bounded cylinders.

\begin{theorem}
  Let $\mathcal{M}$ be an SMP with slow residence-time distributions.
  For any state $s$, $\varepsilon > 0$, time bound $b \in \mathbb{R}_{\geq 0}$, and scheduler $\sigma$,
  there exists $N \in \mathbb{N}$ such that
  \[\prob^\sigma(s)(\cylinder(a_1 \dots a_n, b)) \leq \varepsilon \quad \text{for all } n \geq N.\]
\end{theorem}

Based on this result, we can prove the following theorem using the decidability of the first-order theory of the reals.

\begin{theorem}
  The time-bounded additive approximation problem is decidable for SMPs with effective and slow residence-time distributions.
\end{theorem}

\section{Unambiguous Processes}\label{sec:unambiguous-intro}
If we consider only generative processes,
then we can also recover decidability by restricting to unambiguous processes.
Intuitively, an unambiguous process is one in which a given output label uniquely
identifies the successor state.

\begin{definition}
  A generative SMP is \emph{unambiguous} if for every $s \in S$ and $a \in \aout$
  there exists at most one $s' \in S$ such that $\tau(s)(s',a) \neq 0$.
  
  For an unambiguous SMP, we write $T(s,w)$ for the \emph{unique} state reached from $s$
  after outputting the word $w$.
\end{definition}

Now consider the set of ``loops'' reachable from $s$ and $s'$,
which we denote by $L(s_1,s_2) \subseteq S^2 \times \aout^{\leq S^2}$ and defined by
\[L(s_1,s_2) = \left\{(p_1,p_2,v) \mid \exists w \in \aout^{\leq S^2},
  \begin{array}{c}
    T(s_1,w) = p_1,\ T(s_2,w) = p_2,\\
    T(p_1,v) = p_1,\ T(p_2,v) = p_2
  \end{array}
  \right\}.\]
Intuitively, $(p_1,p_2,v) \in L(s_1,s_2)$ means that there exists a word $w$
such that $s_1$ goes to $p_1$ and $s_2$ goes to $p_2$ when outputting $w$,
and furthermore, whenever $p_1$ and $p_2$ output the word $v$,
they end up back in $p_1$ and $p_2$.
Thus, $v$ makes $p_1$ and $p_2$ loop back to themselves.
We can then prove the following lemma.

\begin{lemma}\label{lem:unambiguous-intro}
  $s_1 \ft s_2$ if and only if
  \begin{itemize}
    \item $\prob(s_1)(\cylinder(w,t)) \geq \prob(s_2)(\cylinder(w,t))$ for all $w \in \aout^{\leq S^2}$ and $t \in \mathbb{R}_{\geq 0}$ and
    \item $\prob(p_1)(\cylinder(v,t)) \geq \prob(p_2)(\cylinder(v,t))$ for all $(p_1,p_2,v) \in L(s_1,s_2)$ and $t \in \mathbb{R}_{\geq 0}$.
  \end{itemize}
\end{lemma}

Since for a given word, the inequalities in Lemma~\ref{lem:unambiguous-intro} can be checked for effective distributions,
and since there are only finitely many words check, we obtain the following theorem.

\begin{theorem}
  For unambiguous generative SMPs with effective residence-time distributions,
  the faster-than problem is decidable in $\coNP$.
\end{theorem}

\section{Logical Characterisation of the Faster-Than Relation}
In this section we give a logical characterisation of the faster-than relation
when we restrict to generative SMPs.
The language $\mathcal{L}$ that we use for this consists of path formulas
\[\varphi ::= \top \mid \diam{a} \varphi\]
and state formulas
\[\psi ::= \mathcal{P}^{\leq t}_{\geq p} (\varphi)\]
where $t,p \in \mathbb{Q}_{\geq 0}$.

The semantics of $\mathcal{L}$ are given in terms of paths $\pi = a_1 a_2 \dots \in \aout^{*}$
where we let $\pi[i] = a_i$ be the $i$th term of of $\pi$.
Then the semantics of $\mathcal{L}$ are as follows.

\[\begin{array}{l l l}
  \pi \models \top & & \text{always} \\
  \pi \models \diam{a} \varphi & \text{iff} & \pi[1] = a \text{ and } \pi|_2 \models \varphi \\
  s \models \mathcal{P}^{\leq t}_{\geq p} (\varphi) & \text{iff} & \prob(s)(\cylinder(\mathfrak{W}(\varphi),t)) \geq p 
\end{array}\]
where $\pi|_2$ is the tail of $\pi$,
and $\mathfrak{W}(\varphi)$ is the longest common prefix of all paths which satisfy $\varphi$.

The language $\mathcal{L}$ characterises the faster-than relation in the following sense.

\begin{theorem}
  For generative SMPs, it holds that
  \[s_1 \ft s_2 \quad \text{ if and only if } \quad s_2 \models \psi \text{ implies } s_1 \models \psi \text{ for all } \psi \in \mathcal{L}.\]
\end{theorem}

Unfortunately, it is not clear to us how to make the logical characterisation work for the reactive case.
The issue is that the definition of faster-than for the reactive case has an asymmetry in the quantifiers:
For all schedulers $\sigma$ there must exist a scheduler $\sigma'$.
However, when defining the semantics of the operator, we must choose whether
\[\prob^\sigma(s)(\cylinder(\mathfrak{W}(\varphi),t)) \geq p\]
should hold for all $\sigma$ or just for some $\sigma$.

Apart from characterising the faster-than relation,
the language $\mathcal{L}$ turns out to be quite simple.
In particular, \emph{every} formula in $\mathcal{L}$ is satisfiable,
and even satisfiable by a finite model.

\begin{theorem}
  Any formula $\psi \in \mathcal{L}$ is satisfiable by a finite SMP.
\end{theorem}

This can be easily seen by considering a path formula $\varphi = \diam{a_1} \dots \diam{a_n} \top$
and letting $\mathcal{M}_\varphi$ be an SMP with $n+1$ states such that state number $i$
has an $a_i$-transition with probability $1$ to state number $i+1$,
and each state has a Dirac distribution at $0$ as residence-time distribution.

As a corollary, this immediately implies that the satisfiability is trivially decidable.

\begin{corollary}
  The satisfiability problem for $\mathcal{L}$ is decidable.
\end{corollary}

Finally, by making use of the existential theory of the reals,
we obtain a $\PSPACE$ model checking algorithm for some commonly used residence-time distributions.

\begin{theorem}
  The model checking problem for $\mathcal{L}$ is decidable for SMPs with residence-time distributions that are either
  \begin{itemize}
    \item exponential,
    \item piecewise polynomial,
    \item piecewise affine,
    \item uniform, or
    \item Dirac.
  \end{itemize}
\end{theorem}

\section{Compositionality}
Let us now return to the picture in Figure~\ref{fig:ft-intro},
where a context $\mathcal{M}$ is operating in parallel with a component $\mathcal{M}_2$,
and we wish to replace $\mathcal{M}_2$ by a faster component $\mathcal{M}_1$.
We will first give some examples of when this picture can lead to parallel timing anomalies,
meaning that even though $\mathcal{M}_1$ is faster than $\mathcal{M}_2$,
the system $\mathcal{M}_1 \comp{\star} \mathcal{M}$ is actually slower than $\mathcal{M}_2 \comp{\star} \mathcal{M}$.
We will give timing anomalies for the following types of composition function:

\begin{description}
  \item[Product composition:] $F_{\star(\mu,\nu)} = \Exp{\theta \cdot \theta'}$ if $F_\mu = \Exp{\theta}$ and $F_\nu = \Exp{\theta'}$.
  \item[Minimum composition:] $F_{\star(\mu,\nu)}(t) = \min\{F_\mu(t), F_\nu(t)\}$.
  \item[Maximum composition:] $F_{\star(\mu,\nu)}(t) = \max\{F_\mu(t), F_\nu(t)\}$.
\end{description}

\begin{figure}
  \centering
  \hfill
  \begin{tikzpicture}\tikzset{every state/.style={minimum size=35pt}}
    %Nodes
    \node[state, circle split] (0) {$\mu'$ \nodepart{lower} $s$};
    \node[state, circle split] (1) [right = 1.5cm of 0]{$\nu'$ \nodepart{lower} $s'$};
    \node[state, circle split] (2) [right = 1.5cm of 1]{$\eta$ \nodepart{lower} $s''$};
    
    \node[state, circle split] (3) [below = of 0]{$\mu$ \nodepart{lower} $s_1$};
    \node[state, circle split] (4) [right = 1.5cm of 3]{$\nu$ \nodepart{lower} $s_1'$};
    \node[state, circle split] (5) [right = 1.5cm of 4]{$\eta$ \nodepart{lower} $s_1''$};
    
    \node[state, circle split] (6) [below = of 3]{$\nu$ \nodepart{lower} $s_2$};
    \node[state, circle split] (7) [right = 1.5cm of 6]{$\mu$ \nodepart{lower} $s_2'$};
    \node[state, circle split] (8) [right = 1.5cm of 7]{$\eta$ \nodepart{lower} $s_2''$};
    
    \node[xshift=-0.5cm] at (0.west) {\large{$\mathcal{M}$}};
    \node[xshift=-0.5cm] at (3.west) {\large{$\mathcal{M}_1$}};
    \node[xshift=-0.5cm] at (6.west) {\large{$\mathcal{M}_2$}};
    
    % Edges
    \path[thick, ->] (0) edge [above] node {$a:1$} (1);
    \path[thick, ->] (1) edge [above] node {$a:1$} (2);
    \path[thick, loop right, ->] (2) edge [right] node {$a:1$} (2);
    
    \path[thick, ->] (3) edge [above] node {$a:1$} (4);
    \path[thick, ->] (4) edge [above] node {$a:1$} (5);
    \path[thick, loop right, ->] (5) edge [right] node {$a:1$} (5);
    
    \path[thick, ->] (6) edge [above] node {$a:1$} (7);
    \path[thick, ->] (7) edge [above] node {$a:1$} (8);
    \path[thick, loop right, ->] (8) edge [right] node {$a:1$} (8);
  \end{tikzpicture}
  \hfill \
  \caption{For different instantiations of $\mu$, $\nu$, $\mu'$, $\nu'$ and $\eta'$,
    the context $\mathcal{M}$ together with the components $\mathcal{M}_1$ and $\mathcal{M}_2$ lead to parallel timing anomalies.}
  \label{fig:anomalies-intro}
\end{figure}
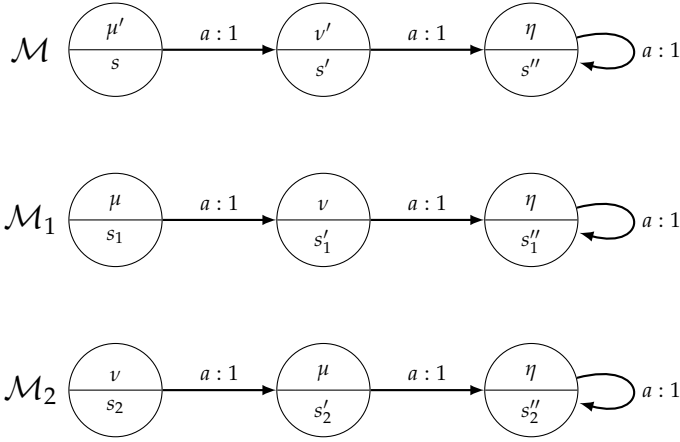

Consider the pointed SMPs $(\mathcal{M},s)$, $(\mathcal{M}_1,s_1)$, and $(\mathcal{M}_2,s_2)$
as depicted in Figure~\ref{fig:anomalies-intro},
and let $F_\mu = \Exp{2}$, $F_\nu = \Exp{0.5}$, and $F_\eta = \Exp{1}$.
Notice that this immediately implies that $\mathcal{M}_1 \ft \mathcal{M}_2$.

\begin{example}[Product composition]
  Let $F_{\mu'} = \Exp{10}$ and $F_{\nu'} = \Exp{0.1}$.
  We then get
  \[\prob(s_1 \comp{\star} s)(\cylinder(aa,2)) \approx 0.09\]
  and
  \[\prob(s_2 \comp{\star} s)(\cylinder(aa,2)) \approx 0.30,\]
  meaning that $\mathcal{M}_1 \comp{\star} \mathcal{M} \not\ft \mathcal{M}_2 \comp{\star} \mathcal{M}$.
\end{example}

\begin{example}[Minimum composition]
  Let $F_{\mu'} = \Exp{1}$ and $F_{\nu'} = \Exp{2}$.
  We then get
  \[\prob(s_1 \comp{\star} s)(\cylinder(aa,2)) \approx 0.40\]
  and
  \[\prob(s_2 \comp{\star} s)(\cylinder(aa,2)) \approx 0.51,\]
  so also in this case $\mathcal{M}_1 \comp{\star} \mathcal{M} \not\ft \mathcal{M}_2 \comp{\star} \mathcal{M}$.
\end{example}

\begin{example}[Maximum composition]
  Let $F_{\mu'} = \Exp{2}$ and $F_{\nu'} = \Exp{1}$.
  We then get
  \[\prob(s_1 \comp{\star} s)(\cylinder(aa,2)) \approx 0.75\]
  and
  \[\prob(s_2 \comp{\star} s)(\cylinder(aa,2)) \approx 0.91,\]
  so once again we get $\mathcal{M}_1 \comp{\star} \mathcal{M} \not\ft \mathcal{M}_2 \comp{\star} \mathcal{M}$.
\end{example}

\subsection{Avoiding Parallel Timing Anomalies}
We have now seen that parallel timing anomalies can occur for many standard ways of composing systems.
Furthermore, none of the examples we showed made use of non-determinism or probabilistic branching,
showing that parallel timing anomalies can occur purely as a consequence of the real-time behaviour of the systems.

We therefore wish to understand under which conditions we can ensure that parallel timing anomalies do not occur.
We provide a first step toward such an understanding by identifying a set of conditions which over-approximate the faster-than relation,
and show that this set of conditions is decidable.
Hence, we can algorithmically verify that a system satisfies the conditions,
and give guarantees that the system can not lead to parallel timing anomalies.

In order to over-approximate the faster-than relation,
we require that $\mathcal{M}_1 \comp{\star} \mathcal{M}$
is pointwise faster than $\mathcal{M}_1$ along all paths (from the initial states).
Likewise, we require that $\mathcal{M}_2$ is pointwise faster than $\mathcal{M}_2 \comp{\star} \mathcal{M}$ along all paths.
Since we already know that $\mathcal{M}_1$ is faster than $\mathcal{M}_2$,
this will imply by transitivity that $\mathcal{M}_1 \comp{\star} \mathcal{M}$
is faster than $\mathcal{M}_2 \comp{\star} \mathcal{M}$,
thus ensuring that parallel timing anomalies can not occur.

\begin{definition}
  A reactive SMP $\mathcal{M} = (S, \tau, \rho, \ell)$ has a \emph{deterministic Markov kernel}
  if for all $s \in S$ and $a \in \ain$ there is at most one state $s' \in S$
  such that $\tau(s,a)(s') > 0$.
\end{definition}

\begin{definition}\label{def:mon-intro}
  Let $n \in \mathbb{N}$.
  We say that $\star$ is \emph{$n$-monotonic} in $\mathcal{M}_1$, $\mathcal{M}_2$, $\mathcal{M}$, and $\mathcal{M}'$
  and write $(\mathcal{M}_1, \mathcal{M}) \mon{n}{\star} (\mathcal{M}_2, \mathcal{M}')$
  if $\mathcal{M}'$ has a deterministic Markov kernel and
  the following holds pointwise along all paths of length up to $n$:
  \begin{enumerate}
    \item The CDF of $\mathcal{M}_1 \comp{\star} \mathcal{M}$ is pointwise greater than that of $\mathcal{M}$.
    \item The CDF of $\mathcal{M}_2$ is pointwise greater than that of $\mathcal{M}_2 \comp{\star} \mathcal{M}'$.
    \item For all schedulers $\sigma$ there exists a scheduler $\sigma'$ such that \label{item:mon-cond-3}
      the transition probability of $\mathcal{M}_1 \comp{\star} \mathcal{M}$ under $\sigma'$
      is greater than that of $\mathcal{M}$ under $\sigma$.
    \item For all schedulers $\sigma$ there exists a scheduler $\sigma'$ such that \label{item:mon-cond-4}
      the transition probability of $\mathcal{M}_2$ under $\sigma'$ is greater than that of
      $\mathcal{M}_2 \comp{\star} \mathcal{M}'$ under $\sigma$.
  \end{enumerate}
  
  We will say that $\star$ is \emph{monotonic} in $\mathcal{M}_1$, $\mathcal{M}_2$, $\mathcal{M}$, and $\mathcal{M}'$
  and write $(\mathcal{M}_1, \mathcal{M}) \mon{}{\star} (\mathcal{M}_2, \mathcal{M}')$
  if $(\mathcal{M}_1, \mathcal{M}) \mon{n}{\star} (\mathcal{M}_2, \mathcal{M}')$ for all $n$.
\end{definition}

\begin{theorem}
  If $\mathcal{M}_1 \ft \mathcal{M}_2$, $\mathcal{M} \ft \mathcal{M}'$, and $(\mathcal{M}_1, \mathcal{M}) \mon{}{\star} (\mathcal{M}_2, \mathcal{M}')$,
  then $\mathcal{M}_1 \comp{\star} \mathcal{M} \ft \mathcal{M}_2 \comp{\star} \mathcal{M}'$.
\end{theorem}

We do not know whether the conditions of monotonicity are decidable,
but if we strength the existential quantifiers of items \ref{item:mon-cond-3} and \ref{item:mon-cond-4} in Definition~\ref{def:mon-intro} to universal quantifiers,
then we arrive at a notion of \emph{strong} $n$-monotonicity, respectively \emph{strong} monotonicity,
denoted by $(\mathcal{M}_1, \mathcal{M}) \smon{n}{\star} (\mathcal{M}_2, \mathcal{M}')$,
respectively $(\mathcal{M}_1, \mathcal{M}) \smon{}{\star} (\mathcal{M}_2, \mathcal{M}')$.

Since clearly $(\mathcal{M}_1, \mathcal{M}) \smon{}{\star} (\mathcal{M}_2, \mathcal{M}')$
implies $(\mathcal{M}_1, \mathcal{M}) \mon{}{\star} (\mathcal{M}_1, \mathcal{M}')$,
we get the following corollary.

\begin{corollary}
  If $\mathcal{M}_1 \ft \mathcal{M}_2$, $\mathcal{M} \ft \mathcal{M}'$, and $(\mathcal{M}_1, \mathcal{M}) \smon{}{\star} (\mathcal{M}_2, \mathcal{M}')$,
  then $\mathcal{M}_1 \comp{\star} \mathcal{M} \ft \mathcal{M}_2 \comp{\star} \mathcal{M}'$.
\end{corollary}

The first step in order to decide whether $(\mathcal{M}_1, \mathcal{M}) \smon{}{\star} (\mathcal{M}_2, \mathcal{M}')$
is to notice that it is enough to consider paths up to a specific length for finite systems.

\begin{lemma}
  Let $\mathcal{M}_1$, $\mathcal{M}_2$, $\mathcal{M}$, and $\mathcal{M}'$ be finite SMPs
  and let
  \[m = \max\{|\mathcal{M}_1| \cdot |\mathcal{M}|, |\mathcal{M}_2| \cdot |\mathcal{M}'|\} + \max\{|\mathcal{M}_1|, |\mathcal{M}_2|, |\mathcal{M}|, |\mathcal{M}'|\} + 1.\]
  If $(\mathcal{M}_1, \mathcal{M}) \smon{m}{\star} (\mathcal{M}_2, \mathcal{M}')$,
  then $(\mathcal{M}_1, \mathcal{M}) \smon{}{\star} (\mathcal{M}_2, \mathcal{M}')$.
\end{lemma}

We can then make use of the first-order theory of the reals to decide strong monotonicity.

\begin{theorem}\label{thm:smon-decidable-intro}
  Let $\mathcal{M}_1$, $\mathcal{M}_2$, $\mathcal{M}$, and $\mathcal{M}'$ be finite, reactive SMPs.
  If for all paths $\pi_1$ in $\mathcal{M}_1 \comp{\star} \mathcal{M}$, $\pi_2$ in $\mathcal{M}$,
  $\pi_3$ in $\mathcal{M}_2$, and $\pi_4$ in $\mathcal{M}_2 \comp{\star} \mathcal{M}'$,
  the sets
  \[\{t \in \mathbb{R}_{\geq 0} \mid F_{\pi_1\sproj{i}}(t) \geq F_{\pi_2\sproj{i}}\}\]
  and
  \[\{t \in \mathbb{R}_{\geq 0} \mid F_{\pi_3\sproj{i}}(t) \geq F_{\pi_4\sproj{i}}\}\]
  are semialgebraic for all $1 \leq i \leq m$,
  then it is decidable whether $(\mathcal{M}_1, \mathcal{M}) \smon{}{\star} (\mathcal{M}_2, \mathcal{M}')$.
\end{theorem}

The sets described in Theorem~\ref{thm:smon-decidable-intro} are semialgebraic for common distributions
such as exponential and uniform distributions,
and for common composition functions such as product, minimum, and maximum composition.

\defaultbib

  \setcounter{figure}{0}
  \setcounter{subfigure}{0}
  \setcounter{section}{0}
  \setcounter{subsection}{0}
  \setcounter{table}{0}
  \setcounter{equation}{0}
  \newcommand{\sfttitle}{Simulation-Based Faster-Than Relation}
\chapter{\sfttitle}\label{chap:sft}

This chapter summarises Paper D ``A Hemimetric Extension of Simulation for Semi-Markov Decision Processes''~\cite{PBLM18}.
For the full paper, see Part~\ref{chap:papers}.

We have seen in Chapter~\ref{chap:tft} how to compare the real-time behaviour of SMPs by considering their traces.
In this chapter we will instead compare processes through the notion of \emph{simulation}.
Roughly speaking, a process $s_1$ simulates a process $s_2$ if anything that $s_2$ can do, $s_1$ can also do.
However, $s_1$ may be able to do more than what $s_2$ can do.
When considering the real-time behaviour of systems,
we in addition require that $s_1$ must be faster than $s_2$ in order for $s_1$ to simulate $s_2$.

Since the real-time behaviour of processes is sensitive to the exact type of distribution and parameters
used to specify the real-time behaviour in each state,
such processes are subject to the approximate modeling problem,
which we discussed in the introduction.
We therefore develop a notion of \emph{simulation distance},
which quantifies how close a process is to simulating another process in terms of its real-time behaviour.
In order to do this, we first consider how to quantitatively compare the residence-time distributions of processes.
In this chapter, we only consider \emph{reactive} processes.

\section{Comparing Residence-Time Distributions}
In order to compare the residence-time distributions of processes,
we will take as our starting point the \emph{usual stochastic order} from the theory of stochastic orders~\cite{shaked2007}.
A distribution $\mu$ is smaller than another distribution $\nu$ in the usual stochastic order
if the CDF of $\mu$ is point-wise greater than the CDF of $\nu$, i.e. when
\[F_\mu(t) \geq F_\nu(t) \quad \text{for all } t \in \mathbb{R}_{\geq 0}.\]
We extend this order to a quantitative notion,
and while doing so, we also shift the focus from the distributions to the CDFs,
since we are only interested in comparing CDFs.

\begin{definition}
  Let $F$ and $G$ be CDFs and let $\varepsilon \in \mathbb{R}_{> 0}$.
  We say that $F$ is \emph{$\varepsilon$-faster than} $G$ and write $F \fft_\varepsilon G$
  if
  \[F(\varepsilon \cdot t) \geq G(t) \quad \text{for all } t \in \mathbb{R}_{\geq 0}. \qedhere\]
\end{definition}

The name $\varepsilon$-faster-than comes from the fact that if $F \fft_\varepsilon G$,
then at every point in time $t$, $F$ will have a higher probability of having fired a transition than $G$,
if we accelerate the real-time behaviour of $F$ by the factor $\varepsilon$.

\begin{figure}
  \centering
  \begin{subfigure}{0.45\textwidth}
    \centering
    \includegraphics[scale=0.37]{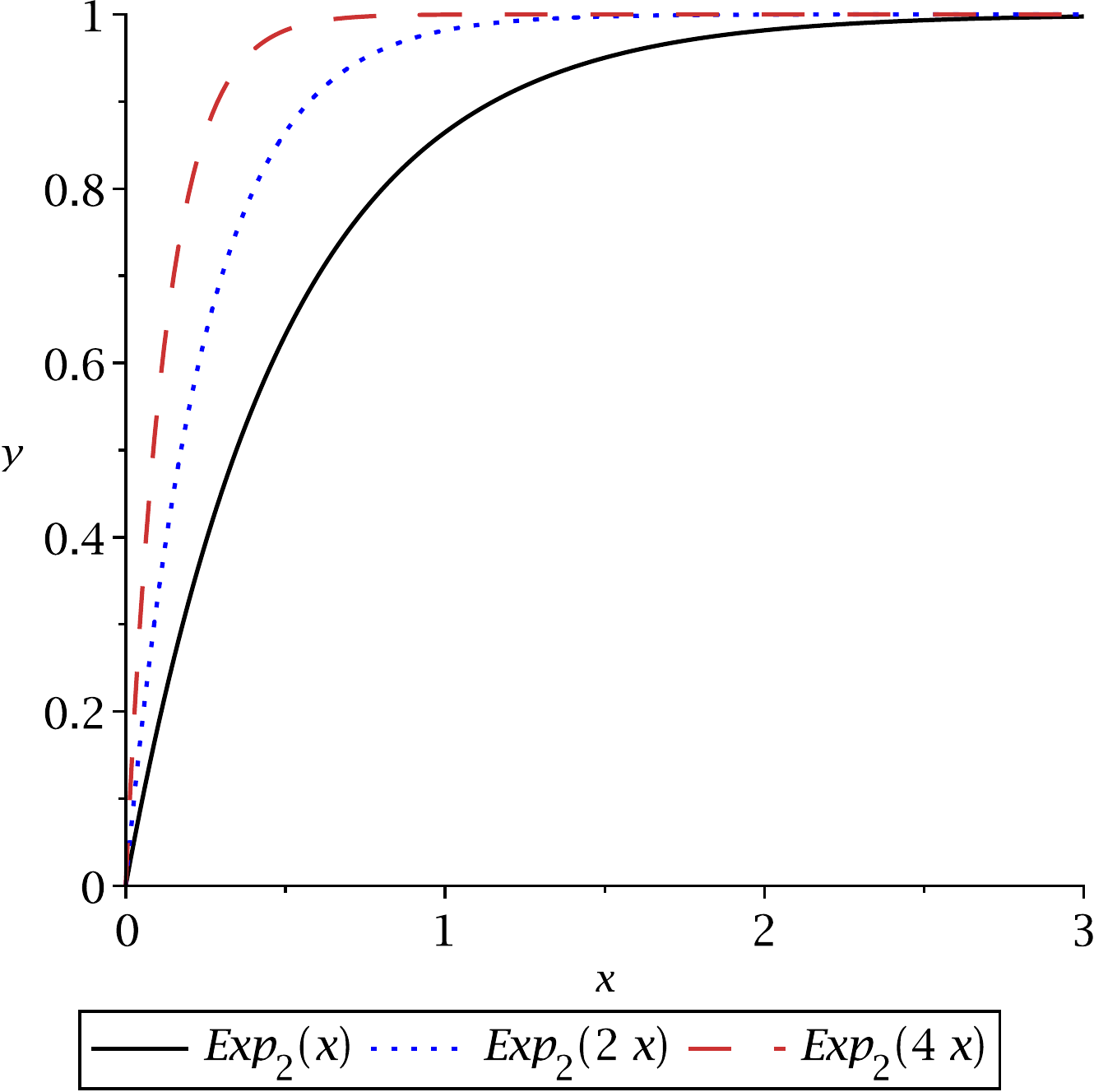}
    \caption{Exponential}
    \label{fig:ft-exp-intro}
  \end{subfigure}%
  \begin{subfigure}{0.45\textwidth}
    \centering
    \includegraphics[scale=0.37]{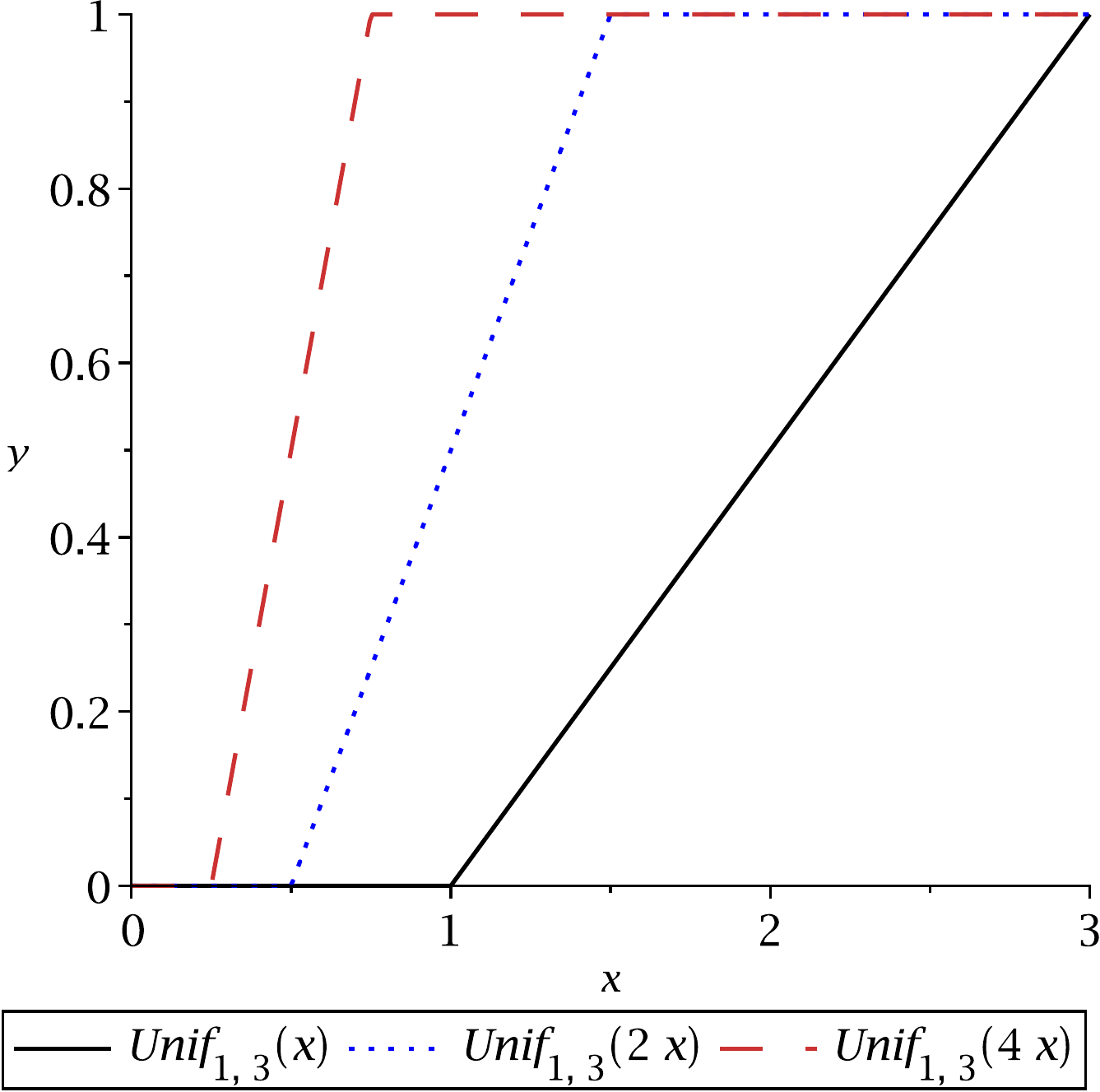}
    \caption{Uniform}
    \label{fig:ft-unif-intro}
  \end{subfigure}
  \caption{When accelerating by a factor $\varepsilon \geq 1$, the CDF becomes faster.
    Here the CDFs of an exponential and a uniform distribution are plotted where the acceleration factor $\varepsilon$ takes the values $1$, $2$, and $4$.}
  \label{fig:eps-ft-intro}
\end{figure}

\begin{example}
  To get a feeling for the significance of the acceleration factor $\varepsilon$,
  consider the plots in Figure~\ref{fig:eps-ft-intro}.
  
  In Figure~\ref{fig:ft-exp-intro}, we see the CDF of an exponential distribution with rate $2$.
  When accelerating this CDF by a factor $2$, we see that the shape of an exponential distribution is preserved,
  but the resulting CDF is faster. The same happens with acceleration factor $4$,
  except the resulting CDF is even faster.
  Thus the net result of accelerating an exponential distribution is to increase its rate.
  
  In Figure~\ref{fig:ft-unif-intro}, we see the CDF of a uniform distribution between $1$ and $3$.
  When accelerating this CDF by a factor $2$, the shape of a uniform distribution is preserved,
  but the resulting uniform distribution is between two values that are less than the original values,
  thus making the CDF faster.
  The same happens with acceleration factor $4$,
  except now the uniform distribution is between values that are even smaller,
  resulting in an even faster CDF.
  The result of accelerating a uniform distribution is therefore to decrease the parameters of the uniform distribution. 
\end{example}

The next three propositions show for what values of $\varepsilon$ the $\varepsilon$-faster-than relation holds
between Dirac, uniform, and exponential distributions.

\begin{proposition}\label{prop:dirac-intro}
  Let $F$ be any CDF. The following holds for any $\varepsilon \in \mathbb{R}_{> 0}$.
  \begin{enumerate}
    \item $\dirac{0} \fft_{\varepsilon} F$.
    \item If $F \neq \dirac{0}$, then $F \not \fft_{\varepsilon} \dirac{0}$.
  \end{enumerate}
\end{proposition}

\begin{proposition}\label{prop:expandunif-intro}
  \leavevmode
  \begin{enumerate}
    \item $\Exp{\theta_1} \fft_{\varepsilon} \Exp{\theta_2}$,
      where $\varepsilon = \frac{\theta_2}{\theta_1}$.
    \item If $c = 0$ and $a > 0$, then $\unif{a}{b} \not\fft_{\varepsilon} \unif{c}{d}$
      for any $\varepsilon \in \mathbb{R}_{> 0}$.
    \item If $c = 0$ and $a = 0$, then $\unif{a}{b} \fft_{\varepsilon} \unif{c}{d}$,
      where $\varepsilon = \frac{b}{d}$.
    \item If $c > 0$, then $\unif{a}{b} \fft_{\varepsilon} \unif{c}{d}$,
      where $\varepsilon = \max\left\{\frac{a}{c},\frac{b}{d}\right\}$.
  \end{enumerate}
  In all cases, the given $\varepsilon$ is the least such that
  the $\varepsilon$-faster than relation holds.
\end{proposition}

\begin{proposition}\label{prop:unifexp-intro}
  \leavevmode
  \begin{enumerate}
    \item $\Exp{\theta} \not \fft_{\varepsilon} \unif{a}{b}$
      for all $\varepsilon \in \mathbb{R}_{> 0}$.
    \item If $a > 0$, then $\unif{a}{b} \not \fft_{\varepsilon} \Exp{\theta}$
      for all $\varepsilon \in \mathbb{R}_{> 0}$.
    \item If $a = 0$, then $\unif{a}{b} \fft_{\varepsilon} \Exp{\theta}$,
      where $\varepsilon = \theta \cdot b$.
      Furthermore, this is the least $\varepsilon$ such that the $\varepsilon$-faster-than relation holds.
  \end{enumerate}
\end{proposition}

\begin{figure}
  \centering
  \begin{subfigure}{0.45\textwidth}
    \centering
    \includegraphics[scale=0.37]{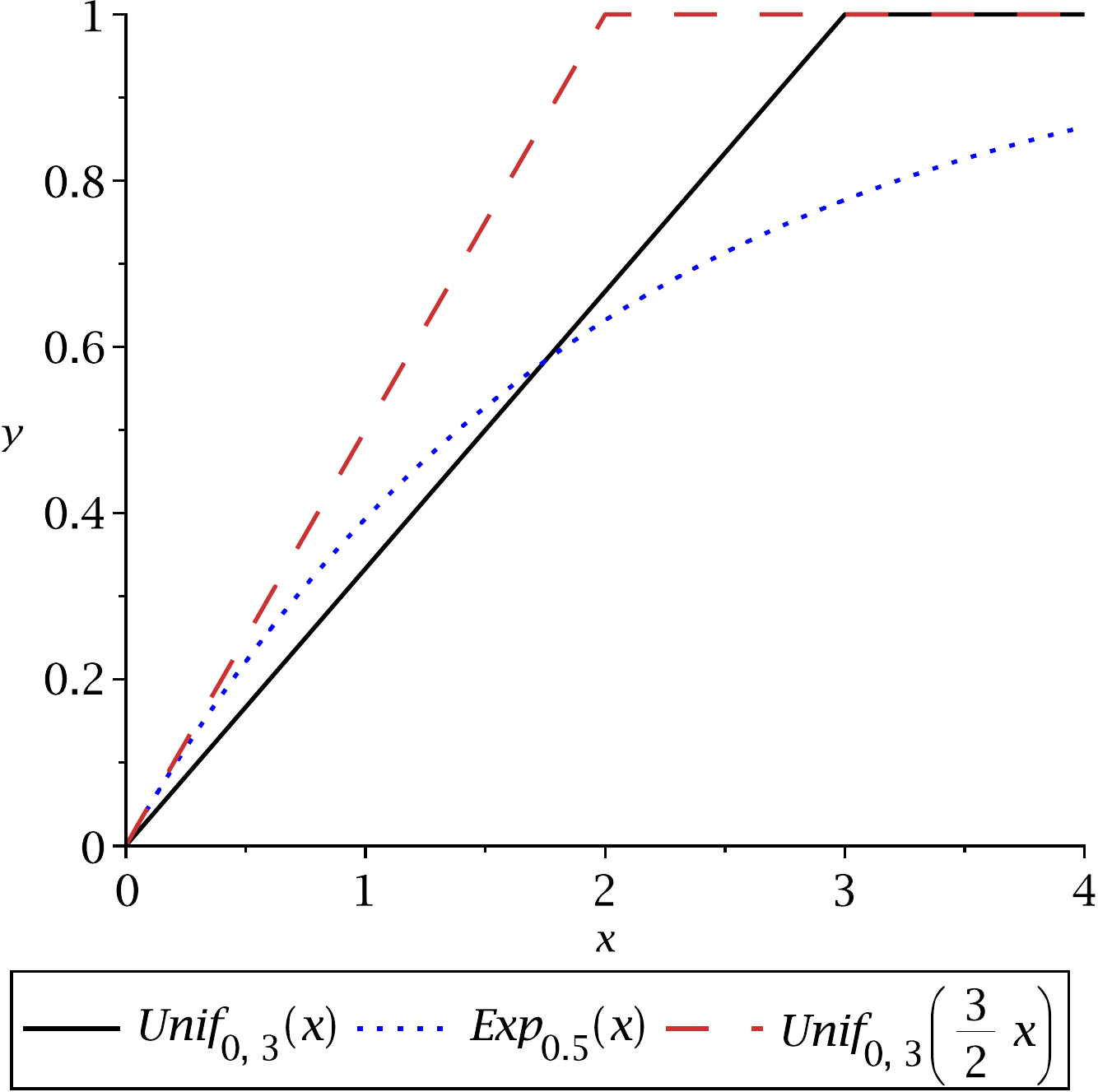}
    \caption{Exponential}
    \label{fig:compare1-intro}
  \end{subfigure}%
  \begin{subfigure}{0.45\textwidth}
    \centering
    \includegraphics[scale=0.37]{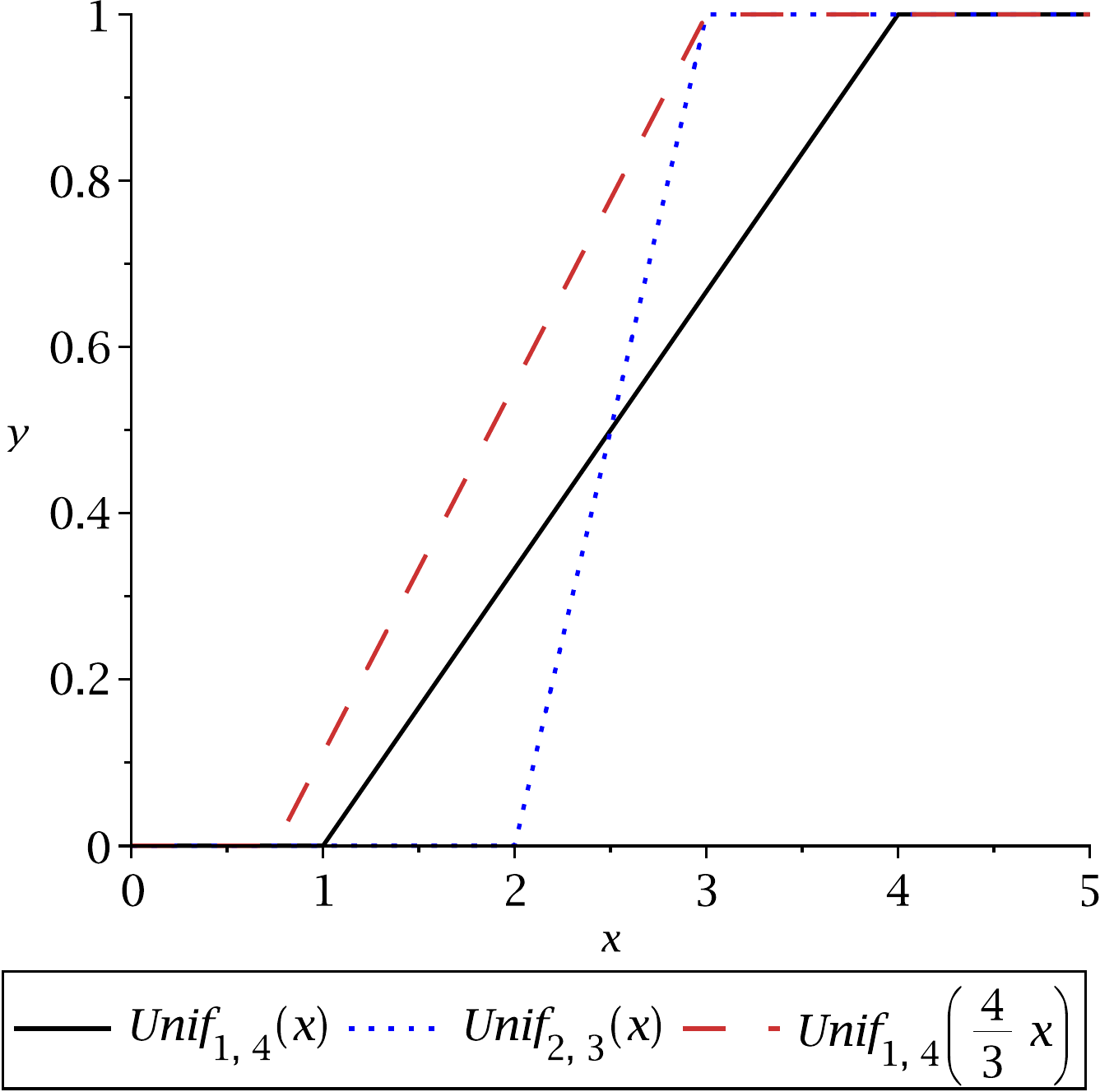}
    \caption{Uniform}
    \label{fig:compare2-intro}
  \end{subfigure}
  \caption{Accelerating the real-time behaviour of a uniform distribution to make it faster than another uniform distribution or an exponential distribution.}
  \label{fig:compare-intro}
\end{figure}

\begin{example}
  Consider the plots in Figure~\ref{fig:compare-intro}.
  In Figure~\ref{fig:compare1-intro}, we see an exponential distribution with rate $0.5$
  and a uniform distribution between $0$ and $3$.
  Clearly, neither of them is faster than the other, since they cross.
  However, Proposition~\ref{prop:unifexp-intro} tells us that if we accelerate the uniform distribution
  by a factor $\varepsilon = 0.5 \cdot 3 = 1.5$,
  then the resulting uniform distribution is faster than the exponential distribution,
  and this can also be seen in Figure~\ref{fig:compare1-intro}.
  
  In Figure~\ref{fig:compare2-intro} we see two different uniform distributions,
  one between $1$ and $4$ and another between $2$ and $3$.
  Again, neither of these is faster than the other,
  but by Proposition~\ref{prop:expandunif-intro},
  accelerating the uniform distribution between $1$ and $4$ by
  a factor $\varepsilon = \max\{\frac{1}{3},\frac{4}{3}\} = \frac{4}{3}$
  results in a uniform distribution that is faster than the one between $2$ and $3$,
  as can be seen in Figure~\ref{fig:compare2-intro}.
\end{example}

In addition, the $\varepsilon$-faster-than relation enjoys the following properties.
The first property is that the relation is monotonic in $\varepsilon$,
and the second property is that the relation is a congruence with respect to convolution.

\begin{lemma}\label{lem:mono-intro}
  Let $\varepsilon \leq \varepsilon'$ and assume that $F \fft_{\varepsilon} G$.
  Then $F \fft_{\varepsilon'} G$.
\end{lemma}

\begin{lemma}\label{lem:conv-intro}
  If $F_{\mu_1} \fft_{\varepsilon} F_{\mu_2}$ and $F_{\nu_1} \fft_\varepsilon F_{\nu_2}$,
  then $F_{(\mu_1 * \nu_1)} \fft_{\varepsilon} F_{(\mu_2 * \nu_2)}$.
\end{lemma}

\section{Simulation Distance}
We will now use the $\varepsilon$-faster-than relation to extend simulation for SMPs to a distance between SMPs,
which intuitively measures how much one process needs to be accelerated in order to simulate another process.
To do this, first note that condition~\ref{enum:s2} in Definition~\ref{def:simul-intro} of simulation for SMPs
is actually an instance of the usual stochastic order.
We can therefore naturally extend this definition using our notion of $\varepsilon$-faster-than.

\begin{definition}
  Let $\mathcal{M} = (S, \tau, \rho, \ell)$ be an SMP.
  A \emph{$\varepsilon$-simulation relation} is a relation $\mathcal{R} \subseteq S \times S$ such that $s_1 \mathcal{R} s_2$ implies
  \begin{enumerate}
    \item $\ell(s_1) = \ell(s_2)$,
    \item $F_{s_2} \fft_\varepsilon F_{s_1}$, and
    \item for all $a \in \ain$ there exists a \emph{weight function} $\Delta_a \in \dist(S \times S)$ such that
      \begin{enumerate}
        \item $\Delta_a(s,s') > 0$ implies $s \mathcal{R} s'$,
        \item $\tau(s_1,a)(s) = \sum_{s' \in S} \Delta_a(s,s')$, and
        \item $\tau(s_2,a)(s') = \sum_{s \in S} \Delta_a(s,s')$.
      \end{enumerate}
  \end{enumerate}
  
  If there exists an $\varepsilon$-simulation relation $\mathcal{R}$ such that $s_1 \mathcal{R} s_2$,
  then we say that $s_2$ \emph{$\varepsilon$-simulates} $s_1$ and write $s_1 \simul_\varepsilon s_2$.
\end{definition}

\begin{figure}
  \centering
  \hfill
  \begin{tikzpicture}
    \node[state, circle split] (1) {$\Exp{4}$ \nodepart{lower} $s_1$};
    
    \path[thick, loop left, ->] (1) edge [left] node {$a : 1$} (1);
  \end{tikzpicture}
  \hfill
  \begin{tikzpicture}
    \node[state, circle split] (2) {$\Exp{2}$ \nodepart{lower} $s_2$};
    
    \path[thick, loop right, ->] (2) edge [right] node {$a : 1$} (2);
  \end{tikzpicture}
  \hfill \
  
  \caption{Two states of an SMP $\mathcal{M}$.}
  \label{fig:eps-simul-intro}
\end{figure}
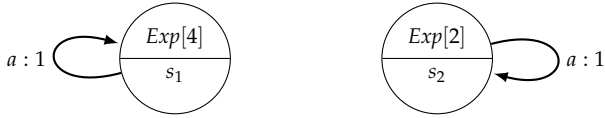

\begin{example}
  Consider the two states in Figure~\ref{fig:eps-simul-intro}.
  Since both states have exponential residence-time distributions,
  Proposition~\ref{prop:expandunif-intro} tells us that
  \[s_1 \simul_2 s_2 \quad \text{and} \quad s_2 \simul_{\frac{1}{2}} s_1. \qedhere\]
\end{example}

If $s_1 \simul_\varepsilon s_2$ and $\varepsilon \leq 1$,
then this means that $s_2$ already simulates $s_1$,
and $\varepsilon$ gives a quantitative measure of how much $s_2$ is faster than $s_1$.
On the other hand, if $\varepsilon > 1$,
then $s_2$ does not simulate $s_1$, but if we accelerate the real-time behaviour of the entire process by $\varepsilon$,
and consider $s_2$ in this accelerated process but $s_1$ still in the original process,
then $s_2$ does simulate $s_1$.
This is made precise by the following proposition.

\begin{proposition}\label{prop:simulpert-intro}
  For any $\varepsilon \in \mathbb{R}_{> 0}$,
  \[s_1 \simul_{\varepsilon} s_2 \quad \text{if and only if} \quad s_1 \simul \pert{s_2}{\varepsilon}\]
  where $\pert{s_2}{\varepsilon}$ is a copy of $s_2$ where the entire process is accelerated by $\varepsilon$.
\end{proposition}

With the notion of $\varepsilon$-faster-than in hand,
it is natural to ask what the smallest $\varepsilon$ is such that the $\varepsilon$-faster-than relation holds,
or in other words, what the smallest acceleration factor is that makes one process simulate another process.
This motivates the definition of the simulation distance.

\begin{definition} \label{def:simuldist-intro}
  The \emph{simulation distance} $\simdist : S \times S \to [1,\infty]$ 
  from a state $s_1$ to a state $s_2$ is given by
  \[\simdist(s_1,s_2) = \inf\{ \varepsilon \geq 1 \mid s_1 \simul_{\varepsilon} s_2\} . \qedhere\]
\end{definition}

For SMPs whose transition function is finitely supported,
the simulation distance has the property that it is a generalisation of simulation in the sense that $s_1 \simul s_2$ if and only if $d(s_1,s_2) = 1$.

The simulation distance itself is not a hemimetric, since it does not satisfy the triangle inequality.
However, it satisfies a multiplicative version of the triangle inequality,
which means that if we take the logarithm of the simulation distance,
then we obtain a hemimetric.

\begin{theorem}
  $\log \simdist$ is a hemimetric.
\end{theorem}

\section{Computing the Simulation Distance}\label{sec:sft-compute-intro}
Our first result on the simulation distance is that it can be computed,
and the algorithm for computing it has polynomial complexity for the residence-time distributions we have considered so far.
A key part of the algorithm is to consider, for two CDFs $F$ and $G$, the value given by
\[c(F,G) = \inf\{\varepsilon \geq 1 \mid F \fft_\varepsilon G\}.\]
Intuitively, $c(F,G)$ denotes the least acceleration factor required for $F$ to be faster than $G$.
Furthermore, given an SMP $\mathcal{M}$, let
\[\mathcal{C}(\mathcal{M}) = \{c(F_s, F_{s'}) \mid s,s' \in S\}.\]
We are interested in those SMPs for which we can actually compute $c(F,G)$
when $F$ and $G$ are residence-time distributions of the SMP.

\begin{definition}
  An SMP $\mathcal{M}$ is \emph{$c$-effective} if $\mathcal{C}(\mathcal{M})$ is computable.
\end{definition}

Given an SMP $\mathcal{M} = (S, \tau, \rho, \ell)$, we will denote by $f(l)$ the complexity of computing $c(F_s,F_{s'})$
for $s,s' \in S$, where $l$ is the length of the representation of the residence-time distributions of $\mathcal{M}$.

Note that by Propositions~\ref{prop:dirac-intro}-\ref{prop:unifexp-intro},
any SMP whose residence-time distributions are Dirac, uniform or exponential is $c$-effective and $f(l)$ is polynomial.

\begin{lemma}\label{lem:correctness-intro}
  Let $\mathcal{M}$ be a finite SMP.
  If $\simdist(s_1,s_2) \neq \infty$, then
  \begin{itemize}
    \item $s_1 \simul_c s_2$, for some $c \in \mathcal{C}(\mathcal{M})$ and
    \item $\simdist(s_1,s_2) = \min\{c \in \mathcal{C}(\mathcal{M}) \mid s_1 \simul_c s_2\}$.
  \end{itemize}
\end{lemma}

Lemma~\ref{lem:correctness-intro} gives a strategy for computing $\simdist(s_1,s_2)$.
First we compute $\mathcal{C}(\mathcal{M})$,
and then we check for each $c \in \mathcal{C}(\mathcal{M})$ whether $s_1 \simul_c s_2$.
If there is no such $c$, then $\simdist(s_1,s_2) = \infty$,
otherwise $\simdist(s_1,s_2)$ will be the smallest $c$ such that $s_1 \simul_c s_2$.
Hence we first need an algorithm to decide whether $s_1 \simul_c s_2$.
We do this by adapting to our setting the classic algorithm for deciding simulation for Markov chains~\cite{baier2000,zhang09}.
Given an SMP $\mathcal{M} = (S, \tau, \rho, \ell)$,
let $n = |S|$ be the number of states, $m = |\ain|$ the number of input actions,
and $k = |\ap|$ the number of atomic propositions.

\begin{theorem}\label{thm:decidable-intro}
  Let $\mathcal{M}$ be a finite and $c$-effective SMP.
  Given $s_1,s_2 \in S$ and $\varepsilon \geq 1$, deciding whether 
  $s_1 \simul_{\varepsilon} s_2$ can be done in time $\bigo(n^2(f(l) + k) + m^2n)$.
\end{theorem}

Using Theorem~\ref{thm:decidable-intro},
we obtain the algorithm shown in Algorithm~\ref{alg:distance-intro},
which uses a bisection method to search through the elements of $\mathcal{C}(\mathcal{M})$
and test them, rather than simply testing them all.

\begin{algorithm}[t]
  \SetAlgoLined
    Order the elements of $\mathcal{C}(M) \setminus \{\infty\}$
    such that $c_1 < c_2 < \dots < c_n$ \;
    \lIf{$s_1 \simul_{c_1} s_2$}{
      \Return $c_1$
    }
    \lElseIf{$s_1 \not\simul_{c_n} s_2$}{
      \Return $\infty$
    }
    \Else{ 
      $i \leftarrow 1, j \leftarrow n$ \;
      \While{$i < j$}{
        $h \leftarrow \left\lceil\frac{j-i}{2}\right\rceil$ \;
        \lIf{$s_1 \simul_{c_{j-h}} s_2$}{
          $j \leftarrow j - h$
        }
        \lElse{
          $i \leftarrow i + h$
        }
      }
      \Return $c_j$ \;
    }
  \caption{Computing the simulation distance between two states $s_1$ and $s_2$.}
  \label{alg:distance-intro}
\end{algorithm}

\begin{theorem}\label{thm:compute-intro}
  Let $\mathcal{M}$ be a finite and $c$-effective SMP. The simulation distance between any two states 
  can be computed in time $\bigo(n^2 (f(l) + k) + m^2n \cdot \log n)$.
\end{theorem}

\section{Compositionality}
The simulation distance turns out to behave nicely with respect to composition.
More concretely, we prove that, under mild assumptions,
composition is non-expansive with respective to the simulation distance.
This result is a quantitative generalisation of the fact that simulation is a precongruence
with respect to composition.

In order to obtain this result,
we restrict our attention to those residence-time composition functions that are monotonic
in the following sense.

\begin{definition}
  A residence-time composition function $\star$ is \emph{monotonic} if
  \[F_\mu \fft_\varepsilon F_\nu \quad \text{implies} \quad F_{\star(\mu,\eta)} \fft_\varepsilon F_{\star(\nu,\eta)}\]
  for all $\varepsilon \geq 1$ and $\mu,\nu,\eta \in \dist(\mathbb{R}_{\geq 0})$.
\end{definition}

This is not a significant restriction,
since most of the residence-time composition functions that are found in the literature are indeed monotonic.
For monotonic residence-time composition functions we then have the promised non-expansiveness result.

\begin{theorem}
  For finite SMPs and monotonic $\star$,
  \[\simdist(s_1,s_2) \leq \varepsilon \quad \text{implies} \quad \simdist(s_1 \comp{\star} s_3, s_2 \comp{\star} s_3) \leq \varepsilon.\]
\end{theorem}

The final aspect of compositionality that we will consider is how to compute the distance between composed states.
We saw in Section~\ref{sec:sft-compute-intro} that we can compute the distance if we know how to compute the constants $c(F,G)$.
For a residence-time composition function on exponential distributions where we take e.g. the product of the rates,
composition poses no problem, since the CDF of two exponential residence-time distributions composed in this manner is still an exponential distribution,
and we know how to compute $c(F,G)$ for those.

However, when composing residence-time distributions using the point-wise maximum of their CDFs,
the composition of two uniform distributions need not be a uniform distribution.
Likewise, composing a uniform distribution and an exponential distribution in this manner yields a CDF that is neither uniform nor exponential.
Hence we must consider how to compute $c(F,G)$ for these composed distributions.

\begin{proposition}\label{prop:comp-intro}
  Let $\star$ be maximum composition.
  The constants $c(F_\mu, F_{\star(\nu,\eta)})$ and $c(F_{\star(\mu,\eta)},F_\nu)$ are computable
  whenever $\mu$, $\nu$, and $\eta$ are taken from the set of exponential, uniform, and Dirac distributions.
\end{proposition}

The proof of Proposition~\ref{prop:comp-intro} is laborious and tedious,
because many combinations and special cases must be considered.
Hence we have not extended the result to a higher number of compositions
or to composition on both sides,
although we strongly believe that such a result will hold for many other kinds of distributions also.

\section{Logical Properties}
If the simulation distance tells us that two processes are close,
then we would also expect them to satisfy almost the same properties.
In this section we make this idea precise by introducing a logical specification language for specifying properties of SMPs.
We show that this language characterises $\varepsilon$-simulation
and that the simulation distance from $s_1$ to $s_2$ is less than $\varepsilon$
if and only if whenever $s_1$ satisfies a formula, $s_2$ satisfies a slight perturbation of the same formula.

The language we use is a slight extension of Markovian logic~\cite{KMP13}
which we call \emph{timed Markovian logic} ($\mlp$).
$\mlp$ has the following syntax,
where $\alpha \in \ap$, $a \in \ain$, $p \in \mathbb{Q} \cap [0,1]$,
and $t \in \mathbb{Q}_{\geq 0}$.

\[\mlp: \quad \varphi ::= \alpha \mid \neg \alpha \mid \ell_p t \mid m_p t \mid L_p^a \varphi \mid M_p^a \varphi \mid \varphi \land \varphi' \mid \varphi \lor \varphi'\]

$\alpha$ and $\neg \alpha$ speak about atomic propositions,
$\land$ and $\lor$ are the usual conjunction and disjunction,
$\ell_p t$ and $m_p t$ speak about the timing behaviour of processes,
and $L_p^a$ and $M_p^a$ speak about the branching behaviour of processes.
This is made precise by the semantics of $\mlp$,
which are given as follows.

\[\begin{array}{l l l l l l l}
  s \models \alpha & \; \text{iff} \; & \alpha \in \ell(s) & \quad & s \models \ell_p t & \; \text{iff} \; & F_s(t) \geq p \\
  s \models \neg \alpha & \; \text{iff} \; & \alpha \notin \ell(s) & \quad & s \models m_p t & \; \text{iff} \; & F_s(t) \leq p \\
  s \models \varphi \land \varphi' & \; \text{iff} \; & s \models \varphi \text{ and } s \models \varphi' & \quad & s \models L_p^a \varphi           & \; \text{iff} \; & \tau(s,a)(\sat{\varphi}) \geq p \\
  s \models \varphi \lor \varphi' & \; \text{iff} \; & s \models \varphi \text{ or } s \models \varphi' & \quad & s \models M_p^a \varphi           & \; \text{iff} \; & \tau(s,a)(\sat{\varphi}) \leq p \\
\end{array}\]
where $\sat{\varphi}$ is the set of states satisfying $\varphi$.

Furthermore, we will consider the following fragments of $\mlp$.

\[\mlpgeq: \quad \varphi ::= \alpha \mid \neg \alpha \mid \ell_p t \mid L_p^a \varphi \mid \varphi \land \varphi' \mid \varphi \lor \varphi'\]
\[\mlpleq: \quad \varphi ::= \alpha \mid \neg \alpha \mid m_p t \mid M_p^a \varphi \mid \varphi \land \varphi' \mid \varphi \lor \varphi'\]

In order to connect $\mlp$ to our simulation distance,
we introduce the notion of \emph{$\varepsilon$-perturbance} of a formula $\varphi \in \mlp$,
denoted by $\pert{\varphi}{\varepsilon}$,
which is given by replacing all occurrences of $\ell_p t$ or $m_p t$ in $\varphi$ by $\ell_p \varepsilon \cdot t$ or $m_p \varepsilon \cdot t$, respectively.
We then get the following result, which shows that the fragments $\mlpleq$ and $\mlpgeq$ characterise $\varepsilon$-simulation.

\begin{theorem}\label{thm:eps-logic-intro}
  Let $\varepsilon \in \mathbb{Q}_{\geq 0}$ with $\varepsilon \geq 1$.
  Then the following holds.
  \begin{itemize}
    \item $s_1 \simul_\varepsilon s_2 \quad$ if and only if $\quad \forall \varphi \in \mlpgeq. s_1 \models \varphi \implies s_2 \models \pert{\varphi}{\varepsilon}$.
    \item $s_1 \simul_\varepsilon s_2 \quad$ if and only if $\quad \forall \varphi \in \mlpleq. s_2 \models \pert{\varphi}{\varepsilon} \implies s_1 \models \varphi$.
  \end{itemize}
\end{theorem}

From this result we get the following corollary,
which connects the fragments of $\mlp$ directly to our simulation distance.

\begin{corollary}
  Let $\varepsilon \in \mathbb{Q}_{\geq 0}$ with $\varepsilon \geq 1$.
  For finite SMPs the following holds.
  \begin{itemize}
    \item $\simdist(s_1,s_2) \leq \varepsilon \quad$ if and only if $\quad \forall \varphi \in \mlpgeq. s_1 \models \varphi \implies s_2 \models \pert{\varphi}{\varepsilon}$.
    \item $\simdist(s_1,s_2) \leq \varepsilon \quad$ if and only if $\quad \forall \varphi \in \mlpleq. s_2 \models \pert{\varphi}{\varepsilon} \implies s_1 \models \varphi$.
  \end{itemize}
\end{corollary}

Another property that is often of interest is that of \emph{reachability}:
Starting from a given state, can we reach a state which satisfies some property?
For SMPs which have both probabilistic branching and real-time behaviour,
a more interesting reachability problem is:

\begin{quote}
  Starting from a given state $s$, can we with probability at least $p$ and before time $t$ reach a state which satisfies property $\varphi$?
\end{quote}

We will now show that the $\varepsilon$-simulation relation also preserves reachability properties.
In order to do this, we consider the following kind of events.
Let $X \subseteq S$ and $t \in \mathbb{R}_{\geq 0}$.
Then we define
\[\eventually{t} X = \{\pi \in \Pi(M) \mid \exists i \in \mathbb{N}. \pi\sproj{i} \in X \text{ and } \sum_{j = 1}^{i - 1} \pi\tproj{j} \leq t\}\]
as the set of paths that eventually reach a state in $X$ within time $t$.
We can then prove the following theorem.

\begin{theorem}\label{thm:linear-intro}
  Let $\beta$ be a Boolean combination of atomic propositions.
  If we have $s_1 \simul_\varepsilon s_2$, then
  for any scheduler $\sigma$ there exists a scheduler $\sigma'$
  such that
  \[\prob^\sigma_{s_1}(\eventually{t} \sat{\beta}) \leq \prob^{\sigma'}_{s_2}(\eventually{\varepsilon \cdot t} \sat{\beta}).\]
\end{theorem}

In other words, $s_1 \simul_\varepsilon s_2$ guarantees that
whenever $s_1$ can reach a state in $\sat{\beta}$ within time $t$ under some scheduler $\sigma$,
then we can find a scheduler $\sigma'$ such that $s_2$ reaches a state in $\sat{\beta}$ within time $\varepsilon \cdot t$,
and with at least as high probability.

\section{Topology of the Simulation Distance}
In this section we investigate the topology of the simulation distance,
in particular with respect to properties expressible in $\mlp$.
The question we wish to answer is whether,
given a sequence of states $\{s_k\}$ that converges to a state $s$ such that $s_i \models \varphi$ for each state $s_i$ in the sequence,
we can be sure that also the state $s$ satisfies $\varphi$.
This is the same as asking whether the set $\sat{\varphi}$ is a closed set in the topology induced by the simulation distance.
Hence, if $\sat{\varphi}$ is closed, reasoning in the limit about properties expressible by $\varphi$ is sound.

Recall that the \emph{right-centered} topology is generated by the open balls
\[\mathcal{B}^R_r(s) = \{s' \mid \simdist(s',s) < r\}\]
and the \emph{left-centered} topology is generated by the open balls
\[\mathcal{B}^L_r(s) = \{s' \mid \simdist(s,s') < r\}.\]

\begin{lemma}\label{lem:right-intro}
  The following holds in the right-centered topology.
  \begin{enumerate}
    \item $\sat{\ell_p t}$ is closed.
    \item If $p = 0$, then $\sat{\ell_p t}$ is open.
    \item If $p > 0$, then $\sat{\ell_p t}$ is not open.
    \item If $p = 1$, then $\sat{m_p t}$ is closed.
    \item If $p < 1$, then $\sat{m_p t}$ is not closed.
  \end{enumerate}
\end{lemma}

We can then use Lemma~\ref{lem:right-intro} to show that reasoning in the limit is sound for the right-centered topology.

\begin{theorem}
  For any $\varphi \in \mlpgeq$, $\sat{\varphi}$ is closed in the right-centered topology.
\end{theorem}

\begin{lemma}\label{lem:left-intro}
  The following holds in the left-centered topology.
  \begin{enumerate}
    \item If $p = 1$, then $\sat{m_p t}$ is open.
    \item If $p < 1$, then $\sat{m_p t}$ is not open.
    \item If $p = 0$, $\sat{\ell_p t}$ is closed.
    \item If $p > 0$, $\sat{\ell_p t}$ is not closed.
  \end{enumerate}
\end{lemma}

However, we have not been able to determine whether $\sat{m_p t}$ is closed in the left-centered topology or not.
If it were closed, then we could prove that reasoning in the limit is also sound for the left-centered topology.
Our intuition leads us to conjecture that this is the case.

\begin{conjecture}
  For any $\varphi \in \mlpleq$, $\sat{\varphi}$ is closed in the left-centered topology.
\end{conjecture}

\defaultbib

  \setcounter{figure}{0}
  \setcounter{subfigure}{0}
  \setcounter{section}{0}
  \setcounter{subsection}{0}
  \setcounter{table}{0}
  \setcounter{equation}{0}
  \newcommand{\conctitle}{Conclusion}
\chapter{\conctitle}\label{chap:conc}

In this thesis we have investigated how to reason about the real-time behaviour of stochastic systems,
both by expressing and verifying properties in a given specification language,
and by comparing systems through different notions of behavioural relations.
Through this we have developed various logical specification languages as well as new ways of comparing systems,
and studied the properties of these.

We have introduced weighted logic with bounds (WLWB) as a specification language,
which allows one to reason about upper and lower bounds on weights in a weighted transition system.
Since many requirements of interest speak about upper and lower bounds,
we argue that this is a useful language for specifying requirements.
Furthermore, this language is less susceptible to the approximate modelling problem,
since it does not speak about exact weights.
We have studied the properties of this language and shown that it characterises a kind of
behavioural equivalence that looks at the upper and lower bounds of transitions
rather than matching each weight exactly.
We have also given a complete axiomatisation of the language,
and developed algorithms for deciding both the model checking and the satisfiability problem for WLWB.

We have defined a notion of a faster-than relation on traces for semi-Markov processes
by requiring that on each trace, the fast process must have a higher probability
of completing that trace within any given time bound than the slow process.
Such a notion is useful, since it allows one to reason about incremental improvement of a system in the design phase,
where the speed of the system can be gradually improved.
It is also useful from a compositional perspective.
From this perspective, one may identify a component that is working too slowly,
and replace it with a component that is faster.
However, in this case one has to be careful about timing anomalies.
We have shown that such timing anomalies can occur,
and we have taken first steps toward avoiding them, by identifying conditions
that are sufficient for guaranteeing the absence of timing anomalies.
We have also shown that deciding the faster-than relation is a difficult problem that is undecidable in general.
Moreover, it can not be approximated up to a multiplicative constant,
due to a close connection to probabilistic automata.
Despite this, we have given an algorithm for approximating a time-bounded variant
of the faster-than problem up to an additive constant,
as well as an algorithm for deciding the faster-than relation exactly
when restricting to unambiguous processes.

Since the qualitative answers of the classical notion of simulation for semi-Markov processes
are too rigid for the quantitative nature of these processes,
we have extended the concept of simulation to a quantitative simulation distance.
This distance gives information about how much one should increase
the speed of a process in order for it to become as fast as another process.
This is useful when you have a system which does not satisfy a given property,
but you have a model of a system which does.
Then the distance tells you how much you need to increase the speed of the original system
in order for it to also satisfy the property.
We have shown how to efficiently compute this distance for some commonly used residence-time distributions,
and that composition is non-expansive with respect to the distance,
meaning that timing anomalies can not occur.
Furthermore, we have shown that the distance can be characterised by a logical specification language
which we call timed Markovian logic,
and that it preserves reachability properties.
We have also shown that, in the topology induced by the simulation distance,
properties expressed in timed Markovian logic are preserved in the limit,
in the sense that if a sequence of states has converges to a certain state,
and if each state in the sequence satisfies some property expressed in timed Markovian logic,
then the state to which the sequence converges will also satisfy that property.

The research presented here has been carried out as part of the research project
Approximate Reasoning for Stochastic Markovian Systems,
which is funded by The Danish Council for Independent Research.
The content of this thesis contributes to the project
by introducing formalisms and algorithms for approximating the behaviour of stochastic system,
and studying the properties of these from a logical, topological, as well as computational point of view.

\section{Future Work}

The work presented here has contributed to our understanding
of how to reason about and compare the stochastic behaviour of systems
from a logical, topological, as well as computational point of view.
However, there still remain many open problems
that we intend to investigate in future work.
We discuss here the most important of these.

\vspace{\baselineskip}

\textbf{Strong completeness and Stone duality.}
The completeness result we have shown for WLWB
is a weak completeness result, showing that
\[\models \varphi \quad \text{implies} \quad \vdash \varphi\]
for any formula $\varphi$ in WLWB.
The notion of strong completeness asks that
\[\Phi \models \varphi \quad \text{implies} \quad \Phi \vdash \varphi\]
for any formula $\varphi$ and set of formulas $\Phi$ in WLWB.
Proving strong completeness would require additional, infinitary axioms
such as the rules
\[\{L_q \varphi \mid q < r\} \vdash L_r \varphi \quad \text{and} \quad \{M_q \varphi \mid q < r\} \vdash M_r \varphi\]
for a given $r$, which describe the Archimedean property of the reals.

A strong completeness result may also point the way to
a Stone duality result~\cite{halmos2009,stone36,johnstone86},
which sheds light on the connection between logic and topology.
Such results have already been developed for Markovian logics~\cite{6571564,FKLMP17,KMP14},
which bear some similarity to WLWB.

\vspace{\baselineskip}

\textbf{Extend logical specification languages with temporal operators.}
The logical specification languages we have introduced and studied are all quite parsimonious,
although our results show that they are expressive enough to characterise the relations under consideration.
However, when specifying requirements in an actual engineering situation,
it may be useful to introduce additional constructs in the language
in order to allow for more expressivity.
In particular, it would be interesting to add temporal operators like those found in LTL~\cite{pnueli77} and CTL~\cite{CE81},
such as ``until'' and ``eventually'', or even fix-point operators like those found in the $\mu$-calculus~\cite{kozen83},
and investigate how many of our results carry over to this more expressive language.

\vspace{\baselineskip}

\textbf{Better understanding of timing anomalies.}
Although we shown that timing anomalies can occur when reasoning with the faster-than relation,
and we have given conditions that are sufficient to guarantee that no timing anomalies occur,
a complete understanding of when and how timing anomalies occur is still missing.
First of all, the conditions that we have given are very restrictive,
and requires that the processes in question are fully deterministic.
Secondly, the conditions impose requirements on all the involved processes,
and not just the context.
It would be preferable to have conditions that only look at the context,
since then one could verify the context once and for all,
and then be guaranteed that no timing anomalies occur when swapping components in and out.
One way to try and gain a better understanding of timing anomalies
may be to consider networks of priced timed automata~\cite{DLLMPVW11}
instead of semi-Markov processes, since the former are more well-suited for compositional reasoning.

Another aspect that is missing from our understanding of timing anomalies
is what happens in the generative case.
We have only considered reactive processes here,
since defining composition for these is more natural than for generative processes~\cite{SV04}.
However, several notions of composition have been defined for generative processes~\cite{BBS95,CSZ92,DAHK99,GSS95},
and we are interested in seeing how timing anomalies behave and can be avoided in this setting.

\vspace{\baselineskip}

\textbf{Develop algorithms for deciding the faster-than relation for reactive systems.}
For both reactive and generative processes, we have shown that the faster-than relation is undecidable in general.
However, for generative processes, we have nonetheless been able to develop algorithms
for two special cases, namely those of time-bounded additive approximation and unambiguous processes.
These cases do not immediately carry over to the setting of reactive systems,
where the main challenge is that of handling the schedulers involved,
of which there may be uncountably many,
depending on the kind of scheduler under consideration~\cite{WJ06}.
We have extended the result on time-bounded additive approximation from generative processes to reactive processes,
but only for the case where there are countably many schedulers,
meaning that the schedulers may not take time into account.
It is therefore still unclear to us whether there exists an algorithm
for the case of uncountably many schedulers.
The same comments also apply to the algorithm for unambiguous processes.

\vspace{\baselineskip}

\textbf{Take probabilistic branching into account in the simulation distance.}
One weakness of the simulation distance is that it only considers differences in the residence-time distributions
and not the differences in the probabilistic branching.
This is because we have chosen to focus on the real-time behaviour of systems.
However, there are some cases where this is not completely satisfactory.

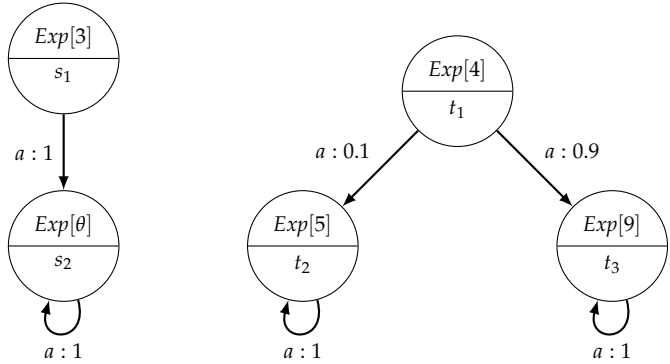
\begin{figure}
  \centering
  \hfill
  \begin{tikzpicture}
    \node[state, circle split] (s1) {$\Exp{3}$ \nodepart{lower} $s_1$};
    \node[state, circle split, below = of s1] (s2) {$\Exp{\theta}$ \nodepart{lower} $s_2$};
    
    \path[thick, ->] (s1) edge [left] node {$a:1$} (s2);
    \path[thick, ->, loop below, looseness=4] (s2) edge [below] node {$a:1$} (s2);
  \end{tikzpicture}
  \hfill
  \begin{tikzpicture}
    \node[state, circle split] (t1) {$\Exp{4}$ \nodepart{lower} $t_1$};
    \node[state, circle split, below left = of t1] (t2) {$\Exp{5}$ \nodepart{lower} $t_2$};
    \node[state, circle split, below right = of t1] (t3) {$\Exp{9}$ \nodepart{lower} $t_3$};
    
    \path[thick, ->] (t1) edge [above left] node {$a:0.1$} (t2);
    \path[thick, ->, loop below, looseness=4] (t2) edge [below] node {$a:1$} (t2);
    \path[thick, ->] (t1) edge [above right] node {$a:0.9$} (t3);
    \path[thick, ->, loop below, looseness=4] (t3) edge [below] node {$a:1$} (t3);
  \end{tikzpicture}
  \hfill \
  
  \caption{A semi-Markov process where $s_1 \simul t_1$ if $\theta \leq 5$ and $s_1 \not\simul t_1$ if $\theta > 5$.}
  \label{fig:conclusion-intro}
\end{figure}

Consider for example the semi-Markov process depicted in Figure~\ref{fig:conclusion-intro}.
In this case, we have $s_1 \simul t_1$ whenever $\theta \leq 5$.
However, if $\theta$ is perturbed just a little bit above $5$,
then we get $s_1 \not\simul t_1$.
For example, consider what happens when $\theta = 5.00001$.
Then it is not difficult to see that the simulation distance will be
\[\simdist(s_1,t_1) = \frac{5.00001}{5} = 1.000002.\]
However, for so small perturbations of $\theta$,
one may argue in the following way that $t_1$ should be considered to be faster than $s_1$.
While it is true that $t_2$ is slower than $s_2$,
and that $t_1$ has a non-zero probability of transitioning to $t_2$,
$t_2$ is only marginally slower than $s_2$,
and furthermore, $t_1$ transitions only to $t_2$ $10\%$ of the time.
The remaining $90\%$ of the time,
$t_1$ will transition to $t_3$, which is significantly faster than $s_2$.
Hence, the greater probability of going to a much faster state
should somehow outweigh the small probability of going to a slightly slower state.

We believe that the kind of reasoning just described
will be possible by taking into consideration also the difference
between the probabilistic branching of the processes in defining the simulation distance.
For this, an obvious possibility would be to incorporate the (non-symmetric) Kantorovich distance (see e.g.~\cite{chatzikokolakis18}).
This is somewhat similar to what has been done for bisimulation distances on continuous-time Markov processes
by combining the Kantorovich distance and the total variation distance~\cite{BBLM17}.
However, it is not clear to us at present how the distance between the real-time behaviour in the states
and the distance between the transition distributions should interact.

\vspace{\baselineskip}

\textbf{Consider behavioural distances starting from the topological point of view.}
In this thesis and in other works such as~\cite{LMP12,KMP14},
topological issues of simulation and bisimulation distances have been
investigated, in particular how properties of the system behave with respect to the topology.
However, in these cases, the distance comes before the topology
in the sense that the distance is defined and the topology is then investigated.
We believe that in order to understand better the interplay between (bi)simulation,
distances, topology, and logical properties,
it will be beneficial to define first the topology that characterises (bi)simulation,
and then see what distances can metrise this topology.

\section{Summary}
The research presented in this thesis
makes a novel contribution to challenging problems encountered when dealing with stochastic systems.
This research has deepened our understanding of how to compare and express properties about stochastic systems,
as well as opened new lines of research that the author intends to pursue in the future.

\defaultbib

%backmatter
\appendix
\part{Papers}\label{chap:papers}
\titleformat{%command
  \chapter
}[%shape
display%
]{%format
  \normalfont\huge
}{%label
  \begin{center}\color{aaublue}\chaptertitlename\ \thechapter\end{center}
}{%style
  1cm
}{%code before title
  \thispagestyle{empty}\begin{center}\Large
}[%code after title
  \end{center}
]

  \cleardoublepage
  \setcounter{enumiii}{0}
  \setcounter{enumii}{0}
  \setcounter{enumiv}{0}
  \setcounter{enumi}{0}
  \setcounter{equation}{0}
  \setcounter{figure}{0}
  \setcounter{footnote}{0}
  \setcounter{mpfootnote}{0}
  \setcounter{paragraph}{0}
  \setcounter{parentequation}{0}
  \setcounter{part}{0}
  \setcounter{section}{0}
  \setcounter{subfigure}{0}
  \setcounter{subparagraph}{0}
  \setcounter{subsection}{0}
  \setcounter{subsubsection}{0}
  \setcounter{table}{0}
  \papertitlepage{%
  Reasoning About Bounds in Weighted Transition Systems
}{paper:paperA}{%
  Mikkel Hansen, Kim Guldstrand Larsen, Radu Mardare, and Mathias Ruggaard Pedersen
}{%
  The paper is under submission to Logical Methods in Computer Science.
  The paper is an extended version of~\cite{hansen2016}.
}{%
%  \noindent\copyright\ 201X IEEE
%  
%  \noindent{\em The layout has been revised and the content extended.}
}

%% required for running head on odd and even pages, use suitable
%% abbreviations in case of long titles and many authors:

%%%%%%%%%%%%%%%%%%%%%%%%%%%%%%%%%%%%%%%%%%%%%%%%%%%%%%%%%%%%%%%%%%%%%%%%%%%

%% the abstract has to PRECEDE the command \maketitle:
%% be sure not to issue the \maketitle command twice!

%% Abstract
\begin{abstract}
  We propose a way of reasoning about minimal and maximal values of
  the weights of transitions in a weighted transition system (WTS). This perspective induces a notion of bisimulation that is coarser than the classic bisimulation: it
  relates states that exhibit transitions to bisimulation classes with the weights within the same boundaries.
  We propose a customised modal logic that expresses these numeric boundaries for transition weights by means of particular modalities.
  We prove that our logic is invariant under the proposed notion of bisimulation.
  We show that the logic enjoys the finite model property
  and we identify a complete axiomatisation for the logic.
  Last but not least, we use a tableau method to show that the satisfiability problem for the logic is decidable.
\end{abstract}

%% Introduction 
\section{Introduction}
Weighted transition systems (WTSs) are used to model concurrent and distributed systems
in the case where some resources are involved, such as time, bandwidth, fuel, or energy consumption.
Recently, the concept of a cyber-physical system (CPS),
which considers the integration of computation and the physical world has become relevant in modeling various real-life situations. 
In these models, sensor feedback affects computation, and through machinery, computation can further affect physical processes.
The quantitative nature of weighted transition systems is well-suited for the quantifiable inputs and sensor measurements of CPSs,
but their rigidity makes them less well-suited for the uncertainty inherent in CPSs.
In practice, there is often some uncertainty attached to the
resource cost, whereas weights in a WTS are precise.
Thus, the model may be too restrictive and unable to capture the uncertainties
inherent in the domain that is being modeled.

In this paper, we attempt to remedy this shortcoming by introducing
a modal logic for WTSs that allows for approximate
reasoning by speaking about upper and lower bounds for the weights of the transitions.
The logic has two types of modal operators that reason about the minimal and maximal weights on transitions, respectively.
This allows reasoning about models where the quantitative information may be imprecise 
(e.g. due to imprecisions introduced when gathering real data), but where we can establish a lower and upper bound for transitions.

In order to provide the semantics for this logic, we use the set of possible transition weights from
one state to a set of states as an abstraction of the actual transition weights.
The logic is expressive enough to characterise WTSs up to a relaxed notion of weighted bisimilarity,
where the classical conditions are replaced with conditions requiring that the minimal
and maximal weights on transitions are matched.

Our main contribution is a complete axiomatisation of our logic,
showing that any validity in this logic can be proved as a theorem from the axiomatic system.
Completeness allows us to transform any validity checking problem into a theorem proving one that can
be solved automatically by modern theorem provers,
thus bridging the gap to the theorem proving community.
The completeness proof adapts the classical filtration method,
which allows one to construct a (canonical) model using maximal consistent sets of formulas.
The main difficulty of adapting this method to our setting
is that we must establish both lower and upper bounds for the transitions in this model.
To achieve this result, we demonstrate that our logic enjoys the finite model property.

Our second significant contribution is a decision procedure for determining the satisfiability of formulas in our logic.
This decision procedure makes use of the tableau method to construct a tableau for a given formula.
If the constructed tableau is successful,
then the formula is satisfiable, and a finite model for the formula can be generated from the tableau.

%% Related work
\subsection*{Related Work.}
In \cite{esik2014}, Zolt\'{a}n \'{E}sik also considered the issue of bisimulation for weighted transition systems,
although in the more general setting of synchronisation trees with weights in an arbitrary monoid or semiring.
Synchronisation trees arise by unfolding the transitions of a weighted transition system starting in some state
which will become the root of the tree.
Both \'{E}sik's and our notion of bisimilarity bears some resemblance
to probabilistic bisimulation \cite{LS91},
by considering not only single transitions but transitions to equivalence classes of states.
However, while we require that the upper and lower bounds of these transitions should match,
the bisimilarity of \'{E}sik requires that the sum of the transitions should be the same.
This is motivated by the fact that the synchronisation trees do not form a category
which respects the additive structure of a semiring.
However, as \'{E}sik proves, if one takes the quotient with respect to
his version of weighted bisimilarity,
then the category one obtains does respect the additive structure.
Thus, the semiring structure of the weights is of vital importance to \'{E}sik's work,
but is an aspect that we have not considered in our work. 

Several logics have been proposed in the past to express properties of quantified (weighted, probabilistic or stochastic) systems.
They typically use modalities indexed with real numbers to express properties such as \textit{``$\varphi$ holds with at least probability $b$''},
\textit{``we can reach a state satisfying $\varphi$ with a cost at least $r$''}, etc.

In the context of weighted automata, weighted monadic second order logic
has been introduced by Droste and Gastin \cite{droste2005}
to capture the behaviour of weighted automata for commutative semirings.
This work has been extended to many closely related systems \cite{babari2016,droste2006a,droste2006b,meinecke2006,fichtner2011}.
There has also been work on connecting weighted monadic second order logic with probabilistic CTL \cite{bollig2009}.
For weighted transition systems, weighted modal logic has been introduced by Larsen and Mardare \cite{larsen1}
to reason about the consumption of resources in such a system.
This logic has been extended to handle recursion \cite{larsen2014a,larsen2014b}
as well as parallel composition and concurrency \cite{LMX15}.
For both the original weighted modal logic and its concurrent extension,
complete axiomatisations were developed.
A weighted extension of the $\mu$-calculus was introduced by Larsen et al. in \cite{larsen2015},
where a complete axiomatisation for this extension was also given.

While our setting is that of weighted transition systems,
our logic and the development of its theory has more in common with Markovian logic
than with the previously mentioned work on weighted systems.
Markovian logic was introduced by Mardare et al. \cite{MCL12,cardelli2011a}
building on previous work on probability logics \cite{Zhou09,Fagin,Heifetz200131}.
Markovian logic reasons about probabilistic and stochastic systems
using operators $L_r$ and $M_r$ which mean that a property hold with \emph{at least} probability $r$
or \emph{at most} probability $r$, respectively.
Much of the work on Markovian logic has focused on giving a complete axiomatisation for the logic \cite{KMP13},
culminating in a Stone duality for Markov processes \cite{6571564}.
However, compositional aspects have been considered in \cite{cardelli2011b},
where also an axiomatisation was given for Markovian logic with an operator for parallel composition.

While our logical syntax resembles that of Markovian logic,
our semantics is different in the sense that we argue not about probabilities,
but about an interval of possible weights.
For instance, in the aforementioned logics we have a validity of type $\vdash\neg L_r\phi\to M_r\phi$
saying that the value of the transition from the current state to $\phi$ is either at least $r$ or at most $r$;
on the other hand, in our logic the formula $\neg L_r\phi\land \lnot M_r\phi$ might have a model since $L_r\phi$ and $M_r\phi$
express the fact that the lower cost of a transition to $\phi$ is at least $r$ and the highest cost is at most $r$ respectively.

Our completeness proof uses a technique similar to the one used for weighted modal logic \cite{larsen1} and Markovian logic \cite{KMP13,MCL12,cardelli2011a}.
It is however different from these related constructions since our axiomatisation is finitary, while the aforementioned ones require infinitary proof rules.
Our axiomatic systems are related to the ones mentioned above and the mathematical structures revealed by this work are also similar to the related ones.
This suggest a natural extension towards a Stone duality result along the lines of \cite{6571564}, which we will consider in a future work. 

Decidability results regarding satisfiability have also been given for some related logics,
such as weighted modal logic \cite{LMX18} and probabilistic versions of CTL and the $\mu$-calculus \cite{katoen:sat}.
However, the satisfiability problem is known to be undecidable for
other related logics, in particular timed logics such as TCTL \cite{ACD93}
and timed modal logic \cite{JLMX14}.
This fact suggests that our logic is an interesting one which, despite its expressivity, remains decidable.

Our approach of considering upper and lower bounds is related
to work on interval-based formalisms such as interval Markov chains (IMCs) \cite{JL91}
and interval weighted modal transition systems (WMTSs) \cite{Juhl2012408}.
Much like our approach, IMCs consider upper and lower bounds on transitions
in the probabilistic case.
WMTSs add intervals of weights to individual transitions
of modal transition systems, in which there can be both may- and must-transitions.
A main focus of the work both on IMCs and WMTSs have been
a process of refinement, making the intervals progressively smaller
until an implementation is obtained.
However, none of these works have explored the logical perspective up to the level of axiomatisation or satisfiability results,
which is the focus of our paper.

%% Model
\section{Model}
The models addressed in this paper are weighted transition systems,
in which transitions are labelled with numbers to specify the cost of the corresponding transition.
In order to specify and reason about properties regarding imprecision, such as
``the maximum cost of going to a safe state is $10$''
and ``the minimum cost of going to a halting state is $5$'',
we will abstract away the individual transitions
and only consider the minimum and maximum costs from a state to another.
We will do this by constructing for any two states the set of weights that are allowed
from one to the other.

First we recap the definition of a weighted transition system.
Let $\mathcal{AP}$ be a countable set of atomic propositions.
A WTS is formally defined as follows:
\begin{definition}
  A \emph{weighted transition system (WTS)} is a tuple $\mathcal{M} = (S, \rightarrow, \ell)$, where
  \begin{itemize}
    \item $S$ is a non-empty set of \emph{states},
    \item $\rightarrow \subseteq S \times \mathbb{R}_{\geq 0} \times S$ is the \emph{transition relation}, and
    \item $\ell : S \to 2^{\mathcal{AP}}$ is a \emph{labelling function} mapping to each state a set of atomic propositions. \qedhere
  \end{itemize}
\end{definition}
Note that we impose no restrictions on the state space $S$; it can be uncountable.
We write $s \xrightarrow{r} t$ to mean that $(s,r,t) \in \rightarrow$.
We will say that a WTS is \emph{image-finite} if for any $s \in S$
there are only finitely many $t \in S$ such that $s \xrightarrow{r} t$
for some $r \in \mathbb{R}_{\geq 0}$.

When modeling cyber-physical systems,
it is often unreasonable to expect one to know the
exact weights for transitions.
However, it is often the case that one has some bounds
on the actual weights, e.g. one might know
that the cost of taking some transition is between $5$ and $25$.
In order to reason about these bounds, we abstract away
the individual transitions, and instead consider the set
of weights between a state and a set of states.

\begin{definition}\label{def:theta}
  For an arbitrary WTS $\mathcal{M} = (S,\rightarrow,\ell)$, the function
  $\theta_{\mathcal{M}} : S \to \left(2^S \to 2^{\mathbb{R}_{\geq 0}}\right)$
  is defined for any state $s \in S$ and set of states $T \subseteq S$ as
  \[
  \transm{s}{T}
  =
  \{r \in \mathbb{R}_{\geq 0} \mid \exists t \in T \text{ such that } s \xrightarrow{r} t\}. \qedhere
  \]
\end{definition}

Thus $\transm{s}{T}$ is the set of all possible weights of going from $s$ to a state in $T$.
We will sometimes refer to $\trans{s}{T}$ as the \emph{image from $s$ to $T$}
or simply as an \emph{image set}.
In the rest of the paper, we will use the notation
\[\transl{s}{T} = \begin{cases} -\infty & \text{if } \trans{s}{T} = \Emptyset \\ \inf\trans{s}{T} & \text{otherwise} \end{cases}\]
and
\[\transr{s}{T} = \begin{cases} \infty & \text{if } \trans{s}{T} = \Emptyset \\ \sup\trans{s}{T} & \text{otherwise.} \end{cases}\]
Thus $\transl{s}{T}$ is a lower bound on the weights from $s$ to $T$
and $\transr{s}{T}$ is an upper bound.

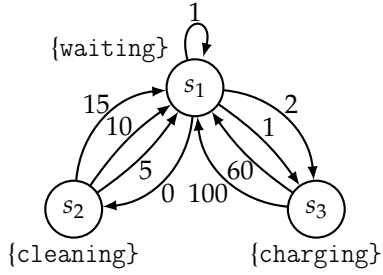
\begin{figure}
  \centering
  \begin{tikzpicture}[WTS, node distance=3cm]
    \node[state, label=above left:{\{\tt waiting\}}] (0) {$s_1$};
    \node[state, label=below:{\{\tt cleaning\}}] (1) [below left = of 0] {$s_2$};
    \node[state, label=below:{\{\tt charging\}}] (2) [below right = of 0] {$s_3$};
    
    \path[->] (0) edge [loop above] node {$1$} (0);
    
    \path[->] (0) edge [bend left = 10, above] node [above, xshift=1.5mm] {$1$} (2);
    \path[->] (0) edge [bend left = 40, above] node [above, xshift=1.5mm] {$2$} (2);
    \path[->] (2) edge [bend left = 10, below] node [below, xshift=-2mm] {$60$} (0);
    \path[->] (2) edge [bend left = 40, below] node [below left, xshift=2mm] {$100$} (0);
    
    \path[->] (0) edge [bend left = 40, below] node [below, xshift=1mm] {$0$} (1);
    \path[->] (1) edge [bend right = 10, below] node [below, xshift=1mm] {$5$} (0);
    \path[->] (1) edge [bend left = 10, above] node [above, xshift=-2mm] {$10$} (0);
    \path[->] (1) edge [bend left = 40, above] node [above, xshift=-2mm] {$15$} (0);
  \end{tikzpicture}
  \caption{A simple model of a robot vacuum cleaner.}
  \label{fig:wts-example}
\end{figure}

\begin{example}
  Figure \ref{fig:wts-example} shows a simple model of a robot vacuum cleaner that
  can be in a waiting state, a cleaning state, or a charging state.
  This is an example of a cyber-physical system where
  the costs of transitions are necessarily imprecise.
  The time it takes to recharge the batteries depends on the
  condition of the batteries as well as that of the charger;
  the time it takes to clean the room depends on how dirty the room is,
  and how free the floor is from obstacles;
  and the time it takes to reach the charger depends on where in the room
  the robot is when it needs to be recharged.
  By constructing the image sets, we can abstract away from the individual transitions.
  For example, we have $\trans{s_2}{\{s_1\}} = \{5,10,15\}$,
  so $\transl{s_2}{\{s_1\}} = 5$ and $\transr{s_2}{\{s_1\}} = 15$.
\end{example}

We will now establish some useful properties of image sets.
In particular, the transition function is monotonic with respect to set inclusion,
and union distributes over image sets as one might expect.

\begin{lemma}[Monotonicity of $\theta$]\label{lem:thetamono}
  Let $\mathcal{M} = (S,\rightarrow,\ell)$ be a WTS
  and let $T_1$ and $T_2$ be subsets of $S$.
  If $T_1 \subseteq T_2$, then $\trans{s}{T_1} \subseteq \trans{s}{T_2}$.
\end{lemma}

\begin{lemma}\label{lem:thetaunion}
  Let $\mathcal{M} = (S,\rightarrow,\ell)$ be a WTS.
  For any $s \in S$ and $T_1,T_2 \subseteq S$, it holds that
  \begin{enumerate}
    \item $\trans{s}{T_1 \cup T_2} = \trans{s}{T_1} \cup \trans{s}{T_2}$ and
    \item $\trans{s}{T_1 \cap T_2} \subseteq \trans{s}{T_1} \cap \trans{s}{T_2}$.
  \end{enumerate}
\end{lemma}

As usual we would like some way of relating model states with equivalent behavior.
To this end we define the notion of a bisimulation relation.
The classical notion of a bisimulation relation for weighted transition systems \cite{blackburn},
which we term weighted bisimulation, is defined as follows.

\begin{definition}
  Given a WTS $\mathcal{M} = (S, \rightarrow, \ell)$,
  an equivalence relation $\mathcal{R} \subseteq S \times S$ on $S$
  is called a \emph{weighted bisimulation relation}
  if and only if for all $s,t \in S$, $s \mathcal{R} t$ implies
  \begin{itemize}
    \item (Atomic harmony) $\ell(s) = \ell(t)$,
    \item (Zig) if $s \xrightarrow{r} s'$ then there exists $t' \in S$ such that $t \xrightarrow{r} t'$ and $s' \mathcal{R} t'$, and
    \item (Zag) if $t \xrightarrow{r} t'$ then there exists $s' \in S$ such that $s \xrightarrow{r} s'$ and $s' \mathcal{R} t'$. \qedhere
  \end{itemize}
\end{definition}

We say that $s,t \in S$ are weighted bisimilar, written $s \sim_W t$,
if and only if there exists a weighted bisimulation relation $\mathcal{R}$ such that $s \mathcal{R} t$.
Weighted bisimilarity, $\sim_W$, is the largest weighted bisimulation relation.

Since it is our goal to abstract away from the exact weights on the transitions,
the bisimulation that we will now introduce
does not impose the classical zig-zag conditions \cite{blackburn} of a bisimulation relation,
but instead require that bounds be matched for any bisimulation class.

\begin{definition}\label{def:bisim}
  Given a WTS $\mathcal{M} = (S, \rightarrow, \ell)$,
  an equivalence relation $\mathcal{R} \subseteq S \times S$ on $S$
  is called a \emph{generalised weighted bisimulation relation}
  if and only if for all $s,t \in S$, $s \mathcal{R} t$ implies
  \begin{itemize}
    \item (Atomic harmony) $\ell(s) = \ell(t)$,
    \item (Lower bound) $\transl{s}{T} = \transl{t}{T}$, and
    \item (Upper bound) $\transr{s}{T} = \transr{t}{T}$
  \end{itemize}
  for any $\mathcal{R}$-equivalence class $T \subseteq S$.
\end{definition}

Given $s,t \in S$ we say that $s$ and $t$
are generalized weighted bisimilar, written $s \sim t$,
if and only if there exists a generalised weighted bisimulation relation
$\mathcal{R}$ such that $s \mathcal{R} t$.
We let $\sim$ denote generalised weighted bisimilarity
which is defined as
\[
\mathord{\sim} = \bigcup \left\{\mathcal{R} \mid \mathcal{R} \text{ is a generalised weighted bisimulation relation} \right\}.
\]

We will now show that generalised weighted bisimilarity, $\sim$, is the largest generalised weighted bisimulation relation.
To this end, we first need to show that $\sim$ is an equivalence relation.

\begin{lemma}\label{lem:simequiv}
  Generalised weighted bisimilarity, $\sim$, is an equivalence relation.
\end{lemma}
\begin{proof}
  In order to prove that generalised weighted bisimilarity is an equivalence relation, we have to show that it is reflexive, symmetric and transitive.
  \begin{description}
  \item[Reflexivity] Consider the identity relation
    \[
    \mathcal{I} = \left\{(s,s) \mid s \in S \text{ for some WTS } \mathcal{M} = (S, \rightarrow, \ell) \right\}.
    \]
    It is trivial to verify that $\mathcal{I}$ is a generalised weighted bisimulation relation, and therefore $\mathcal{I} \subseteq \mathord{\sim}$.

  \item[Symmetry] Let $\mathcal{M} = (S, \rightarrow, \ell)$ be a WTS and $s,t \in S$ states such that $s \sim t$.
  Because $s \sim t$ there must exist a generalised weighted bisimulation relation $\mathcal{R}$ such that $s \mathcal{R} t$.
  Let $\mathcal{R}' = \left\{(t,s) \mid (s,t) \in \mathcal{R} \right\}$.
  $\mathcal{R}'$ is clearly also a generalised weighted bisimulation relation implying $\mathcal{R}' \subseteq \mathord{\sim}$ and therefore $t \sim s$. 

  \item[Transitivity] Let $\mathcal{M} = (S, \rightarrow, \ell)$ be a WTS and $s,t,u \in S$ states such that $s \sim t$ and $t \sim u$.
  There must exist generalised weighted bisimulation relations $\mathcal{R}$ and $\mathcal{R}'$ such that $s \mathcal{R} t$ and $t \mathcal{R}' u$.
  Let $\mathcal{R}'' = (\mathcal{R} \cup \mathcal{R}')^{+}$ be the transitive closure of the union of $\mathcal{R}$ and $\mathcal{R}'$.
  Since $\mathcal{R}$ and $\mathcal{R}'$ are both equivalence relations, $\mathcal{R} \cup \mathcal{R}'$ is reflexive and symmetric,
  and since the transitive closure of a symmetric and reflexive relation is symmetric and reflexive,
  we get that $\mathcal{R}''$ is an equivalence relation.
  We need to show that $\mathcal{R}''$ is a generalised weighted bisimulation relation.
  Atomic harmony is trivially satisfied.

  Suppose that $\trans{u}{T''} \neq \Emptyset$ for some $T'' \in S/\mathcal{R}''$ implying the existence of a state $u' \in T''$ such that $\trans{u}{\{u'\}} \neq \Emptyset$,
  further implying the existence of an equivalence class $T' \in S/\mathcal{R}'$ such that $u' \in T'$ and thus $\trans{u}{T'} \neq \Emptyset$.
  $t \mathcal{R}' u$ implies $\trans{t}{T'} \neq \Emptyset$ which further implies the existence of a state $t' \in T'$ such that $\trans{t}{\{t'\}} \neq \Emptyset$.
  There must exist an equivalence class $T \in S/\mathcal{R}$ such that $t' \in T$ implying $\trans{t}{T} \neq \Emptyset$.
  Because $s \mathcal{R} t$ we must have $\trans{s}{T} \neq \Emptyset$ implying the existence of a state $s' \in T$ such that $\trans{s}{\{s'\}} \neq \Emptyset$.
  $s',t' \in T$ implies $s' \mathcal{R} t'$, $t',u' \in T'$ implies $t' \mathcal{R}' u'$,
  and therefore $s' \mathcal{R}'' u'$ implying $s' \in T''$ which further implies $\trans{s}{T''} \neq \Emptyset$.
  Therefore $\trans{u}{T''} \neq \Emptyset$ implies $\trans{s}{T''} \neq \Emptyset$ for all $T'' \in S/\mathcal{R}''$.
  Symmetric arguments show that $\trans{s}{T''} \neq \Emptyset$ implies $\trans{u}{T''} \neq \Emptyset$ for all $T'' \in S/\mathcal{R}''$,
  and therefore $\trans{s}{T''} = \Emptyset$ if and only if $\trans{u}{T''} = \Emptyset$ for all $T'' \in S/\mathcal{R}''$.

  Suppose towards a contradiction that $\transl{s}{T''} \neq \transl{u}{T''}$ for some $T'' \in S/\mathcal{R}''$.
  We have two cases to consider, namely
  \[\transl{s}{T''} < \transl{u}{T''} \quad \text{and} \quad \transl{s}{T''} > \transl{u}{T''}.\]
  If $\transl{s}{T''} < \transl{u}{T''}$ there must exist a rational number $q \in \mathbb{Q}$ such that $\transl{s}{T''} < q < \transl{u}{T''}$,
  implying the existence of a state $s' \in T''$ such that $\transl{s}{T''} \leq \transl{s}{\{s'\}} < q$.
  There must exist $T \in S/\mathcal{R}$ such that $s' \in T$ implying $\transl{s}{T} < q$.
  Because $s \mathcal{R} t$ we must have $\transl{s}{T} = \transl{t}{T}$ implying the existence of a state $t' \in T$ such that $\transl{t}{\{t'\}} < q$.
  There must exist $T' \in S/\mathcal{R}'$ such that $t' \in T'$ implying $\transl{t}{T'} < q$.
  Because $t \mathcal{R}' u$ we must have $\transl{t}{T'} = \transl{u}{T'}$ implying the existence of a state $u' \in T'$ such that $\transl{u}{\{u'\}} < q$.
  $s',t' \in T$ implies $s' \mathcal{R} t'$, $t',u' \in T'$ implies $t' \mathcal{R} u'$,
  and therefore $s' \mathcal{R}'' u'$, implying $u' \in T''$ and therefore $\transl{u}{T''} < q$, leading to a contradiction.
  Symmetric arguments show that also $\transl{s}{T''} > \transl{u}{T''}$ leads to a contradiction and therefore $\transl{s}{T} = \transl{u}{T}$ for any $T \in S/\mathcal{R}''$.

  Similar arguments show that $\transr{s}{T} = \transr{u}{T}$ for any $T \in S/\mathcal{R}''$ thus showing that $\mathcal{R}''$
  is a generalised weighted bisimulation relation implying $\mathcal{R}'' \subseteq \mathord{\sim}$ and therefore $s \sim t$ and $t \sim u$ implies $s \sim u$. \qedhere
  \end{description}
\end{proof}

Having established that $\sim$ is an equivalence relation, we will now show that it is indeed the largest generalised weighted bisimulation relation.

\begin{theorem}
  Generalised weighted bisimilarity, $\sim$, is the largest generalised weighted bisimulation relation.
\end{theorem}
\begin{proof}
  We first show that $\sim$ is a generalised weighted bisimulation relation.
  By Lemma \ref{lem:simequiv} we know that $\sim$ is an equivalence relation.
  Let $\mathcal{M} = (S, \rightarrow, \ell)$ be a WTS and $s,t \in S$ states such that $s \sim t$.
  There must exist a generalised weighted bisimulation relation $\mathcal{R}$ such that $s \mathcal{R} t$,
  which trivially verifies atomic harmony.
  
  Suppose that $\trans{t}{T} \neq \Emptyset$ for some $T \in S/\mathord{\sim}$, implying the existence of a state $t' \in T$ such that $\trans{t}{\{t'\}} \neq \Emptyset$.
  There must exist an equivalence class $T' \in S/\mathcal{R}$ such that $t' \in T'$, which implies that $\trans{t}{T'} \neq \Emptyset$.
  Because $s \mathcal{R} t$ we must have $\trans{s}{T'} \neq \Emptyset$, implying the existence of a state $s' \in T'$ such that $\trans{s}{\{s'\}} \neq \Emptyset$.
  Because $s',t' \in T'$ we must have $s' \mathcal{R} t'$ and hence $s' \sim t'$,
  so $s' \in T$ and thus $\trans{s}{T} \neq \Emptyset$.
  Symmetric arguments show that $\trans{s}{T} \neq \Emptyset$ implies $\trans{t}{T} \neq \Emptyset$
  and therefore $\trans{s}{T} = \Emptyset$ if and only if $\trans{t}{T} = \Emptyset$ for all $T \in S/\mathord{\sim}$.

  Suppose $\transl{s}{T} \neq \transl{t}{T}$ for some $T \in S/\mathord{\sim}$.
  We have two cases to consider, namely $\transl{s}{T} < \transl{t}{T}$ and $\transl{s}{T} > \transl{t}{T}$.
  If $\transl{s}{T} < \transl{t}{T}$ there must exist a rational number $q \in \mathbb{Q}$ such that $\transl{s}{T} < q < \transl{t}{T}$,
  implying the existence of a state $s' \in T$ such that $\transl{s}{T} \leq \transl{s}{\{s'\}} < q$.
  There must exist $T' \in S/\mathcal{R}$ such that $s' \in T'$ and hence $\transl{s}{T'} < q$.
  Because $s \mathcal{R} t$ we have $\transl{s}{T'} = \transl{t}{T'}$, which means that there exists a state $t' \in T'$ such that $\transl{t}{\{t'\}} < q$.
  $s',t' \in T'$ implies $s' \mathcal{R} t'$ which further implies $s' \sim t'$ and therefore $\transl{t}{T} < q$, leading to a contradiction.
  Symmetric arguments show that also $\transl{s}{T} > \transl{t}{T}$ leads to a contradiction, and therefore $\transl{s}{T} = \transl{t}{T}$ for all $T \in S/\mathord{\sim}$.

  Similar arguments show that $\transr{s}{T} = \transr{t}{T}$ for any $T \in S/\mathord{\sim}$, thus showing that $\sim$ is a generalised weighted bisimulation relation.
  
  $\sim$ was defined as the union of all generalised weighted bisimulation relations,
  so for any generalised weighted bisimulation relation $\mathcal{R}$
  we must have $\mathcal{R} \subseteq \mathord{\sim}$,
  and hence we conclude that $\sim$ is the largest generalised weighted bisimulation relation.
\end{proof}

In what follows, we will use bisimulation to mean generalised weighted bisimulation and bisimilarity to mean generalised weighted bisimilarity.

\begin{example}\label{ex:sim_nwsim}
  Consider the WTS depicted in Figure \ref{fig:ex_bisim_nwbisim}.
  It is easy to see that $\{s',t'\}$ is a $\sim$-equivalence class,
  and in fact it is the only $\sim$-equivalence class with in-going transitions.
  Since $\transl{s}{\{s',t'\}} = \transl{t}{\{s',t'\}} = 1$
  and $\transr{s}{\{s',t'\}} = \transr{t}{\{s',t'\}} = 3$ we must have $s \sim t$,
  but because $s \xrightarrow{2} s'$ and $t \not \xrightarrow{2}$ it cannot be the case that $s \sim_W t$.
\end{example}

\begin{figure}
  \centering
  \begin{tikzpicture}[WTS, node distance=2cm]
    \node[state, label=left:{$\{a\}$}]  (s0)               {$s$};
    \node[state, label=left:{$\{b\}$}]  (s1) [below=of s0] {$s'$};
    \node[state, label=right:{$\{a\}$}] (t0) [right=of s0] {$t$};
    \node[state, label=right:{$\{b\}$}] (t1) [below=of t0] {$t'$};

    \path (s0) edge[bend right=45] node[left] {$1$} (s1);
    \path (s0) edge                node[left] {$2$} (s1);
    \path (s0) edge[bend left=45]  node[left] {$3$} (s1);

    \path (t0) edge[bend right=45] node[left] {$1$} (t1);
    \path (t0) edge[bend left=45]  node[left] {$3$} (t1);
  \end{tikzpicture}
  \captionof{figure}{$s \sim t$ but $s \not \sim_W t$.}
  \label{fig:ex_bisim_nwbisim}
\end{figure}

The following lemma shows that if two states are weighted bisimilar,
then their image sets match exactly for any weighted bisimulation class.

\begin{lemma}\label{lem:bisim}
  Let $\mathcal{M} = (S, \rightarrow, \ell)$ be a WTS and let $s,t \in S$.
  $s \sim_W t$ implies that $\trans{s}{T} = \trans{t}{T}$
  for any $\sim_W$-equivalence class $T \subseteq S$.
\end{lemma}
\begin{proof}
  Assume $s \sim_W t$ and let $T \subseteq S$ be a $\sim_W$-equivalence class.
  If $r \in \trans{s}{T}$, then there exists some $s' \in T$ such that $s \xrightarrow{r} s'$.
  Because $s \sim_W t$, there must exist some $t' \in T$ such that $t \xrightarrow{r} t'$
  and $s' \sim_W t'$. Since $T$ is a $\sim_W$-equivalence class,
  this means that $r \in \trans{t}{T}$.
  A similar argument shows that if $r \in \trans{t}{T}$,
  then $r \in \trans{s}{T}$.
\end{proof}

We can now show the following relationship between $\sim$ and $\sim_W$.

\begin{theorem}\label{thm:bisimcoarse}
  Generalised weighted bisimilarity is coarser than weighted bisimilarity, i.e.
  \[
  \sim_W \mathbin{\subseteq} \sim \quad\text{and}\quad \sim_W \mathbin{\neq} \sim.
  \]
\end{theorem}
\begin{proof}
  Assume that $s \sim_W t$. We have that $\ell(s) = \ell(t)$,
  and by Lemma \ref{lem:bisim}, we have that $\trans{s}{T} = \trans{t}{T}$
  for any $\sim_W$-equivalence class $T \subseteq S$.
  This implies in particular that
  $\transl{s}{T} = \transl{t}{T}$ and $\transr{s}{T} = \transr{t}{T}$.
  Hence $\sim_W$ is a bisimulation relation.
  
  By Example \ref{ex:sim_nwsim}, the inclusion is strict.
\end{proof}

This result is not surprising,
as our bisimulation relation only looks at the extremes
of the transition weights, whereas weighted bisimulation
looks at all of the transition weights.

%% Logic
\section{Logic}\label{sec:logic}
In this section we introduce a modal logic which is inspired by Markovian logic \cite{MCL12}.
Our aim is that our logic should be able to capture the notion of bisimilar states as presented in the previous section,
and as such it must be able to reason about the lower and upper bounds on transition weights.

\begin{definition}
  The formulas of the logic $\mathcal{L}$ are induced by the abstract syntax
  \[\mathcal{L}: \quad \varphi, \psi ::= p \mid \neg \varphi \mid \varphi \land \psi \mid L_r \varphi \mid M_r \varphi\]
  where $r \in \mathbb{Q}_{\geq 0}$ is a non-negative rational number and $p \in \mathcal{AP}$ is an atomic proposition.
\end{definition}

$L_r$ and $M_r$ are modal operators. An illustration of how $L_r$ and $M_r$ are interpreted can be seen in Figure \ref{fig:semantics}.
Intuitively, $L_r \varphi$ means that the cost of transitions to where $\varphi$
holds is \emph{at least} $r$ (see Figure \ref{fig:Lsemantics}), and $M_r \varphi$ means that
the cost of transitions to where $\varphi$ holds is \emph{at most} $r$ (see Figure \ref{fig:Msemantics}).
We now give the precise semantics interpreted over WTSs.

\begin{definition}\label{def:semantics}
  Given a WTS $\mathcal{M} = (S, \rightarrow, \ell)$,
  a state $s \in S$ and a formula $\varphi \in \mathcal{L}$,
  the satisfiability relation $\models$ is defined inductively as
  \[
  \begin{array}{l l l}
    \mathcal{M},s \models p                   & \text{ iff } & p \in \ell(s), \\
    \mathcal{M},s \models \neg \varphi        & \text{ iff } & \mathcal{M},s \not\models \varphi, \\
    \mathcal{M},s \models \varphi \wedge \psi & \text{ iff } & \mathcal{M},s \models \varphi \;\text{and}\; \mathcal{M},s \models \psi, \\
    \mathcal{M},s \models L_r \varphi         & \text{ iff } & \transl{s}{\sat{\varphi}_{\mathcal{M}}} \geq r,\\
    \mathcal{M},s \models M_r \varphi         & \text{ iff } & \transr{s}{\sat{\varphi}_{\mathcal{M}}} \leq r,\\
  \end{array}
  \]
  where $\sat{\varphi}_{\mathcal{M}} = \left\{s \in S \mid \mathcal{M},s \models \varphi \right\}$
  is the set of all states of $\mathcal{M}$ having the property $\varphi$.
\end{definition}

\begin{figure}
  \centering
  \begin{subfigure}{0.45\textwidth}
    \centering
    \begin{tikzpicture}[scale=.5]
      %% X-axis
      \draw [->,thick] (0,0) -- (8,0);

      %% Arc
      \draw [-,semithick] (6,0) arc (0:180:2cm) node[above, xshift=10mm, yshift=10mm] {$\trans{s}{\sat{\varphi}}$};

      %% r
      \draw[shift={(1,0)},-] (0pt,5pt) -- (0pt,-5pt) node[below] {$r$};

      %% Lower bound
      \draw[->] (2,0) ++ (0,-.5) -- (2,0) node[below,at start] {$\theta^{-}$};
      %% Upper bound
      \draw[->] (6,0) ++ (0,-.5) -- (6,0) node[below,at start] {$\theta^{+}$};
    \end{tikzpicture}
    \caption{$\mathcal{M},s \models L_r \varphi$}
    \label{fig:Lsemantics}
  \end{subfigure}
  \begin{subfigure}{0.45\textwidth}
    \centering
    \begin{tikzpicture}[scale=.5]
      %% X-axis
      \draw [->,thick] (0,0) -- (8,0);

      %% Arc
      \draw [-,semithick] (6,0) arc (0:180:2cm) node[above, xshift=10mm, yshift=10mm] {$\trans{s}{\sat{\varphi}}$};

      %% r
      \draw[shift={(7,0)},-] (0pt,5pt) -- (0pt,-5pt) node[below] {$r$};

      %% Lower bound
      \draw[->] (2,0) ++ (0,-.5) -- (2,0) node[below,at start] {$\theta^{-}$};
      %% Upper bound
      \draw[->] (6,0) ++ (0,-.5) -- (6,0) node[below,at start] {$\theta^{+}$};
    \end{tikzpicture}
    \caption{$\mathcal{M},s \models M_r \varphi$}
    \label{fig:Msemantics}
  \end{subfigure}
  \caption{The semantics of $L_r$ and $M_r$. If $\mathcal{M},s \models L_r \varphi$,
           then $r$ is to the left of $\transl{s}{\sat{\varphi}}$, and if $\mathcal{M},s \models M_r \varphi$,
           then $r$ is to the right of $\transr{s}{\sat{\varphi}}$.}
  \label{fig:semantics}
\end{figure}

We will omit the subscript ${\mathcal{M}}$ from $\sat{\varphi}_{\mathcal{M}}$ whenever the model is clear from the context.
If $\mathcal{M},s \models \varphi$ we say that $\mathcal{M}$ is a model of $\varphi$. 
A formula is said to be \emph{satisfiable} if it has at least one model.
We say that $\varphi$ is a \emph{validity} and write $\models \varphi$ if $\neg \varphi$ is not satisfiable.
In addition to the operators defined by the syntax of $\mathcal{L}$,
we also have the derived operators such as $\bot$, $\to$, etc.
defined in the usual way.
A \emph{literal} is a formula that is of the form $p$ or $\neg p$ where $p \in \mathcal{AP}$.

The formula $L_0 \varphi$ has special significance in our logic,
as this formula means that there exists some transition to where $\varphi$ holds.
In fact, it follows in a straightforward manner from the semantics that
$\mathcal{M}, s \models L_0 \varphi$ if and only if $\trans{s}{\sat{\varphi}} \neq \Emptyset$.
We can therefore encode the usual box and diamond modalities in our logic in the following way.
\[\Diamond \varphi = L_0 \varphi \quad \Box \varphi = \neg \Diamond \neg \varphi.\]
Notice also that in general, the following schemes \emph{do not hold}.
\begin{align*}
  L_r \varphi \land L_r \psi &\rightarrow L_r(\varphi \land \psi) \\
  M_r \varphi \land M_r \psi &\rightarrow M_r(\varphi \land \psi)
\end{align*}
The reason that they do not hold in general is that there may be no transition to where $\varphi \land \psi$ holds,
i.e. $\neg L_0 (\varphi \land \psi)$. If we assume $L_0 (\varphi \land \psi)$,
then both schemes hold, as we show in Lemma \ref{lem:theorems}.
Another thing to note about the logic is that the formulas $L_r \varphi$ and $L_r \neg \varphi$
can both hold in the same model.
To see this, simply construct a state that
has two transitions with weight $x \geq r$
to two different states,
one where $\varphi$ holds and one where $\varphi$ does not hold.

\begin{example}\label{ex:logic}
  Consider again our model of a robot vacuum cleaner depicted in Figure \ref{fig:wts-example}.
  Perhaps we want a guarantee that it takes no more than one time unit to go
  from a waiting state to a charging state.
  This can be expressed by the formula ${\tt waiting} \to M_1{\tt charging}$,
  but since we know the only waiting state in our model is $s_1$
  this can be simplified to simply checking whether
  $\mathcal{M},s_1 \models M_1{\tt charging}$.
  We thus have to check that $\transr{s_1}{\sat{{\tt charging}}} \leq 1$.
  We do this by constructing the image set $\trans{s_1}{\sat{{\tt charging}}}$.
  Since we have $\sat{{\tt charging}} = \{s_3\}$,
  it follows that
  \[\trans{s_1}{\sat{{\tt charging}}} = \{1,2\}.\]
  Hence
  \[\transr{s_1}{\sat{{\tt charging}}} = 2 \not\leq 1,\]
  so $\mathcal{M},s_1 \not\models M_1{\tt charging}$.
\end{example}

\begin{lemma}\label{lem:invariance}
  Let $\mathcal{M} = (S, \rightarrow, \ell)$ be an image-finite WTS and $s \in S$.
  Let $T \subseteq S$ be a set such that all elements of $T$ satisfy exactly the same formulas,
  and furthermore for any $t \in T$ and $t' \notin T$,
  there exists a formula $\varphi$ such that $t \models \varphi$ and $t' \not\models \varphi$.
  Then there exists a formula $\varphi \in \mathcal{L}$ such that
  $\trans{s}{T} = \trans{s}{\sat{\varphi}}$.
\end{lemma}
\begin{proof}
  The idea of the proof is to repeatedly use the observation that if $t' \notin T$,
  then there exists a formula $\varphi$ such that $t' \not\models \varphi$ and $t \models \varphi$ for all $t \in T$.
  First pick some formula $\varphi_1$ such that $t \models \varphi_1$ for all $t \in T$.
  Then $T \subseteq \sat{\varphi_1}$, so $\trans{s}{T} \subseteq \trans{s}{\sat{\varphi_1}}$.
  If $\trans{s}{T} \subsetneq \trans{s}{\sat{\varphi_1}}$,
  then there must exist some $t_1 \notin T$
  such that $s \xrightarrow{r} t_1$ and $t_1 \models \varphi_1$.
  Since $t_1 \notin T$, there must exist some formula $\varphi_2$
  such that $t_1 \not \models \varphi_2$ and $t \models \varphi_2$ for all $t \in T$.
  We then get $\trans{s}{T} \subseteq \trans{s}{\sat{\varphi_1 \land \varphi_2}}$.
  Again, if $\trans{s}{T} \subsetneq \trans{s}{\sat{\varphi_1 \land \varphi_2}}$,
  then there must exist some $t_2 \notin T$ such that
  $s \xrightarrow{r} t_2$ and $t_2 \models \varphi_2$.
  Since $t_2 \notin T$, there must exist some formula $\varphi_3$
  such that $t_1 \not \models \varphi_3$ and $t \models \varphi_3$ for all $t \in T$.
  Since $\mathcal{M}$ is image-finite, there can only be finitely many states $t_i \notin T$ with $s \xrightarrow{r} t_i$,
  so continuing in the same way, we will eventually get a formula $\varphi_1 \land \dots \land \varphi_n$
  such that $\trans{s}{T} = \trans{s}{\sat{\varphi_1 \land \dots \land \varphi_n}}$.
\end{proof}

Next we show that our logic $\mathcal{L}$ is invariant under bisimulation,
which is also known as the Hennessy-Milner property.
In order to prove this result, we have to restrict our models to only those that are image-finite, as shown by the following example.

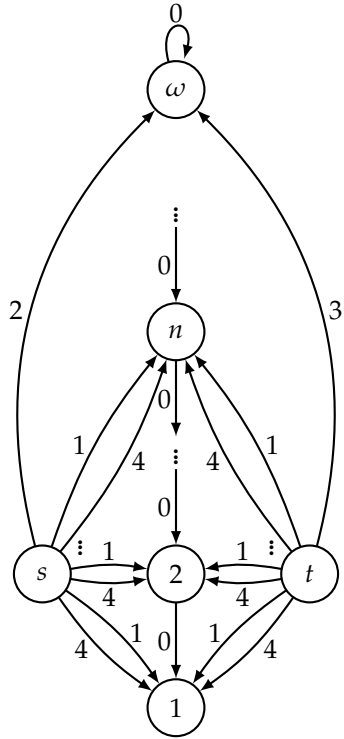
\begin{figure}
  \centering
  \begin{tikzpicture}[WTS, node distance=2cm]
    % States
    \node[state] (omega)                          {$\omega$};
    \node        (dots1) [below = of omega]       {\Huge $\vdots$};
    \node[state] (n) [below = of dots1]           {$n$};
    \node        (dots2) [below = of n]           {\Huge $\vdots$};
    \node[state] (2) [below = of dots2]           {$2$};
    \node[state] (1) [below = of 2]               {$1$};
    \node[state] (s) [left = 2cm of 2]            {$s$};
    \node[state] (t) [right = 2cm of 2]           {$t$};
    
    \coordinate (c1) at ($(s) + (1,0.8)$);
    \coordinate (c2) at ($(s) + (1.7,-0.7)$);
    \coordinate (c3) at ($(t) + (-1,0.8)$);
    \coordinate (c4) at ($(t) + (-1.7,-0.7)$);
    
    \node (dots3) at (c1) {\Huge $\vdots$};
    \node (dots5) at (c3) {\Huge $\vdots$};

    % Natural numbers
    \path (omega) edge[loop above] node[above] {$0$}  (omega);
    \path (dots1) edge node[left] {$0$}               (n);
    \path (n) edge node[left] {$0$}                   (dots2);
    \path (dots2) edge node[left] {$0$}               (2);
    \path (2) edge node[left] {$0$}                   (1);

    % s
    \path (s) edge[bend left] node[left = 0.1cm] {$2$} (omega);
    
    \path (s) edge[bend left = 10] node[left = 0.1cm] {$1$} (n);
    \path (s) edge[bend right = 10] node[right = 0.1cm] {$4$} (n);
    
    \path (s) edge[bend left = 10] node[above] {$1$} (2);
    \path (s) edge[bend right = 10] node[below] {$4$} (2);
    
    \path (s) edge[bend left = 10] node[right = 0.2cm] {$1$} (1);
    \path (s) edge[bend right = 10] node[left = 0.2cm] {$4$} (1);
    
    % t
    \path (t) edge[bend right] node[right = 0.1cm] {$3$} (omega);
    
    \path (t) edge[bend right = 10] node[right = 0.1cm] {$1$} (n);
    \path (t) edge[bend left = 10] node[left = 0.1cm] {$4$} (n);
    
    \path (t) edge[bend right = 10] node[above] {$1$} (2);
    \path (t) edge[bend left = 10] node[below] {$4$} (2);
    
    \path (t) edge[bend right = 10] node[left = 0.2cm] {$1$} (1);
    \path (t) edge[bend left = 10] node[right = 0.2cm] {$4$} (1);
  \end{tikzpicture}
  \captionof{figure}{$s$ and $t$ satisfy the same logical formulas, but $s \not\sim t$.}
  \label{fig:ex_non_invariance}
\end{figure}

\begin{example}
  Consider the WTS depicted in Figure \ref{fig:ex_non_invariance}
  with state space $S = \mathbb{N} \cup \{\omega, s, t\}$
  and $\ell(s') = \Emptyset$ for all $s' \in S$.
  The transition relation is given by $\omega \xrightarrow{0} \omega$,
  $s \xrightarrow{2} \omega$, $t \xrightarrow{3} \omega$,
  and $n+1 \xrightarrow{0} n$, $s \xrightarrow{1} n$, and $t \xrightarrow{1} n$
  for all $n \in \mathbb{N}$.
  
  Then we have that $s_1 \sim s_2$ if and only if $s_1 = s_2$,
  since any states in $\mathbb{N} \cup \{\omega\}$ can be distinguished by
  the number of steps they can take,
  and $s$ and $t$ can be distinguished by the fact that
  $\transl{s}{\{\omega\}} = 2 \neq 3 = \transl{t}{\{\omega\}}$.
  However, $s$ and $t$ satisfy all the same formulas,
  since any formula that holds in $\omega$ will also hold in $n$ for some $n \in \mathbb{N}$,
  and the weights on the transitions to $\omega$
  will therefore be masked by the bounds $1$ and $4$,
  and hence any formula can not distinguish between $s$ and $t$.
\end{example}

The proof strategy follows a classical pattern: The left to right direction is shown by induction on $\varphi$ for $\varphi \in \mathcal{L}$.
The right to left direction is shown by constructing a relation $\mathcal{R}$ relating those states that satisfy the same formulas and showing that this relation is a bisimulation relation.

\begin{theorem}[Bisimulation invariance]\label{thm:bisiminvariance}
  For any WTS $\mathcal{M} = (S, \rightarrow, \ell)$ and states $s,t \in S$ it holds that
  \[
  s \sim t \quad\text{implies}\quad \left[\forall \varphi \in \mathcal{L}.\; \mathcal{M},s \models \varphi \;\;\text{iff}\;\; \mathcal{M},t \models \varphi\right].
  \]
  Furthermore, if $\mathcal{M}$ is image-finite, then it also holds that
  \[
  \left[\forall \varphi \in \mathcal{L}.\; \mathcal{M},s \models \varphi \;\;\text{iff}\;\; \mathcal{M},t \models \varphi\right] \quad\text{implies}\quad s \sim t.
  \]
\end{theorem}
\begin{proof}
  We first show that $s \sim t$ implies $\mathcal{M},s \models \varphi$ if and only if
  $\mathcal{M},t \models \varphi$ for all $\varphi \in \mathcal{L}$ by induction on $\varphi$.
  The Boolean cases are trivial.
  If $\varphi = L_r \psi$, then
  we have $\transl{s}{\sat{\psi}} \geq r$,
  which implies that $\transl{s}{\sat{\psi}} \neq - \infty$.
  Assume towards a contradiction that $\transl{t}{\sat{\psi}} < r$.
  It can not be the case that $\transl{t}{\sat{\psi}} = - \infty$,
  hence it follows that $\sat{\psi}$ and $\trans{t}{\sat{\psi}}$ are non-empty,
  so there must exist some element $t' \in \sat{\psi}$
  such that $\transl{t}{\sat{\psi}} \leq \transl{t}{\{t'\}} < r$.
  Since $\sim$ is an equivalence relation,
  there must exists some $\sim$-equivalence class $T$ such that $t' \in T$.
  This means that $\{t'\} \subseteq T$,
  so that also $\transl{t}{T} \leq \transl{t}{\{t'\}} < r$.
  By the induction hypothesis we have that $T \subseteq \sat{\psi}$.
  Because $s \sim t$, we have that $\transl{s}{T} = \transl{t}{T} < r$,
  so by monotonicity we get $\transl{s}{\sat{\psi}} \leq \transl{s}{T} < r$,
  which is a contradiction.
  The $M_r$ case is handled similarly.

  For the reverse direction, assume that $\mathcal{M}$ is image-finite.
  We have to show that if for all $\varphi \in \mathcal{L}$,
  $\mathcal{M},s \models \varphi$ if and only if $\mathcal{M},t \models \varphi$ then $s \sim t$.
  To this end, we define a relation $\mathcal{R}$ on $S$ as
  \[
  \mathcal{R} = \left\{(s,t) \in S \times S \mid \forall \varphi \in \mathcal{L}.\; \mathcal{M},s \models \varphi \;\text{iff}\; \mathcal{M},t \models \varphi \right\}  .
  \]
  $\mathcal{R}$ is clearly an equivalence relation and $s \mathcal{R} t$.

  It is clear that $\ell(s) = \ell(t)$.
  Next we show that $\transl{s}{T} = \transl{t}{T}$ and $\transr{s}{T} = \transr{t}{T}$ for any $\mathcal{R}$-equivalence class $T$.
  Let $T \subseteq S$ be an $\mathcal{R}$-equivalence class.
  We first show that $\trans{s}{T} = \Emptyset$ if and only if $\trans{t}{T} = \Emptyset$.
  Assume that $\trans{s}{T} = \Emptyset$.
  By Lemma \ref{lem:invariance} there exists a formula $\varphi$
  such that $\trans{s}{T} = \trans{s}{\sat{\varphi}} = \Emptyset$,
  and therefore $s \not \models L_0 \varphi$.
  Now assume towards a contradiction that $\trans{t}{T} \neq \Emptyset$.
  Since $\mathcal{M}$ is image-finite,
  there must be a finite subset $T' \subseteq T$ such that
  $\trans{t}{T} = \trans{t}{T'}$.
  By Lemma \ref{lem:thetaunion}, we then get
  $\trans{t}{T} = \bigcup_{t' \in T'}\trans{t}{\{t'\}} \neq \Emptyset$,
  from which it follows that there must be some $t' \in T'$
  such that $\trans{t}{\{t'\}} \neq \Emptyset$.
  Since $t' \in T$, we must have $t' \models \varphi$,
  and therefore $t \models L_0 \varphi$,
  which contradicts the fact that $s \mathcal{R} t$
  and $s \not \models L_0 \varphi$.
  
  Now assume that $\trans{s}{T} \neq \Emptyset$ and $\trans{t}{T} \neq \Emptyset$.
  We need to show that $\transl{s}{T} = \transl{t}{T}$
  and $\transr{s}{T} = \transr{t}{T}$.
  We do this by contradiction, which gives us four cases to consider:
  $\transl{s}{T} < \transl{t}{T}$, $\transl{s}{T} > \transl{t}{T}$,
  $\transr{s}{T} < \transr{t}{T}$, and $\transr{s}{T} > \transr{t}{T}$.
  
  For the case of $\transl{s}{T} < \transl{t}{T}$,
  there exists $q \in \mathbb{Q}_{\geq 0}$ such that
  \[
  \transl{s}{T} < q < \transl{t}{T} .
  \]
  By Lemma \ref{lem:invariance}, there exists a formula $\varphi$
  such that $\transl{t}{T} = \transl{t}{\sat{\varphi}}$.
  Since $T \subseteq \sat{\varphi}$, we then obtain
  \[
  \transl{s}{\sat{\varphi}} \leq \transl{s}{T} < q < \transl{t}{T} = \transl{t}{\sat{\varphi}} ,
  \]
  which implies that $s \not \models L_q \varphi$ but $t \models L_q \varphi$,
  and thus we get a contradiction.
  The other cases are handled similarly.
\end{proof}

%% Metatheory
\section{Metatheory}
In this section we propose an axiomatisation for our logic that we prove not only sound,
but also complete with respect to the proposed semantics.

\subsection{Axiomatic System}\label{sec:axioms}
Let $r,s \in \mathbb{Q}_{\geq 0}$.
Then the deducibility relation $\vdash \, \subseteq 2^{\mathcal{L}} \times \mathcal{L}$
is a classical conjunctive deducibility relation,
and is defined as the smallest relation which satisfies
the axioms of propositional logic in addition to the axioms given in Table \ref{tab:axioms}.
We will write $\vdash \varphi$ to mean $\Emptyset \vdash \varphi$,
and we say that a formula or a set of formulas is \emph{consistent} if it can not derive $\bot$.

\begin{table}
  \centering
  \begin{tabular}{l l l}
    \hline
    (A$1$):   & $\vdash \neg L_0 \bot$ & \\
    (A$2$):   & $\vdash L_{r + q}\varphi \rightarrow L_r \varphi$ & if $q > 0$ \\
    (A$2'$):  & $\vdash M_r\varphi \rightarrow M_{r + q} \varphi$ & if $q > 0$ \\
    (A$3$):   & $\vdash L_r \varphi \land L_q \psi \rightarrow L_{\min\{r,q\}}(\varphi \lor \psi)$ & \\
    (A$3'$):  & $\vdash M_r \varphi \land M_q \psi \rightarrow M_{\max\{r,q\}}(\varphi \lor \psi)$ & \\
    (A$4$):   & $\vdash L_r(\varphi \lor \psi) \rightarrow L_r \varphi \lor L_r \psi$ & \\
    (A$5$):   & $\vdash \neg L_0 \psi \rightarrow (L_r \varphi \rightarrow L_r(\varphi \lor \psi))$ & \\
    (A$5'$):  & $\vdash \neg L_0 \psi \rightarrow (M_r \varphi \rightarrow M_r(\varphi \lor \psi))$ & \\
    (A$6$):   & $\vdash L_{r + q}\varphi \rightarrow \neg M_r\varphi$ & if $q > 0$ \\
    (A$7$):   & $\vdash M_r \varphi \rightarrow L_0 \varphi$ & \\
    (R$1$):   & $\vdash \varphi \rightarrow \psi \implies \vdash (L_r \psi \land L_0 \varphi) \rightarrow L_r \varphi$ & \\
    (R$1'$):  & $\vdash \varphi \rightarrow \psi \implies \vdash (M_r \psi \land L_0 \varphi) \rightarrow M_r \varphi$ & \\
    (R$2$):   & $\vdash \varphi \rightarrow \psi \implies \vdash L_0 \varphi \rightarrow L_0 \psi$ & \\
    \hline
  \end{tabular}
  \caption{The axioms for our axiomatic system, where $\varphi, \psi \in \mathcal{L}$ and $q,r \in \mathbb{Q}$.}
  \label{tab:axioms}
\end{table}

The axioms presented in Table \ref{tab:axioms} bear some resemblance to the axiomatic systems of \cite{MCL12} and \cite{cardelli2011a}. 
Notably, our axiom A2 is almost identical to A2 of these works and capture similar properties about the systems being studied,
with the major difference being that we reason about transition weights whereas the aforementioned works reason about rates or probabilities of transitions.
Also worth noting here is the similarity between the rule R1 of these works and R1 of our axiomatic system.
A notable difference is that we do not have the additive properties of measures for disjoint sets (since we are not working with probability measures),
as is captured by the axioms A3 and A4 of these works.
Also, in one of the axiomatisations of \cite{MCL12}, the axioms A2 and A2$'$
are not axioms, but can be derived from the axioms.

Rules R2 and R3 of \cite{MCL12} and \cite{cardelli2011a} reflect the Archimedean property of rationals,
and while similar axioms can be proven sound in our setting,
these were not needed to show our completeness result.
We suspect, however, that if we were to pursue strong completeness,
infinitary axioms similar to these would be needed.

Axiom A1 captures the notion that since $\bot$ is never satisfied,
we can never take a transition to where $\bot$ holds.
Axiom A2 says that if we know some value is the lower bound for going to
where $\varphi$ holds, then any lower value is also a lower bound for going to where
$\varphi$ holds. Axiom A2$'$ is the analogue for upper bounds.
Axioms A3-A4 show how $L_r$ and $M_r$ distribute over conjunction and disjunction.
The version of axiom A4 where $L_r$ is replaced with $M_r$
is also sound, but as we show in Lemma \ref{lem:theorems},
it can be proven from the other axioms.
Axioms A5 and A5$'$ say that if it is not possible
to take a transition to where $\psi$ holds,
then including the states where $\psi$ holds does not change the bounds.
Axioms A6 and A7 show the relationship between $L_r$ and $M_r$.
In particular, A6 ensures that all bounds are well-formed.
Notice also that the contrapositive of axiom A2 and A7
together gives us that $\neg L_0 \varphi$
implies $\neg L_r \varphi$ and $\neg M_r \varphi$ for any $r \in \mathbb{Q}_{\geq 0}$.
The rules R1 and R1$'$ give a sort of monotonicity for $L_r$ and $M_r$,
and rule R2 says that if $\psi$ follows from $\varphi$,
then if it is possible to take a transition to where $\varphi$ holds,
it is also possible to take a transition to where $\psi$ holds.

We now show some of the theorems which can be deduced from the axioms.
T1, T1$'$, and T5 together complete the distributivity properties for conjunction and disjunction.
T2 and T2$'$ make precise the intuitively clear idea that if two formulas are equivalent,
then their upper and lower bounds should also be the same.
T3 extends axiom A1 to hold for any $r \geq 0$,
and T4 then extends this to any $\varphi$ which implies $\bot$.
\begin{lemma}\label{lem:theorems}
  From the axioms listed in Table \ref{tab:axioms} we can derive the following theorems:\\
  \begin{tabular}{l l}
    \emph{(T1):}    & $\vdash (L_r \varphi \land L_q \psi \land L_0(\varphi \land \psi)) \to L_{\max\{r,q\}}(\varphi \land \psi)$ \\
    \emph{(T1$'$):} & $\vdash (M_r \varphi \land M_q \psi \land L_0(\varphi \land \psi)) \to M_{\min\{r,q\}}(\varphi \land \psi)$ \\
    \emph{(T2):}    & $\vdash \varphi \leftrightarrow \psi \implies \vdash L_r \varphi \leftrightarrow L_r \psi$ \\
    \emph{(T2$'$):} & $\vdash \varphi \leftrightarrow \psi \implies \vdash M_r \varphi \leftrightarrow M_r \psi$ \\
    \emph{(T3):}    & $\vdash \neg L_r \bot, \quad r \geq 0$ \\
    \emph{(T4):}    & $\vdash \varphi \to \bot \implies \vdash \neg L_r \varphi,\quad r \geq 0$ \\
    \emph{(T5):}    & $\vdash M_r(\varphi \lor \psi) \rightarrow M_r \varphi \lor M_r \psi$
  \end{tabular}
\end{lemma}
\begin{proof}
  \hfill
  \begin{description}
  \item[T1]
    Rule R1 implies
    \[
    \vdash \neg L_q (\varphi \land \psi) \rightarrow (\neg L_q \varphi \lor \neg L_0 (\varphi \land \psi))  ,
    \]
    so also
    \[
    \vdash \neg L_q (\varphi \land \psi) \rightarrow (\neg L_q \varphi \lor \neg L_0 (\varphi \land \psi) \lor \neg L_r \psi)  .
    \]
    This is equivalent to
    \[
      \vdash (L_r \varphi \land L_q \psi \land L_0 (\varphi \land \psi)) \rightarrow L_q (\varphi \land \psi)  .
      \]

  \item[T1$'$]
    Similar to T1.

  \item[T2]
    Suppose $\vdash \varphi \leftrightarrow \psi$.
    We have that $\vdash L_r \varphi \rightarrow L_0 \varphi$ by A2
    and $\vdash L_0 \varphi \rightarrow L_0 \psi$ by R2.
    Hence $\vdash L_r \varphi \rightarrow (L_r \varphi \land L_0 \psi)$,
    so $\vdash L_r \varphi \rightarrow L_r \psi$ by R1.
    A similar argument shows that $\vdash L_r \psi \rightarrow L_r \varphi$,
    so $\vdash L_r \varphi \leftrightarrow L_r \psi$.
    
  \item[T2$'$]
    Similar to T2.
    
  \item[T3]
    From axiom A1 we know that $\vdash \neg L_0 \bot$ which,
    by the contrapositive of A2, implies $\vdash \neg L_r \bot$ for any $r > 0$.
    
  \item[T4]
    Suppose $\vdash \varphi \to \bot$.
    We know for any $\psi \in \mathcal{L}$ that $\vdash \bot \to \psi$ and therefore $\vdash \varphi \to \bot \implies \vdash \varphi \leftrightarrow \bot$.
    From A1 we know that $\vdash \neg L_0 \bot$ and from T3 that $\vdash \neg L_r \bot$ for any $r > 0$ implying, by T2, that $\vdash \neg L_r \varphi$ for any $r \geq 0$.
    
  \item[T5]
    By axiom A7 we get $\vdash M_r (\varphi \lor \psi) \rightarrow L_0(\varphi \lor \psi)$
    and A4 gives
    \[\vdash L_0(\varphi \lor \psi) \rightarrow L_0 \varphi \lor L_0 \psi.\]
    Hence we get $\vdash M_r(\varphi \lor \psi) \rightarrow (M_r(\varphi \lor \psi) \land L_0 \varphi) \lor (M_r(\varphi \lor \psi) \land L_0 \psi)$.
    Since $\vdash \varphi \rightarrow (\varphi \lor \psi)$
    and $\vdash \psi \rightarrow (\varphi \lor \psi)$,
    rule R1$'$ then gives
    \[\vdash M_r(\varphi \lor \psi) \rightarrow M_r \varphi \lor M_r \psi. \qedhere\]
  \end{description}
\end{proof}

Next we prove that our axioms are indeed sound.

\begin{theorem}[Soundness]\label{thm:soundness}\hfill
  \[
  \vdash \varphi \quad\text{implies}\quad \models \varphi.
  \]
\end{theorem}
\begin{proof}
  The soundness of each axiom is easy to show,
  and many of them use the distributive property from Lemma \ref{lem:thetaunion}.
  Here we prove the soundness for a few of the more interesting axioms.
  \begin{description}
  \item[A3]
    Suppose $\mathcal{M},s \models L_r \varphi \land L_q \psi$
    implying that $\mathcal{M},s \models L_r \varphi$ and $\mathcal{M},s \models L_q \psi$,
    implying further that $\transl{s}{\sat{\varphi}} \geq r$ and $\transl{s}{\sat{\psi}} \geq q$.
    
    By Lemma \ref{lem:thetaunion} we must have that 
    \[
    \trans{s}{\sat{\varphi \lor \psi}}
    = \trans{s}{\sat{\varphi} \cup \sat{\psi}}
    = \trans{s}{\sat{\varphi}} \cup \trans{s}{\sat{\psi}}
    \]
    and because $\transl{s}{\sat{\varphi}} \geq r$ and $\transl{s}{\sat{\psi}} \geq q$
    we must have
    \[
    \transl{s}{\sat{\varphi \lor \psi}} = \inf \trans{s}{\sat{\varphi}} \cup \trans{s}{\sat{\psi}}\geq \min\left\{r,q\right\}
    \]
    implying $\mathcal{M},s \models L_{\min\{r,q\}} (\varphi \lor \psi)$.
    
  \item[A4]
    Suppose $\mathcal{M},s \models L_r (\varphi \lor \psi)$ implying that
    \[
    \transl{s}{\sat{\varphi \lor \psi}}
    = \inf \trans{s}{\sat{\varphi}} \cup \trans{s}{\sat{\psi}}
    \geq r  .
    \]
    This implies that at least one of $\trans{s}{\sat{\varphi}}$
    and $\trans{s}{\sat{\psi}}$ is non-empty.
    If $\trans{s}{\sat{\varphi}} \neq \Emptyset$, then $\transl{s}{\sat{\varphi}} \geq r$,
    and also if $\trans{s}{\sat{\psi}} \neq \Emptyset$, then $\transl{s}{\sat{\psi}} \geq r$,
    so at least one of $\mathcal{M},s \models L_r \varphi$
    and $\mathcal{M},s \models L_r \psi$ must hold.
    Hence $\mathcal{M},s \models L_r \varphi \lor L_r \psi$.
    
  \item[A6]
    Suppose $\mathcal{M},s \models L_{r+q} \varphi$ implying that
    \[
    \transl{s}{\sat{\varphi}} = \inf \trans{s}{\sat{\varphi}} \geq r+q  .
    \]
    It is clear that $\inf \trans{s}{\sat{\varphi}} \leq \sup \trans{s}{\sat{\varphi}}$, so
    \[
    \transr{s}{\sat{\varphi}} = \sup \trans{s}{\sat{\varphi}} \geq \inf \trans{s}{\sat{\varphi}} \geq r + q > r  .
    \]
    Therefore, it cannot be the case that $\mathcal{M},s \models M_r \varphi$
    and thus $\mathcal{M},s \models \neg M_r \varphi$.
    
  \item[R1]
    Suppose $\models \varphi \to \psi$ implying that $\sat{\varphi} \subseteq \sat{\psi}$,
    implying further, by the monotonicity of $\theta$,
    that $\trans{s}{\sat{\varphi}} \subseteq \trans{s}{\sat{\psi}}$. 
    Suppose further that $\mathcal{M},s \models L_r \psi \land L_0 \varphi$
    implying $\mathcal{M},s \models L_r \psi$ and $\mathcal{M},s \models L_0 \varphi$,
    implying further that
    \[
    \transl{s}{\sat{\psi}} = \inf \trans{s}{\sat{\psi}} \geq r \quad \text{and} \quad \trans{s}{\sat{\varphi}} \neq \Emptyset  .
    \]
    Since $\trans{s}{\sat{\varphi}}$ is non-empty, we then get that
    \[
    \inf \trans{s}{\sat{\varphi}} \geq \inf \trans{s}{\sat{\psi}} \geq r  ,
    \]
    which means that $\mathcal{M},s \models L_r \varphi$. \qedhere
  \end{description}
\end{proof}

\subsection{Finite Model Property and Completeness}\label{sec:finitemodel-completeness}
With our axiomatisation proven sound we are now ready to present our main results,
namely that our logic has the finite model property and that our axiomatisation is complete. 

To show the finite model property we will adapt
the classical filtration method to our setting.
Starting from an arbitrary formula $\rho$,
we define a finite fragment of our logic, $\mathcal{L}[\rho]$,
which we then use to construct a finite model for $\rho$.
The main difference from the classical filtration method
is that we must find an upper and a lower bound for the transitions
in the model.
For an arbitrary formula $\rho \in \mathcal{L}$ we define the following based on $\rho$:
\begin{itemize}
  \item Let $Q_{\rho} \subseteq \mathbb{Q}_{\geq 0}$ be the set of all rational numbers
  $r \in \mathbb{Q}_{\geq 0}$ such that $L_r$ or $M_r$ appears in the syntax of $\rho$.

  \item Let $\Sigma_{\rho}$ be the set of all atomic propositions $p \in \mathcal{AP}$
    such that $p$ appears in the syntax of $\rho$.

  \item The \emph{granularity} of $\rho$, denoted as $gr(\rho)$,
    is the least common denominator of all the elements in $Q_{\rho}$.

  \item The \emph{range} of $\rho$, denoted as $R_{\rho}$, is defined as
      \[R_{\rho} =
      \begin{cases}
        \Emptyset & \text{if}\; Q_{\rho} = \Emptyset\\
        I_\rho \cup \{0\}
        & \text{otherwise},
      \end{cases}\]
    where $I_\rho = \left\{
          q \in \mathbb{Q}_{\geq 0} \mid
          \exists j \in \mathbb{N}.\; q = \frac{j}{gr(\rho)} \;\text{and}\;
          \min Q_{\rho} \leq q \leq \max Q_{\rho} 
        \right\}$.
    Here the granularity is used to pick out finitely many numbers in the interval.
    Note that we need to add $0$ to $R_{\rho}$
    whether or not $\rho$ actually contains $0$ in any of its modalities.
    This is because, as we have pointed out before, formulas involving $L_0$
    have special significance in our logic.
  \item The \emph{modal depth} of $\rho$, denoted as $md(\rho)$, is defined inductively as:
    \[
    md(\rho) =
    \begin{cases}
      0 & 
      \text{if}\; \rho = p \in \mathcal{AP}\\

      md(\varphi) & 
      \text{if}\; \rho = \neg \varphi\\

      \max\left\{md(\varphi_1),md(\varphi_2)\right\} & 
      \text{if}\; \rho = \varphi_1 \land \varphi_2\\
      1 + md(\varphi) &
      \text{if}\; \rho = L_r \varphi \;\text{or}\; \rho = M_r \varphi  .
    \end{cases}
    \]
\end{itemize}
Since all formulas are finite, the modal depth is always a non-negative integer.
The \emph{language} of $\rho$, denoted by $\mathcal{L}[\rho]$, is defined as
\[
\mathcal{L}[\rho] = \{\varphi \in \mathcal{L} \mid R_\varphi \subseteq R_{\rho} , md(\varphi) \leq md(\rho) \;\text{and}\; \Sigma_\varphi \subseteq \Sigma_{\rho} \},
\]
and we take $\mathcal{L}_{\leftrightarrow}[\rho]$ to be the Lindenbaum algebra
of $\mathcal{L}[\rho]$, i.e. the quotient with respect to logical equivalence.
The Lindenbaum algebra is a Boolean algebra with equivalence classes as elements.
Note that the quotient
\[\quot : \mathcal{L}[\rho] \rightarrow \mathcal{L}_{\leftrightarrow}[\rho]\]
is a homomorphism between Boolean algebras,
and therefore preserves the structure of $\mathcal{L}[\rho]$.
For each element $x \in \mathcal{L}_{\leftrightarrow}[\rho]$,
we fix now a formula $\varphi \in x$ to be the representative of that equivalence class,
and we write $\hat{\varphi}$ for $x$.
The order $\leq$ in $\mathcal{L}_{\leftrightarrow}[\rho]$
is then given by $\hat{\varphi} \leq \hat{\psi}$
if and only if $\vdash \varphi \rightarrow \psi$.
The join and meet in $\mathcal{L}_\leftrightarrow[\rho]$ are given by
\[\hat{\varphi} \lor \hat{\psi} = \quot(\varphi \lor \psi) \quad \hat{\varphi} \land \hat{\psi} = \quot(\varphi \land \psi),\]
and complement is given by
\[\neg \hat{\varphi} = \quot(\neg \varphi).\]

Note here the difference between $\quot(\varphi)$ and $\hat{\varphi}$.
The quotient $h$ sends $\varphi$ to its equivalence class $x \in \mathcal{L}_{\leftrightarrow}[\rho]$.
However, it may be the case that $\varphi$ is not the representative for $x$,
but some other formula $\psi$ is. In that case we have $\quot(\varphi) = x = \hat{\psi}$.
On the other hand, $\hat{\varphi}$ denotes both that $\varphi \in \hat{\varphi}$,
and also that $\varphi$ is the chosen representative of its equivalence class,
which ensures that in this case we have $\quot(\varphi) = \hat{\varphi}$.

The idea is that $\Sigma_\rho$ ensures that only finitely many atomic propositions are used,
$R_\rho$ ensures that only finitely many weights on the modalities are used,
and $md(\rho)$ puts a bound on the modal depth of formulas.
The language $\mathcal{L}[\rho]$ itself is not finite,
but contains only finitely many logically non-equivalent formulas.
Hence $\mathcal{L}_{\leftrightarrow}[\rho]$ must be finite,
and as we shall see, it contains all the information necessary to
construct a model for $\rho$.

\begin{proposition}
  The language $\mathcal{L}_{\leftrightarrow}[\rho]$ is finite.
\end{proposition}
\begin{proof}
  Let $\mathcal{L}_{\leftrightarrow}^n[\rho]$ be the subset of $\mathcal{L}_{\leftrightarrow}[\rho]$
  which only contains formulas of modal depth $n$.
  Then it is clear that
  \[\mathcal{L}_{\leftrightarrow}[\rho] = \bigcup_{i = 0}^{md(\rho)} \mathcal{L}^i_{\leftrightarrow}[\rho].\]
  We will now prove by induction on the modal depth that for each $i$,
  $\mathcal{L}_{\leftrightarrow}^i[\rho]$ is finite.
  
  $i = 0$: In this case, each element of $\mathcal{L}_{\leftrightarrow}^0[\rho]$
    is a Boolean combination of atomic propositions in $\Sigma_\rho$.
    There are $2^{2^{|\Sigma_\rho|}}$ non-equivalent such formulas,
    so this set is finite.
  
  $i > 0$: Each element of $\mathcal{L}_{\leftrightarrow}^i[\rho]$
    is a Boolean combination of formulas of the form $L_r \varphi$ and $M_r \varphi$,
    where $\varphi \in \mathcal{L}_\leftrightarrow^{j}[\rho]$ for some $j < i$ and $r \in R_\rho$.
    By induction hypothesis, we know that there are only finitely many such $\varphi$.
    We know from Lemma \ref{lem:theorems} that if $\varphi$ and $\psi$ are logically equivalent,
    then $L_r \varphi$ and $L_r \psi$ as well as $M_r \varphi$ and $M_r \psi$ are also logically equivalent.
    Since $R_\rho$ is finite, we conclude that $\mathcal{L}_\leftrightarrow^i[\rho]$ is finite.
\end{proof}

In order to define the model,
we need the standard notions of filters and ultrafilters on Boolean algebras \cite{halmos2009}.
A non-empty subset of a Boolean algebra $B$ is called a \emph{filter}
if it is upward-closed with respect to the order,
and closed under finite meets.
A filter $F$ is \emph{proper} if $F \neq B$.
An \emph{ultrafilter} is a proper filter which is maximal
in the sense of set inclusion.

The following property of ultrafilters is often useful.

\begin{lemma}
  For an ultrafilter $F$ of $\mathcal{L}_{\leftrightarrow}[\rho]$ it holds that for any
  $\varphi \in \mathcal{L}[\rho]$,
  either $\quot(\varphi) \in F$ or $\neg \quot(\varphi) \in F$,
  but not both.
\end{lemma}

We let $\mathcal{U}[\rho]$ denote the set of all ultrafilters on $\mathcal{L}_{\leftrightarrow}[\rho]$.
Since $\mathcal{L}_{\leftrightarrow}[\rho]$ is finite, $\mathcal{U}[\rho]$
is also finite and consequently, any ultrafilter $u \in \mathcal{U}[\rho]$
must be a finite set.
For any set $\Phi \subseteq \mathcal{L}_{\leftrightarrow}[\rho]$,
the characteristic formula of $\Phi$, denoted $\tas{\Phi}$,
is defined as
\[\tas{\Phi} = \bigwedge_{\hat{\varphi} \in \Phi} \varphi.\]
Note that $\tas{\Phi} \in \mathcal{L}[\rho]$ is a finite formula,
and that if $u \in \mathcal{U}[\rho]$, then $\quot(\tas{u}) \in u$.

We will now construct a (finite) model, $\mathcal{M}_{\rho}$,
for $\rho$ with state space $\mathcal{U}[\rho]$.
In order to define the transition relation
$\rightarrow_\rho \subseteq \mathcal{U}[\rho] \times \mathbb{R}_{\geq 0} \times \mathcal{U}[\rho]$,
we consider any two ultrafilters $u,v \in \mathcal{U}[\rho]$ and define two functions
\[L,M : \mathcal{U}[\rho] \times \mathcal{U}[\rho] \to 2^{R_{\rho}}\]
as
\[
L(u,v) = \{r \mid \quot(L_r \tas{v}) \in u\} 
\quad\text{and}\quad
M(u,v) = \{s \mid \quot(M_s \tas{v}) \in u\}  .
\]

The following lemma establishes a relationship between $L$ and $M$,
that we will need to define the transition relation.
The lemma is a straightforward consequence of axiom $A7$.

\begin{lemma}\label{lem:emptynonempty}
  Given any ultrafilters $u,v \in \mathcal{U}[\rho]$,
  it can not be the case that $L(u,v) = \Emptyset$ and $M(u,v) \neq \Emptyset$.
\end{lemma}
\begin{proof}
  Assume towards a contradiction that $L(u,v) = \Emptyset$
  and $M(u,v) \neq \Emptyset$. Then there exists some $r \in Q_\rho$ such that
  \[\quot(\neg L_0 \tas{v}) \in u \quad \text{and} \quad \quot(M_r \tas{v}) \in u.\]
  However, by axiom A7, this implies that $\quot(L_0 \tas{v}) \in u$,
  which is a contradiction.
\end{proof}

We can now define the transition relation in terms of $L(u,v)$ and $M(u,v)$.
In Figure \ref{fig:finitemodel}, we have illustrated the different cases that we must consider.
Here, the area between $\min Q_\rho$ and $\max Q_\rho$
is the only part that the restricted language $\mathcal{L}[\rho]$ can speak about.
The arches represent the interval within which transitions with that weight are possible.
For any of the arches in the figure, we have the following correspondence with $L_r$ and $M_r$.
\begin{itemize}
  \item If a number $r$ on the real line is contained within the arch,
    then we have $\quot(\neg L_r \tas{v}) \in u$ and $\quot(\neg M_r \tas{v}) \in u$.
  \item If a number $r$ on the real line is to the left of the arch,
    then we have $\quot(L_r \tas{v}) \in u$ and $\quot(\neg M_r \tas{v}) \in u$.
  \item If a number $r$ on the real line is to the right of the arch,
    then we have $\quot(M_r \tas{v}) \in u$ and $\quot(\neg L_r \tas{v}) \in u$.
\end{itemize}
In case (a) in Figure \ref{fig:finitemodel},
we therefore have $L(u,v) \neq \Emptyset$ and $M(u,v) \neq \Emptyset$,
so we have all the information we need to define the transition.
In case (b) and (f), we have $L(u,v) \neq \Emptyset$ and $M(u,v) = \Emptyset$,
since there exist numbers within the interval $[\min Q_\rho, \max Q_\rho]$
that are to the left of these arches, but none that are to the right.
This means that we have enough information to define the minimum transition,
but we do not know what the maximum transition is.
Note that we can not simply say that the maximum transition is $\max Q_\rho$,
because that would imply $\quot(M_{\max Q_\rho} \tas{v}) \in u$,
but we know that $M(u,v) = \Emptyset$.
Hence we need to pick a number that is to the right of $\max Q_\rho$ as the maximum.
In case (d), we have both $L(u,v) = \Emptyset$ and $M(u,v) = \Emptyset$.
This implies that $\quot(\neg L_0 \tas{v}) \in u$,
which means that there should be no transition from $u$ to $v$.
In case (c) and (e), we have $L(u,v) = \Emptyset$ and $M(u,v) \neq \Emptyset$,
but according to Lemma \ref{lem:emptynonempty} these cases can never occur.

\begin{figure}
  \centering
  \resizebox{\textwidth}{!}{
    \begin{tikzpicture}
      %% X-axis
      \draw [->,thick] (-3,0) -- (11,0);
      %% Zero
      \draw[shift={(-3,0)}] (0pt,3pt) -- (0pt,-3pt);
      \draw[shift={(-3,0)}] (0pt,3pt) -- (0pt,-3pt) node[below] {$0$};
      
      %% Lower bound
      \draw[semithick] (1.2, -0.5) node[below] {$\min Q_\rho$} |- (1, -0.5) -- (1, 2.5) |- (1.2, 2.5);
      %% Upper bound
      \draw[semithick] (6.8, -0.5) node[below] {$\max Q_\rho$} |- (7, -0.5) -- (7, 2.5) |- (6.8, 2.5);

      %% Case a
      \coordinate (1) at (3,0);
      \coordinate (2) at (5,0);
      \draw (1) to [bend left=60] node[above] {(a)} (2);
      
      %% Case b
      \coordinate (3) at (6,0);
      \coordinate (4) at (7,0.5);
      \coordinate (5) at (8,0);
      \draw (3) to [bend left=25] node[above] {(b)} (4);
      \draw[dashed] (4) to [bend left=25] (5);
      
      %% Case c
      \coordinate (6) at (0,0);
      \coordinate (7) at (1,0.5);
      \coordinate (8) at (2,0);
      \draw[dashed] (6) to [bend left=25] (7);
      \draw (7) to [bend left=25] node[above] {(c)} (8);
      
      %% Case d
      \coordinate (9) at (-0.5,0);
      \coordinate (10) at (8.5,0);
      \coordinate (15) at (1,1.5);
      \coordinate (16) at (7,1.5);
      \draw[dashed] (9) to [bend left=15] (15);
      \draw (15) to [bend left=26] node[above] {(d)} (16);
      \draw[dashed] (16) to [bend left=15] (10);
      
      %% Case e
      \coordinate (11) at (-2.7,0);
      \coordinate (12) at (-0.7,0);
      \draw[dashed] (11) to [bend left=60] node[above] {(e)} (12);
      
      %% Case f
      \coordinate (13) at (8.7,0);
      \coordinate (14) at (10.7,0);
      \draw[dashed] (13) to [bend left=60] node[above] {(f)} (14);
    \end{tikzpicture}
  }
  \caption{When constructing a transition from $u$ to $v$, we will only have information about what happens in the region $Q_\rho$ and at $0$. The line represents the non-negative real line and the arches represent the transitions that would be possible in a full model (i.e. one not restricted to $\mathcal{L}[\rho]$). The dashed part of the arches represent the part of the transition that we do not have information about.}
  \label{fig:finitemodel}
\end{figure}
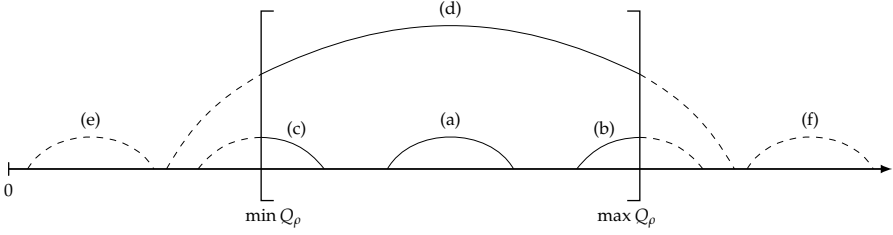

We therefore distinguish the following three cases in order to define the transition relation:
\begin{enumerate}
  \item If $L(u,v) \neq \Emptyset$ and $M(u,v) \neq \Emptyset$,
    then we add the two transitions $u \xrightarrow{r_1} v$ and $u \xrightarrow{r_2} v$
    where $r_1 = \max L(u,v)$ and $r_2 = \min M(u,v)$. \label{item:trans1}
  \item If $L(u,v) \neq \Emptyset$ and $M(u,v) = \Emptyset$,
    then we add the two transitions $u \xrightarrow{r_1} v$ and $u \xrightarrow{r_2} v$
    where $r_1 = \max L(u,v)$ and $r_2 = \max Q_\rho + \frac{1}{gr(\rho)}$. \label{item:trans2}
  \item If $L(u,v) = \Emptyset$ and $M(u,v) = \Emptyset$,
    then there is no transition from $u$ to $v$. \label{item:trans3}
\end{enumerate}
The following lemma tells us that these transitions are well-formed,
i.e. that the lower bound on transitions is less than or equal to the upper bound.

\begin{lemma}\label{lem:leq}
  For any ultrafilters $u,v \in \mathcal{U}[\rho]$,
  if $L(u,v) \neq \Emptyset$ and $M(u,v) \neq \Emptyset$,
  then $\max L(u,v) \leq \min M(u,v)$.
\end{lemma}
\begin{proof}
  Assume towards a contradiction that $\max L(u,v) > \min M(u,v)$.
  Then there exist $q,q' \in Q_\rho$ such that $q > q'$,
  $\quot(L_q \tas{v}) \in u$ and $\quot(M_{q'} \tas{v}) \in u$.
  Since $q > q'$, axiom A6 gives $\quot(\neg M_{q'} \tas{v}) \in u$,
  which is a contradiction.
\end{proof}

Finally we define the labelling function $\ell_\rho : \mathcal{U}[\rho] \rightarrow 2^{\mathcal{AP}}$
for any $u \in \mathcal{U}[\rho]$ as $ \ell_\rho(u) = \{p \in \mathcal{AP} \mid p \in u\} $.
We then have a model $\mathcal{M}_\rho = (\mathcal{U}[\rho], \rightarrow_\rho, \ell_\rho)$, and it is not difficult to prove that $\mathcal{M}_\rho$ is a WTS.
Before we can prove the truth lemma,
we need the following technical lemma.

\begin{lemma}\label{lem:filtration}
  For any consistent formula $\varphi \in \mathcal{L}[\rho]$,
  if $[\mathcal{M}_\rho,u \models \varphi$ iff $\quot(\varphi) \in u]$, then
  \[
  \bigvee_{v \in \sat{\varphi}} \quot(\tas{v}) \in u \quad \text{iff} \quad \quot(\varphi) \in u.
  \]
\end{lemma}
\begin{proof}
  Suppose $\bigvee_{v \in \sat{\varphi}} \quot(\tas{v}) \in u$.
  Assume towards a contradiction that $\quot(\neg \tas{v}) \in u$ for all $v \in \sat{\varphi}$.
  Then, since $u$ is an ultrafilter,
  we must have $\bigwedge_{v \in \sat{\varphi}} \quot(\neg \tas{v}) \in u$,
  which means that $\neg \bigvee_{v \in \sat{\varphi}} \quot(\tas{v}) \in u$,
  which is a contradiction.
  Hence there exists some $v' \in \sat{\varphi}$ such that $\quot(\tas{v'}) \in u$.
  If $\hat{\psi} \in v'$, then $\vdash \tas{v'} \rightarrow \psi$,
  so $\hat{\psi} \in u$ because $u$ is an ultrafilter.
  Since $v' \in \sat{\varphi}$, we have by assumption that $\quot(\varphi) \in v'$,
  so we get $\quot(\varphi) \in u$.
  
  Suppose $\quot(\varphi) \in u$, which by assumption means that $u \in \sat{\varphi}$,
  so
  \[\vdash \tas{u} \rightarrow \bigvee_{v \in \sat{\varphi}} \tas{v}.\]
  Since $u$ is an ultrafilter, we have $\quot(\tas{u}) \in u$,
  and hence $\bigvee_{v \in \sat{\varphi}} \quot(\tas{v}) \in u$.
\end{proof}

We are now in a position to state and prove the truth lemma,
which says that an ultrafilter satisfies a formula in our model
if and only if that formula is included in the ultrafilter.

\begin{lemma}[Truth lemma]\label{lem:truth}
  If $\rho \in \mathcal{L}$ is a consistent formula,
  then for all $\varphi \in \mathcal{L}[\rho]$
  and $u \in \mathcal{U}[\rho]$ we have
  \[\mathcal{M}_\rho,u \models \varphi \quad \text{iff} \quad \quot(\varphi) \in u.\]
\end{lemma}
\begin{proof}%[Proof of Lemma \ref{lem:truth}]
  The proof is by induction on the structure of $\varphi$.
  The Boolean cases are trivial.
  For the case $\varphi = L_r \psi$, we proceed as follows.

    ($\implies$)
      Assume $\mathcal{M}_\rho, u \models L_r \psi$,
      meaning that $\transl{u}{\sat{\psi}} \geq r$.
      It can not be the case that $\trans{u}{\sat{\psi}} = \Emptyset$,
      because otherwise $\transl{u}{\sat{\psi}} = - \infty$,
      and we have assumed $\transl{u}{\sat{\psi}} \geq r$.
      It also can not be the case that $\sat{\psi} = \Emptyset$,
      because otherwise $\trans{u}{\sat{\psi}} = \Emptyset$.
      We can partition all the ultrafilters $v \in \sat{\psi}$
      as follows.
      Let $E = \{v \in \sat{\psi} \mid L(u,v) = \Emptyset\}$
      and $N = \{v \in \sat{\psi} \mid L(u,v) \neq \Emptyset\}$.
      We then get that $E \cap N = \Emptyset$, $E \cup N = \sat{\psi}$,
      $\quot(\neg L_0 \tas{v}) \in u$ for all $v \in E$,
      and $\quot(L_r \tas{v}) \in u$ for all $v \in N$.
      Because $u$ is an ultrafilter, we then have
      \[
      \quot\left(\bigwedge_{v \in E} \neg L_0 \tas{v} \land \bigwedge_{v \in N} L_r \tas{v}\right) \in u  .
      \]
      By axiom A3, this implies
      \[
      \quot\left(\bigwedge_{v \in E} \neg L_0 \tas{v} \land L_r \bigvee_{v \in N} \tas{v}\right) \in u  .
      \]
      Then axiom A5 gives
      \[
      \quot\left(L_r \bigvee_{v \in \sat{\psi}} \tas{v}\right) \in u .
      \]
      By the induction hypothesis, T2, and Lemma \ref{lem:filtration},
      we then get $\quot(L_r \psi) \in u$.
    
    ($\impliedby$)
      Let $\quot(L_r \psi) \in u$. It follows from A1, A2, and R2 that $\psi$ is consistent. 
      Hence, by the induction hypothesis, $\sat{\psi}$ is non-empty.
      We first show that $\trans{u}{\sat{\psi}} \neq \Emptyset$.
      Assume therefore towards a contradiction that $\trans{u}{\sat{\psi}} = \Emptyset$.
      Then for all $v \in \sat{\psi}$, we must have that case \ref{item:trans3} holds,
      and hence $L(u,v) = \Emptyset$,
      meaning $\quot(\neg L_r \tas{v}) \in u$ for all $v \in \sat{\psi}$.
      Since there are finitely many $v \in \sat{\psi}$,
      we can enumerate them as $v_1,v_2,\dots,v_n$.
      Then, since $u$ is an ultrafilter, we have
      \[
      \quot\left(\neg L_r \tas{v_1} \land \neg L_r \tas{v_2} \land \dots \land \neg L_r \tas{v_n}\right) \in u  .
      \]
      By De Morgan's law, this is equivalent to
      \[
      \quot\left(\neg (L_r \tas{v_1} \lor L_r \tas{v_2} \lor \dots \lor L_r \tas{v_n})\right) \in u  .
      \]
      The contrapositive of axiom A4 then gives that
      \[
      \quot\left(\neg L_r (\tas{v_1} \lor \tas{v_2} \lor \dots \lor \tas{v_n})\right) \in u  ,
      \]
      and by the induction hypothesis, T2, and Lemma \ref{lem:filtration},
      this is equivalent to $\neg \quot(L_r \psi) \in u$, which is a contradiction.
      
      Now assume towards a contradiction that $\transl{u}{\sat{\psi}} < r$. Then
      there exists some $v \in \sat{\psi}$ such that $\transl{u}{\{v\}} < r$ and
      case \ref{item:trans1} or case \ref{item:trans2} holds.
      In either case we have $\max L(u,v) < r$ and hence
      there exists some $q \in Q_\rho$ such that
      $\quot(L_q \tas{v}) \in u$, which implies $\quot(L_0 \tas{v}) \in u$ by axiom A2.
      By the induction hypothesis, $\quot(\psi) \in v$,
      which means that $\vdash \tas{v} \rightarrow \psi$.
      rule R1 then gives $\quot(L_r \tas{v}) \in u$,
      but this is a contradiction since $\max L(u,v) < r$.
    
    The $M_r$ case is similar, using axiom A7 instead of A2 to derive $\quot(L_0 \psi) \in u$.
\end{proof}

Having established the truth lemma, we can now show that any consistent formula is satisfied by some finite model.

\begin{theorem}[Finite model property]\label{thm:finitemodel}
  For any consistent formula $\varphi \in \mathcal{L}$,
  there exists a finite WTS $\mathcal{M} = (S, \rightarrow, \ell)$ and a state $s \in S$
  such that $\mathcal{M},s \models \varphi$.
\end{theorem}
\begin{proof}%[Proof of Theorem \ref{thm:finitemodel}]
  Since $\varphi \in \mathcal{L}$ is consistent,
  $\quot(\varphi) \neq \quot(\bot)$,
  and since $\mathcal{L}_{\leftrightarrow}[\rho]$ is finite,
  there must exist an ultrafilter $u \in \mathcal{U}[\rho]$ such that $\quot(\varphi) \in u$.
  By the truth lemma, this means that $\mathcal{M}_\varphi, u \models \varphi$,
  and by construction, $\mathcal{M}_\varphi$ is a finite model.
\end{proof}

We are now able to state and prove our main result, namely that our axiomatisation is complete.

\begin{theorem}[Completeness]\label{thm:completeness}
  For any formula $\varphi \in \mathcal{L}$, it holds that
  \[
  \models \varphi \quad \text{implies} \quad \vdash \varphi  .
  \]
\end{theorem}
\begin{proof}%[Proof of Theorem \ref{thm:completeness}]\hfill
  \[
  \models \varphi \quad \text{implies} \quad \vdash \varphi
  \]
  is equivalent to
  \[
  \not \vdash \varphi \quad \text{implies} \quad \not \models \varphi  ,
  \]
  which is equivalent to
  \[
  \text{the consistency of } \neg \varphi \text{ implies the existence of a model for } \neg \varphi  ,
  \]
  and this is guaranteed by the finite model property.
\end{proof}

We have thus established completeness for our logic.
There is also a stronger notion of completeness, often called strong completeness,
which asserts that $\Phi \models \varphi$ implies $\Phi \vdash \varphi$
for any set of formulas $\Phi \subseteq \mathcal{L}$.
Completeness is a special case of strong completeness where $\Phi = \Emptyset$.
In the case of compact logics, strong completeness follows directly from completeness.
However, our logic is non-compact.

\begin{theorem}\label{thm:noncompact}
  Our logic is non-compact, meaning that there exists an infinite set $\Phi \subseteq \mathcal{L}$
  such that each finite subset of $\Phi$ admits a model, but $\Phi$ does not.
\end{theorem}
\begin{proof}
  Consider the set
  \[\Phi = \{L_q \varphi \mid q < r \} \cup \{\neg L_r \varphi\}.\]
  For any finite subset of $\Phi$, it is easy to construct a model.
  However, if $\mathcal{M},s \models L_q \varphi$ for all $q < r$
  where $q,r \in \mathbb{Q}_{\geq 0}$,
  then by the Archimedean property of the rationals,
  we also have $\mathcal{M},s \models L_r \varphi$.
  Hence there can be no model for $\Phi$. 
\end{proof}

%% Satisfiability
\section{Model Checking and Satisfiability}\label{sec:sat}
We now turn our attention to decision problems related to our logic.
First we consider the model checking problem,
which asks us to decide whether $\mathcal{M},s \models \varphi$
for a given model $\mathcal{M}$, state $s$, and formula $\varphi$.
We will develop a polynomial time algorithm for this problem
by adapting the classical model checking algorithm of Clarke et al. \cite{CES86} to our setting.
In what follows, we will assume that all models have a finite state space,
and that each state has finitely many outgoing transitions.
Furthermore, we will assume that the set $\ap$ of atomic propositions is finite.

Given a formula $\varphi$ and a model $\mathcal{M} = (S, \rightarrow, \ell)$,
we construct a function $F_\varphi : S \rightarrow 2^{\mathcal{L}}$
which assigns a set of formulas to each state.
Intuitively, $F_\varphi(s)$ will be the set of subformulas of $\varphi$ that are true in $s$.

In order to do this, we first introduce the following terminology.
The \emph{closure} of a formula $\varphi$, denoted $\cl(\varphi)$,
is given by
\[\cl(\varphi) =  \begin{cases}
                    \{p\} & \text{if } \varphi = p \\
                    \cl(\varphi') \cup \{\varphi\} & \text{if } \varphi = \neg \varphi', \varphi = L_r \varphi', \text{ or } \varphi = M_r \varphi' \\
                    \cl(\varphi_1) \cup \cl(\varphi_2) \cup \{\varphi\} & \text{if } \varphi = \varphi_1 \land \varphi_2
                  \end{cases}\]

\begin{definition}
  A formula $\varphi'$ is said to be a \emph{subformula} of $\varphi$ if $\varphi' \in \cl(\varphi)$.
  $\varphi'$ is said to be a \emph{proper} subformula of $\varphi$ if it is a subformula and $\varphi' \neq \varphi$.
\end{definition}

\begin{definition}
  For a formula $\varphi$, we define the length of $\varphi$ as follows.
  \begin{itemize}
    \item $\varphi$ has length $1$ if it has no proper subformulas.
    \item $\varphi$ has length $i$ if its longest proper subformula has length $i - 1$.
  \end{itemize}
  We will denote the length of $\varphi$ by $\len(\varphi)$.
\end{definition}

We can now construct the function $F_\varphi$ by means of Algorithm \ref{alg:model-checking}.

\begin{algorithm}
  \SetAlgoLined
    Let $F_\varphi(s) = \Emptyset$ for all $s \in S$ \;
    Let $\Phi_i$ be the set of all subformulas of $\varphi$ of length $i$ \;
    
    \For{$i = 1, \dots, \len(\varphi)$}{
      \For{$\varphi' \in \Phi_i$}{
        \For{$s \in S$}{
          \If{$\varphi' = p$}{
            \If{$p \in \ell(s)$}{
              $F_\varphi(s) := F_\varphi(s) \cup \{\varphi'\}$ \;
            }
          }
          \If{$\varphi' = \neg \psi$}{
            \If{$\psi \notin F_\varphi(s)$}{
              $F_\varphi(s) := F_\varphi(s) \cup \{\varphi'\}$ \;
            }
          }
          \If{$\varphi' = \psi_1 \land \psi_2$}{
            \If{$\psi_1 \in F_\varphi(s)$ and $\psi_2 \in F_\varphi(s)$}{
              $F_\varphi(s) := F_\varphi(s) \cup \{\varphi'\}$ \;
            }
          }
          \If{$\varphi' = L_r \psi$}{
            Let $S_\psi = \{s' \in S \mid \psi \in F_\varphi(s')\}$ \;
            \If{$S_\psi \neq \Emptyset$ and $\min\{r' \mid s \xrightarrow{r'} s' \text{ for some } s' \in S_\psi\} \geq r$}{
              $F_\varphi(s) := F_\varphi(s) \cup \{\varphi'\}$ \;
            }
          }
          \If{$\varphi' = M_r \psi$}{
            Let $S_\psi = \{s' \in S \mid \psi \in F_\varphi(s')\}$ \;
            \If{$S_\psi \neq \Emptyset$ and $\max\{r' \mid s \xrightarrow{r'} s' \text{ for some } s' \in S_\psi\} \leq r$}{
              $F_\varphi(s) := F_\varphi(s) \cup \{\varphi'\}$ \;
            }
          }
        }
      }
    }
    
    \Return $F_\varphi$ \;
  \caption{Constructing the function $F_\varphi$ for a given formula $\varphi$.}
  \label{alg:model-checking}
\end{algorithm}

\begin{lemma}\label{lem:model-checking}
  Let $\mathcal{M} = (S, \rightarrow, \ell)$ be a model, $s \in S$ a state, and $\varphi$ a formula.
  For any subformula $\varphi'$ of $\varphi$ it holds that
  \[\mathcal{M},s \models \varphi' \quad \text{if and only if} \quad \varphi' \in F_\varphi(s).\]
\end{lemma}
\begin{proof}
  We will prove this by structural induction on $\varphi'$.
  
  \begin{description}
    \item[($\varphi' = p$):]
      \begin{align*}
        \mathcal{M},s \models p &\text{ iff } p \in \ell(s) && \text{(Definition \ref{def:semantics})}\\
                                &\text{ iff } p \in F_\varphi(s) && \text{(Algorithm \ref{alg:model-checking})}.
      \end{align*}
    \item[($\varphi' = \neg \psi$):]
      \begin{align*}
        \mathcal{M},s \models \neg \psi &\text{ iff } \mathcal{M},s \not\models \psi && \text{(Definition \ref{def:semantics})}\\
                                        &\text{ iff } \psi \notin F_\varphi(s) && \text{(ind. hyp.)} \\
                                        &\text{ iff } \neg \psi \in F_\varphi(s) && \text{(Algorithm \ref{alg:model-checking})}.
      \end{align*}
    \item[($\varphi' = \psi_1 \land \psi_2$):]
      \begin{align*}
        \mathcal{M},s \models \psi_1 \land \psi_2 &\text{ iff } \mathcal{M},s \models \psi_1 \text{ and } \mathcal{M},s \models \psi_2 && \text{(Definition \ref{def:semantics})}\\
                                                  &\text{ iff } \psi_1 \in F_\varphi(s) \text{ and } \psi_2 \in F_\varphi(s) && \text{(ind. hyp.)} \\
                                                  &\text{ iff } \psi_1 \land \psi_2 \in F_\varphi(s) && \text{(Algorithm \ref{alg:model-checking})}.
      \end{align*}
    \item[($\varphi' = L_r \psi$):]
      \begin{align*}
        \mathcal{M},s \models L_r \psi  &\text{ iff } \transl{s}{\sat{\psi}} \geq r && \text{(Definition \ref{def:semantics})} \\
                                        &\text{ iff } \transl{s}{S_\psi} \geq r && \text{(ind. hyp.)} \\
                                        &\text{ iff } S_\psi \neq \Emptyset \text{ and } \\
                                        &\phantom{\text{ iff }} \min\{r' \mid \exists s' \in S_\psi. s \xra{r'} s'\} \geq r \\
                                        &\text{ iff } L_r \psi \in F_\varphi(s) && \text{(Algorithm \ref{alg:model-checking})}.
      \end{align*}
    \item[($\varphi' = M_r \psi$):]
      \begin{align*}
        \mathcal{M},s \models M_r \psi  &\text{ iff } \transr{s}{\sat{\psi}} \leq r && \text{(Definition \ref{def:semantics})} \\
                                        &\text{ iff } \transr{s}{S_\psi} \leq r && \text{(ind. hyp.)} \\
                                        &\text{ iff } S_\psi \neq \Emptyset \text{ and } \\
                                        &\phantom{\text{ iff }} \max\{r' \mid \exists s' \in S_\psi.s \xra{r'} s'\} \leq r \\
                                        &\text{ iff } M_r \psi \in F_\varphi(s) && \text{(Algorithm \ref{alg:model-checking})}. \qedhere
      \end{align*}
  \end{description}
\end{proof}

We can now prove that, given a model $\mathcal{M} = (S, \rightarrow, \ell)$ and formula $\varphi$,
the model checking problem is decidable in polynomial time.

\begin{theorem}
  The model checking problem for our logic is decidable in time
  \[\bigo(|\cl(\varphi)| \cdot |S| \cdot (n \cdot \log n + |\ap| + |\cl(\varphi)|)),\]
  where $n$ is the degree of $\mathcal{M}$, i.e. the maximum number of outgoing transitions from a state in $S$.
\end{theorem}
\begin{proof}
  It follows from Lemma \ref{lem:model-checking} that the model checking problem
  is decidable by constructing the function $F_\varphi$ and checking whether $\varphi \in F_\varphi(s)$.
  It therefore remains to argue that Algorithm \ref{alg:model-checking} runs in the claimed time.
  
  The first two loops in Algorithm \ref{alg:model-checking} at line $3$ and $4$ iterate over all elements of $\cl(\varphi)$,
  and the third loop iterates over all elements of $S$.
  Inside the third loop, the algorithm enters one of the if-statements depending on the structure of the current subformula.
  If it enters the first if-statement at line $7$, then we must search $\ell(s)$ for $p$.
  This takes at most time $|\ap|$.
  If it enters the second or third if-statement at line $11$ and $16$,
  then we must search $F_\varphi(s)$ for one of the formulas.
  This takes at most time $|\cl(\varphi)|$,
  since the function $F_\varphi$ can only label states with subformulas of $\varphi$.
  Lastly, if the algorithm enters the fourth of fifth if-statement at line $21$ and $27$,
  then we must find the corresponding minimum and maximum value.
  This can be done using e.g. mergesort in time $n \cdot \log n$.
  
  Together, this analysis of Algorithm \ref{alg:model-checking} gives the claimed run-time.
\end{proof}

Next we will consider the satisfiability problem,
which asks us to decide whether a given formula $\varphi$ has a model or not.
The finite model property gives us a way of
deciding this problem.
An algorithm would be to enumerate all finite models
and all theorems derivable from the axioms,
which can be done since there are countably many of each of these.
If $\varphi$ is satisfiable, it has a model,
and by the finite model property, it has a finite one.
So we can check one by one whether a finite model satisfies $\varphi$
by using the model checking algorithm described previously.
On the other hand, if $\varphi$ is not satisfiable,
then $\neg \varphi$ is a theorem,
so we can search through all theorems to see whether $\neg \varphi$ is one of them.
Since $\varphi$ is either satisfiable or its negation is a theorem,
one of these two algorithms must eventually halt.
By running these two algorithms in parallel,
we have shown that the problem of deciding satisfiability for a given formula is decidable.

In what follows we do more:
We propose an algorithm that constructs a tableau syntactically from a given formula.
By inspecting this tableau, we can decide whether or not the formula is satisfiable,
and if it is satisfiable, we can construct a model for the formula from the tableau.

As in the previous section, we impose an order on formulas
given by $\varphi \leq \psi$ if and only if $\models \varphi \rightarrow \psi$.
Given a finite set of formulas $\Gamma = \{\varphi_1, \dots, \varphi_n\}$,
we denote by $\min(\Gamma)$ the set of minimal elements of $\Gamma$, i.e.
\[\min(\Gamma) = \{\varphi_i \in \Gamma \mid \text{there is no } \varphi_j \text{ such that } \varphi_j \leq \varphi_i\},\]
and we let
\[\mathcal{L}(\Gamma) = \{\varphi_i \in \Gamma \mid \text{there is no } j < i \text{ such that } \models \varphi_j \leftrightarrow \varphi_i\}.\]
Furthermore, we let $\upw{\Gamma}(\varphi)$ be the upward closure of $\varphi$ in $\Gamma$, i.e.
\[\upw{\Gamma}(\varphi) = \{\varphi' \in \Gamma \mid \varphi \leq \varphi'\}.\]

\begin{table}
  \centering
  \begin{tabular}{c}
    \hline
    
    \\
    
    {\begin{prooftree}
      \hypo{\langle \Gamma \cup \{\varphi \land \psi\}, \mathcal{I}^L, \mathcal{I}^M \rangle}
      \infer[left label = {($\land$)}]1{\langle \Gamma \cup \{\varphi, \psi\}, \mathcal{I}^L, \mathcal{I}^M \rangle}
    \end{prooftree}} \\
    
    \\
    
    {\begin{prooftree}
      \hypo{\langle \Gamma \cup \{\neg (\varphi \land \psi)\}, \mathcal{I}^L, \mathcal{I}^M \rangle}
      \infer[left label = {($\neg \land$)}]1{\langle \Gamma \cup \{\neg \varphi\}, \mathcal{I}^L, \mathcal{I}^M \rangle \quad \langle \Gamma \cup \{\neg \psi\}, \mathcal{I}^L, \mathcal{I}^M \rangle}
    \end{prooftree}} \\
    
    \\
  
    {\begin{prooftree}
      \hypo{\langle \Gamma \cup \{\neg \neg \varphi\}, \mathcal{I}^L, \mathcal{I}^M \rangle}
      \infer[left label = {($\neg\neg$)}]1{\langle \Gamma \cup \{\varphi\}, \mathcal{I}^L, \mathcal{I}^M \rangle}
    \end{prooftree}} \\
    
    \\
    
    {\begin{prooftree}
      \hypo{\langle \Gamma \cup \{N^1_{r_1} \varphi_1, \dots, N^n_{r_n} \varphi_n\} \cup \{\neg O^1_{r_1'} \varphi_1', \dots, \neg O^{n'}_{r_{n'}'} \varphi_{n'}'\}, \mathcal{I}^L, \mathcal{I}^M \rangle}
      \infer[left label = {(mod)}]1{\langle \{\psi_1\}, \mathcal{I}^L_1, \mathcal{I}^M_1 \rangle \quad \cdots \quad \langle \{\psi_k\}, \mathcal{I}^L_k, \mathcal{I}^M_k \rangle}
    \end{prooftree}} \\
    
    \\
    
    \parbox{10cm}{
      if $N^i \in \{L,M\}$ for all $1 \leq i \leq n$, $O^j \in \{L,M\}$ for all $1 \leq j \leq n'$,
      and no formula in $\Gamma$ is of the form $N_r \varphi$ or $\neg N_r \varphi$ where $N \in \{L,M\}$.
    } \\
    
    \\
    
    \hline
  \end{tabular}
  \caption{Tableau rules.}
  \label{tab:rules}
\end{table}

A \emph{tableau} is a tree with nodes of the form $\langle \Gamma, \mathcal{I}^L, \mathcal{I}^M \rangle$
that is constructed from the rules of Table \ref{tab:rules},
where the (mod) rule may only be used when no other rule can be used.
For each node $\langle \Gamma, \mathcal{I}^L, \mathcal{I}^M \rangle$,
$\Gamma$ is a set of formulas, and $\mathcal{I}^L$ and $\mathcal{I}^M$
are intervals of the form $\lbag a, b \rbag$ where
$a \in \mathbb{R}_{\geq 0} \cup \{-\infty\}$, $b \in \mathbb{R}_{\geq 0} \cup \{\infty\}$, $\lbag \in \{[,(\}$, and $\rbag \in \{], )\}$,
subject to the constraint that $\lbag = ($ if $a = -\infty$ and $\rbag = \; )$ if $b = \infty$.
We will say that an interval $\lbag a, b \rbag$ is \emph{consistent} if $a < b$ or $a = b$ and the interval is closed.

For the rule (mod), the objects $\psi_i$, $\mathcal{I}^L_i$ and $\mathcal{I}^M_i$ in the conclusion are constructed as follows.
The $\psi_i$ are given by
\[\{\psi_1, \dots, \psi_k\} = \min(\mathcal{L}(\{\varphi_1, \dots, \varphi_n\})).\]
Let $\Gamma' = \{\varphi_1, \dots, \varphi_n\}$ and
\[\mathbb{L}^+_i = \{r \mid L_r \varphi_j = N^j_{r_j} \varphi_j \text{ for some } j \text{ and } \varphi_j \in \upw{\Gamma'}(\psi_i)\}\]
\[\mathbb{M}^+_i = \{r \mid M_r \varphi_j = N^j_{r_j} \varphi_j \text{ for some } j \text{ and } \varphi_j \in \upw{\Gamma'}(\psi_i)\}\]
as well as
\[\mathbb{L}^-_i = \{r \mid L_r \varphi_j' = O^j_{r_j} \varphi_j' \text{ for some } j \text{ and } \models \psi_i \rightarrow \varphi_j'\}\]
\[\mathbb{M}^-_i = \{r \mid M_r \varphi_j' = O^j_{r_j} \varphi_j' \text{ for some } j \text{ and } \models \psi_i \rightarrow \varphi_j'\}.\]

Then the intervals $\mathcal{I}^L_i$ and $\mathcal{I}^M_i$ are given by
\[\mathcal{I}^L_i = \begin{cases}
                      [ \max \mathbb{L}^+_i, \min \mathbb{L}^-_i ) & \text{if } \mathbb{L}^+_i \neq \Emptyset \text{ and } \mathbb{L}^-_i \neq \Emptyset \\
                      [0, \min \mathbb{L}^-_i) & \text{if } \mathbb{L}^+_i = \Emptyset \text{ and } \mathbb{L}^-_i \neq \Emptyset \\
                      [ \max \mathbb{L}^+_i, \infty) & \text{if } \mathbb{L}^+_i \neq \Emptyset \text{ and } \mathbb{L}^-_i = \Emptyset \\
                      [0, \infty) & \text{if } \mathbb{L}^+_i = \Emptyset \text{ and } \mathbb{L}^-_i = \Emptyset
                    \end{cases}\]
\[\mathcal{I}^M_i = \begin{cases}
                      ( \max \mathbb{M}^-_i, \min \mathbb{M}^+_i ] & \text{if } \mathbb{M}^-_i \neq \Emptyset \text{ and } \mathbb{M}^+_i \neq \Emptyset \\
                      [0, \min \mathbb{M}^+_i ] & \text{if } \mathbb{M}^-_i = \Emptyset \text{ and } \mathbb{M}^+_i \neq \Emptyset \\
                      ( \max \mathbb{M}^-_i, \infty) & \text{if } \mathbb{M}^-_i \neq \Emptyset \text{ and } \mathbb{M}^+_i = \Emptyset \\
                      [0, \infty) & \text{if } \mathbb{M}^-_i = \Emptyset \text{ and } \mathbb{M}^+_i = \Emptyset
                    \end{cases}\]

Informally, one should think of a node $m = \langle \Gamma, \mathcal{I}^L, \mathcal{I}^M \rangle$ as satisfying all the formulas in $\Gamma$.
Moreover, the (mod)-rule signifies a state transition,
where the new states are given by the nodes in the conclusion,
and any transition to $m$ must have a minimum weight that lies in the interval $\mathcal{I}^L$,
and a maximum weight that lies in the interval $\mathcal{I}^M$.

\begin{example}
  We now illustrate the use of the (mod) rule through an example.
  Consider the node
  \[m = \langle \{p_1, p_2, L_2 p_1, L_4(p_1 \land p_2), L_0 p_3, \neg L_5 p_2, \neg M_6 p_3\}, \mathcal{I}^L, \mathcal{I}^M \rangle.\]
  We group the formulas as
  \[\Gamma = \{p_1, p_2\}, \Gamma' = \{L_2 p_1, L_4 (p_1 \land p_2), L_0 p_3\}, \text{ and } \Gamma'' = \{\neg L_5 p_2, \neg M_6 p_3\},\]
  so that $m = \langle \Gamma \cup \Gamma' \cup \Gamma'', \mathcal{I}^L, \mathcal{I}^M \rangle$.
  Since $\Gamma$ only includes literals, it is clear that we can use no other rules,
  so we are allowed to use (mod) on $m$.
  
  We see that $\models (p_1 \land p_2) \rightarrow p_1$,
  and hence $\{\psi_1, \psi_2\} = \{p_1 \land p_2, p_3\}$,
  so there are two children of $m$. For the first child, we find
  \begin{align*}
    &\mathbb{L}^+_1 = \{2,4\} &&\mathbb{M}^+_1 = \Emptyset \\
    &\mathbb{L}^-_1 = \{5\} &&\mathbb{M}^-_1 = \Emptyset,
  \end{align*}
  and for the second child we find
  \begin{align*}
    &\mathbb{L}^+_2 = \{0\} &&\mathbb{M}^+_2 = \Emptyset \\
    &\mathbb{L}^-_2 = \Emptyset &&\mathbb{M}^-_2 = \{6\}.
  \end{align*}
  Hence the intervals become
  \begin{align*}
    &\mathcal{I}^L_1 = [4,5) &&\mathcal{I}^M_1 = [0,\infty) \\
    &\mathcal{I}^L_2 = [0,\infty) &&\mathcal{I}^M_2 = (6,\infty),
  \end{align*}
  and our application of the rule becomes
  \[
    \begin{prooftree}
      \hypo{\langle \{p_1, p_2, L_2 p_1, L_4(p_1 \land p_2), L_0 p_3, \neg L_5 p_2, \neg M_6 p_3\}, \mathcal{I}^L, \mathcal{I}^M \rangle}
      \infer[left label = {(mod)}]1{\langle \{p_1 \land p_2\}, [4,5), [0,\infty) \rangle \quad \langle \{p_3\}, [0,\infty) (6,\infty) \rangle}
    \end{prooftree} \qedhere
  \]
\end{example}

Given a formula $\varphi$, we will say that a tableau $\mathcal{T}$ is a \emph{tableau for $\varphi$}
if $\langle \{\varphi\}, [0,0], [0,0] \rangle$ is the root of $\mathcal{T}$.
A node $m$ in a tableau is called a \emph{modal node} if the (mod)-rule was applied to $m$.
We will say that a node is a \emph{terminal node} if it is either a modal node or a leaf node.

\begin{definition}\label{def:consistent}
  A node $m = \langle \Gamma, \lbag_1 a,b \rbag_1, \lbag_2 c,d \rbag_2 \rangle$ is \emph{consistent} if
  \begin{itemize}
    \item for any $p \in \ap$ we do not have both $p \in \Gamma$ and $\neg p \in \Gamma$,
    \item $\lbag_1 a,b \rbag_1$ and $\lbag_2 c,d \rbag_2$ are consistent, and
    \item either $a < d$ or $a = d$, $\lbag_1 = [$, and $\rbag_2 = \; ]$. \qedhere
  \end{itemize}
\end{definition}

\begin{definition}\label{def:success}
  A tableau $\mathcal{T}$ is \emph{successful} if there exists a subtree $\mathcal{T}'$ of $\mathcal{T}$ such that
  \begin{itemize}
    \item every leaf in $\mathcal{T}'$ is also a leaf in $\mathcal{T}$,
    \item if a modal node $m$ is included in $\mathcal{T}'$,
      then every child of $m$ is also included in $\mathcal{T}'$, and
    \item every terminal node in $\mathcal{T}'$ is consistent. \qedhere
  \end{itemize}
\end{definition}

Given a successful tableau $\mathcal{T}$,
we construct the WTS $\mathcal{M}(\mathcal{T})$ with state $s_{\mathcal{T}}$
using Algorithm~\ref{alg:model}.

\begin{algorithm}
  \SetAlgoLined
    Let $\mathcal{T}'$ be a witness for the fact that $\mathcal{T}$ is successful \;
    $S := \{s_\mathcal{T}\}, \rightarrow := \Emptyset, \ell := \Emptyset$ \;
    Let $X$ be a stack and $X := \Emptyset$ \;
    $X.push((s_\mathcal{T},r))$ where $r$ is the root of $\mathcal{T}'$ \;
    
    \While{$X \neq \Emptyset$}{
      $(s,m) := X.pop$ \;
      Let $m = \langle \Gamma, \Delta, (a,b) \rangle$ \;
      \If{$m$ is not a terminal node}{
        Let $m'$ be the left-most child of $m$ in $\mathcal{T}'$ \;
        $X.push((s,m'))$ \;
      }
      \If{$m$ is a leaf node}{
        $\ell := \ell \cup \{(s,p) \mid p \in \ap \text{ and } p \in \Gamma\}$ \;
      }
      \If{$m$ is a modal node}{
        $\ell := \ell \cup \{(s,p) \mid p \in \ap \text{ and } p \in \Gamma\}$ \;
        Let $m_1 = \langle \Gamma_1, \mathcal{I}^L_1, \mathcal{I}^M_1 \rangle, \dots, m_n = \langle \Gamma_n, \mathcal{I}^L_n, \mathcal{I}^M_n \rangle$ be the children of $m$ in $\mathcal{T}'$ \;
        \For{$i = 1, \dots, n$}{
          Let $\mathcal{I}^L_i = \lbag a_i,b_i \rbag$ and $\mathcal{I}^M_i = \lbag c_i,d_i \rbag$ \;
          $x_i := a_i$ \;
          $y_i := \begin{cases} \max\{a_i, \frac{d_i - c_i}{2} + c_i\} & \text{if } d_i \neq \infty \\ \max\{a_i, c_i + 1\} & \text{if } d_i = \infty\end{cases}$ \;
          $S := S \cup \{s_i\}$ \;
          $\rightarrow := \rightarrow \cup \{(s,x_i,s_i), (s,y_i,s_i)\}$ \;
          $X.push((s_i,m_i))$ \;
        }
      }
    }
    
    $\mathcal{M}(\mathcal{T}) := (S, \rightarrow, \ell)$ \;
    
    \Return $(\mathcal{M}(\mathcal{T}), s_{\mathcal{T}})$ \;
  \caption{Constructing the model $\mathcal{M}(\mathcal{T})$ for a successful tableau $\mathcal{T}$.}
  \label{alg:model}
\end{algorithm}

\begin{lemma}\label{lem:tableaumodel}
  If $\mathcal{T}$ is a successful tableau for $\varphi$, then $\mathcal{M}(\mathcal{T}),s_{\mathcal{T}} \models \varphi$.
\end{lemma}
\begin{proof}
  Let $Y$ be the set of all pairs $(s,m)$ that are added to the stack $X$ by Algorithm \ref{alg:model} at some point during the construction of $\mathcal{M}(\mathcal{T})$.
  We wish to prove that for any $(s,\langle \Gamma, \mathcal{I}^L, \mathcal{I}^M\rangle) \in Y$
  we have $\mathcal{M}(\mathcal{T}),s \models \Gamma$,
  where we write $\mathcal{M}(\mathcal{T}),s \models \Gamma$ to mean $\mathcal{M}(\mathcal{T}),s \models \varphi$ for all $\varphi \in \Gamma$.
  Note that if we can prove this, then it follows that $\mathcal{M}(\mathcal{T}),s_\mathcal{T} \models \varphi$
  since $(s_\mathcal{T}, \langle \{\varphi\}, \Emptyset, (0,0) \rangle) \in Y$.
  
  Let $(s, m)$ be an arbitrary element of $Y$ and let $l$ be the length of the longest path from $m$ to a leaf.
  We will prove, by induction on $l$, that $\mathcal{M}(\mathcal{T}), s \models \Gamma$ where $m = \langle \Gamma, \mathcal{I}^L, \mathcal{I}^M \rangle$.
  
  $l = 0$: In this case, $m$ is a leaf. Hence $\Gamma$ only contains literals,
    and by construction we have $p \in \ell(s)$ if and only if $p \in \Gamma$.
    Since $m$ is consistent, we thus get $\mathcal{M}(\mathcal{T}),s \models \Gamma$.
    
  $l > 0$: In this case we consider the different rules that may be applied to $m$.
  \begin{description}
    \item[($\land$)] We have 
      \[
        \begin{prooftree}
          \hypo{m = \langle \Gamma \cup \{\varphi_1 \land \varphi_2\}, \mathcal{I}^L, \mathcal{I}^M \rangle}
          \infer[left label = {($\land$)}]1{m' = \langle \Gamma \cup \{\varphi_1, \varphi_2\}, \mathcal{I}^L, \mathcal{I}^M \rangle}
        \end{prooftree}
      \]
      By induction hypothesis we get $\mathcal{M}(\mathcal{T}),s \models \Gamma \cup \{\varphi_1,\varphi_2\}$.
      This implies that $\mathcal{M}(\mathcal{T}),s \models \varphi_1$ and $\mathcal{M}(\mathcal{T}),s \models \varphi_2$,
      so $\mathcal{M}(\mathcal{T}),s \models \Gamma \cup \{\varphi_1 \land \varphi_2\}$.
      
    \item[($\neg \land$)] We have
      \[
        \begin{prooftree}
          \hypo{m = \langle \Gamma \cup \{\neg (\varphi_1 \land \varphi_2)\}, \mathcal{I}^L, \mathcal{I}^M \rangle}
          \infer[left label = {($\neg \land$)}]1{m_1 = \langle \Gamma \cup \{\neg \varphi_1\}, \mathcal{I}^L, \mathcal{I}^M \rangle \quad m_2 = \langle \Gamma \cup \{\neg \varphi_2\}, \mathcal{I}^L, \mathcal{I}^M \rangle}
        \end{prooftree}
      \]
      We have three cases to consider;
      either $m_1$ is included in $\mathcal{T}'$,
      $m_2$ is included in $\mathcal{T}'$,
      or both $m_1$ and $m_2$ are included in $\mathcal{T}'$.
      If $m_1$ is included in $\mathcal{T}'$ we get, by the induction hypothesis,
      that $\mathcal{M}(\mathcal{T}),s \models \Gamma \cup \{\neg \varphi_1 \}$
      implying that $\mathcal{M}(\mathcal{T}),s \not \models \varphi_1$.
      If $m_2$ is included in $\mathcal{T}'$ we get, by the induction hypothesis,
      that $\mathcal{M}(\mathcal{T}),s \models \Gamma \cup \{ \neg \varphi_2 \}$
      implying that $\mathcal{M}(\mathcal{T}),s \not \models \varphi_2$.
      In either case we get that $\mathcal{M}(\mathcal{T}),s \not \models \varphi_1 \land \varphi_2$ and $\mathcal{M}(\mathcal{T}),s \models \Gamma$,
      and therefore $\mathcal{M}(\mathcal{T}),s \models \Gamma \cup \{\neg (\varphi_1 \land \varphi_2)\}$.
      The last case follows trivially from the preceding arguments.
    
    \item[($\neg\neg$)] We have
      \[
        \begin{prooftree}
          \hypo{m = \langle \Gamma \cup \{\neg \neg \varphi'\}, \mathcal{I}^L, \mathcal{I}^M \rangle}
          \infer[left label = {($\neg\neg$)}]1{m' = \langle \Gamma \cup \{\varphi'\}, \mathcal{I}^L, \mathcal{I}^M \rangle}
        \end{prooftree}
      \]
      By induction hypothesis we know that $\mathcal{M}(\mathcal{T}),s \models \Gamma \cup \{\varphi'\}$,
      and hence $\mathcal{M}(\mathcal{T}),s \models \Gamma \cup \{\neg \neg \varphi'\}$.
      
    \item[(mod)] We have
      \[
        \begin{prooftree}
          \hypo{m = \langle \Gamma \cup \{N^1_{r_1} \varphi_1, \dots, N^n_{r_n} \varphi_n\} \cup \{\neg O^1_{r_1'} \varphi_1', \dots, \neg O^{n'}_{r_{n'}'} \varphi_{n'}'\}, \mathcal{I}^L, \mathcal{I}^M \rangle}
          \infer[left label = {(mod)}]1{m_1 = \langle \{\psi_1\}, \mathcal{I}^L_1, \mathcal{I}^M_1 \rangle \quad \cdots \quad m_k = \langle \{\psi_k\}, \mathcal{I}^L_k, \mathcal{I}^M_k \rangle}
        \end{prooftree}
      \]
      $\Gamma$ must consist only of literals, because otherwise the (mod) rule could not be used.
      As in the case for $l = 0$, we then get $\mathcal{M}(\mathcal{T}),s \models \Gamma$ since $m$ is consistent.
      Let $\Psi = \{\psi_1, \ldots, \psi_k\}$,
      and for any $1 \leq j \leq k$, let $\mathcal{I}^L_j = \lbag a_j,b_j \rbag$
      and $\mathcal{I}^M_j = \lbag c_j, d_k \rbag$.
      By the induction hypothesis, we know that $\mathcal{M}(\mathcal{T}), s_j \models \psi_j$ for all $j \in \{1,\ldots,k\}$,
      and, by construction, $s_j$ is the only successor of $s$ that satisfies $\psi_j$.    
      Now consider a formula $N^i_{r_i} \varphi_i$.
      There must exist a subset $\Psi_{\varphi_i} \subseteq \Psi$ such that
      $\trans{s}{\sat{\varphi_i}} = \trans{s}{\bigcup_{\psi' \in \Psi_{\varphi_i}} \sat{\psi'}}$.   
      We first consider the case where $N^i = L$.
      Because $\Psi_{\varphi_i}$ is finite,
      there exists $\psi_j' \in \Psi_{\varphi_i}$ such that
      $\transl{s}{\sat{\varphi_i}} = \transl{s}{\sat{\psi_j'}}$,
      implying the existence of $\psi_j \in \Psi$ such that $\transl{s}{\sat{\varphi_i}} = \transl{s}{\sat{\psi_j}} = a_j$.
      We must have $a_j \geq r_i$ implying $\transl{s}{\sat{\varphi_i}} \geq r_i$, and thus $\mathcal{M}(\mathcal{T}),s \models L_{r_i} \varphi_i$.
      In the case where $N^i = M$ we can, similarly to the previous case,
      find $\psi_j \in \Psi$ such that
      $\transr{s}{\sat{\varphi_i}} = \transr{s}{\sat{\psi_j}}$,
      and we know that $d_i \neq \infty$ implying
      \[\transr{s}{\sat{\psi_j}} = \max\left\{ a_j, \frac{d_j - c_j}{2} + c_j \right\} \leq d_j \leq r_i .\]
      Therefore, $\transr{s}{\sat{\varphi}} \leq r_i$ and thus $\mathcal{M}(\mathcal{T}),s \models M_{r_i} \varphi_i$.
      
      Lastly we consider a formula $\neg O^i_{r_i'}\varphi_i'$.
      If there is no $\psi_j \in \Psi$ such that $\models \psi_j \to \varphi_i'$, then,
      by the construction of $\mathcal{M}(\mathcal{T})$, there is no successor
      $s'$ of $s$ such that $\mathcal{M}(\mathcal{T}),s \models \varphi_i'$.
      Therefore, $\transl{s}{\sat{\varphi_i'}} = \infty$ and $\transr{s}{\sat{\varphi_i'}} = - \infty$,
      and thus $\mathcal{M}(\mathcal{T}),s \models \neg O^i_{r_i'}\varphi_i'$
      is trivially satisfied for $O^i \in \{L, M\}$.
      Suppose $\models \psi_j' \to \varphi_i'$ for some $\psi_j' \in \Psi$.
      We first consider the case where $O^i = L$.
      There must exist $\psi_j \in \Psi$ such that
      $\transl{s}{\sat{\varphi_i'}} = \transl{s}{\sat{\psi_j}} = a_j$.
      By the assumption that $\mathcal{T}$ is successful,
      we must have that $m_j$ is consistent.
      Therefore, $a_j < b_j \leq r_{i'}$ implying
      $\transl{s}{\sat{\varphi_i'}} < r_i'$,
      and thus $\mathcal{M}(\mathcal{T}),s \models \neg L_{r_i'}\varphi_i'$.
      In the case where $O^i = M$ we must be able to find $\psi_j \in \Psi$
      such that $\transr{s}{\sat{\varphi_i'}} = \transr{s}{\sat{\psi_j}}$.
      We have to consider $d_j = \infty$ and $d_j \neq \infty$ separately.
      If $d_j = \infty$ we have
      \[\transr{s}{\sat{\psi_j}} = \max\left\{a_j, c_j + 1\right\} > c_j \geq r_i' .\]
      If $d_j \neq \infty$ we have
      \[\transr{s}{\sat{\psi_j}} = \max\left\{a_j, \frac{d_j - c_j}{2} + c_j\right\} > c_j \geq r_i' .\]
      In either case we have that $\trans{s}{\sat{\varphi_i'}} > r_i'$, so $\mathcal{M}(\mathcal{T}),s \models \neg M_{r_i'}\varphi_i'$. \qedhere
  \end{description}
\end{proof}

\begin{lemma}\label{lem:tableaux}
  Let $\mathcal{T}_1$ and $\mathcal{T}_2$ be tableaux for $\varphi$.
  Then it holds that $\mathcal{T}_1$ is successful if and only if $\mathcal{T}_2$ is successful.
\end{lemma}
\begin{proof}
  Assume that $\mathcal{T}_1$ is a successful tableau.
  Let $\mathcal{T}_1'$ be a subtree of $\mathcal{T}_1$
  which witnesses the fact that $\mathcal{T}_1$ is successful.
  If $\mathcal{T}_1'$ is also a subtree of $\mathcal{T}_2$,
  then we are done. If not, let $d$ be the smallest number such that
  $\mathcal{T}_1'$ differs at depth $d$ from any subtree of $\mathcal{T}_2$ with the same root as $\mathcal{T}_2$.
  Note that we must have $d > 0$ because $\mathcal{T}_1'$ and $\mathcal{T}_2$ have the same root.
  Denote by $\mathcal{T}_1' |_n$ the restriction of $\mathcal{T}_1'$ to depth $n$.
  Then $\mathcal{T}_1' |_{d-1}$ is a subtree of $\mathcal{T}_2$.
  
  Let $X$ be the set of all terminal nodes that are in $\mathcal{T}_1'$ at depth $d$ or below.
  Then every node in $X$ is also a node in $\mathcal{T}_2$,
  and furthermore, every node in $X$ is reachable in $\mathcal{T}_2$ from $\mathcal{T}_1' |_{d-1}$.
  Hence, if we extend $\mathcal{T}_1' |_{d-1}$ to include all paths leading to an element in $X$,
  then this extension is a subtree of $\mathcal{T}_2$ that witnesses the fact that $\mathcal{T}_2$ is successful.
\end{proof}

\begin{lemma}\label{lem:sat}
  $\varphi$ is satisfiable if and only if there exists a successful tableau for $\varphi$.
\end{lemma}
\begin{proof}
  ($\implies$) Assume $\varphi$ is satisfiable, meaning that $\mathcal{M},s \models \varphi$
  for some $\mathcal{M} = (S, \rightarrow, \ell)$ and $s \in S$.
  
  Let $\mathcal{T}$ be a tableau for $\varphi$,
  and note that such a tableau always exists by applying the tableau rules to $\langle \{\varphi\}, [0,0], [0,0] \rangle$.
  Now construct a marking $\mathfrak{M} \subseteq S \times \mathcal{T}$ as follows.
  \begin{itemize}
    \item $(s,r) \in \mathfrak{M}$ where $r$ is the root of $\mathcal{T}$.
    \item If $(s',m) \in \mathfrak{M}$ and ($\land$) or ($\neg\neg$) was applied to $m$,
      add $(s',m')$ to $\mathfrak{M}$, where $m'$ is the child of $m$.
    \item If $(s',m) \in \mathfrak{M}$ and ($\neg \land$) was applied to $m$, meaning that
      \[
        \begin{prooftree}
          \hypo{m = \langle \Gamma \cup \{\neg (\varphi_1 \land \varphi_2)\}, \mathcal{I}^L, \mathcal{I}^M \rangle}
          \infer[left label = {($\neg \land$)}]1{m_1 = \langle \Gamma \cup \{\neg \varphi_1\}, \mathcal{I}^L, \mathcal{I}^M \rangle \quad m_2 = \langle \Gamma \cup \{\neg \varphi_2\}, \mathcal{I}^L, \mathcal{I}^M \rangle}
        \end{prooftree}
      \]
      then add $(s',m_1)$ to $\mathfrak{M}$ if $s' \in \sat{\neg \varphi_1}$ and add $(s',m_2)$ to $\mathfrak{M}$ if $s' \in \sat{\neg \varphi_2}$.
    \item If $(s',m) \in \mathfrak{M}$ and (mod) was applied to $m$, meaning that
      \[
        \begin{prooftree}
          \hypo{m = \langle \Gamma \cup \{N^1_{r_1} \varphi_1, \dots, N^n_{r_n} \varphi_n\} \cup \{\neg O^1_{r_1'} \varphi_1', \dots, \neg O^{n'}_{r_{n'}'} \varphi_{n'}'\}, \mathcal{I}^L, \mathcal{I}^M \rangle}
          \infer[left label = {(mod)}]1{m_1 = \langle \{\psi_1\}, \mathcal{I}^L_1, \mathcal{I}^M_1\rangle \quad \cdots \quad m_k = \langle \{\psi_k\}, \mathcal{I}^L_k, \mathcal{I}^M_k\rangle}
        \end{prooftree}
      \]
      then add $(t',m_i)$ to $\mathfrak{M}$ if $t' \in \sat{\psi_i}$ and $s' \xrightarrow{r} t'$ for some $r \in \mathbb{R}_{\geq 0}$.
  \end{itemize}
  We will first argue that for any $(s', \langle \Gamma, \mathcal{I}^L, \mathcal{I}^M \rangle) \in \mathfrak{M}$ we have $\mathcal{M},s' \models \Gamma$,
  meaning $\mathcal{M},s' \models \varphi'$ for all $\varphi' \in \Gamma$.
  We prove this by induction on the depth $d$ of $m$.
  
  $d = 0$: We have $(s', \langle \Gamma, \mathcal{I}^L, \mathcal{I}^M \rangle) = (s,r) = (s, \langle \{\varphi\}, [0,0], [0,0] \rangle)$,
    and by assumption we get $\mathcal{M},s \models \varphi$.
    
  $d > 0$: We consider which rule was applied to the parent of $m$.
  
    ($\land$):
      \[
        \begin{prooftree}
          \hypo{m' = \langle \Gamma \cup \{\varphi_1 \land \varphi_2\}, \mathcal{I}^L, \mathcal{I}^M \rangle}
          \infer[left label = {($\land$)}]1{m = \langle \Gamma \cup \{\varphi_1, \varphi_2\}, \mathcal{I}^L, \mathcal{I}^M \rangle}
        \end{prooftree}
      \]
      By induction hypothesis, we have $\mathcal{M},s' \models \Gamma \cup \{\varphi_1 \land \varphi_2\}$,
      so $\mathcal{M},s' \models \varphi_1$ and $\mathcal{M},s' \models \varphi_2$,
      and hence $\mathcal{M},s' \models \Gamma \cup \{\varphi_1, \varphi_2\}$.
  
    ($\neg \land$):
      \[
        \begin{prooftree}
          \hypo{m' = \langle \Gamma \cup \{\neg (\varphi_1 \land \varphi_2)\}, \mathcal{I}^L, \mathcal{I}^M \rangle}
          \infer[left label = {($\neg \land$)}]1{m_1 = \langle \Gamma \cup \{\neg \varphi_1\}, \mathcal{I}^L, \mathcal{I}^M \rangle \quad m_2 = \langle \Gamma \cup \{\neg \varphi_2\}, \mathcal{I}^L, \mathcal{I}^M \rangle}
        \end{prooftree}
      \]
      If $m = m_1$, then by the way $\mathfrak{M}$ was constructed we get $\mathcal{M},s' \models \neg \varphi_1$,
      and hence by induction hypothesis, $\mathcal{M},s' \models \Gamma \cup \{\neg \varphi_1\}$.
      Likewise we get $\mathcal{M},s' \models \Gamma \cup \{\neg \varphi_2\}$ if $m = m_2$.
  
    ($\neg\neg$):
      \[
        \begin{prooftree}
          \hypo{m' = \langle \Gamma \cup \{\neg \neg \varphi'\}, \mathcal{I}^L, \mathcal{I}^M \rangle}
          \infer[left label = {($\neg\neg$)}]1{m = \langle \Gamma \cup \{\varphi'\}, \mathcal{I}^L, \mathcal{I}^M \rangle}
        \end{prooftree}
      \]
      By induction hypothesis we have $\mathcal{M},s' \models \Gamma \cup \{\neg \neg \varphi'\}$,
      which is equivalent to $\mathcal{M},s' \models \Gamma \cup \{\varphi'\}$.
  
    (mod):
      \[
        \begin{prooftree}
          \hypo{m' = \langle \Gamma \cup \{N^1_{r_1} \varphi_1, \dots, N^n_{r_n} \varphi_n\} \cup \{\neg O^1_{r_1'} \varphi_1', \dots, \neg O^{n'}_{r_{n'}'} \varphi_{n'}'\}, \mathcal{I}^L, \mathcal{I}^M \rangle}
          \infer[left label = {(mod)}]1{m_1 = \langle \{\psi_1\}, \mathcal{I}^L_1, \mathcal{I}^M_1 \rangle \quad \cdots \quad m_k = \langle \{\psi_k\}, \mathcal{I}^L_k, \mathcal{I}^M_k \rangle}
        \end{prooftree}
      \]
      We must have $m = m_i$ for some $1 \leq i \leq k$.
      By construction of $\mathfrak{M}$ we know that $\mathcal{M},m_i \models \psi_i$.
    
    Now let $\mathcal{T}'$ be the subtree of $\mathcal{T}$
    consisting of those nodes $m$ where there exists a state $s'$ such that $(s',m) \in \mathfrak{M}$.
    We will now prove that $\mathcal{T}'$
    satisfies the three conditions in Definition~\ref{def:success}.
    
    For the first condition we prove the contrapositive:
    If $m$ is not a leaf in $\mathcal{T}$, then it is not a leaf in $\mathcal{T}'$.
    Hence we assume that $m$ is not a leaf in $\mathcal{T}$.
    If $m$ is not a node in $\mathcal{T}'$,
    then it is also not a leaf node in $\mathcal{T}'$.
    If $m$ is a node in $\mathcal{T}'$,
    then there must exist some state $s'$
    such that $(s',m) \in \mathfrak{M}$.
    We now consider which rule was applied to $m$ in $\mathcal{T}$.
    
    \begin{description}
      \item[($\land$) or ($\neg\neg$)] In these cases, $m$ has a child $m'$ in $\mathcal{T}$,
        and by construction of $\mathfrak{M}$, we get $(s',m') \in \mathfrak{M}$, so $m'$ is a child of $m$ in $\mathcal{T}'$.
      \item[($\neg \land$)]
        \[
          \begin{prooftree}
            \hypo{m = \langle \Gamma \cup \{\neg (\varphi_1 \land \varphi_2)\}, \mathcal{I}^L, \mathcal{I}^M\rangle}
            \infer[left label = {($\neg \land$)}]1{m_1 = \langle \Gamma \cup \{\neg \varphi_1\}, \mathcal{I}^L, \mathcal{I}^M \rangle \quad m_2 = \langle \Gamma \cup \{\neg \varphi_2\}, \mathcal{I}^L, \mathcal{I}^M \rangle}
          \end{prooftree}
        \]
        We know that $\mathcal{M},s' \models \Gamma \cup \{\neg (\varphi_1 \land \varphi_2)\}$,
        so we must have $\mathcal{M},s' \models \neg \varphi_1$ or $\mathcal{M},s' \models \neg \varphi_2$.
        By construction of $\mathfrak{M}$, this means that $(s',m_1) \in \mathfrak{M}$ or $(s',m_2) \in \mathfrak{M}$,
        and hence $m_1$ or $m_2$ must be a child of $m$ in $\mathcal{T}'$.
      \item[(mod)]
        \[
          \begin{prooftree}
            \hypo{m = \langle \Gamma \cup \{N^1_{r_1} \varphi_1, \dots, N^n_{r_n} \varphi_n\} \cup \{\neg O^1_{r_1'} \varphi_1', \dots, \neg O^{n'}_{r_{n'}'} \varphi_{n'}'\}, \mathcal{I}^L, \mathcal{I}^M \rangle}
            \infer[left label = {(mod)}]1{m_1 = \langle \{\psi_1\}, \mathcal{I}^L_1, \mathcal{I}^M_1 \rangle \quad \cdots \quad m_k = \langle \{\psi_k\}, \mathcal{I}^L_k, \mathcal{I}^M_k \rangle}
          \end{prooftree}
        \]
        For each $m_i$ there must exist some $j$ such that $N^j_{r_j} \varphi_j = N^j_{r_j} \psi_i$.
        Then we know that $\mathcal{M},s' \models N^j_{r_j} \psi_i$,
        and hence
        \[\transl{s'}{\sat{\psi_i}} \geq r_j \quad \text{or} \quad \transr{s'}{\sat{\psi_i}} \leq r_j.\]
        In either case there must exist some $t' \in \sat{\psi_i}$
        such that $s' \xrightarrow{r} t'$ for some $r$.
        Hence $(t',m_i) \in \mathfrak{M}$ and $m_i$ is a child of $m$ in $\mathcal{T}'$.
    \end{description}
    
    For the second condition, let $(s',m) \in \mathfrak{M}$ where $m$ is a modal node, meaning that
    \[
      \begin{prooftree}
        \hypo{m = \langle \Gamma \cup \{N^1_{r_1} \varphi_1, \dots, N^n_{r_n} \varphi_n\} \cup \{\neg O^1_{r_1'} \varphi_1', \dots, \neg O^{n'}_{r_{n'}'} \varphi_{n'}'\}, \mathcal{I}^L, \mathcal{I}^M \rangle}
        \infer[left label = {(mod)}]1{m_1 = \langle \{\psi_1\}, \mathcal{I}^L_1, \mathcal{I}^M_1 \rangle \quad \cdots \quad m_k = \langle \{\psi_k\}, \mathcal{I}^L_k, \mathcal{I}^M_k \rangle}
      \end{prooftree}
    \]
    For every $\psi_i$ we must have $N^j_{r_j} \varphi_j = N^j_{r_j} \psi_i$ for some $j$,
    so $\mathcal{M},s' \models N^j_{r_j} \psi_i$,
    which implies that there exists $t' \in \sat{\psi_i}$ such that $s' \xrightarrow{r} t'$ for some $r$.
    Hence we get $(t',m_i) \in \mathfrak{M}$.
    Since this holds for any $i$, we get that every $m_i$ is included in $\mathcal{T}'$.
    
    For the third condition, let $m = \langle \Gamma, \mathcal{I}^L, \mathcal{I}^M \rangle$ be a terminal node in $\mathcal{T}'$.
    We check the conditions of Definition \ref{def:consistent}.
    There must exist a state $s'$ such that $(s',m) \in \mathfrak{M}$,
    which means that $\mathcal{M},s' \models \Gamma$.
    Hence $s'$ satisfies all the literals in $\Gamma$,
    which can only happen if the first condition is satisfied.
    The second and third condition are satisfied because of the way
    the intervals of the children are constructed in the (mod) rule.
    
  ($\impliedby$) This follows from Lemma \ref{lem:tableaumodel}.
\end{proof}

\begin{theorem}
  The satisfiability problem for our logic is decidable.
\end{theorem}
\begin{proof}
  By Lemma \ref{lem:sat}, to decide whether a formula $\varphi$ is satisfiable,
  it is enough to check whether there exists a successful tableau for $\varphi$.
  Furthermore, by Lemma \ref{lem:tableaux} it is enough to only check a single tableau for $\varphi$:
  If the tableau is successful, then all tableaux for $\varphi$ are successful,
  and if it is not successful, then no tableau for $\varphi$ is successful.
  
  One can construct such a tableau for $\varphi$
  by applying the tableau rules of Table \ref{tab:rules} to the tuple $\langle \{\varphi\}, \Emptyset, (0,0) \rangle$
  until no more rules can be applied.
  We will now argue that there is an effective procedure for constructing such a tableau by induction on the modal depth of $\varphi$.
  
  $md(\varphi) = 0$: In this case, the (mod) rule is never used when constructing the tableau.
    Hence the procedure proceeds by syntactically checking which rules can be used at a given moment,
    and choosing a valid rule to apply.
  
  $md(\varphi) > 0$: In this case we proceed as for the case where $md(\varphi) = 0$,
    except that now the (mod) rule may also be applied,
    in which case we need to be able to compute the $\psi_i$, $\Delta_i$ and $(a_i,b_i)$.
    The difficulty lies in computing the set $\{\psi_1, \dots, \psi_k\} = \min(\mathcal{L}(\{\varphi_1, \dots, \varphi_n\}))$
    and the sets
    \[\mathbb{L}^-_i = \{r \mid L_r\varphi_j' = O^j_{r_j} \varphi_j' \text{ for some } j \text{ and } \models \psi_i \rightarrow \varphi_j'\}\]
    \[\mathbb{M}^-_i = \{r \mid M_r \varphi_j' = O^j_{r_j} \varphi_j' \text{ for some } j \text{ and } \models \psi_i \rightarrow \varphi_j'\}.\]
    However, note that all $\varphi_i$ and $\varphi_i'$ and  have modal depth less than $md(\varphi)$.
    Therefore, by induction hypothesis, we have an effective procedure to decide whether
    $\models \varphi_i \rightarrow \varphi_j$ and $\models \varphi_i \leftrightarrow \varphi_j$,
    which is exactly what we need to compute the aforementioned sets.
    Given this we can compute the values needed for the intervals $\mathcal{I}^L_i$ and $\mathcal{I}^M_i$.
\end{proof}

\begin{example}\label{ex:sat}
  Consider the formula $\varphi = \neg(\neg (L_2 p_1 \land M_5L_1 p_1) \land \neg M_2 p_2 ))$.
  Using the tableau rules, we get the following tableau $\mathcal{T}$ for $\varphi$.
  \[
    \begin{prooftree}[proof style=downwards]
      \hypo{\langle \{p_1\}, [1,\infty), [0,\infty) \rangle}
      \infer[left label = {(mod)}]1{\langle \{p_1, L_1 p_1\}, [2, \infty), [5, \infty) \rangle}
      \infer[left label = {(mod)}]1{\langle \{L_2 p_1, M_5L_1 p_1\}, [0,0], [0,0] \rangle}
      \infer[left label = {($\land$)}]1{\langle \{L_2 p_1 \land M_5L_1 p_1\}, [0,0], [0,0] \rangle}
      \infer[left label = {($\neg \neg$)}]1{\langle \{\neg\neg (L_2 p_1 \land M_5L_1 p_1)\}, [0,0], [0,0] \rangle}
      \hypo{\langle \{p_2\}, [0,\infty), [0, 2] \rangle}
      \infer[left label = {(mod)}]1{\langle \{M_2 p_2\}, [0,0], [0,0] \rangle}
      \infer[left label = {($\neg\neg$)}]1{\langle \{\neg\neg M_2 p_2\}, [0,0], [0,0] \rangle}
      \infer[left label = {($\neg \land$)}]2{\langle \{\neg (\neg (L_2 p_1 \land M_5L_1 p_1) \land M_2 p_2)\}, [0,0], [0,0] \rangle}
    \end{prooftree}
  \]
  In this case the tableau is successful,
  since all terminal nodes are consistent.
  In fact, there are three distinct subtrees witnessing this fact:
  one that chooses the left branch, one that chooses the right branch,
  and one that chooses both branches.
  In Figure \ref{fig:sat-ex} we show the resulting model $\mathcal{M}(\mathcal{T})$
  for the witness that chooses the left branch.
\end{example}

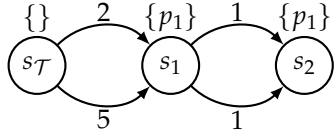
\begin{figure}
  \centering
  \begin{tikzpicture}[WTS, node distance=2cm]
    \node[state, label=above:{$\{\}$}]    (st)               {$s_\mathcal{T}$};
    \node[state, label=above:{$\{p_1\}$}] (s1) [right=of st] {$s_1$};
    \node[state, label=above:{$\{p_1\}$}] (s2) [right=of s1] {$s_2$};

    \path (s0) edge[bend right=45] node[below] {$5$} (s1);
    \path (s0) edge[bend left=45]  node[above] {$2$} (s1);

    \path (s1) edge[bend right=45] node[below] {$1$} (s2);
    \path (s1) edge[bend left=45]  node[above] {$1$} (s2);
  \end{tikzpicture}
  \captionof{figure}{The model $\mathcal{M}(\mathcal{T})$ for the successful tableau $\mathcal{T}$ in Example \ref{ex:sat}.}
  \label{fig:sat-ex}
\end{figure}

\begin{example}
  Consider the formula $\varphi = p_1 \land L_4 p_1 \land \neg L_3 p_1 \land L_2 p_2$.
  Using the tableau rules, we get the following tableau $\mathcal{T}$ for $\varphi$.
  \[
    \begin{prooftree}
      \hypo{\langle \{p_1 \land L_4 p_1 \land \neg L_3 p_1 \land L_2 p_2\}, [0,0], [0,0] \rangle}
      \infer[left label = {($\land$)}]1{\langle \{p_1, L_4 p_1 \land \neg L_3 p_1 \land L_2 p_2\}, [0,0], [0,0] \rangle}
      \infer[left label = {($\land$)}]1{\langle \{p_1, L_4 p_1, \neg L_3 p_1 \land L_2 p_2\}, [0,0], [0,0] \rangle}
      \infer[left label = {($\land$)}]1{\langle \{p_1, L_4 p_1, \neg L_3 p_1, L_2 p_2\}, [0,0], [0,0] \rangle}
      \infer[left label = {(mod)}]1{\langle \{p_1\}, [4,3), [0,\infty] \rangle \quad \langle \{p_2\}, [2, \infty), [0,\infty) \rangle}
    \end{prooftree}
  \]
  In this case the interval $[4,3)$ is not consistent,
  and hence the tableau is not successful,
  so we can conclude that $\varphi$ is not satisfiable.
\end{example}

%% Conclusion
\section{Concluding Remarks}
Our contributions in this paper have been to define a new bisimulation relation
for weighted transition systems (WTSs),
which relates those states that have similar behavior with respect to their minimum
and maximum weights on transitions,
as well as an accompanying modal logic to reason about the
upper and lower bounds of weights on transitions.
We have shown that this logic characterises exactly
those states that are bisimilar for image-finite systems.
Furthermore, we have provided a complete axiomatisation of our logic,
and we have shown that it enjoys the finite model property.
Based on this finite model property,
we have developed an algorithm which decides the satisfiability of a formula in our logic
and constructs a finite model for the formula if it is satisfiable.

This work could be extended in different ways.
Since our logic is non-compact, strong completeness does not follow directly from weak completeness,
and hence it would be interesting to explore a strong-complete axiomatisation of the proposed logic.
Such an axiomatisation would need additional, infinitary axioms.
Examples of such axioms would be
\[\{L_q \varphi \mid q < r\} \vdash L_r \varphi \quad \text{and} \quad \{M_q \varphi \mid q < r\} \vdash M_r \varphi,\]
which are easily proven sound and describe the Archimedean property discussed in Theorem \ref{thm:noncompact}.

Although we have shown that our logic is expressive enough to capture bisimulation,
it would also be of interest to extend our logic with a kind of fixed-point operator or
standard temporal logic operators such as until in order to increase its expressivity,
and hence its practical use.
We envisage two ways in which such a logic could be given semantics:
either by accumulating weights or by taking the maximum or minimum of weights.
In the accumulating case in particular,
one could also allow negative weights to model that the system gains resources.

\defaultbib

  \cleardoublepage
  \setcounter{enumiii}{0}
  \setcounter{enumii}{0}
  \setcounter{enumiv}{0}
  \setcounter{enumi}{0}
  \setcounter{equation}{0}
  \setcounter{figure}{0}
  \setcounter{footnote}{0}
  \setcounter{mpfootnote}{0}
  \setcounter{paragraph}{0}
  \setcounter{parentequation}{0}
  \setcounter{part}{0}
  \setcounter{section}{0}
  \setcounter{subfigure}{0}
  \setcounter{subparagraph}{0}
  \setcounter{subsection}{0}
  \setcounter{subsubsection}{0}
  \setcounter{table}{0}
  \papertitlepage{%
  Timed Comparisons of Semi-Markov Processes
}{paper:paperB}{%
  Mathias R. Pedersen, Nathanaël Fijalkow, Giorgio Bacci, Kim G. Larsen, and Radu Mardare
}{%
  The paper has been published in the\\
  \textit{Proceedings of the 12th International Conference on Language and Automata Theory and Applications}, pp.~271--283, 2018.
}{%
  \noindent\copyright\ 2018 Springer
  
  \noindent{\em The layout has been revised and the content extended.}
}

%% Abstract
\begin{abstract}
  Semi-Markov processes are Markovian processes
  in which the firing time of the transitions is modelled by probabilistic distributions over positive reals
  interpreted as the probability of firing a transition at a certain moment in time.

  In this paper we consider the trace-based semantics of semi-Markov processes, 
  and investigate the question of how to compare two semi-Markov processes with respect to their time-dependent behaviour.
  To this end, we introduce the relation of being ``faster than'' between processes and study its algorithmic complexity.
  Through a connection to probabilistic automata we obtain hardness results showing in particular that this relation is undecidable.
  However, we present an additive approximation algorithm for a time-bounded variant of the faster-than problem 
  over semi-Markov processes with slow residence-time functions, and a $\coNP$ algorithm 
  for the exact faster-than problem over unambiguous semi-Markov processes.
  Finally, we give a logical characterisation of the faster-than relation
  and show that satisfiability and model checking are decidable for this logic.
\end{abstract}

%% Introduction
\section{Introduction}
Semi-Markov processes are Markovian stochastic systems
that model the firing time of transitions as probabilistic distribution over positive reals;
thus, one can encode the probability of firing a certain transition within a certain time interval.
For example, continuous-time Markov processes are particular case of semi-Markov processes where the timing distributions are always exponential.

Semi-Markov processes have been used extensively to model real-time systems such as power plants~\cite{PST97} and power supply units~\cite{PTV04}.
For such real-time systems, non-functional requirements are becoming increasingly important.
Many of these requirements, such as response time and throughput, depend heavily on the timing behaviour of the system in question.
It is therefore natural to understand and be able to compare the timing behaviour of different systems. 
 
Moller and Tofts~\cite{MT91} proposed the notion of a \emph{faster-than} relation for systems with discrete-time in the context of process algebras.
Their goal was to be able to compare processes that are functionally behaviourally equivalent,
except that one process may execute actions faster than the other.
This line of study was continued by L{\"{u}}ttgen and Vogler~\cite{LV01}, who moreover considered
upper bounds on time, in order to allow for reasoning about worst-case timing behaviours.
For timed automata, Guha et al.~\cite{GNA12} introduced a bisimulation-like faster-than relation
and studied its compositional properties.
For continuous-time probabilistic systems, Baier et al.~\cite{BKHW05} considered a simulation relation
where the timing distribution on each state is required to stochastically dominate the other.
They introduced both a weak and a strong version of their simulation relation,
and gave a logical characterisation of these in terms of the logic CSL.

In the literature, less attention has been drawn to trace-based notions of faster-than relations
although trace equivalence and inclusion are important concepts when considering linear-time properties such as liveness or safety~\cite{BK08}.
In this paper we propose a simple and intuitive notion of trace 
inclusion for semi-Markov processes, which we call \emph{faster-than} relation, that compares the relative speed of
processes with respect to the execution of arbitrary sequences of actions.

Differently from trace inclusion, 
our relation does not make a step-wise comparison of the timing delays for each individual action in a sequence, but over the overall execution time of the sequence. 
As an example, consider the semi-Markov process in Fig. \ref{fig:faster-than-b}.
The states $s$ and $s'$, although performing the same sequences of actions,
are not related by trace inclusion because the first two actions in any sequence
are individually executed at opposite order of speeds (here governed by exponential-time distributions).
Instead, according to our relation, $s$ \emph{is} faster-than $s'$ (but not vice versa) because it executes single-action sequences at a faster rate than $s'$,
and action sequences of length greater than one at the same speed -- this is due to the fact that the execution time of each action is governed by random variables
that are independent of each other and the sum of independent random variables is commutative.

\begin{figure}
  \centering
  \begin{tikzpicture}
    \node[state, pin={[pin distance=0.2cm]$\text{Exp}(4)$}] (s) {$s$};
    \node[state, right = of s, pin={[pin distance=0.2cm]$\text{Exp}(2)$}] (s1) {};
    \node[state, right = of s1, pin={[pin distance=0.2cm]$\text{Exp}(1)$}] (s2) {};
    \node[state, right = of s2, pin={[pin distance=0.2cm]$\text{Exp}(4)$}] (s3) {};
    \node[state, right = of s3, pin={[pin distance=0.2cm]$\text{Exp}(2)$}] (s') {$s'$};
    
    \path[->] (s) edge node[above] {$a, 1$} (s1);
    \path[->] (s1) edge node[above] {$a, 1$} (s2);
    \path[->] (s3) edge node[above] {$a, 1$} (s2);
    \path[->] (s') edge node[above] {$a, 1$} (s3);
  \end{tikzpicture}
  \caption{A semi-Markov process where $s$ is faster than $s'$. The states of the process are annotated with their timing distributions 
           and each action-labelled transition is decorated with its probability to be executed.}
  \label{fig:faster-than-b}
\end{figure}
 
In this paper we investigate the algorithmic complexity of various problems regarding the faster-than relation,
emphasising their connection with classical algorithmic problems over Rabin's probabilistic automata.
In particular, we prove that the faster-than problem over generic semi-Markov processes is undecidable and that it is 
Positivity-hard when restricted to processes with only one action label. The reduction from the Positivity
problem is important because it relates the faster-than problem to the Skolem problem,
an important problem in number theory, whose decidability status has been an open problem for 
at least 80 years~\cite{OW14,AAOW15}.

We show that undecidability for the faster-than problem can not be tackled even by approximation techniques:
via the same connection with probabilistic automata we are able to prove that the faster-than 
problem can not be approximated up to a multiplicative constant.
However, as a positive result, we show that a time-bounded variant of the faster-than problem,
which compares processes up to a given finite time bound, although still undecidable,
admits approximated solutions up to an \emph{additive} constant over semi-Markov processes with slow residence-time distributions.
These include the important cases of uniform and exponential distributions.
As a second positive result, we present a $\coNP$ algorithm for solving the faster-than problem exactly over unambiguous semi-Markov processes,
where a process is unambiguous if every transition to a next state is unambiguously determined by the label that it outputs.

Finally, we give a logical characterisation of the faster-than relation in terms of a very simple logic.
Every formula in this logic is satisfiable in some finite model,
and hence the satisfiability problem for the logic is trivially decidable.
Furthermore, we show that the model checking problem for the logic is also decidable
for many common residence-time distributions.

%% Definitions
\section{Definitions}
For a finite set $S$ we let $\dist(S)$ denote the set of (sub)distributions over $S$, i.e. functions $\delta : S \to [0,1]$
such that $\sum_{s \in S} \delta(s) \le 1$.
The subset of total distributions is $\distone(S)$.

We let $\mathbb{N}$ denote the natural numbers and $\mathbb{R}_{\ge 0}$ denote the non-negative real numbers.
We equip $\mathbb{R}_{\ge 0}$ with the Borel $\sigma$-algebra $\B$,
so that $(\mathbb{R}_{\ge 0}, \B)$ is a measurable space.
Let $\dist(\mathbb{R}_{\ge 0})$ denote the set of (sub)distributions over $(\mathbb{R}_{\ge 0}, \B)$,
i.e. measures $\mu : \B \to [0,1]$ such that $\mu(\R_{\ge 0}) \le 1$.
Throughout the paper we will write $\mu(t)$ for $\mu([0,t])$.

To avoid confusion we will refer to $\mu$ in $\dist(\R_{\ge 0})$ as timing distributions,
and to $\delta$ in $\dist(S)$ as distributions.

\begin{definition}[Semi-Markov process]
A \emph{semi-Markov process}, usually written $\M$, is given by:
  \begin{itemize}
    \item $S$ is a (finite) set of \emph{states},
    \item $\aout$ is a (finite) set of \emph{output labels},
    \item $\Delta : S \to \dist(S \times \aout)$ is a \emph{transition function},
    \item $\rho : S \to \dist(\R_{\ge 0})$ is a \emph{residence-time function}. \qedhere
  \end{itemize}
\end{definition}

The operational behaviour of a semi-Markov process can be described as follows.
In a given state $s \in S$,
the process fires a transition within time $t$ with probability $\rho(s)(t)$,
leading to the state $s' \in S$ while outputting the label $a \in \aout$ with probability $\Delta(s)(s',a)$.

We aim at defining $\P_\M(s,w,t)$, 
the probability that from the state~$s$,
the output of the semi-Markov process $\M$ within time $t$ \emph{starts with} the word $w$.
It is important to note here that time is accumulated:
we sum together the time spent in all states along the way,
and ask that this total time is less than the specified bound $t$.
A full and formal definition of the probability can be done through the usual cylinder construction.
However, we will spare the reader this well-known construction and give seemingly ad-hoc definitions in this conference version.

In order to account for the accumulated time in the probability,
we need the notion of convolution. The convolution of two timing distributions $\mu$ and $\nu$ is $\mu * \nu$ defined by
\[
(\mu * \nu)(E) = \int_0^\infty \nu(E - x) \mu(\wrt{x})
\]
for any Borel set $E$.
Convolution is both associative and commutative.
Let $X$ and $Y$ be two independent random variables with timing distributions $\mu$ and $\nu$,
i.e. $\P(X \in E) = \mu(E)$ and $\P(Y \in E) = \nu(E)$,
then 
\[
\P(X + Y \in E) = (\mu * \nu)(E) .
\]

\begin{definition}[Probability]
  Consider a semi-Markov process $\M$. 
  We define the timing distribution $\P_\M(s,w)$ inductively:
  $\P_\M(s,\varepsilon) = \mathbbm{1}$ for the empty word $\varepsilon$,
  where $\mathbbm{1}$ is the function such that $\mathbbm{1}(t) = 1$ for all $t$ in $\R_{\ge 0}$,
  and for a word $w$ in $\aout^*$, a letter $a$ in $\aout$ and a state $s$,
  \[\P_\M(s,aw) = \sum_{s' \in S} \Delta(s)(s',a) \cdot \left(\rho(s) * \P_\M(s',w)\right) .\]
  We will then write $\P_\M(s,w,t)$ to mean $\P_\M(s,w)(t)$.
\end{definition}

\subsection{Timed Comparisons}
We introduce the following relation which will be the focus of our paper.

\begin{definition}[Faster-than relation]
  Consider a semi-Markov process $\M$ and two states $s$ and $s'$.
  We say that $s$ is \emph{faster than} $s'$, denoted $s \ft s'$,
  if for all $w$, for all $t$,
  \[
  \P_\M(s,w,t) \ge \P_\M(s',w,t) . \qedhere
  \]
\end{definition}

The algorithmic problem we consider in this paper is the \emph{faster-than problem}: 
given a semi-Markov process and two states $s$ and $s'$, determine whether $s \ft s'$.

\subsection{Algorithmic Considerations}
The definition we use for semi-Markov processes is very general, because we allow for any residence-time function.
The aim of the paper is to give generic algorithmic results which apply to \emph{effective} classes of timing distributions,
a notion we define now.
Recall that a residence-time function associates with each state a timing distribution.
We first give some examples of classical timing distributions.
\begin{itemize}
	\item The prime example is exponential distributions, defined by the timing distribution
	$\mu(t) = 1 - e^{-\lambda t}$ for some parameter $\lambda > 0$ usually called the rate.
	\item Another interesting example is that of piecewise polynomial distributions. Consider finitely many polynomials $P_1,\dots,P_n$
	and a finite set of pairwise disjoint intervals $I_1 \cup I_2 \cup \cdots \cup I_n$ covering $[0,\infty)$ such that 
	for every $k$, $P_k$ is non-negative over $I_k$ and $\sum_k \int_{I_k} P_k = 1$.
	This induces the timing distribution
	\[
	\mu(t) = \sum_k \int_{I_k \cap [0,t]} P_k(t) .
	\]
	\item A special case of the previous example is given by piecewise affine distributions, where the polynomials are affine functions.
	\item Another important special case of piecewise polynomial distributions are the uniform distributions with parameters $0 \leq a < b$ 
	defining the timing distribution
	\[
	\mu(t) = 
	\begin{cases}
	1 & \text{if } t < a, \\
	\frac{t-a}{b-a} & \text{if } t \in [a,b) \\ 
	0 & \text{if } x \ge b .
	\end{cases}
	\]
	\item The simplest example is given by Dirac distributions defined for the parameter $a$ by 
	$\mu(E) = 1$ if $a$ is in $E$, and $0$ otherwise.
\end{itemize}

The following definition captures these examples, and more.
For a class $\C$ of timing distributions, we let $\Convex(\C)$ be the smallest class of timing distributions containing $\C$ 
and closed under convex combinations, 
and similarly $\Conv(\C)$ adding closure under convolutions.

\begin{lemma}
  \label{lem:convC}
  Let $\C$ be a class of timing distributions.
  Consider a semi-Markov process $\M$ whose residence-time function uses timing distributions from $\C$,
  a state $s$ and a word $w$, then $\P_\M(s,w) \in \Conv(\C)$.
\end{lemma}

Lemma~\ref{lem:convC} is established by a straightforward induction on the word $w$ using the definition of $\P_\M(s,w)$.

In the rest of the paper we will consider only distributions that are suitable for algorithmic manipulation. Clearly, we must be able to give them as input to a computational device, so we assume they can be described by finitely many rational parameters. Moreover, we require that testing inequalities between them is decidable, since this is essential for determining the 
faster-than relation. The next definition formalises this intuition.

\begin{definition}[Effective timing distributions]
  A class $\C$ of timing distributions is \emph{effective} if, for any 
  $\varepsilon \ge 0$, $b \in \R_{\ge 0} \cup \set{\infty}$, and $\mu_1,\mu_2 \in \Conv(\C)$, 
  it is decidable whether $\mu_1(t) \ge \mu_2(t) - \varepsilon$, for all $t \le b$.
\end{definition}

Many common classes of timing distributions are effective,
and can be decided using the existential theory of the reals.
To show this, we make use of the following lemma.

\begin{lemma}\label{lem:transform}
  Let $F_1,F_2 : \mathbb{R}_{\geq 0} \rightarrow [0,1]$ be given timing distributions.
  If there exists a surjective function $T : [0,1] \rightarrow \mathbb{R}_{\geq 0}$ such that
  the functions $F_1 \circ T$ and $F_2 \circ T$ are semialgebraic,
  then it is decidable whether $F_1(t) \geq F_2(t)$ for all $t \in \mathbb{R}_{\geq 0}$.
\end{lemma}
\begin{proof}
  Since $F_1 \circ T$ and $F_2 \circ T$ are semialgebraic,
  the formula $\varphi$ defined by
  \[\varphi = \forall t. ((0 \leq t \leq 1) \implies F_1(T(t)) \geq F_2(T(t)))\]
  is expressible in the existential theory of the reals (or rather, its negation is),
  so we can decide whether $\varphi$ is true by exploiting the decidability of the existential theory of the reals \cite{Tarski51}.
  We now claim that $F_1(t) \geq F_2(t)$ for all $t$ is true if and only if $\varphi$ is true.
  
  Assume that $F_1(t) \geq F_2(t)$ for all $t \in \mathbb{R}_{\geq 0}$ is true.
  Pick an arbitrary $t' \in [0,1]$.
  Then $T(t') \in \mathbb{R}_{\geq 0}$, so we know that $F_1(T(t')) \geq F_2(T(t'))$,
  and hence $\varphi$ is true.
  
  Next assume that $\varphi$ is true and pick an arbitrary $t' \in \mathbb{R}_{\geq 0}$.
  Because $T$ is surjective, there must be some $y \in [0,1]$ such that $t' = T(y)$.
  Hence we know that 
  \[F_1(t') = F_1(T(y)) \geq F_2(T(y)) = F_2(t'). \qedhere\]
\end{proof}

In Lemma~\ref{lem:transform}, $T$ is a \emph{transformation} or \emph{variable change}
which turns the given functions into piecewise polynomial functions.
The requirement that the transformation be surjective ensures that deciding
the inequality between the transformed functions is equivalent to deciding it between the original functions.

\begin{proposition}\label{thm:effective_classes}
  The following classes of timing distributions are effective:
  \begin{itemize}
	  \item exponential distributions with rational rates,
	  \item piecewise polynomial distributions,
	  \item piecewise affine distributions,
	  \item uniform distributions,
	  \item Dirac distributions.
  \end{itemize}
\end{proposition}
\begin{proof}
  Let $\mathcal{C}$ be a class of timing distributions.
  We want to decide whether
  \[(\mu_1 * \cdots * \mu_n)(t) \geq (\nu_1 * \cdots * \nu_n)(t) \quad \text{for all } t \in \mathbb{R}_{\geq 0}\]
  whenever $\mu_1, \dots, \mu_n, \nu_1, \dots, \nu_n \in \mathcal{C}$.
  If we let $F_1(t) = (\mu_1 * \cdots * \mu_n)(t)$ and $F_2(t) = (\nu_1 * \cdots * \nu_n)(t)$,
  then Lemma~\ref{lem:transform} tells us that we can decide whether $F_1(t) \geq F_2(t)$ for all $t \in \mathbb{R}_{\geq 0}$
  by finding an appropriate surjective function $T : [0,1] \rightarrow \mathbb{R}_{\geq 0}$.
  
  When $\mathcal{C}$ is either the class of Dirac distributions, the class of piecewise affine distributions,
  or the class of piecewise polynomial distributions, this is trivial,
  since these are all already semialgebraic.
  Hence we can simply take $T$ to be the identity function.
  Although it is perhaps less obvious, the same is also true when $\mathcal{C}$ is the class of uniform distributions \cite{KvC10}.
  
  This leaves the case when $\mathcal{C}$ is the class of exponential distributions with rational rates.
  Assume that each $\mu_i$ has rate $\lambda_i$ and each $\nu_i$ has rate $\lambda_i'$.
  It was shown in \cite{bon1999} that $F_1$ has the following closed form.
  Assume that there are $m$ distinct rates among $\lambda_1, \dots, \lambda_n$
  and reorder $\lambda_1, \dots, \lambda_n$ such that $\lambda_1, \dots, \lambda_{r_1}$ are identical,
  $\lambda_{r_1 + 1}, \dots, \lambda_{r_1 + r_2}$ are identical, and so forth.
  This reordering does not change the values of $F_1$ because convolution is both associative and commutative.
  Now let $\alpha_1 = \lambda_{r_1}, \alpha_2 = \lambda_{r_1 + r_2}, \dots, \alpha_m = \lambda_{r_1 + \dots + r_m}$.
  The closed form of $F_1$ is then given by
  \[F_1(t) = 1 - \sum_{k = 1}^m \sum_{l = 1}^{r_k} C_1(k,l) \cdot e^{-\lambda_k \cdot t},\]
  where $C_1(k,l)$ is an expression that depends on $k$ and $l$, but not on $t$.
  Note also that $C_1(k,l)$ is expressible in the existential theory of the reals.
  Likewise, $F_2$ will have a closed form
  \[F_2(t) = 1 - \sum_{k = 1}^{m'} \sum_{l = 1}^{r_k'} C_2(k,l) \cdot e^{-\beta_k \cdot t}.\]
  
  Now let $T : [0,1] \rightarrow \mathbb{R}_{\geq 0}$ be given by
  \[T(x) = \begin{cases} 0 & \text{if } x = 0 \\ - \frac{1}{\gcd(\lambda_1, \dots, \lambda_n, \lambda_1', \dots, \lambda_n')} \cdot \ln(x) & \text{otherwise.}\end{cases}\]
  For convenience, we let $\theta = \gcd(\lambda_1, \dots, \lambda_n, \lambda_1', \dots, \lambda_n')$.
  Then $T$ is surjective, for if $y \in \mathbb{R}_{\geq 0}$,
  then $x = e^{-\theta \cdot y} \in [0,1]$ and $T(x) = y$.
  Furthermore,
  \[F_1(T(x)) = 1 - \sum_{k = 1}^m \sum_{l = 1}^{r_k} C_1(k,l) \cdot x^{\frac{\alpha_k}{\theta}},\]
  and because $\theta$ divides $\alpha_k$ for each $k$, it follows that $F_1 \circ T$ is a polynomial and hence semialgebraic.
  In a similar fashion we can show that $F_2 \circ T$ is also semialgebraic.
\end{proof}

Proposition~\ref{thm:effective_classes} relies on decidability results 
for the existential theory of the reals~\cite{Canny88,Tarski51},
implying that the most demanding operations above can be performed in polynomial space.

An effective class $\C$ of timing distributions induces the set of semi-Markov processes whose residence-time functions
use timing distributions from $\C$.
Furthermore, a given semi-Markov process has only finitely many states,
and hence can only use finitely many timing distributions.
For our decidability results we will therefore focus on finite classes of timing distributions.
This paper gives algorithmic results for generic effective classes of timing distributions.
In our complexity analyses, we will always assume that the operations on the timing distributions have a unit cost.

%% Hardness results
\section{Hardness Results}\label{sec:hardness}
We start the technical part of this article by hardness results inherited from Markov processes.
A Markov process is a semi-Markov process without the residence-time function,
and for a Markov process $\M = (S, \aout, \Delta)$,
we define the probability
\[\P_\M(s,aw) = \sum_{s' \in S} \Delta(s)(s',a) \cdot \P_\M(s')(w)\]
and
\[\P_\M(s,\varepsilon) = 1\]
for the empty word.
The faster-than relation for Markov processes is then $s \preceq s'$ if for all $w$ we have $\P_\M(s,w) \geq \P_\M(s',w)$.

We show that the faster-than problem for Markov processes, and hence also for semi-Markov processes, is undecidable in general,
can not be multiplicatively approximated, and relates to an open problem in number theory even in a restricted case.
These limitations shape and motivate our positive results,
which will be the topic of the remaining sections.

We first explain how hardness results for Markov processes directly imply hardness results for semi-Markov processes.
The following lemma formalises the two ways semi-Markov processes subsume Markov processes.

\begin{lemma}\label{lem:Markov_process_reduction}
  Consider a semi-Markov process $\M = (S,\aout,\Delta,\rho)$ and its induced Markov process $\M' = (S,\aout,\Delta)$.
  \begin{itemize}
	  \item If $\rho$ is constant, i.e. for all $s,s'$ we have $\rho(s) = \rho(s')$,
	  then for all $w$, for all $t$, we have $\P_\M(s,w,t) = \P_{\M'}(s,w) \cdot 
	  (\underbrace{\rho(s) * \cdots * \rho(s)}_{|w| \text{ times}})(t)$.
	  \item If for all $s$, $\rho(s)$ is the Dirac distribution for $0$, 
	  then for all $w$, for all $t$, we have $\P_\M(s,w,t) = \P_{\M'}(s,w)$.
  \end{itemize}
  In particular in both cases, the following holds:
  for $s,s'$ two states, we have $s \ft s'$ in $\M$ if, and only if, $s \ft s'$ in $\M'$.
\end{lemma}

We will use Lemma~\ref{lem:Markov_process_reduction} to draw corollaries about semi-Markov processes from hardness results of Markov processes.

The hardness results of this section will be based on a connection to probabilistic automata.
A probabilistic automaton is given by 
\[
\A = (Q,A,q_0,\Delta : Q \times A \to \D_{=1}(Q),F) ,
\]
where $Q$ is the state space, $A$ is the alphabet, $q_0$ is an initial state, $\Delta$ is the transition function,
and $F$ is a set of final or accepting states.
Any probabilistic automaton $\A$ induces the probability $\P_\A(w)$ that a run over $w \in A^*$ is accepting,
i.e. starts in $q_0$ and ends in $F$.
The key property of probabilistic automata that we will exploit is the undecidability of the universality problem, 
which was proved in~\cite{Paz71}, see also~\cite{GO10}.
The universality problem is as follows: given a probabilistic automaton $\A$, determine whether for all words $w$ in $A^+$ we have 
$\P_\A(w) \ge \frac{1}{2}$.

We describe a construction which given a probabilistic automaton $\A$, constructs the \emph{derived} Markov process $\M(\A)$.
The set of states of $\M(\A)$ is $Q \times \set{\ell,r} \cup \set{\top}$, where $\top$ is a new state.
Let $s = (q_0,\ell)$ and $s' = (q_0,r)$, where $q_0$ is the initial state of $\A$.
The set of output labels is $A$, and the transition function $\Delta'$ is defined as follows:
\begin{align*}
  \Delta'((p,\ell))((q,\ell),a) &= \frac{1}{2|A|} \Delta(p,a)(q)  &\quad \quad
  \Delta'((p,\ell))(\top,a)     &= \frac{1}{2|A|} \text{ if } p \in F \\
  \Delta'((p,r))((q,r),a)       &= \frac{1}{2|A|} \Delta(p,a)(q)  &\quad \quad
  \Delta'((p,r))(\top,a)        &= \frac{1}{4|A|} .
\end{align*}

\begin{lemma}\label{lem:equalities}
  \[
  \P_{\M(\A)}(s,wa)	= \frac{1}{(2 |A|)^{|w| + 1}} \left(1 + \P_\A(w) \right)
  \]
  and
  \[
  \P_{\M(\A)}(s',wa)= \frac{1}{(2 |A|)^{|w| + 1}} \left(1 + \frac{1}{2} \right).
  \]
\end{lemma}
\begin{proof}
  First observe that it can easily be proven by induction that
  \[\prob_{\M(\A)}(s,w) = \sum_{s_1 \in S} \dots \sum_{s_n \in S} \Delta'(s)(s_1,w_1) \cdots \Delta'(s_{n-1})(s_n,w_n)\]
  where $w = w_1 \dots w_n$ by simply unfolding the inductive definition of $\prob$.
  
  For the first equality, we therefore have
  \begin{align*}
    &\phantom{{}={}}\prob_{\M(\A)}(s,wa) \\
    &= \sum_{s_1 \in S} \dots \sum_{s_{n+1} \in S} \Delta'(s)(s_1,w_1) \cdots \Delta'(s_{n-1})(s_n,w_n) \cdot \Delta'(s_n)(s_{n+1},a) \\
    &= \sum_{s_1 \in Q \times \{\ell\}} \dots \sum_{s_{n+1} \in Q \times \{\ell\}} \frac{1}{2|A|} \Delta(s,w_1)(s_1) \cdots \frac{1}{2|A|} \Delta(s_n,a)(s_{n+1}) \\
    &\phantom{{}={}} + \sum_{s_1 \in Q \times \{\ell\}} \dots \sum_{s_n \in Q \times \{\ell\}} \frac{1}{2|A|} \Delta(s,w_1)(s_1) \cdots \Delta'(s_n)(\top,a) \\
    &= \frac{1}{(2|A|)^{|w| + 1}} + \frac{1}{(2|A|)^{|w| + 1}} \cdot \prob_{\A}(w) \\
    &= \frac{1}{(2|A|)^{|w| + 1}}\left(1 + \prob_{\A}(w)\right).
  \end{align*}
  
  For the second equality, we get
  
  \begin{align*}
    &\phantom{{}={}}\prob_{\M(\A)}(s',wa) \\
    &= \sum_{s_1 \in S} \dots \sum_{s_{n+1} \in S} \Delta'(s)(s_1,w_1) \cdots \Delta'(s_{n-1})(s_n,w_n) \cdot \Delta'(s_n)(s_{n+1},a) \\
    &= \sum_{s_1 \in Q \times \{r\}} \dots \sum_{s_{n+1} \in Q \times \{r\}} \frac{1}{2|A|} \Delta(s,w_1)(s_1) \cdots \frac{1}{2|A|} \Delta(s_n,a)(s_{n+1}) \\
    &\phantom{{}={}} + \sum_{s_1 \in Q \times \{r\}} \dots \sum_{s_n \in Q \times \{r\}} \frac{1}{2|A|}\Delta(s,w_1)(s_1) \cdots \frac{1}{4|A|} \\
    &= \frac{1}{(2|A|)^{|w| + 1}} + \frac{1}{(2|A|)^{|w|}} \cdot \frac{1}{4|A|} \\
    &= \frac{1}{(2|A|)^{|w| + 1}}\left(1 + \frac{1}{2}\right). \qedhere
  \end{align*}
\end{proof}

\begin{theorem}\label{thm:fasterthan_undecidability}
  The faster-than problem is undecidable for Markov processes.
\end{theorem}
\begin{proof}
  Given a probabilistic automaton $\A$, we construct
  the derived Markov process $\M(\A)$.
  Thanks to the equalities in Lemma~\ref{lem:equalities}, $\A$ is universal if, and only if, $s \ft s'$.
\end{proof}

We discuss three approaches to recover decidability.

A first approach is to look for \emph{structural restrictions} on the underlying graph.
However, the undecidability result above for probabilistic automata is quite robust in this aspect,
as it already applies when the underlying graph is acyclic, meaning that the only loops are self-loops.
In spite of this, we present in Section~\ref{sec:unambiguous} an algorithm to solve the faster-than problem
for \emph{unambiguous} semi-Markov processes.

A second approach is to restrict the \emph{observations}.
The undecidability result above holds already when there are two different output letters,
hence a natural question is to look at what happens when we only have one output letter.
Interestingly, specialising the construction above yields a reduction from the Positivity problem.
This problem appears in various contexts, prominently in number theory,
and its decidability status has been an open problem for at least 30 years~\cite{OW14}.
Formally, the Positivity problem reads: given a linear recurrence sequence, are all terms of the sequence non-negative?
It has been shown that the universality problem for probabilistic automata with one letter alphabet is equivalent 
to the Positivity problem~\cite{AAOW15}.
Thus, using again the derived Markov process $\M(\A)$ for a probabilistic automaton $\A$ with only one label, we obtain the following result.

\begin{theorem}
  The faster-than problem is Positivity-hard over Markov processes with one output label.
\end{theorem}

A third approach is \emph{approximations}.
However, we can exploit further the connection we made with probabilistic automata, %in Theorem~\ref{thm:fasterthan_undecidability},
obtaining an impossibility result for \emph{multiplicative approximation}.
We rely on the following celebrated theorem for probabilistic automata due to Condon and Lipton \cite{CL89}.
The following formulation of their theorem is described in detail in \cite{Fijalkow17}.

\begin{theorem}[\hspace{1sp}\cite{CL89}]
\label{thm:inapproximability_automata}
Let $0 < \alpha < \beta < 1$ be two constants.
There is no algorithm which, given a probabilistic automaton $\A$,
  \begin{itemize}
    \item if for all $w$ we have $\P_\A(w) \geq \beta$, returns YES,
    \item if there exists $w$ such that $\P_\A(w) \leq \alpha$, returns NO.
  \end{itemize}
\end{theorem}

\begin{theorem}\label{thm:impossibility_approximation}
  Let $0 < \varepsilon < \frac{1}{3}$ be a constant.
  There is no algorithm which, given a Markov process $\M$ and two states $s,s'$,
  \begin{itemize}
    \item if for all $w$ we have $\P_\M(s,w) \geq \P_\M(s',w)$, returns YES,
    \item if there exists $w$ such that $\P_\M(s,w) \leq \P_\M(s',w) \cdot (1 - \varepsilon)$, returns NO.
  \end{itemize}
\end{theorem}
\begin{proof}
  Assume towards a contradiction that there exists an algorithm as described in the theorem.
  We then construct an algorithm satisfying the specifications of Theorem~\ref{thm:inapproximability_automata}.

  Let $\alpha = \frac{1}{2} - \frac{3\varepsilon}{2}$ and $\beta = \frac{1}{2}$,
  and let $\A$ be a probabilistic automaton.
  We now run the algorithm on the derived Markov process $\M(\A)$.
  \begin{itemize}
      \item If for all $w$ we have $\P_{\M(\A)}(s,w) \geq \P_{\M(\A)}(s',w)$, then the algorithm returns YES. 
      Indeed, this is equivalent to $\P_\A(w) \ge \beta$.
      \item If there exists $w$ such that $\P_{\M(\A)}(s,w) \leq \P_{\M(\A)}(s',w) \cdot (1 - \varepsilon)$, then the algorithm returns NO.
      Indeed, this is equivalent to $\P_\A(w) \le \alpha$.
  \end{itemize}
  Hence we constructed an algorithm satisfying the specifications of Theorem~\ref{thm:inapproximability_automata},
  a contradiction.
\end{proof}

These hardness results for Markov processes together with Lemma~\ref{lem:Markov_process_reduction},
gives us the following hardness results for semi-Markov processes.

\begin{corollary}
  The following holds for semi-Markov processes for any class of timing distributions.
  \begin{itemize}
    \item The faster-than problem is undecidable.
    \item The faster-than problem with only one output label is Positivity-hard.
    \item The faster-than problem can not be multiplicatively approximated.
  \end{itemize}
\end{corollary}

%% Approximation
\section{Time-Bounded Additive Approximation}\label{sec:approximation}
Instead of considering multiplicative approximation,
we can also consider additive approximation,
meaning that we want to decide whether for all $w$ and $t$ we have $\P_\M(s,w,t) \geq \P_\M(s',w,t) - \varepsilon$
for some constant $\varepsilon > 0$.
In this section, we present an algorithm to solve the problem of approximating additively the faster-than relation with two assumptions:
\begin{itemize}
	\item \emph{time-bounded}: we only look at the behaviours up to a given bound $b$ in $\mathbb{R}_{\ge 0}$,
	\item \emph{slow residence-time functions}: each transition takes \emph{some} time to fire.
\end{itemize}
As we will show, the combination of these two assumptions imply that the relevant words have bounded length.
This is in contrast to the impossibility of approximating the faster-than relation multiplicatively
that we showed in Sect. \ref{sec:hardness}.
More precisely, we consider the \emph{time-bounded} variant of the faster-than problem:
given a time bound $b$ in $\R_{\ge 0}$, a semi-Markov process, and two states $s$ and $s'$,
determine whether for all $t \leq b$ and $w$ it holds that $\P_\M(s,w,t) \geq \P_\M(s',w,t)$.

We first observe that this restriction of the faster-than problem
does not make any of the problems in Sect. \ref{sec:hardness} easier for semi-Markov processes. 
Indeed, if the residence-time functions are all Dirac distributions on $0$, then all transitions are fired instantaneously, and
the time-bounded restriction is immaterial.
Thus we focus on distributions that do not fire instantaneously,
as made precise by the following definition.

\begin{definition}[Slow distributions]
  We say that a class $\C$ of timing distributions is \emph{slow} if 
  for all finite subset $\C_0$ of $\C$, 
  there exists a computable function $\varepsilon \colon \N \times \R_{\ge 0} \to [0,1]$ such that 
  for all $n$, $t$, and $\mu_1, \dots,\mu_n \in \Convex(\C_0)$ we have 
  $(\mu_1 * \dots *\mu_n)(t) \leq \varepsilon(n,t)$
  and $\lim_{n \to \infty} \varepsilon(n,t) = 0$.
\end{definition}

Given a slow and effective class $\C$ of timing distributions,
we can do additive approximation of the time-bounded faster-than problem in the following way.
We introduce the following notation.
Fix a semi-Markov process $\M$.
Let $\C_\M = \Convex(\set{\rho(s) \mid s \in S})$, and $n$ in $\N$.
We define the timing distribution $F_{\M,n}$ by $F_{\M,n}(t) = 1$ if $n = 0$ and otherwise
\[
  F_{\M,n}(t) = \sup \set{(\mu_1 * \cdots * \mu_n)(t) \mid \mu_1,\ldots,\mu_n \in \C_\M} .
\]

\begin{lemma}\label{lem:upperbound_f}
  For all $s$ and all $w$, we have $\P_\M(s,w) \le F_{\M,|w|}$.
\end{lemma}
\begin{proof}
  We proceed by induction on the length of $w$.
  It is clear for $|w| = 0$.
  \begin{align*}
    \P_\M(s,aw) 
    &= \sum_{s' \in S} \Delta(s)(s',a) \cdot \rho(s) * \P_\M(s',w)  \\
    &\le \underbrace{\sum_{s' \in S} \Delta(s)(s',a) \cdot \rho(s)}_{\in \C_\M} * F_{\M,|w|} \\
    &\le F_{\M,|w|+1} .
  \end{align*}
  This concludes.
\end{proof}

\begin{theorem}\label{thm:approx_faster}
  There exists an additive approximation algorithm for the time-bounded faster-than problem over semi-Markov processes
  for all slow and effective classes of timing distributions.

  In other words, for a constant $\varepsilon > 0$, there exists an algorithm which, given a semi-Markov process $\M$,
  two states $s,s'$, and a bound $b$ in $\R_{\geq 0}$,
  determines whether
  \[
  \forall w, \forall t \le b,\ \P_\M(s,w,t) \ge \P_\M(s',w,t) - \varepsilon .
  \]
\end{theorem}
\begin{proof}
Let $\C_\M = \Convex(\set{\rho(s) \mid s \in S})$, since $S$ is finite there exists 
a computable function $\varepsilon \colon \N \times \R_{\ge 0} \to [0,1]$ such that 
for all $n$, $t$, and $\mu_1, \dots,\mu_n \in \C_\M$ we have 
$(\mu_1 * \dots *\mu_n)(t) \leq \varepsilon(n,t)$
and $\lim_{n \to \infty} \varepsilon(n,t) = 0$.
Given $\varepsilon > 0$, there exists $N$ such that $\varepsilon(N,b) < \varepsilon$.
Let $n \ge N$. By assumption
\[
(\mu_1 * \dots * \mu_n)(b) \leq \varepsilon(n,b) \le \varepsilon(N,b) < \varepsilon
\] 
for all $\mu_1, \dots, \mu_n \in \C_\M$.
Taking the supremum over $\mu_1, \dots, \mu_n$, we then get $F_{\M, n}(b) < \varepsilon$,
and by Lemma \ref{lem:upperbound_f}, this means that for all $w$ of length at least $N$, we have $\P_\M(s',w,b) < \varepsilon$.
Hence it holds trivially that for all $t \leq b$ and $w$ of length at least $N$, we have $\P_\M(s,w,t) \geq \P_\M(s',w,t) - \varepsilon$.

Thus the algorithm checks whether for all words of length less than $N$, for all $t \le b$, we have 
$\P_\M(s,w,t) \geq \P_\M(s',w,t) - \varepsilon$, which is decidable thanks to the effectiveness of $\C$.
\end{proof}

Next we show that there are interesting classes of timing distributions that are indeed slow.
For this we introduce a class of timing distributions that are not just slow,
but furthermore are guaranteed to converge to zero rapidly.
We say that a timing distribution $\mu$ is \emph{very slow} if there exists a computable function $\varepsilon : \R_{\ge 0} \to [0,1]$ 
such that $\lim_{t \to 0} \frac{\varepsilon(t)}{t} = 0$
and for all $t$, we have $\mu(t) \le \varepsilon(t)$.
In order to show that very slow timing distributions are slow,
we need the following lemma.

\begin{lemma}\label{lem:bound_sum}
  Let $\mu_1,\ldots,\mu_n$ be timing distributions.
  Then 
  \[
  (\mu_1 * \mu_2 * \cdots * \mu_n)(t) \leq \sum_{i = 1}^n \mu_i \left(\frac{t}{n}\right) .
  \]
\end{lemma}
\begin{proof}
  We proceed by induction on $n$.
  The case of $n = 1$ is trivial.
  Recall that for any non-negative function $f$ and measure $\mu$ we have
  \begin{equation}\label{eq:ineq}
    \int_E f(x) \mu(\wrt{x}) \leq \mu(E) \cdot (\sup_E f(x)) .
  \end{equation}
  Let $\mu = \mu_1 * \dots * \mu_n$.
  \begin{align*}
		     & (\mu_1 * \cdots * \mu_{n+1})(t) \\
             &= \int_0^t \mu(t-x) \mu_{n+1}(\wrt{x}) \\
             &= \int_0^{\frac{nt}{n+1}} \mu(t-x) \mu_{n+1}(\wrt{x}) 
             + \int_{\frac{nt}{n+1}}^t \mu(t-x) \mu_{n+1}(\wrt{x}) \\
             &= \int_0^{\frac{nt}{n+1}} \mu(t-x) \mu_{n+1}(\wrt{x}) 
             + \int_0^{\frac{t}{n+1}} \mu \left(\frac{t}{n+1} - u \right) \mu_{n+1}(\wrt{u}) \\
             &\leq \mu \left(\frac{nt}{n+1}\right) 
             + \mu_{n+1}\left(t-\frac{nt}{n+1}\right) \\
             &\leq \sum_{i=1}^n \mu_i\left(\frac{n}{n+1} \frac{t}{n}\right) 
             + \mu_{n+1} \left(\frac{t}{n+1}\right) 
             = \sum_{i=1}^{n+1} \mu_i\left(\frac{t}{n+1}\right) .
  \end{align*}
  The third equality is the change of variable $u = x - \frac{nt}{n+1}$.
  The first inequality uses for each summand the inequality~\eqref{eq:ineq}.
  The second inequality is by induction hypothesis.
\end{proof}

We can now prove the following theorem.

\begin{theorem}
  The following classes of timing distributions are slow:
  \begin{itemize}
    \item very slow distributions,
    \item uniform distributions, and
    \item exponential distributions.
  \end{itemize}
\end{theorem}
\begin{proof}
  Let $\C$ be a class of very slow timing distributions, and 
  \[\C_0 = \set{\mu_1,\ldots,\mu_n}\] a finite subset of $\C$.
  Since every timing distribution in $\C$ is very slow,
  for every $i \in \set{1,\ldots,n}$ there exists a function $\varepsilon_i$ such that $\mu_i(t) \leq \varepsilon_i(t)$ for all $t$.
  Let $\varepsilon(n,t) = n \cdot \max \set{\varepsilon_i\left(\frac{t}{n} \right) \mid i \in \set{1,\ldots,n}}$.
  Note that $\lim_{n \to \infty} \varepsilon(n,t) = 0$.
  Let $\nu_1,\ldots,\nu_n$ in $\Convex(\C_0)$, we have $(\nu_1 * \cdots * \nu_n)(t) \le \sum_{i = 1}^n \nu_i \left(\frac{t}{n} \right)$
  thanks to Lemma~\ref{lem:bound_sum}.
  This implies that $(\nu_1 * \cdots * \nu_n)(t) \le \varepsilon(n,t)$, 
  which concludes.

  For exponential distributions, we proceed as follows.
  Let $\C_0$ be a finite class of exponential distributions.
  Let $\lambda > 0$ be the rate of the slowest exponential distributions appearing in $\C_0$,
  and let $\mu(t) = 1 - e^{-\lambda t}$.
  Then for any $\mu_1, \dots, \mu_n$ in $\Convex(\C_0)$ we have
  \[(\mu_1 * \dots * \mu_n)(t) \leq (\underbrace{\mu * \dots * \mu}_{n \text{ times}})(t).\]
  The distribution $\mu * \dots * \mu$ is called the Gamma (or more precisely, Erlang) distribution,
  and there is a computable closed form for it.
  In particular, if we let 
  \[\varepsilon(n,t) = (\underbrace{\mu * \dots * \mu}_{n \text{ times}})(t),\]
  we have $\lim_{n \to \infty} \varepsilon(n,b) = 0$,
  so exponential distributions are slow.

  Uniform distributions can be handled using a similar way as for exponential distributions.
  Let $\C_0$ be a finite class of uniform distributions with parameters $a_i$ and $b_i$ for $i \in \set{1,\ldots,n}$.
  Let $a$ be the smallest $a_i$ and $b$ the smallest $b_i$,
  and let $\mu$ be the uniform distribution with parameters $a$ and $b$.
  Then it follows that
  \[(\mu_1 * \dots * \mu_n)(t) \leq (\underbrace{\mu * \dots * \mu}_{n \text{ times}})(t) = \varepsilon(n,t).\]
  Then $(\mu * \dots * \mu)$ also has a nice closed form \cite{KvC10}
  and $\lim_{n \to \infty} \varepsilon(n,b) = 0$.
\end{proof}

%% Unambiguous processes
\section{Unambiguous Semi-Markov Processes}\label{sec:unambiguous}
In order to regain decidability of the faster-than relation,
we can look at structurally simpler special cases of semi-Markov processes.
Here we will focus on semi-Markov processes such that each output word induces
at most one trace of states.
More precisely, we will say that a semi-Markov process is \emph{unambiguous} if 
for every $s$ in $S$ and $a$ in $\aout$, there exists at most one $s'$ in $S$ such that $\Delta(s)(s',a) \neq 0$.
A related notion of bounded ambiguity has been utilised to obtain decidability results in the context of probabilistic automata \cite{FRW17}.
We introduce the following notation for unambiguous semi-Markov processes:
$T(s,w)$ is the state reached after emitting $w$ from $s$.

\begin{example}
  Figure \ref{fig:unambiguous} gives an example of an unambiguous semi-Markov process.
  For each of the three states, there is at most one state that can be reached by a given output label.
  However, there need not be a transition for each output label from every state.
  In this example, the state $s_2$ has no $b$-transition, so for instance $T(s_1,ab) = s_2$, but $T(s_1,abb)$ is undefined.
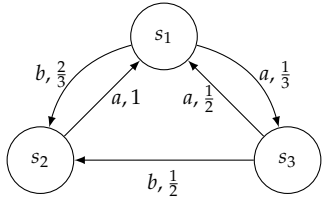
\begin{figure}
  \centering
  \begin{tikzpicture}
    \node[state] (s1) {$s_1$};
    \node[state, below left = of s1] (s2) {$s_2$};
    \node[state, below right = of s1] (s3) {$s_3$};
    
    \path[->] (s3) edge node[left] {$a, \frac{1}{2}$} (s1);
    \path[->] (s3) edge node[below] {$b, \frac{1}{2}$} (s2);
    \path[->] (s2) edge node[right] {$a, 1$} (s1);
    \path[->] (s1) edge[bend left] node[right] {$a, \frac{1}{3}$} (s3);
    \path[->] (s1) edge[bend right] node[left] {$b, \frac{2}{3}$} (s2);
  \end{tikzpicture}
  \caption{An example of an unambiguous semi-Markov process.}
  \label{fig:unambiguous}
\end{figure}
\end{example}

\begin{theorem}\label{thm:unambiguous}
  The faster-than problem is decidable in $\coNP$ over unambiguous semi-Markov processes for all effective classes of timing distributions.
\end{theorem}

Theorem~\ref{thm:unambiguous} follows from the next proposition.

\begin{proposition}\label{prop:unambiguous}
  Consider an unambiguous semi-Markov process $\M$ and two states $s,s'$.
  Let $L(s,s')$ be the set of loops reachable from $(s,s')$:
  \[
  \set{(p,p',v) \in S^2 \times \aout^{\le S^2} 
  \left|
  \ \exists w \in \aout^{\le S^2},\ 
  \begin{array}{c}
  T(s,w) = p,\ T(s',w) = p',\\
  T(p,v) = p,\ T(p',v) = p'
  \end{array}
  \right.} .
  \]

  We have $s \ft s'$ if, and only if
  \begin{itemize}
	  \item for all $w$ in $\aout^{\le S^2}$, we have $\P_\M(s,w) \ge \P_\M(s',w)$, and
	  \item for all $(p,p',v)$ in $L(s,s')$, we have $\P_\M(p, v) \ge \P_\M(p',v)$.
  \end{itemize} 
\end{proposition}

Before going into the proof, we explain how to use Proposition~\ref{prop:unambiguous}
to construct an algorithm solving the faster-than problem over unambiguous semi-Markov processes.
\begin{enumerate}
	\item The first step is to compute $L(s,s')$, which can be done in polynomial time using a simple graph analysis,
	\item The second step is to check the two properties, which both can be reduced 
	to exponentially many queries of the form: $\mu_1 \ge \mu_2$ for $\mu_1,\mu_2$ in $\Conv(\C)$.
\end{enumerate}
To obtain a $\coNP$ algorithm, in the second step we guess which of the two properties is not satisfied and a witness of polynomial length, 
which is either a word of quadratic length for the first property,
or two states and a word of quadratic length for the second property.

We split the proof of Proposition~\ref{prop:unambiguous} into two lemmas, each proving one direction of the proposition.
The following lemma gives the first direction.

\begin{lemma}
  \label{lem:hard_direction_unambiguous}
  If $s \ft s'$, then, for all $(p,p',v) \in L(s,s')$, $\P_\M(p, v) \ge \P_\M(p', v)$.
\end{lemma}

\begin{proof}
  Assume that $s$ is faster than $s'$ and let $(p,p')$ be in $L(s,s')$.
  There exist $w,v$ in $\aout^*$ such that 
  $T(s,w) = p,\ T(s',w) = p',\ T(p,v) = p,\ T(p',v) = p'$.
  Let $n$ in $\N$.
  Since $s$ is faster than $s'$, we have $\P_\M(s,w v^n) \ge \P_\M(s',w v^n)$.
  We have 
  \begin{align*}
    \P_\M(s, w v^n) &= \P_\M(s,w) * \underbrace{\P_\M(p,v) * \dots * \P_\M(p,v)}_{n \text{ times}} \\
    \P_\M(s',w v^n) &= \P_\M(s',w) * \underbrace{\P_\M(p',v) * \dots * \P_\M(p',v)}_{n \text{ times}} .
  \end{align*}
  Let $X_{s,w}$ be the random variable measuring the time elapsed from $s$ emitting $w$.
  Similarly, we define $X_{p,v}, Y_{s',w}$ and $Y_{p',v}$.
  We have: for all $n$ in $\N$, for all $t$,
  \[\P_\M(X_{s,w} + n X_{p,v} \leq t) \geq \P_\M(Y_{s',w} + n Y_{p',v} \leq t) ,\]
  Dividing both sides by $n$ yields
  \[\P_\M \left(\frac{X_{s,w}}{n} + X_{p,v} \leq \frac{t}{n} \right) \geq \P_\M \left(\frac{Y_{s',w}}{n} + Y_{p',v} \leq \frac{t}{n} \right) .\]
  We make the change of variables $x = \frac{t}{n}$: for all $n$ in $\N$, for all $x$ we have
  \[\P_\M \left(\frac{X_{s,w}}{n} + X_{p,v} \leq x\right) \geq \P_\M \left(\frac{Y_{s',w}}{n} + Y_{p',v} \leq x\right) .\]
  Letting $n \rightarrow \infty$, we then obtain, for all $x$
  \[\P_\M(X_{p,v} \leq x) \geq \P_\M(Y_{p',v} \leq x) ,\]
  which is equivalent to $\P_\M(p,v) \ge \P_\M(p',v)$.
\end{proof}

The following lemma gives the converse implication of Proposition~\ref{prop:unambiguous}.

\begin{lemma}\label{lem:easy_direction_unambiguous}
  Assume that
  \begin{itemize}
	  \item for all $w$ in $\aout^{\le S^2}$, we have $\P_\M(s,w) \ge \P_\M(s',w)$, and
	  \item for all $(p,p',v)$ in $L(s,s')$, we have $\P_\M(p, v) \ge \P_\M(p', v)$.
  \end{itemize} 
  Then $s \ft s'$.
\end{lemma}

\begin{proof}
  We prove that for all $w$, we have $\P_\M(s,w) \ge \P_\M(s',w)$ by induction on the length of $w$.

  For $w$ of length at most $S^2$, this is ensured by the first assumption.
  Let $w$ be a word longer than $S^2$.
  There exist two states $p,p'$ such that $p$ is reached by $s$ and $p'$ by $s'$
  after emitting $i$ letters of $w$ and again after emitting $j$ letters of $w$,
  with $j$ at most $S^2$.
  Let $w = w_1\ v\ w_2$ where $v$ starts at position $i$ and ends at position $j$. 
  By construction $(p,p',v)$ is in $L(s,s')$.
  We have
  \begin{align*}
  \P_\M(s,w) &= \P_\M(s,w_1) * \P_\M(p,v) * \P_\M(p,w_2) \\
		         &= \P_\M(s,w_1) * \P_\M(p,w_2) * \P_\M(p,v) \\
		         &= \P_\M(s,w_1 w_2) * \P_\M(p,v) \\
		         &\ge \P_\M(s',w_1 w_2) * \P_\M(p',v) \\
		         &= \P_\M(s',w_1) * \P_\M(p',w_2) * \P_\M(p',v) \\
		         &= \P_\M(s',w_1) * \P_\M(p',v) * \P_\M(p',w_2) \\
             &= \P_\M(s',w) .
  \end{align*}
  The equalities use the associativity and commutativity of the convolution.
  The inequality $\P_\M(s,w_1 w_2) \ge \P_\M(s',w_1 w_2)$ holds by induction hypothesis, 
  because $w_1 w_2$ is shorter than $w$.
  The inequality $\P_\M(p,v) \ge \P_\M(p',v)$ holds thanks to the second assumption.
\end{proof}

%% Logic
\section{Logic}
In this section we give a logical characterisation of the faster-than relation.
The logic needed for this turns out to be quite simple,
and it therefore possesses many nice properties.
In particular, every formula is satisfiable by a finite model.

The logic $\mathcal{L}$ consists of path formulas
\[\varphi ::= \top \mid \diam{a} \varphi\]
and state formulas
\[\psi ::= \mathcal{P}^{\leq t}_{\geq p} (\varphi)\]
where $t,p \in \mathbb{Q}_{\geq 0}$.

For the semantics of $\mathcal{L}$,
we consider paths $\pi = a_1 a_2 \dots \in \aout^{*}$ to be infinite sequences of output labels,
and we let $\pi[i] = a_i$ be the $i$th label of $\pi$.
The semantics are then given by
\[\begin{array}{l l l}
  \pi \models \top & & \text{always} \\
  \pi \models \diam{a} \varphi & \text{iff} & \pi[1] = a \text{ and } \pi|_2 \models \varphi \\
  \mathcal{M},s \models \mathcal{P}^{\leq t}_{\geq p} (\varphi) & \text{iff} & \prob_{\mathcal{M}}(s, \mathfrak{W}(\varphi))(t) \geq p 
\end{array}\]
where $\pi|_2$ is the tail of $\pi$,
and $\mathfrak{W}(\varphi)$ is the longest common prefix of all paths which satisfy $\varphi$.

\begin{theorem}\label{thm:logic}
  $s \ft s'$ if and only if $\mathcal{M},s' \models \psi$ implies $\mathcal{M},s \models \psi$, for all $\psi \in \mathcal{L}$.
\end{theorem}
\begin{proof}
  (${\implies}$)
  Let $s \ft s'$ and assume $\mathcal{M},s' \models \mathcal{P}^{\leq t}_{\geq p} (\varphi)$.
  One can easily prove by structural induction on $\varphi$
  that $\prob_\mathcal{M}(s',\mathfrak{W}(\varphi)) = \prob_\mathcal{M}(s',a_1 \dots a_n)$ for some $a_1, \dots, a_n$.
  Hence we know that
  \[\prob_\mathcal{M}(s',\mathfrak{W}(\varphi))(t) = \prob_\mathcal{M}(s', a_1 \dots a_n)(t) \geq p,\]
  and since $s \ft s'$, this implies that
  \[\prob_\mathcal{M}(s,a_1 \dots a_n)(t) \geq \prob_\mathcal{M}(s',a_1 \dots a_n)(t) \geq p,\]
  so $\prob_\mathcal{M}(s,\mathfrak{W}(\varphi))(t) \geq p$.
  
  (${\impliedby}$)
  We show the contrapositive.
  Assume that $s \not \ft s'$,
  meaning that there exists $a_1 \dots a_n$ and $t$ such that
  \[\prob_\mathcal{M}(s,a_1 \dots a_n)(t) < \prob_\mathcal{M}(s',a_1 \dots a_n)(t).\]
  Then we can find a rational $q$ such that
  \[\prob_\mathcal{M}(s,a_1 \dots a_n)(t) < q < \prob_\mathcal{M}(s',a_1 \dots a_n)(t).\]
  Now let $\varepsilon = q - \prob(s,a_1 \dots a_n)(t) > 0$.
  By right-continuity, there exists some $\delta > 0$ such that
  $t < x < t + \delta$ implies
  \[\prob_\mathcal{M}(s,a_1 \dots a_n)(x) - \prob_\mathcal{M}(s,a_1 \dots a_n)(t) < \varepsilon.\]
  Choose a rational $q'$ such that $t < q' < t + \delta$ in order to obtain
  \[\prob_\mathcal{M}(s,a_1 \dots a_n)(q') < q \leq \prob_\mathcal{M}(s',a_1 \dots a_n)(q').\]
  But then we have
  \[\mathcal{M},s' \models \mathcal{P}^{\leq q'}_{\geq q} (\diam{a_1} \dots \diam{a_n} \top) \quad \text{and} \quad \mathcal{M},s \not \models \mathcal{P}^{\leq q'}_{\geq q} (\diam{a_1} \dots \diam{a_n} \top). \qedhere\]
\end{proof}

Next we show that every formula has a finite model,
which also implies that every formula is satisfiable.

\begin{theorem}[Finite model property]
  Any formula $\psi \in \mathcal{L}$
  has a finite semi-Markov process satisfying it.
\end{theorem}
\begin{proof}
  For any path formula $\varphi = \diam{a_1} \dots \diam{a_n} \top$,
  we construct the model $\mathcal{M}_\varphi = (S, \tau, \rho)$ as follows.
  Let $S = \{s_1, \dots s_{n+1}\}$, and let
  \[\tau(s_i)(s_{i+1},a) = \begin{cases} 1 & \text{if } a = a_i \\ 0 & \text{otherwise.}\end{cases}\]
  Finally, let $\rho(s) = \delta_0$ be the Dirac distribution at $0$ for all states.
  
  Now it is easy to see that $\mathcal{M}_\varphi, s_1 \models \mathcal{P}^{\leq t}_{\geq p} (\varphi)$
  for any $t$ and $q$.
\end{proof}

\begin{corollary}
  Every formula $\psi$ is satisfiable,
  and hence the satisfiability problem is trivially decidable.
\end{corollary}

Lastly we consider the model checking problem for $\mathcal{L}$.
This problem can be solved by once more making use of the existential theory of the reals,
thus giving a $\PSPACE$ algorithm.

\begin{theorem}
  The model checking problem is decidable for any semi-Markov process with residence-time distributions
  from one of the following classes of timing distributions.
  \begin{itemize}
    \item Exponential distributions,
    \item piecewise polynomial distributions,
    \item piecewise affine distributions,
    \item uniform distributions, or
    \item Dirac distributions.
  \end{itemize}
\end{theorem}
\begin{proof}
  This essentially follows from Proposition \ref{thm:effective_classes},
  by letting the right-hand side of the inequality be a constant.
\end{proof}

%% Conclusion
\section{Conclusion and Open Problems}
We studied the model of semi-Markov processes where the timing behaviour can be described by arbitrary timing distributions.
We have introduced a trace-based relation called the faster-than relation which asks that for any prefix and any time bound,
the probability of outputting a word with that prefix within the time bound is higher in the faster process than in the slower process.
We have shown through a connection to probabilistic automata that the faster-than relation is highly undecidable. 
It is undecidable in general, and remains Positivity-hard even for one output label.
Furthermore, approximating the faster-than relation up to a multiplicative constant is impossible.

However, we constructed algorithms for special cases of the faster-than problem.
We have shown that if one considers approximating up to an additive constant rather than a multiplicative constant,
and if one gives a bound on the time up to which one is interested in comparing the two processes,
then approximation can be done for timing distributions in which we are sure to spend some amount of time to take a transition.
In addition, we have shown that the faster-than relation is decidable and in $\coNP$ for unambiguous processes,
in which there is a unique successor state for every output label.
Furthermore, we have given a logical characterisation of the faster-than relation
and shown that both the satisfiability and the model checking problem for this logic are decidable.

In this paper, we have focused on the generative model,
where the labels are treated as outputs.
An alternative viewpoint is the reactive model,
where the labels are instead treated as inputs~\cite{GSS95}.
While all the undecidability and hardness results we have shown
can also easily be shown to hold for the reactive case,
the same is not true for the algorithms we have constructed.
It is non-trivial to extend these algorithms to the reactive case,
and the main obstacle in doing so is that for reactive systems,
one has to also handle schedulers, often uncountably many.

\defaultbib

  \cleardoublepage
  \setcounter{enumiii}{0}
  \setcounter{enumii}{0}
  \setcounter{enumiv}{0}
  \setcounter{enumi}{0}
  \setcounter{equation}{0}
  \setcounter{figure}{0}
  \setcounter{footnote}{0}
  \setcounter{mpfootnote}{0}
  \setcounter{paragraph}{0}
  \setcounter{parentequation}{0}
  \setcounter{part}{0}
  \setcounter{section}{0}
  \setcounter{subfigure}{0}
  \setcounter{subparagraph}{0}
  \setcounter{subsection}{0}
  \setcounter{subsubsection}{0}
  \setcounter{table}{0}
  \papertitlepage{%
  A Faster-Than Relation for Semi-Markov Decision Processes
}{paper:paperC}{%
  Mathias R. Pedersen, Giorgio Bacci, and Kim G. Larsen
}{%
  The paper is based on an unpublished manuscript.
}{%
  %\noindent\copyright\ 201X IEEE
}

%% Abstract
\begin{abstract}
  When modeling concurrent or cyber-physical systems,
  non-functional requirements such as time
  are important to consider. In order to improve the timing aspects of a model,
  it is necessary to have some notion of what it means
  for a process to be faster than another,
  which can guide the stepwise refinement of the model.
  To this end we study a \emph{faster-than relation} for
  semi-Markov decision processes
  and compare it to standard notions for relating systems.
  We show that checking whether a system is faster than another one is undecidable,
  but as a positive result we give a decision procedure for approximating it.
  Furthermore, we consider the compositional aspects of this relation, 
  and show that the faster-than relation is not a precongruence
  with respect to parallel composition, hence 
  giving rise to so-called parallel timing anomalies.
  We take the first steps toward understanding this problem
  by identifying decidable conditions sufficient to avoid parallel timing anomalies 
  in the absence of non-determinism.
\end{abstract}

%% Introduction
\section{Introduction}
Timing aspects are important when considering real-time or cyber-physical systems.
For example, they are of interest in real-time embedded systems when one wants to verify the worst-case execution time for guaranteeing minimal system performance
or in safety-critical systems when one needs to ensure that unavoidable rigid deadlines will always be met~\cite{lee2008}. 

Semi-Markov decision processes are continuous-time Markov decision processes where the residence-time on states is governed by generic distributions on the positive real line.
These systems have been extensively used to model real-time cyber-physical systems~\cite{SSR15,TS16}. 

For reasoning about timing aspects it is important to understand what it formally means 
for a real-time or cyber-physical system to operate faster than another. 
To this end we define the notion of \emph{faster-than relation} for semi-Markov decision processes.
The definition of faster-than relation we propose in this paper is a reactive version of an 
analogous notion of faster-than relation previously introduced in~\cite{PFBLM18} for the case 
of generative systems.
According to our relation, a semi-Markov decision process is faster than another one when
it reacts to any sequence of inputs with equal or higher probability than the slower process, within the same time bound.

Similarly to~\cite{PFBLM18}, we show that also the faster-than relation on semi-Markov decision
processes is undecidable. However, by extending the approximation algorithm from~\cite{PFBLM18},
we obtain an approximation algorithm for the case where we only consider timed events within some fixed time bound.
The extension of the algorithm in~\cite{PFBLM18} is not a trivial task,
because the definition of faster-than relation on semi-Markov decision processes
requires us to deal with universal and existential quantifications over schedulers,
which were not present in the original definition in~\cite{PFBLM18} for the case of generative systems.

Often, complex cyber-physical systems are organised as concurrent systems 
of multiple components running in parallel and interacting with each other.
Such systems are better analysed \emph{compositionally}, that is, by breaking them into smaller 
components that are more easily examined~\cite{CLM89}.
However, it is not always the case that an analysis on the components
carries over to the full composite system. A well known example of this, occurring in
real-time systems such as scheduling for processors~\cite{cassez2012,lundqvist1999}, are 
\emph{timing anomalies}, that is, when locally faster behaviour leads to a globally slower behaviour~\cite{kirner2009}.

In this paper we study the compositional aspects of the faster-than relation for semi-Markov
decision processes. The situation we are interested in is depicted in Figure~\ref{fig:ft}
where we have a composite system consisting of a context $W$ and a component $V$,
and we want to understand what happens when we replace $V$ with another component $U$ that
is faster than $V$. We consider some common notions of parallel composition, and show that 
timing anomalies can occur using our faster-than relation, even in the absence of
non-determinism. This shows that
timing anomalies are not caused by non-determinism, but arise from 
the linear timing behaviour of processes.

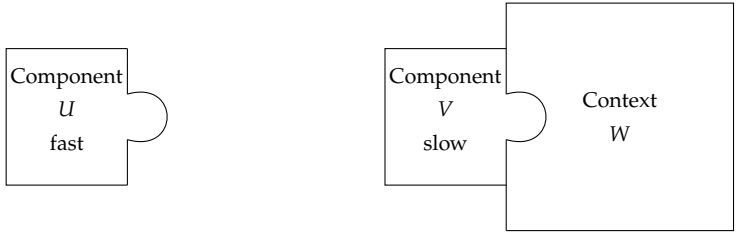
\begin{figure}
  \center
  \begin{tikzpicture}
    % Context
    \draw (0,0) -- (3,0) -- (3,3) -- (0,3) -- (0,1.8);
    \draw (0,1.2) -- (0,0);
    \draw (0,1.8) .. controls  (0.7,2.0) and (0.7,1.0) .. (0,1.2);
    \node[align=center] at (1.5,1.5) {Context \\ $W$};
    
    % Component 1
    \draw (0,0.6) -- (-1.6,0.6) -- (-1.6,2.4) -- (0,2.4);
    \node[align=center] at (-0.8,1.6) {Component \\ $V$ \\ slow};
    
    % Component 2
    \draw (-5,0.6) -- (-6.6,0.6) -- (-6.6,2.4) -- (-5,2.4);
    \draw (-5,2.4) -- (-5,1.8);
    \draw (-5,1.2) -- (-5,0.6);
    \draw (-5,1.8) .. controls (-4.3,2.0) and (-4.3,1.0) .. (-5,1.2);
    \node[align=center] at (-5.8,1.6) {Component \\ $U$ \\ fast};
    
  \end{tikzpicture}
  \caption{The context $W$ operates in parallel with the component $V$.
           If the component $U$ is faster than $V$, then if we replace $V$ with $U$,
           we would expect the overall behaviour to also be faster.}
  \label{fig:ft}
\end{figure}

We then take a first step toward recovering compositional reasoning
for the faster-than relation, by identifying conditions sufficient for avoiding timing anomalies,
which we call \emph{monotonicity}.
Presently we do not know whether these conditions are decidable,
however we introduce another set of conditions, called \emph{strong monotonicity}, which are decidable.
Unfortunately, strong monotonicity only applies to processes which have no non-determinism.

\subsubsection{Related Work.}
The notion of a faster-than relation has been studied in many different contexts throughout the literature.
The work most closely related to ours is that of Pedersen et al. \cite{PFBLM18},
which considers a generative version of the faster-than relation,
whereas we study the reactive version.
The focus of \cite{PFBLM18} is on decidability issues,
and the faster-than relation is proved undecidable.
However, positive results are also given
in the form of an approximation algorithm,
and a decidability result for unambiguous processes.
Baier et al. \cite{BKHW05} define, among other relations,
a simulation relation for continuous-time Markov chains
which can be interpreted as a faster-than relation,
and study its logical characterisation.
However, none of these works consider compositional aspects.

For process algebras, discrete-time faster-than relations have been defined for variations of Milner's CCS,
and shown to be precongruences with respect to parallel composition \cite{corradini1995,LV01,MT91,satoh1994}.
L{\"u}ttgen and Vogler \cite{luttgen2006} attempt to unify some of these process algebraic approaches
and also consider the issue of parallel timing anomalies.
For Petri nets, Vogler \cite{vogler1995a,vogler1995b} considers a testing preorder as a faster-than relation
and shows that this is a precongruence with respect to parallel composition.
Geilen et al. \cite{geilen2011} introduces a refinement principle for timed actor interfaces
under the slogan ``the earlier, the better'', which can also be seen as an example of a faster-than relation.

Work on timing anomalies date back to at least 1969 \cite{graham1969},
but the most influential paper in the area is probably that of Lundqvist and Stenstr\"{o}m \cite{lundqvist1999},
in which they show that timing anomalies can occur in dynamically scheduled processors,
contrary to what most people assumed at the time.
More recent work has focused on compositional aspects \cite{kirner2009}
and defining timing anomalies formally, using transition systems as the formalism \cite{cassez2012,reineke2006}.

%% Preliminaries
\tikzset{->, >=latex}

\section{Notation and Preliminaries}
In this section we fix some notation and recall concepts
that are used throughout the rest of the paper.
Let $\mathbb{N}$ denote the natural numbers and
let $\mathbb{R}_{\geq 0}$ denote the non-negative real numbers,
which we equip with the standard Borel $\sigma$-algebra $\mathbb{B}$.
For any set $X$,
let $\dist(X)$ denote the set of probability measures on $X$,
and let $\subdist(X)$ denote the set of subprobability measures on $X$.
For an element $x \in X$ of some set $X$,
we will use $\delta_x$ to denote the Dirac measure at $x$ defined as
$\delta_x(y) = 1$ if $x = y$ and $\delta_x(y) = 0$ otherwise.
We fix a non-empty, countable set $L$ of \emph{labels} or \emph{actions}
and equip them with the discrete $\sigma$-algebra $\Sigma_L$.

For a probability measure $\mu \in \dist(\mathbb{R}_{\geq 0})$,
we denote by $F_\mu$ its \emph{cumulative distribution function (CDF)} defined as $F_{\mu}(t) = \mu([0,t])$, for all $t \in \mathbb{R}_{\geq 0}$.
We will denote by $\Exp{\theta}$ the CDF of an exponential distribution with rate $\theta > 0$.
The \emph{convolution} of two probability measures $\mu, \nu \in \dist(\mathbb{R}_{\geq 0})$, 
written $\mu * \nu$, is the probability measure on $\mathbb{R}_{\geq 0}$ given by $(\mu * \nu)(B) = \int_{-\infty}^{\infty} \nu(B - x) \;\mu(\wrt{x})$,
for all $B \in \mathbb{B}$ \cite{billingsley1995}.
Convolution is associative, i.e., $\mu * (\nu * \eta) = (\mu * \nu) * \eta$, and commutative, i.e., $\mu * \nu = \nu * \mu$. 

%% Semi-Markov decision processes
\section{Semi-Markov Decision Processes}
In this section we recall the definition of semi-Markov decision processes.
\begin{definition}
  A \emph{semi-Markov decision process (SMDP)} is a tuple
  $M = (S,\tau,\rho)$ where
  \begin{itemize}
    \item $S$ is a non-empty, countable set of \emph{states},
    \item $\tau : S \times L \rightarrow \subdist(S)$ is a \emph{transition probability function}, and
    \item $\rho : S \rightarrow \dist(\mathbb{R}_{\geq 0})$ is a \emph{residence-time probability function}. \qedhere
  \end{itemize}
\end{definition}

The operational behaviour of an SMDP $M = (S,\tau,\rho)$ is as follows.
The process in the state $s \in S$ reacts to an external input $a \in L$ provided by the environment 
by changing its state to $s' \in S$ within time $t \in \mathbb{R}_{\geq 0}$ with probability 
$\tau(s,a)(s') \cdot \rho(s)([0, t])$.

Notice that Markov decision processes are a special case of SMDPs
where for all $s \in S$, $\rho(s) = \delta_0$ (i.e. transitions happen instantaneously), and that
continuous-time Markov decision processes are also a special case of SMDPs
where, for all states $s \in S$, $F_{\rho(s)} = \Exp{\theta_s}$ for some rate $\theta_s \in \mathbb{R}_{\geq 0}$.

The executions of an SMDP $M = (S,\tau,\rho)$ are infinite timed transition sequences of the form
$\pi = (s_1,t_1,a_1)(s_2,t_2,a_2)\dots \in (S \times \mathbb{R}_{\geq 0} \times L)^\omega$, representing
the fact that $M$ waited in state $s_i$ for $t_i$ time units after the action $a_i$ was input.
We will refer to executions of an SMDP as \emph{timed action paths}.
For $i \in \mathbb{N}$, let $\pi\sproj{i} = s_i$, $\pi\tproj{i} = t_i$, $\pi\aproj{i} = a_i$,
$\pi |_i = (s_1,t_1,a_1)\dots(s_i,t_i,a_i)$, and $\pi |^i = (s_i,t_i,a_i)(s_{i+1},t_{i+1},a_{i+1})\dots$.
We let $\paths(M)$ denote the set of all timed action paths in $M$,
and denote by $\paths_n(M) = \{\pi|_n \mid \pi \in \paths(M)\}$ the set of all prefixes of length $n$.
Hereafter, we refer to timed action paths simply as paths,
unless we wish to distinguish between different kinds of paths.

Next we recall the standard construction of the measurable space of paths. 
A \emph{cylinder set} of rank $n \geq 1$ is the set of all paths whose $n$th prefix is contained in a common subset $E \subseteq \paths_n(M)$,
and is given by
\[\cylinder(E) = \{ \pi \in \paths(M) \mid \pi |_n \in E \}.\]
It will be convenient to denote \emph{rectangular cylinders} 
of the form
\[\cylinder(S_1 \times L_1 \times R_1 \times \dots \times S_n \times L_n \times R_n),\]
for $S_i \subseteq S$, $L_i \subseteq L$, and $R_i \subseteq \mathbb R_{\geq 0}$, as
\[\cylinder(S_1 \dots S_n, L_1 \dots L_n, R_1 \dots R_n).\]

We denote by $(\paths(M), \Sigma)$ the \emph{measurable space of timed action paths}, where 
$\Sigma$ is the smallest $\sigma$-algebra generated by the cylinders of the form
\[\cylinder(S_1 \dots S_n, L_1 \dots L_n, R_1 \dots R_n)\]
for $S_i \in 2^S$, $L_i \in 2^L$, and $R_i \in \mathbb{B}$.

In this paper we assume that external choices are resolved by means of memoryless stochastic schedulers, 
however all the results we present still hold for memoryful schedulers.
\begin{definition}
  Given an SMDP $M = (S, \tau, \rho)$, a \emph{scheduler} for $M$ is a function $\sigma : S \rightarrow \dist(L)$ that assigns to each state a probability distribution over action labels.
\end{definition}

We will use the notation $\tau^\sigma(s,a)(s')$
as shorthand for $\tau(s,a)(s') \cdot \sigma(s)(a)$ to denote the probability of moving from state $s$ to $s'$
under the stochastic choice of $a$ given by $\sigma$.
Given an SMDP $M$ and a scheduler $\sigma$ for it, the probabilistic execution of a path starting from
the state $s$ is governed by the probability $\prob^\sigma_M(s)$ on $(\paths(M),\Sigma)$ defined as follows.

\begin{definition}\label{def:prob}
  Let $M = (S,\tau,\rho)$ be an SMDP. Given a scheduler $\sigma$ for M and a state $s \in S$,
  $\prob^\sigma_M(s)$ is defined as the unique (sub)probability measure\footnote{Existence and uniqueness is guaranteed by the Hahn-Kolmogorov theorem \cite{tao2013}.} on
  $(\paths(M), \Sigma)$ such that for all $S_i \in 2^S$, $L_i \in 2^L$, and $R_i \in \mathbb{B}$, 
  with $1 \leq i \leq n$, we have
  \[\prob_M^\sigma(s)(\cylinder(S_1,L_1,R_1)) = \rho(s)(R_1) \cdot \sum_{a \in L_1} \sum_{s' \in S_1} \tau^\sigma(s,a)(s')\]
  and
  \begin{align*}
    &\phantom{{}={}}\prob_M^{\sigma}(s)(\cylinder(S_1\dots S_n, L_1 \dots L_n, R_1 \dots R_n)) \\
    &= \rho(s)(R_1) \cdot \sum_{a \in L_1} \sum_{s' \in S_1} \tau^\sigma(s,a)(s') \cdot \prob_M^\sigma(s')(\cylinder(S_2 \dots S_n, L_2 \dots L_n, R_2 \dots R_n)).
  \end{align*}
\end{definition}

Intuitively, to get the probability $\prob_M^{\sigma}(s)(\cylinder(S_1 \dots S_n, L_1 \dots L_n, R_1 \dots R_n))$,
we first take the probability that $s$ takes a transition at a time point in $R_1$, given by $\rho(s)(R_1)$,
after which we sum over the probabilities of all the possible transitions that can be taken by choosing
a label $a \in L_1$ and a state $s' \in S_1$, and then the rest of the probability is given inductively by continuing on $s'$.
For the rest of the paper, we will omit the subscript $M$ in $\prob_M^{\sigma}$
whenever it is clear from the context which SMDP is being referred to.

%% Faster-than relation
\section{A Faster-Than Relation}
Our aim is to define a relation that formalises the intuitive idea of an SMDP $U$ being 
``faster than'' another SMDP $V$. 
For a process $U$ to be faster than $V$, it must be able to execute any sequence of actions $a_1, \dots, a_n$ in less time than $V$.
Since we are dealing with probabilistic systems,
we must speak of the probability of executing a sequence of actions within some time bound.

\begin{figure}
  \centering
  \hfill
  \begin{tikzpicture}
    % Nodes
    \node[state, circle split] (0) {$\mu$ \nodepart{lower} $u_0$};
    \node[state, circle split] (1) [right = 1.5cm of 0]{$\nu$ \nodepart{lower} $u_1$};
    \node[state, circle split] (8) [right = 1.5cm of 1]{$\eta$ \nodepart{lower} $u_2$};
    
    \node[xshift=-0.5cm] at (0.west) {\large{$U$}};
    
    % Edges
    \path[thick] (0) edge [above] node {$(1,a)$} (1);
    \path[thick] (1) edge [above] node {$(1,a)$} (8);
    \path[thick, loop right] (8) edge [right] node {$(1,a)$} (8);
    
    % Nodes
    \node[state, circle split] (2) [below = 1cm of 0] {$\nu$ \nodepart{lower} $v_0$};
    \node[state, circle split] (3) [right = 1.5cm of 2] {$\mu$ \nodepart{lower} $v_1$};
    \node[state, circle split] (9) [right = 1.5cm of 3] {$\eta$ \nodepart{lower} $v_2$};
    
    \node[xshift=-0.5cm] at (2.west) {\large{$V$}};
    
    % Edges
    \path[thick] (2) edge [above] node {$(1,a)$} (3);
    \path[thick] (3) edge [above] node {$(1,a)$} (9);
    \path[thick, loop right] (9) edge [right] node {$(1,a)$} (9);
  \end{tikzpicture}
  \hfill \
  \caption{If $F_{\mu}(t) \geq F_{\nu}(t)$ for all $t$, then $U$ is faster than $V$ in the first states, and after that their probabilities are the same,
           so $U$ is faster than $V$.}
  \label{fig:faster-than-c}
\end{figure}
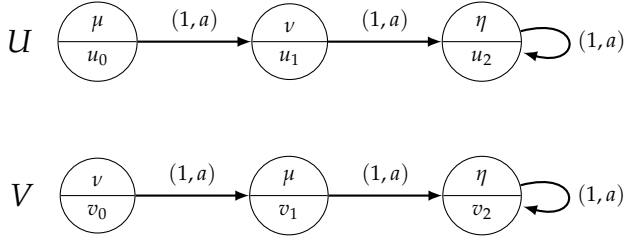

Consider the two simple SMDPs $U$ and $V$ in Figure~\ref{fig:faster-than-c} 
with just a single transition label and initial states $u_0$ and $v_0$, respectively.
Here $\mu,\nu,\eta$ are arbitrary probability measures on $\mathbb{R}_{\geq 0}$, representing
the residence-time distributions at each state.
An arrow with label $(p,a)$ means that when $a$ is chosen as the action,
then the SMDP takes the transition given by the arrow with probability $p$.
The only finite sequences of actions that can be executed in these SMDPs are of the form $a^n$ for $n > 0$.

For $U$ to be faster than $V$, it should be the case that for any time bound $t$ and no matter which 
scheduler $\sigma$ we choose for $V$, we must be able to find a scheduler $\sigma'$ for $U$
such that there is an earlier time bound $t' \leq t$ which allows $U$ to execute
any sequence $a^n$ within time $t'$ with higher or equal probability than that of $V$ executing the same
sequence of actions within time $t$.
Formally, this amounts to saying that 
$\prob^{\sigma'}(u_0)(\cylinder(a^n, t')) \geq \prob^{\sigma}(v_0)(\cylinder(a^n, t))$,
where $\cylinder(a_1 \dots a_n, t)$ denotes the event of executing the sequence of actions $a_1, \dots, a_n$ within time $t$.
Hence, the type of events on which we want to focus are the following.

\begin{definition}\label{def:cylinder}
  For any finite sequence of actions $a_1, \dots, a_n$,
  and $t \in \mathbb{R}_{\geq 0}$, we say that
  \[\cylinder(a_1 \dots a_n, t) = \{ \pi \in \Pi(M) \mid \forall 1 \leq i \leq n,\; \pi\aproj{i} = a_i \text{ and } \sum_{j = 1}^n \pi\tproj{j} \leq t \}\]
  is a \emph{time-bounded cylinder}.
  The \emph{length} of a time-bounded cylinder is the length
  of the sequence of actions in the time-bounded cylinder.
\end{definition}

%Note that time-bounded cylinders $\cylinder(a_1 \dots a_n, x)$ are measurable in $(\paths(M),\Sigma)$,
%since they are given as the measurable pre-image
%\todo{is this really measurable?}
%$(\pi |_n \circ f)^{-1}(S^n \times \{ (a_1, \dots, a_n) \} \times B^n_x)$,
%\todo{projection here not necessary?}
%where $B^n_x$ is the Borel set $\{(r_1, \dots, r_n) \in \mathbb{R}^n_{\geq 0} \mid \sum_{j = 1}^n r_j \in [0,x]\}$
%\todo{is this really Borel?}
%and $f$ is the canonical measurable isomorphism from $\Pi_n(M) = (S \times L \times \mathbb{R}_{\geq 0})^n$ to $S^n \times L^n \times \mathbb{R}^n_{\geq 0}$.

Note that $\cylinder(a_1 \dots a_n, t)$ is measurable in $(\paths(M), \Sigma)$,
since
\[f : \paths_n(\mathcal{M}) \rightarrow S^n \times L^n \times \mathbb{R}_{\geq 0}^n\]
given by
\[f((s_1,o_1,t_1), \dots, (s_n,o_n,t_n)) = (s_1, \dots, s_n), (o_1, \dots, o_n), (t_1, \dots, t_n)\]
and
\[\texttt{res}_n : \paths(\mathcal{M}) \rightarrow \paths_n(\mathcal{M})\]
given by
\[\texttt{res}_n(\pi) = \pi|_n\]
are both measurable,
and hence
\[(f \circ \texttt{res}_n)^{-1}(S^n \times \{(a_1, \dots, a_n)\} \times B^n_t) = \cylinder(a_1 \dots a_n, t)\]
is measurable,
where $B^n_t = \{(r_1, \dots, r_n) \in \mathbb{R}^n_{\geq 0} \mid \sum_{i = 1}^n r_i \leq t\}$.

\begin{example}
  The time-bounded cylinder $\cylinder(aa,2)$ denotes the set of all paths
  where the first two output labels are both $a$'s,
  and the first two steps of the path are completed within $2$ time units.
\end{example}

We will use the notation $(M,s_0)$ to indicate that $M = (S, \tau, \rho)$
is an SMDP with initial state $s_0 \in S$ and call it \emph{pointed SMDP}.
For the rest of the paper, we fix three SMDPs
$M = (S, \tau, \rho)$, $U = (S_U, \tau_U, \rho_U)$, and $V = (S_V, \tau_V, \rho_V)$,
with initial states $s_0 \in S$, $u_0 \in S_U$, $v_0 \in S_V$, respectively. 
Now we are ready to define what it means for an SMDP to be ``faster than'' another one.

\begin{definition}[Faster-than]
  We say that $U$ is \emph{faster than} $V$, written $U \ft V$, if
  for all schedulers $\sigma$ for $V$, time bounds $t$, and sequences of actions $a_1 \dots a_n$,
  there exists a scheduler $\sigma'$ for $U$ and time bound $t' \leq t$, such that 
  $\prob^{\sigma'}(u_0)(\cylinder(a_1 \dots a_n, t')) \geq \prob^{\sigma}(v_0)(\cylinder(a_1 \dots a_n, t))$.
\end{definition}

Clearly, the faster-than relation $\ft$ is a preorder.
The following proposition gives a characterisation of
the faster-than relation that is often easier to work with.

\begin{proposition}\label{prop:alternative}
  $U \ft V$ if and only if for all schedulers $\sigma$ for $V$
  there exists a scheduler $\sigma'$ for $U$ such that
  $\prob^{\sigma'}(u_0)(C) \geq \prob^{\sigma}(v_0)(C)$,
  for all time-bounded cylinders $C$.
\end{proposition}
\begin{proof}%[Proof of Proposition \ref{prop:alternative}]
  Clearly, if for all schedulers $\sigma$ for $V$ there exists a scheduler $\sigma'$ for $U$ such that
  $\prob^{\sigma}(u_0)(C) \geq \prob^{\sigma'}(v_0)(C)$
  for all time-bounded cylinders $C$, then $U \ft V$ by taking $C' = C$.
  If $U \ft V$, then consider an arbitrary scheduler $\sigma$,
  and time-bounded cylinder $C = \cylinder(a_1 \dots a_n, t)$.
  There exists a scheduler $\sigma'$ and
  $t' \in \mathbb{R}_{\geq 0}$ such that $t \geq t'$ and
  \[\prob^{\sigma'}(u_0)(a_1 \dots a_n, t') \geq \prob^{\sigma}(v_0)(a_1 \dots a_n, t).\]
  By monotonicity, $t \geq t'$ implies that
  \[\prob^{\sigma'}(u_0)(a_1 \dots a_n, t) \geq \prob^{\sigma'}(u_0)(a_1 \dots a_n, t'),\]
  and hence $\prob^{\sigma'}(u_0)(C) \geq \prob^{\sigma}(v_0)(C)$.
\end{proof}

Before showing an example of an SMDP being faster than another one, we provide an analytic solution
for computing the probability over time-bounded cylinders in terms of convolutions of the residence time distributions.

\begin{proposition}\label{prop:prodmeasure}
  For any SMDP $M$, scheduler $\sigma$ for $M$, and $s \in S$, we have
  \begin{align*}
    &\phantom{{}={}} \prob^\sigma(s)(\cylinder(S_1 \dots S_n, L_1 \dots L_n, R_1 \dots R_n)) \\
    &= \sum_{s_n \in S_n} \sum_{a_n \in L_n} \dots \sum_{s_1 \in S_1} \sum_{a_1 \in L_1} \tau^\sigma(s,a_1)(s_1) \cdots \tau^\sigma(s_{n-1},a_n)(s_n) \\
    &\phantom{{}={}}\cdot \rho(s) \times \rho(s_1) \times \dots \times \rho(s_{n-1})(R_1 \times R_2 \times \dots \times R_n).
  \end{align*}
\end{proposition}
\begin{proof}%[Proof of Proposition \ref{prop:prodmeasure}]
  The proof is by induction on the length $n$ of the cylinder.
  If the cylinder has length $n = 1$ then
  \[\prob^\sigma(s)(\cylinder(S_1, L_1, R_1)) = \sum_{s_1 \in S_1} \sum_{a_1 \in L_1} \tau^\sigma(s, a_1)(s_1) \cdot \rho(s)(R_1).\]
  
  If the cylinder has length $n = k + 1$, then
  \begin{align*}
    &\phantom{{}={}}\prob^\sigma(s)(\cylinder(S_1 \dots S_{k+1}, L_1 \dots L_{k+1}, R_1 \dots R_{k+1})) \\
    &= \rho(s)(R_1) \cdot \sum_{s_1 \in S_1} \sum_{a_1 \in L_1} \tau^\sigma(s,a_1)(s_1) \cdot \\
    &\phantom{{}={}}\prob^\sigma(s_1)(\cylinder(S_2 \dots S_{k+1}, L_2 \dots L_{k+1}, R_2 \dots R_{k+1})) \\
    &= \rho(s)(R_1) \cdot \sum_{s_1 \in S_1} \sum_{a_1 \in L_1} \tau^\sigma(s,a_1)(s_1) \\
    &\phantom{{}={}}\cdot \sum_{s_{k+1} \in S_{k+1}} \sum_{a_{k+1} \in L_{k+1}} \dots \sum_{s_2 \in S_2} \sum_{a_2 \in L_2} \tau^\sigma(s_1,a_2)(s_2) \cdots \tau^\sigma(s_k, a_{k+1})(s_{k+1}) \\
    &\phantom{{}={}}\cdot \rho(s_1) \times \dots \times \rho(s_{k+1}) (R_2 \times \dots \times R_{k+1}) \\
    &= \sum_{s_{k+1} \in S_{k+1}} \sum_{a_{k+1} \in L_{k+1}} \dots \sum_{s_1 \in S_1} \sum_{a_1 \in L_1} \tau^\sigma(s,a_1)(s_1) \cdots \tau^\sigma(s_k, a_{k+1})(s_{k+1}) \\
    &\phantom{{}={}}\cdot \rho(s) \times \rho(s_1) \times \dots \times \rho(s_{k+1})(R_1 \times \dots \times R_{k+1}). \qedhere
  \end{align*}
\end{proof}

\begin{corollary}\label{cor:prodmeasure}
  For any SMDP $M$, scheduler $\sigma$ for $M$, $s \in S$, and Borel set $B \in \mathbb{R}^n_{\geq 0}$ we have
  \begin{align*}
    &\phantom{{}={}} \prob^\sigma(s)(\cylinder(S \dots S, \{a_1\} \dots \{a_n\}, B)) \\
    &= \sum_{s_n \in S} \dots \sum_{s_1 \in S} \tau^\sigma(s,a_1)(s_1) \cdots \tau^\sigma(s_{n-1},a_n)(s_n) \\
    &\phantom{{}={}}\cdot \rho(s) \times \rho(s_1) \times \dots \times \rho(s_{n-1})(B).
  \end{align*}
\end{corollary}

\begin{proposition}\label{prop:hyp}
  For any SMDP $M = (S,\tau,\rho)$, scheduler $\sigma$ for $M$, and $s \in S$ we have
  \begin{align*}
    &\phantom{{}={}}\prob^\sigma(s)(\cylinder(a_1 \dots a_n, t)) \\
    &=\sum_{s_1 \in S} \dots \sum_{s_n \in S} \tau^\sigma(s,a_1)(s_1) \cdots \tau^\sigma(s_{n-1},a_n)(s_n) \\
    &\phantom{{}={}}\cdot (\rho(s) * \rho(s_1) * \dots * \rho(s_{n-1}))([0,t]) .
  \end{align*}
\end{proposition}
\begin{proof}%[Proof of Proposition \ref{prop:hyp}]
  By Corollary \ref{cor:prodmeasure}, we know that
  \begin{align*}
    &\phantom{{}={}} \prob^\sigma(s)(\cylinder(a_1 \dots a_n, t)) \\
    &= \sum_{s_n \in S} \dots \sum_{s_1 \in S} \tau ^\sigma(s,a_1)(s_1) \cdots \tau^\sigma(s_{n-1},a_n)(s_n) \\
    &\phantom{{}={}} \cdot \rho(s) \times \rho(s_1) \times \dots \times \rho(s_{n-1})(B^n_t).
  \end{align*}
  Hence, if we can show that
  \[\rho(s) \times \rho(s_1) \times \dots \times \rho(s_{n-1}) (B^n_t) = (\rho(s) * \rho(s_1) * \dots * \rho(s_{n-1}))([0,t]),\]
  the proof is done.
  
  The proof now proceeds by induction on the length $n$
  of the time-bounded cylinder $\cylinder(a_1 \dots a_n, t)$.
  If $n = 1$, then
  \[\rho(s)(B^1_t) = \rho(s)([0,t]).\]
  
  If $n = k + 1$, then
  \begin{align*}
    &\phantom{{}={}}(\rho(s) \times \rho(s_1) \times \dots \times \rho(s_k))(B^{k+1}_t) \\
    &= \int_0^t (\rho(s_1) \times \dots \times \rho(s_k))(B^k_{t - x}) \; \rho(s)(\wrt{x}) && \text{(Fubini)} \\
    &= \int_0^t (\rho(s_1) * \dots * \rho(s_k))([0, t - x]) \; \rho(s)(\wrt{x}) && \text{(ind. hyp.)}\\
    &= (\rho(s) * (\rho(s_1) * \dots * \rho(s_k)))([0, t]) && \text{(def. of convolution)}\\
    &= (\rho(s) * \rho(s_1) * \dots * \rho(s_k))([0,t]). && \text{(associativity)} \qedhere
  \end{align*}
\end{proof}

Proposition \ref{prop:hyp} intuitively says that the absorption-time
of any path of length $n$ through the SMDP is distributed
as the $n$-fold convolution of its residence-time probabilities.
Therefore, the probability of doing transitions with labels $a_1, \dots, a_n$
within time $t$ is the sum of the probabilities of taking a path of length $n$
with labels $a_1, \dots, a_n$ through the SMDP,
weighted by the probability of reaching the end of each of these paths within time $t$.
This is similar in spirit to a result on phase-type distributions,
see e.g. \cite[Proposition 2.11]{pulungan2009}.

From Proposition \ref{prop:hyp} we can also derive the following
which gives a more direct inductive definition of the probability on time-bounded cylinders.
If we fix $a_1 \dots a_n$ and let $t$ vary, we get a CDF
\[\prob^\sigma(s)(a_1 \dots a_n)([0,t]) = \prob^\sigma(s)(\cylinder(a_1 \dots a_n, t)).\]

\begin{proposition}\label{prop:inductive}
  The CDF $\prob^\sigma(s)(a_1 \dots a_n)$ can be characterised inductively by
  \[\prob^\sigma(s)(a_1)([0,t]) = \sum_{s' \in S} \tau^\sigma(s,a_1)(s') \cdot \rho(s)([0,t]),\]
  \[\prob^\sigma(s)(a_1 \dots a_n)([0,t]) = \sum_{s' \in S} \tau^\sigma(s,a_1)(s') \cdot (\rho(s) * \prob^\sigma(s')(a_2 \dots a_n))([0,t]).\]
\end{proposition}
\begin{proof}%[Proof of Proposition \ref{prop:inductive}]
  For $n = 1$ we have
  \[\prob^\sigma(s)(a)([0,t]) = \prob^\sigma(s)(\cylinder(a,t)) = \sum_{s' \in S} \tau^\sigma(s,a)(s') \cdot \rho(s)([0,t]).\]
  For $n = k + 1$ we have
  \begin{align*}
    &\phantom{{}={}}\prob^\sigma(s)(a_1 \dots a_n) \\
    &= \sum_{s_1 \in S} \dots \sum_{s_n \in S} \tau^\sigma(s,a_1)(s_1) \cdots \tau^\sigma(s_k, a_n)(s_n) \cdot (\rho(s) * \dots * \rho(s_n))([0,t]) \\
    &= \sum_{s_1 \in S} \dots \sum_{s_n \in S} \tau^\sigma(s,a_1)(s_1) \cdots \tau^\sigma(s_k, a_n)(s_n) \\
    &\phantom{{}={}}\cdot \int_0^t (\rho(s_1) * \dots * \rho(s_n))(t-x) \; \rho(s)(\wrt{x}) \\
    &= \sum_{s_1 \in S} \tau^\sigma(s,a_1)(s_1) \cdot (\rho(s) * \prob^\sigma(s_1)(a_2 \dots a_n))([0,t]). \qedhere
  \end{align*}
\end{proof}

Proposition \ref{prop:inductive} also shows that our definition of faster-than
coincides with the one from \cite{PFBLM18},
except ours is reactive rather than generative.

\begin{example}\label{ex:faster-than}
  Consider the pointed SMDPs $(U,u_0)$ and $(V,v_0)$ that are depicted in 
  Figure~\ref{fig:faster-than-c}. 
  Assuming that $F_\mu(t) \geq F_\nu(t)$ for all $t$, we now show that $U \ft V$.
  To compare $U$ and $V$, first notice that we only need to consider time-bounded cylinders of the form $\cylinder(a^n,t)$, for $n \geq 1$.
  Since the set of actions is $L = \{a\}$, the only possible valid scheduler $\sigma$ for both $U$ and $V$ is 
  the one assigning the Dirac measure $\delta_a$ to all states. We consider two cases.

  \textbf{(Case $n = 1$)} In this case we get
    \[\prob^\sigma(u_0)(\cylinder(a, t)) = F_\mu(t) \quad \text{ and } \quad \prob^\sigma(v_0)(\cylinder(a, t)) = F_\nu(t).\]
    Since we assumed $F_\mu(t) \geq F_\nu(t)$ for all $t$, this implies
    \[\prob^\sigma(u_0)(\cylinder(a, t)) \geq \prob^\sigma(v_0)(\cylinder(a, t)).\]
    
  \textbf{(Case $n > 1$)} By Proposition~\ref{prop:hyp} we have both
    \[\prob^\sigma(u_0)(\cylinder(a^n, t)) = (\mu * \nu * \eta^{*(n-2)})([0,t])\]
    and
    \[\prob^\sigma(v_0)(\cylinder(a^n, t)) = (\nu * \mu * \eta^{*(n-2)})([0,t]),\] 
    where $\eta^{*n}$ is the $n$-fold convolution of $\eta$, defined inductively by 
    $\eta^{*0} = \delta_0$ and $\eta^{*(n+1)} = \eta * \eta^{*n}$. Since convolution is commutative and
    associative, and $\delta_0$ is the identity for convolution, we obtain
    \[\prob^\sigma(u_0)(\cylinder(a^n, t)) = \prob^\sigma(v_0)(\cylinder(a^n, t)).\]
    
  We therefore conclude that $U \ft V$.
\end{example}

\subsection{Comparison With Simulation and Bisimulation}
The standard notions used to compare processes are bisimulation \cite{neuhausser2007} and simulation \cite{BKHW05}.
We next recall their definitions,
naturally extended to our setting of SMDPs.

\begin{definition}
  For an SMDP $M$, a relation $R \subseteq S \times S$ is a \emph{bisimulation relation} (resp. \emph{simulation relation}) on $M$
  if for all $(s_1, s_2) \in R$ we have
  \begin{itemize}
    \item $F_{\rho(s_1)}(t) = F_{\rho(s_2)}(t)$ (resp. $F_{\rho(s_1)}(t) \leq F_{\rho(s_2)}(t)$) for all $t \in \mathbb{R}_{\geq 0}$ and
    \item for all $a \in L$ there exists a weight function $\Delta_a : S \times S \rightarrow [0,1]$ such that
      \begin{itemize}
        \item $\Delta_a(s,s') > 0$ implies $(s,s') \in R$,
        \item $\tau(s_1,a)(s) = \sum_{s' \in S} \Delta_a(s,s')$ for all $s \in S$, and
        \item $\tau(s_2,a)(s') = \sum_{s \in S} \Delta_a(s,s')$ for all $s' \in S$.
      \end{itemize}
  \end{itemize}
  
  If there is a bisimulation relation (resp. simulation relation) $R$ such that $(s_1,s_2) \in R$,
  then we say that $s_1$ and $s_2$ are \emph{bisimilar} (resp. $s_2$ \emph{simulates} $s_1$) and write $s_1 \sim s_2$ (resp. $s_1 \simul s_2$).
\end{definition}

We lift bisimulation and simulation relations to two different SMDPs
by considering the disjoint union of the two and comparing their initial states.
We denote by $\sim$ the largest bisimulation relation and by $\simul$ the largest simulation relation.
Furthermore, we say that $U$ and $V$ are \emph{equally fast} and write $U \eqft V$
if $U \ft V$ and $V \ft U$.

\begin{example}\label{ex:comparison1}
  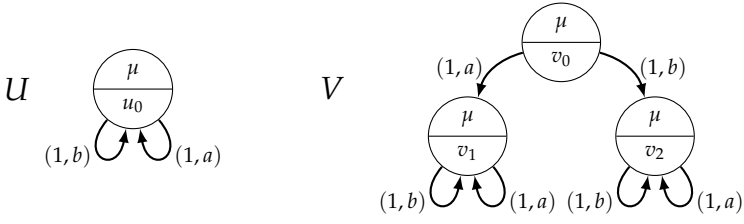
\begin{figure}
    \centering
    % First picture
    \hfill
    \begin{tikzpicture}[->, >=latex, baseline={(current bounding box.center)}]
      % Nodes
      \node[state, circle split] (0) {$\mu$ \nodepart{lower} $u_0$};
      \node[xshift=-1cm] at (0.west) {\large{$U$}};
      %
      % Edges
      \path[thick] (0) edge [out=310,in=280,looseness=6] node[right] {$(1,a)$} (0);
      \path[thick] (0) edge [out=230,in=260,looseness=6] node[left] {$(1,b)$} (0);
    \end{tikzpicture}
    \hfill
    % Second picture
    \begin{tikzpicture}[->, >=latex, baseline={(current bounding box.center)}]
      % Nodes
      \node[state, circle split] (2) [below = 2 cm of 0] {$\mu$ \nodepart{lower} $v_0$};
      \node[state, circle split] (3) [below left = 0.7cm of 2] {$\mu$ \nodepart{lower} $v_1$};
      \node[state, circle split] (4) [below right = 0.7cm of 2] {$\mu$ \nodepart{lower} $v_2$};
      \node[xshift=-2.5cm,yshift=-0.6cm] at (2.west) {\large{$V$}};
      %
      % Edges
      \path[thick] (2) edge [bend right, left] node {$(1,a)$} (3);
      \path[thick] (2) edge [bend left, right] node {$(1,b)$} (4);
      \path[thick] (3) edge [out=310,in=280,looseness=6] node[right] {$(1,a)$} (3);
      \path[thick] (3) edge [out=230,in=260,looseness=6] node[left] {$(1,b)$} (3);
      \path[thick] (4) edge [out=310,in=280,looseness=6] node[right] {$(1,a)$} (4);
      \path[thick] (4) edge [out=230,in=260,looseness=6] node[left] {$(1,b)$} (4);
    \end{tikzpicture}
    \hfill \ 
    
    \caption{Example showing that the faster-than relation and the simulation relation are incomparable.}
    \label{fig:comparison}
  \end{figure}
  
  Consider the two SMDPs $U$ and $V$ in Figure \ref{fig:comparison} with the same probability measure $\mu$ in all states.
  It is easy to see that $U$ is bisimilar to $V$, and hence $V$ also simulates $U$.
  However, we show that $U \not \ft V$ in the following way.
  Construct the scheduler $\sigma$ for $V$ by letting
  \[\sigma(v_0)(a) = 0.5, \; \sigma(v_0)(b) = 0.5, \; \sigma(v_1)(a) = 1, \, \text{ and } \, \sigma(v_2)(b) = 1.\]
  Now, for any scheduler $\sigma'$ for $U$, we must have either
  $\sigma'(u_0)(a) < 1$ or $\sigma'(u_0)(b) < 1$.
  If $\sigma'(u_0)(a) < 1$, then
  \[\sigma'(u_0)(a) > (\sigma'(u_0)(a))^2 > \dots > (\sigma'(u_0)(a))^n.\]
  Furthermore, we see that
  \[\prob^{\sigma}(v_0)(\cylinder(a^n,t)) = 0.5 \cdot \mu^{*n}(t) \text{ and } \prob^{\sigma'}(u_0)(\cylinder(a^n, t)) = (\sigma'(u_0)(a))^n \cdot \mu^{*n}(t)\]
  for $n > 1$.
  Take some $n$ such that $(\sigma'(u_0)(a))^n < 0.5$.
  In that case we get $\prob^{\sigma'}(u_0)(\cylinder(a^n, t)) < \prob^{\sigma}(v_0)(\cylinder(a^n, t))$.
  The same procedure can be used in case $\sigma'(u_0)(b) < 1$.
  Hence we conclude that $U \not \ft V$, and therefore also that $U \not \eqft V$.
\end{example}

Example \ref{ex:comparison1} also works for schedulers with memory,
although the argument has to be modified a bit.
In that case, in each step either the probability of a trace consisting only of $a$'s
or the probability of a trace consisting only of $b$'s must decrease in $U$,
so after some number of steps, the probability of one of these two
must decrease below $0.5$, and then the rest of the argument is the same.

\begin{example}\label{ex:comparison2}
  Consider the SMDPs $U$ and $V$ in Figure \ref{fig:faster-than-c} and
  let $F_\mu = \Exp{\theta_1}$ and $F_\nu = \Exp{\theta_2}$
  be exponential distributions with rates $\theta_1 > \theta_2 > 0$.
  Then, as shown in Example \ref{ex:faster-than},
  it holds that $U \ft V$.
  However, we have both $U \not \simul V$ and $U \not \sim V$.
\end{example}

From Examples \ref{ex:comparison1} and \ref{ex:comparison2},
we get the following theorem.

\begin{theorem}
  $\simul$ and $\ft$ are incomparable, $\sim$ and $\ft$ are incomparable, and we have $\sim \, \not \subseteq \, \eqft$.
\end{theorem}

%% Approximation
\section{Approximation}
It has been shown in \cite{PFBLM18}
that the faster-than relation is undecidable for the generative case.
A small modification of the argument shows that
the same is true for the reactive case.

\begin{theorem}\label{thm:undecidable}
  It is undecidable whether $U \ft V$.
\end{theorem}
\begin{proof}
  The result follows from the fact that $U \ft V$ is undecidable for the generative case.
  Let $U = (S_U, \tau, \rho)$ be a generative Markov process with set of actions $L$
  and construct the reactive Markov process $V = (S_V, \tau', \rho')$ as follows.
  Let $L' = L \cup \{\sharp\}$, where $\sharp$ is a new symbol not in $L$.
  For every state $s \in S_U$, we let $s^{a} \in S_V$ for every symbol $a \in L'$.
  \[
    \tau'\left(s_1^a, a'\right)\!\left(s_2^{a''}\right) = \begin{cases}
                                                          \tau(s_1)(a', s_2) & \text{if } a = a' \in L \text{ and } a'' = \sharp\\
                                                          \frac{1}{|L'|}     & \text{if } a = a' = \sharp \text{ and } a'' \in L.
                                                        \end{cases}\]
  So each state $s^a$ only has one outgoing action, namely $a$,
  and hence controllers play no role in the probabilities of $V$.
  Finally, let $\rho'(s^a) = \rho(s)$.
  Then we have
  \[P_{U}(s)(\cylinder(a_1 \dots a_n, t)) = |L'|^n P_{V}\!\left(s^\sharp\right)\!(\cylinder(\sharp a_1 \sharp \dots \sharp a_n, t)),\]
  and hence
  \[P_{U}(s_1)(\cylinder(a_1 \dots a_n, t)) \geq P_{U}(s_2)(\cylinder(a_1 \dots a_n, t))\]
  if and only if
  \[P_{V}\!\left(s_1^\sharp\right)\!(\cylinder(\sharp a_1 \sharp \dots \sharp a_n, t)) \geq P_{V}\!\left(s_2^\sharp\right)\!(\cylinder(\sharp a_1 \sharp \dots \sharp a_n, t)).\]
  This means that $s_1 \leq s_2$ if and only if $s_1^\sharp \leq s_2^\sharp$.
\end{proof}

In view of Theorem~\ref{thm:undecidable}, we can extend the approximation algorithm
for the generative case from \cite{PFBLM18}
to the reactive case.
In order to do this, we need to also consider
the schedulers that are necessary for reactive systems.
Instead of deciding the faster-than relation,
we consider the time-bounded approximation problem,
which asks the following:
Given $\varepsilon > 0$, a time bound $b \in \mathbb{R}_{\geq 0}$,
and two SMDPs $U$ and $V$, determine whether for all schedulers $\sigma$
there exists a scheduler $\sigma'$ such that
\begin{equation}\label{eq:approx}
  \prob^{\sigma'}(u_0)(C) \geq \prob^{\sigma}(v_0)(C) - \varepsilon
\end{equation}
for all time-bounded cylinders $C = \cylinder(a_1 \dots a_n,t)$ where $t \leq b$.

First we identify the kind of distributions
for which our algorithm will work.
Given a class $\mathcal{C}$ of distributions,
we let $\Convex(\mathcal{C})$ denote the closure of $\mathcal{C}$
under convex combinations, and $\Conv(\mathcal{C})$ denote the closure of $\mathcal{C}$
under both convex combinations and convolutions.

\begin{definition}\label{def:effective}
  A class of distributions $\mathcal{C}$
  is \emph{effective} if for any $\varepsilon > 0$,
  $b \in \mathbb{R}_{\geq 0}$, and $\mu_1,\mu_2 \in \Conv(\mathcal{C})$,
  $\{ t \in \mathbb{R}_{\geq 0} \mid \mu_1([0,t]) \geq \mu_2([0,t]) - \varepsilon \text{ and } t \leq b\}$
  is a semialgebraic set.
\end{definition}

A semialgebraic set is essentially one that can be expressed in the first-order theory of the reals,
and hence membership in such a set can be decided by utilising the decidability of the first-order theory of reals \cite{Tarski51}.
In addition to effectiveness,
we will also require residence-time distributions
to take some non-zero amount of time to fire.
This requirement is made precise by the following definition.

\begin{definition}
  A class $\mathcal{C}$ of distributions is \emph{slow}
  if for any finite subset $\mathcal{C}_0 \subseteq \mathcal{C}$,
  there exists a computable function $\varepsilon : \mathbb{N} \times \mathbb{R}_{\geq 0} \to [0,1]$
  such that for all $n \in \mathbb{N}$, $t \in \mathbb{R}_{\geq 0}$
  and $\mu_1, \dots, \mu_n \in \Convex(\mathcal{C}_0)$ we have
  \[(\mu_1 * \dots * \mu_n)([0,t]) \leq \varepsilon(n,t)\]
  and $\lim_{n \rightarrow \infty} \varepsilon(n,t) = 0$.
\end{definition}

It has been shown in \cite{PFBLM18} that the class of uniform distributions
and the class of exponential distributions are both effective and slow.
The importance of the closure under convex combinations and convolutions
in Definition \ref{def:effective} is explained by the following lemma.

\begin{lemma}\label{lem:conv}
  Let $\mathcal{C}$ be a class of distributions,
  and let $U$ be a SMDP with residence-time distributions taken from $\mathcal{C}$.
  Then $\prob^\sigma(s)(a_1 \dots a_n) \in \Conv(\mathcal{C})$
  for any scheduler $\sigma$, state $s$, and $a_1, \dots, a_n \in L$.
\end{lemma}
\begin{proof}
  The lemma follows essentially from Proposition \ref{prop:inductive}.
  For $n = 1$ we get
  \[\prob^\sigma(s)(a) = \sum_{s' \in S} \tau^\sigma(s,a)(s') \cdot \rho(s) \in \Conv(\mathcal{C}).\]
  
  For $n > 1$ we get
  \[\prob^\sigma(s)(a_1 \dots a_n) = \sum_{s' \in S} \tau^\sigma(s,a)(s') \cdot (\rho(s) * \prob^\sigma(s')(a_2 \dots a_n)),\]
  and since $\prob^\sigma(s')(a_2 \dots a_n) \in \Conv(\mathcal{C})$ by induction hypothesis,
  it follows that $\prob^\sigma(s)(a_1 \dots a_n) \in \Conv(\mathcal{C})$.
\end{proof}

Lemma \ref{lem:conv} shows that, for fixed schedulers $\sigma$ and $\sigma'$,
we can decide whether $\prob^\sigma(u)(C) \geq \prob^{\sigma'}(v)(C)$
whenever $U$ and $V$ have effective residence-time distributions using the first-order theory of reals.

The following theorem shows that, again for fixed schedulers,
we can find an $N \in \mathbb{N}$ such that the probability of any time-bounded cylinder
with length greater than $N$ is less than $\varepsilon$.
Therefore any such time-bounded cylinder trivially satisfies the inequality \eqref{eq:approx}
and can thus be disregarded.

\begin{theorem}[\hspace{1sp}{\cite[Theorem 5]{PFBLM18}}]\label{thm:limit}
  Let $U$ be a SMDP with slow residence-time distributions.
  For any state $s \in S$, $\varepsilon > 0$, $b \in \mathbb{R}_{\geq 0}$, and scheduler $\sigma$,
  there exists $N \in \mathbb{N}$ such that
  $\prob^\sigma(s)(\cylinder(a_1 \dots a_n, b)) \leq \varepsilon$
  for all $n \geq N$.
\end{theorem}

All that is left now is to consider schedulers.
However, since we only need to consider time-bounded cylinders up to some finite length,
we can also represent a scheduler as a collection of finitely many probability distributions over the action labels.
Each such distribution can in turn be represented as a collection of real variables that must sum to no more than $1$.
Hence schedulers can also be represented in the first-order theory of reals.

\begin{theorem}\label{thm:approx}
  Let $U$ be a SMDP with slow residence-time distributions.
  Then the time-bounded approximation problem is decidable.
\end{theorem}
\begin{proof}
  By Theorem \ref{thm:limit},
  we can find some $N \in \mathbb{N}$
  such that $\prob^{\sigma'}(v_0)(C) - \varepsilon \leq 0$
  for any scheduler $\sigma'$ and any time-bounded cylinder
  bounded by $b$ and of length $n \geq N$.
  This means that for any such time-bounded cylinder,
  we trivially have
  \[\prob^\sigma(u_0)(C) \geq \prob^{\sigma'}(v_0)(C) - \varepsilon\]
  for any scheduler $\sigma$.
  It is therefore enough to only consider time-bounded cylinders of length $n \leq N$.
  
  Now let $\sigma$ be a scheduler.
  We can represent $\sigma$ in the first-order theory of reals as follows.
  For each state $s$ and label $a$ (recall there are finitely many of these),
  let $x_{s,a}$ be a real variable.
  Then we interpret $x_{s,a}$ to be the probability $\sigma(s)(a)$,
  and we impose the constraint $\sum_{a \in L} x_{s,a} \leq 1$.
  The whole time-bounded approximation problem
  can therefore be encoded in the first-order theory of reals,
  and is thus decidable.
\end{proof}

%% Composition
\section{Compositionality}\label{sec:comp}
Next we introduce the notion of composition of SMDPs.
As argued in \cite{SV04}, the style of synchronous CSP composition
is the most natural one to consider for reactive probabilistic systems,
so this is the one we will adopt.
However, we leave the composition of the residence-times as a parameter,
so that we can compare different kinds of composition.

\begin{definition}
  A function $\star : \dist(\mathbb{R}_{\geq 0}) \times \dist(\mathbb{R}_{\geq 0}) \rightarrow \dist(\mathbb{R}_{\geq 0})$
  is called a \emph{residence-time composition function} if it is commutative,
  i.e. $\star(\mu,\nu) = \star(\nu,\mu)$ for all $\mu,\nu \in \dist(\mathbb{R}_{\geq 0})$.
\end{definition}

One example of such a composition function is when $\star$ is a coupling,
which is a joint probability measure such that its marginals are $\mu$ and $\nu$.
A simple special case of this is the product measure $\star(\mu,\nu) = \mu \times \nu$,
which is defined by $(\mu \times \nu)(B_1 \times B_2) = \mu(B_1) \cdot \nu(B_2)$
for all Borel $B_1$ and $B_2$.

In order to model the situation in which we want the composite system
only to take a transition when both components can take a transition,
it is natural to take the minimum of the two probabilities,
which corresponds to waiting for the slowest of the two.
In that case, we let
\[F_{\star(\mu,\nu)}(t) = \min\{F_{\mu}(t), F_{\nu}(t)\},\]
and we call this \emph{minimum composition}.
Likewise, if we only require one of the components to be able to take a transition,
then it is natural to take the maximum of the two probabilities by letting
\[F_{\star(\mu,\nu)}(t) = \max\{F_{\mu}(t), F_{\nu}(t)\},\]
which we call \emph{maximum composition}.
A special case of minimum composition is the composition on rates used in PEPA \cite{hillston2005},
and a special case of maximum composition is the composition on rates used in TIPP \cite{gotz1993}.

Further knowledge about the processes that are being composed
lets one define more specific composition functions.
As an example, if we know that the components only have exponential distributions,
then we can define composition functions that work directly on the rates of the distributions.
If $F_\mu = \Exp{\theta}$ and $F_{\nu} = \Exp{\theta'}$,
then one could for example let $\star(\mu,\nu)$ be such that
\[F_{\star(\mu,\nu)} = \Exp{\theta \cdot \theta'}.\]
This corresponds to the composition on rates that is used in SPA \cite{hermanns1998},
and we will call it \emph{product composition}.
Note that product composition is not given by the product measure.

\begin{definition}
  Let $\star$ be a residence-time composition function.
  Then the \emph{$\star$-composition} of $U$ and $V$, denoted by $U \comp{\star} V = (S, \tau, \rho)$, is given by
  \begin{itemize}
    \item $S = U \times V$,
    \item $\tau((u,v),a)((u',v')) = \tau_U(u,a)(u') \cdot \tau_V(v,a)(v')$ for all $a \in L$ and $(u',v') \allowbreak \in S$, and
    \item $\rho((u,v)) = \star(\rho_U(u), \rho_V(v))$. \qedhere
  \end{itemize}
%  for all $(u,v) \in S$.
\end{definition}

When considering the composite SMDP $U \comp{\star} V$
of two SMDPs $U$ and $V$,
we will also write $u \comp{\star} v$
to denote the composite state $(u,v)$ of $U \comp{\star} V$
where $u \in S_U$ and $v \in S_V$.

%%%%%%%%%%%%%%%%%%%%%%%%%%%%%%%%%%%%%%%%%%%%%%%
%% TIMING ANOMALIES
%%%%%%%%%%%%%%%%%%%%%%%%%%%%%%%%%%%%%%%%%%%%%%%
\subsection{Parallel Timing Anomalies}\label{sec:anomalies}
If we  have two components $U$ and $V$,
and we know that $U$ is faster than $V$,
then if $V$ is in parallel with some context $W$,
we would expect this composition to become faster when
we replace the component $V$ with the component $U$.
However, sometimes this fails to happen,
and we will call such an occurrence a \emph{parallel timing anomaly}.

In this section we show that parallel timing anomalies can occur
for some of the kinds of composition discussed in Section \ref{sec:comp}.
We do this by giving different contexts $W$ for the SMDPs $U$ and $V$
from Figure \ref{fig:faster-than-c},
for which it was shown in Example \ref{ex:faster-than} that $U \ft V$.
Our examples of parallel timing anomalies make no use of non-determinism or probabilistic branching,
thus showing that the parallel timing anomalies are caused
inherently by the timing behaviour of the SMDPs.
For ease of presentation, we let the set of labels $L$
consist only of the label $a$ in this section.

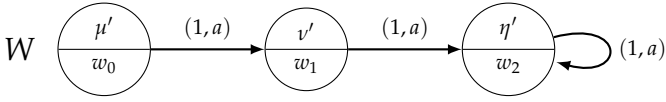
\begin{figure}
  \centering
  \hfill
  \begin{tikzpicture}
    %Nodes
    \node[state, circle split] (0) {$\mu'$ \nodepart{lower} $w_0$};
    \node[state, circle split] (1) [right = 1.5cm of 0]{$\nu'$ \nodepart{lower} $w_1$};
    \node[state, circle split] (8) [right = 1.5cm of 1]{$\eta'$ \nodepart{lower} $w_2$};
    
    \node[xshift=-0.5cm] at (0.west) {\large{$W$}};
    
    % Edges
    \path[thick] (0) edge [above] node {$(1,a)$} (1);
    \path[thick] (1) edge [above] node {$(1,a)$} (8);
    \path[thick, loop right] (8) edge [right] node {$(1,a)$} (8);
  \end{tikzpicture}
  \hfill \
  \caption{For different instantiations of $\mu'$, $\nu'$, and $\eta'$, the context $W$ leads to parallel timing anomalies for product, minimum, and maximum rate composition, respectively.}
  \label{fig:anomalies}
\end{figure}

Consider the two SMDPs $U$ and $V$ depicted in Figure \ref{fig:faster-than-c}.
For the examples in this section,
let $F_\mu = \Exp{2}$, $F_\nu = \Exp{0.5}$,
and let $\eta$ be arbitrary.

\begin{example}[Product composition]\label{ex:prodanomaly}
  Let $\star$ be product composition and
  let the context $(W,w_0)$ be given by Figure \ref{fig:anomalies},
  where $F_{\mu'} = \Exp{10}$, $F_{\nu'} = \Exp{0.1}$ and $\eta = \eta'$.
  In $U \comp{\star} W$, the rates in the first two states will then be $20$ and $0.05$,
  and in $V \comp{\star} W$ they will be $5$ and $0.5$.
  Consider the time-bounded cylinder $\cylinder(aa, 2)$.
  Then we see that
  \[\prob(u_0 \comp{\star} w_0)(\cylinder(aa, 2)) \approx 0.09 \quad \text{and} \quad \prob(v_0 \comp{\star} w_0)(\cylinder(aa, 2)) \approx 0.30,\]
  showing that $U \comp{\star} W \not\ft V \comp{\star} W$.
  Hence we have a parallel timing anomaly.
  What happens is that in the process $V \comp{\star} W$ the probability of taking a transition before time $2$
  with rate $5$ is already very close to $1$,
  so the process $U \comp{\star} W$ does not gain much by having a rate of $20$,
  whereas in the next step, $V \comp{\star} W$ gains a lot of probability by having a rate of $0.5$
  compared to the rate $0.05$ of $U \comp{\star} W$.
\end{example}

\begin{example}[Minimum composition]\label{ex:minanomaly}
  Let $\star$ be minimum composition
  and let the context $(W,w_0)$ be given by Figure \ref{fig:anomalies},
  where $F_{\mu'} = \Exp{1}$, $F_{\nu'} = \Exp{2}$, and $\eta = \eta'$.
  The rates of $U \comp{\star} W$ are then $1$ and $0.5$,
  whereas they are $0.5$ and $2$ in $V \comp{\star} W$.
  Then
  \[\prob(u_0 \comp{\star} w_0)(\cylinder(aa, 2)) \approx 0.40 \quad \text{and} \quad \prob(v_0 \comp{\star} w_0)(\cylinder(aa, 2)) \approx 0.51,\]
  so $U \comp{\star} W \not\ft V \comp{\star} W$.
  What happens in this example is that in the second step,
  $U \comp{\star} W$ has the same rate as $V \comp{\star} W$ had in the first step.
  This means that $U \comp{\star} W$ must be proportionally faster in the second step.
  However, $V \comp{\star} W$ has a rate of $2$ in the second step,
  but $U \comp{\star} W$ only had a rate of $1$ in the first step.
\end{example}

\begin{example}[Maximum composition]\label{ex:maxanomaly}
  Let $\star$ be maximum composition and
  let the context $(W,w_0)$ be given by Figure \ref{fig:anomalies},
  where $F_{\mu'} = \Exp{2}$, $F_{\nu'} = \Exp{1}$, and $\eta = \eta'$.
  $U \comp{\star} W$ then has rates $2$ and $1$,
  and $V \comp{\star} W$ has rates $2$ and $2$.
  Then
  \[\prob(u_0 \comp{\star} w_0)(\cylinder(aa, 2)) \approx 0.75 \quad \text{and} \quad \prob(v_0 \comp{\star} w_0)(\cylinder(aa, 2)) \approx 0.91,\]
  so $U \comp{\star} W_3 \not\ft V \comp{\star} W_3$.
  The reason for the timing anomaly in this case is clear:
  $V \comp{\star} W$ simply has a higher rate in each step than $U \comp{\star} W$ does.
\end{example}

%%%%%%%%%%%%%%%%%%%%%%%%%%%%%%%%%%%%%%%%%%%%%%%
%% AVOIDING TIMING ANOMALIES
%%%%%%%%%%%%%%%%%%%%%%%%%%%%%%%%%%%%%%%%%%%%%%%
\subsection{Avoiding Parallel Timing Anomalies}
We have seen in the previous section that parallel timing anomalies can occur.
We now wish to understand what kind of contexts do not lead to timing anomalies.
In this section we assume that the set $L$ of transition labels is a finite set.
Also, we fix a residence-time composition function $\star$
and two additional SMDPs $(W,w_0) = (S_W, \tau_W, \rho_W)$ and $(W',w'_0) = (S_{W'}, \tau_{W'}, \rho_{W'})$
which should be thought of as contexts.
Next we identify conditions on $(W,w_0)$ such that
$U \ft V$ will imply $U \comp{\star} W \ft V \comp{\star} W$.

We first give conditions that over-approximate the faster-than relation between the composite systems
by requiring that when $U$ and $W$ are put in parallel,
then the composite system is point-wise faster than $U$ along all paths.
Likewise, we require that when $V$ and $W$ are put in parallel,
the composite system is point-wise slower than $V$ along all paths.
If we already know that $U$ is faster than $V$,
this will imply by transitivity that $U \comp{\star} W$ is faster than $V \comp{\star} W$.
We have already seen in Example \ref{ex:faster-than}
that a process $U$ need not be point-wise faster than $V$ along all paths
in order for $U$ to be faster than $V$.
However, by imposing this condition,
we do not need to compare convolutions of distributions,
but can compare the distributions directly.

We first introduce some terminology.
We will say that a SMDP $M$ has a \emph{deterministic Markov kernel}
if for all states $s$ and labels $a$,
there is at most one state $s'$ such that $\tau(s,a)(s') > 0$.

\begin{definition}
  A \emph{state path in $M$} is a sequence of states $s_1,s_2, \dots$
  where for all $i \in \mathbb{N}$ there exists a label $a \in L$
  such that $\tau(s_i, a)(s_{i+1}) > 0$.
  For a state path $\pi = s_1, s_2, \dots$, we let $\pi\sproj{i} = s_i$,
  $\pi|^i = s_i, s_{i+1}, \dots$, $\pi|_i = s_1, s_2, \dots, s_i$,
  and we let $\spaths{}{M}$ denote the set of all state paths in $M$.
  For a state $s \in S$, we let $\spaths{}{s} = \{ \pi \in \spaths{}{M} \mid \pi\sproj{1} = s\}$
  and we let $\spaths{n}{s} = \{\pi|_n \mid \pi \in \spaths{}{s}\}$.
\end{definition}

\begin{definition}\label{def:safe}
  Let $n \in \mathbb{N}$.
  We say that $\star$ is \emph{$n$-monotonic in $U$, $V$, $W$, and $W'$},
  written $(U,W) \mon{n}{\star} (V,W')$,
  if $W'$ has a deterministic Markov kernel and
  \begin{itemize}
    \item $F_{\rho(\pi_U\sproj{i} \comp{\star} \pi_W\sproj{i})}(t) \geq F_{\rho_U(\pi_U\sproj{i})}(t)$ and
      $F_{\rho_V(\pi_V\sproj{i})}(t) \geq F_{\rho(\pi_V\sproj{i} \comp{\star} \pi_{W'}\sproj{i})}(t)$ for all $t \in \mathbb{R}_{\geq 0}$ and $1 \leq i \leq n$,
    \item for all schedulers $\sigma_U$ for $U$ there exists a scheduler $\sigma_{U,W}$ for $U \comp{\star} W$ such that we have
  \end{itemize}
      \[\tau^{\sigma_{U,W}}(\pi_U\sproj{i} \comp{\star} \pi_W\sproj{i}, a)(\pi_U\sproj{i+1} \comp{\star} \pi_W\sproj{i+1}) \geq \tau^{\sigma_U}_U(\pi_U\sproj{i},a)(\pi_U\sproj{i+1}),\]
  \begin{itemize}
    \item[] and
    \item for all schedulers $\sigma_{V,W'}$ for $V \comp{\star} W'$ there exists a scheduler $\sigma_V$ for $V$ such that we have
  \end{itemize}
  \[\tau^{\sigma_V}_V(\pi_V\sproj{i},a)(\pi_V\sproj{i+1}) \geq \tau^{\sigma_{V,W'}}(\pi_V\sproj{i} \comp{\star} \pi_{W'}\sproj{i},a)(\pi_V\sproj{i+1} \comp{\star} \pi_{W'}\sproj{i+1})\]
  for all state paths $\pi_U \in \spaths{n}{u_0}$,
  $\pi_V \in \spaths{n}{v_0}$, $\pi_W \in \spaths{n}{w_0}$, and $\pi_{W'} \in \spaths{n}{w'_0}$,
  and for all $a \in L$ and $1 \leq i < n$.
  Furthermore, we will say that $\star$ is \emph{monotonic in $U$, $V$, $W$, and $W'$}
  and write $(U,W) \mon{}{\star} (V,W')$,
  if it is $n$-monotonic in $U$, $V$, $W$, and $W'$ for all $n \in \mathbb{N}$.
\end{definition}

Clearly, if $(U,W) \mon{n}{\star} (V,W')$,
then $(U,W) \mon{m}{\star} (V,W')$ for all $m \leq n$.
The next result shows that if $(U,W) \mon{}{\star} (V,W')$,
then we are guaranteed to avoid parallel timing anomalies.

\begin{theorem}\label{thm:avoid}
  If $(U,W) \mon{}{\star} (V,W')$ as well as $U \ft V$ and $W \ft W'$,
  then we have $U \comp{\star} W \ft V \comp{\star} W'$.
\end{theorem}
\begin{proof}%[Proof of Theorem \ref{thm:avoid}]
  Let $\cylinder(a_1 \dots a_n, x)$ be an arbitrary time-bounded cylinder,
  and let $\sigma_{V, W'}$ be an arbitrary scheduler for $V \comp{\star} W'$.
  Because $(U,W) \mon{}{\star} (V,W')$,
  there exists a scheduler $\sigma_V$ for $V$ and a path $\pi$ such that 
  \begin{align*}
    &\phantom{{}={}} \prob^{\sigma_{V, W'}}(v_0 \comp{\star} w'_0)(\cylinder(a_1 \dots a_n, t)) \\
    &= \tau^{\sigma_{V,W'}}(\pi\sproj{1},a_1)(\pi\sproj{2}) \cdots \tau^{\sigma_{V,W'}}(\pi\sproj{n},a_n)(\pi\sproj{n+1}) \\
    &\phantom{{}={}}\cdot (\rho(\pi\sproj{1}) * \dots * \rho(\pi\sproj{n}))([0,t]) \\
    &\leq \tau_V^{\sigma_V}(\pi_V\sproj{1},a_1)(\pi_V\sproj{2}) \cdots \tau_V^{\sigma_V}(\pi_V\sproj{n},a_n)(\pi\sproj{n+1}) \\
    &\phantom{{}={}}\cdot (\rho_V(\pi_V\sproj{1}) * \dots * \rho_V(\pi_V\sproj{n}))([0,t])\\
    &\leq \sum_{\pi \in \spaths{n+1}{v_0}} \tau_V^{\sigma_V}(\pi\sproj{1},a_1)(\pi\sproj{2}) \cdots \tau_V^{\sigma_V}(\pi\sproj{n},a_n)(\pi\sproj{n+1}) \\
    &\phantom{{}={}}\cdot (\rho_V(\pi\sproj{1}) * \dots * \rho_V(\pi\sproj{n}))([0,t])\\
    &= \prob^{\sigma_V}(v_0)(\cylinder(a_1 \dots a_n, t)),
  \end{align*}

  Since $U \ft V$, there must exist some scheduler $\sigma_U$ for $U$ such that
  \[\prob^{\sigma_V}(v_0)(\cylinder(a_1 \dots a_n, t))) \leq \prob^{\sigma_U}(u_0)(\cylinder(a_1 \dots a_n, t)).\]
  Again, since $(U,W) \mon{}{\star} (V,W')$,
  there exists a scheduler $\sigma_{U,W}$ for $U \comp{\star} W$ such that
  \begin{align*}
    &\phantom{{}={}} \prob^{\sigma_U}(u_0)(\cylinder(a_1 \dots a_n,t))\\
    &= \sum_{\pi \in \spaths{n+1}{u_0}} \tau_U^{\sigma_U}(\pi\sproj{1},a_1)(\pi\sproj{2}) \cdots \tau_U^{\sigma_U}(\pi\sproj{n},a_n)(\pi\sproj{n+1}) \\
    &\phantom{{}={}}\cdot (\rho_U(\pi\sproj{1}) * \dots * \rho_U(\pi\sproj{n}))([0,t])\\
    &\leq \sum_{\pi_W \in \spaths{n+1}{w_0}}\sum_{\pi \in \spaths{n+1}{u_0}} \tau_U^{\sigma_U}(\pi\sproj{1},a_1)(\pi\sproj{2}) \cdots \tau_U^{\sigma_U}(\pi\sproj{n},a_n)(\pi\sproj{n+1}) \\
    &\phantom{{}={}}\cdot (\rho_U(\pi\sproj{1}) * \dots * \rho_U(\pi\sproj{n}))([0,t])\\
    &\leq \sum_{\pi \in \spaths{n+1}{u_0 \comp{\star} w_0}}\tau^{\sigma_{U,W}}(\pi\sproj{1},a_1)(\sproj{2}) \cdots \tau^{\sigma_{U,W}}(\sproj{n},a_n)(\sproj{n+1}) \\
    &\phantom{{}={}}\cdot (\rho(\pi\sproj{1}) * \dots * \rho(\pi\sproj{n}))([0,t])\\
    &= \prob^{\sigma_{U,W}}(u_0 \comp{\star} w_0)(\cylinder(a_1 \dots a_n, t)). \qedhere
  \end{align*}
\end{proof}

The special case where $W = W'$ shows that this
condition is sufficient to avoid parallel timing anomalies.
We do not know if it is decidable whether $(U,W) \mon{}{\star} (V,W')$.
However, there is a stronger condition which is decidable
in the case of finite SMDPs.
We present it in the next definition.

\begin{definition}\label{def:smon}
  We say that $\star$ is \emph{strongly} $n$-monotonic in $U$, $V$, $W$, and $W'$
  and write $(U,W) \smon{n}{\star} (V,W')$
  if $W'$ has a deterministic Markov kernel and for all state paths $\pi_U \in \spaths{n}{u_0}$,
  $\pi_V \in \spaths{n}{v_0}$, $\pi_W \in \spaths{n}{w_0}$, and $\pi_{W'} \in \spaths{n}{w'_0}$,
  the first condition of Definition \ref{def:safe} is satisfied and
  \begin{itemize}
    \item for all schedulers $\sigma_U$ for $U$ and all
      schedulers $\sigma_{U,W}$ for $U \comp{\star} W$, it is the case that
  \end{itemize}
  \[\tau^{\sigma_{U,W}}(\pi_U\sproj{i} \comp{\star} \pi_W\sproj{i},a)(\pi_U\sproj{i+1} \comp{\star} \pi_W\sproj{i+1}) \geq \tau^{\sigma_U}_U(\pi_U\sproj{i},a)(\pi_U\sproj{i+1}),\]
  \begin{itemize}
    \item[] and
    \item for all schedulers $\sigma_{V,W'}$ for $V \comp{\star} W'$
      and all schedulers $\sigma_V$ for $V$, it is the case that
  \end{itemize}
  \[\tau^{\sigma_V}_V(\pi_V\sproj{i},a)(\pi_V\sproj{i+1}) \geq \tau^{\sigma_{V,W'}}(\pi_V\sproj{i}\comp{\star}\pi_{W'}\sproj{i},a)(\pi_V\sproj{i+1} \comp{\star} \pi_{W'}\sproj{i+1})\]
  for all $a \in L$ and $1 \leq i < n$.
  If $(U,W) \smon{n}{\star} (V,W')$ for all $n \in \mathbb{N}$,
  we say that $\star$ is \emph{strongly} monotonic in $U$, $V$, $W$, and $W'$
  and write $(U,W) \smon{}{\star} (V,W')$.
\end{definition}

The conditions of Definition \ref{def:smon} are
the second and third conditions from Definition \ref{def:safe} with
the existential quantifier strengthened to a universal quantifier.
It is obvious that $(U,W) \smon{}{\star} (V,W')$ implies $(U,W) \mon{}{\star} (V,W')$,
and hence we get the following corollary.

\begin{corollary}
  If $(U,W) \smon{}{\star} (V,W')$ as well as $U \ft V$ and $W \ft W'$,
  then $U \comp{\star} W \ft V \comp{\star} W'$.
\end{corollary}

\begin{example}
  Let $U$ and $V$ be given by Figure \ref{fig:faster-than-c}
  with $F_\mu \geq F_\nu$ as in Example \ref{ex:faster-than}.
  Let $\star$ be minimum rate composition and consider the context $W$ from Figure \ref{fig:anomalies},
  where $\mu' = \mu$, $\nu' = \nu$, and $\eta' = \eta$.
  There is only one possible scheduler $\sigma$, which is the Dirac measure at $a$,
  and hence it is clear that the second and third conditions are satisfied. We also find that
  \begin{align*}
    F_{\rho(u_0 \comp{\star} w_0)}(t) = F_{\rho_U(u_0)}(t) & \quad & F_{\rho_V(v_0)}(t) = F_{\rho(v_0 \comp{\star} w_0)}(t) \\
    F_{\rho(u_1 \comp{\star} w_1)}(t) = F_{\rho_U(u_1)}(t) & \quad & F_{\rho_V(v_1)}(t) = F_{\rho(v_1 \comp{\star} w_1)}(t) \\
    F_{\rho(u_2 \comp{\star} w_2)}(t) = F_{\rho_U(u_2)}(t) & \quad & F_{\rho_V(v_2)}(t) = F_{\rho(v_2 \comp{\star} w_2)}(t)
  \end{align*}
  and hence the first condition is also satisfied, so $(U,W) \smon{}{\star} (V,W)$.
\end{example}

\begin{example}
  All the examples we gave in Section \ref{sec:anomalies} are not monotonic,
  and hence also not strongly monotonic, since they all violate condition 1 of monotonicity.
  
  In Example \ref{ex:prodanomaly}, this is because
  \[F_{\rho(v_0)}(t) = \Exp{0.5}(t) < \Exp{5}(t) = F_{\rho(v_0 \comp{\star} w_0)}(t)\]
  for any $t > 0$. Likewise, in Example \ref{ex:minanomaly} we have
  \[F_{\rho(u_0 \comp{\star} w_0)}(t) = \Exp{1}(t) < \Exp{2}(t) = F_{\rho(u_0)}(t)\]
  for any $t > 0$. Finally, in Example \ref{ex:maxanomaly} we have
  \[F_{\rho(v_0)}(t) = \Exp{0.5}(t) < \Exp{2}(t) = F_{\rho(v_0 \comp{\star} w_0)}(t)\]
  for any $t > 0$.
\end{example}

%\begin{definition}
%  A state $s$ is called \emph{recurrent} if there exists a state path $\pi$
%  such that $\pi\sproj{1} = s$ and $\pi\sproj{k} = s$ for some $k$.
%\end{definition}

%A crucial part of showing decidability is showing that for finite SMDPs,
%paths will eventually start repeating.
%This will allow us to only consider a finite number of paths
%to decide the faster-than relation.

We now wish to show that it is decidable whether $(U,W) \smon{}{\star} (V,W')$ for finite SMDPs,
thereby giving a decidable condition for avoiding timing anomalies.
To do this, we first show that in order to establish strong monotonicity,
it is enough to consider paths up to length
\[m = \max\{|S_U| \cdot |S_W|, |S_V| \cdot |S_{W'}|\} + \max\{|S_U|,|S_V|,|S_W|,|S_{W'}|\} + 1,\]
due to the fact that they start repeating.
%as can be shown by a simple pigeonhole argument.
%as we showed in Lemma \ref{lem:cycles}.

\begin{lemma}\label{lem:cycles}
  Let $U$ and $V$ be two finite, pointed SMDPs.
  For any state paths $\pi_U$ and $\pi_V$
  of length $l > |S_U| \cdot |S_V|$,
  there will be $i < j \leq |S_U| \cdot |S_V| + 1$ such that
  $\pi_U\sproj{i} = \pi_U\sproj{j}$, $\pi_V\sproj{i} = \pi_V\sproj{j}$.
\end{lemma}
\begin{proof}
  Since there are $|S_U| \cdot |S_V|$ ways of choosing a pair $(u_i,v_j) \in S_U \times S_V$
  of states from $U$ and $V$, if we pair the states of $\pi_U$ and $\pi_V$
  such that we get the pairs $(\pi_U\sproj{1},\pi_V\sproj{1})$, $(\pi_U\sproj{2},\pi_V\sproj{2})$, $\dots$, $(\pi_U\sproj{l},\pi_V\sproj{l})$,
  there must be two of these pairs that are the same because $l > |S_U| \cdot |S_V|$.
  Hence we get states $\pi_U\sproj{i} = \pi_U\sproj{j}$
  and $\pi_V\sproj{i} = \pi_V\sproj{j}$ with $i < j \leq n$.
  It also follows that $i$ and $j$ can be chosen so that $i < j \leq |S_U| \cdot |S_V|$,
  because otherwise we would have $j - i > |S_U| \cdot |S_V|$ different pairs
  \[(\pi_U\sproj{i},\pi_V\sproj{i}), (\pi_U\sproj{i+1},\pi_V\sproj{i+1}), \dots, (\pi_U\sproj{j},\pi_V\sproj{j}),\]
  contradicting the fact that there are only $|S_U| \cdot |S_V|$ such different pairs.
\end{proof}

\begin{lemma}\label{lem:m-safe}
  Let $(U,u_0)$, $(V,v_0)$, $(W,w_0)$, and $(W',w'_0)$ be finite, pointed SMDPs.
  If $(U,W) \smon{m}{\star} (V,W')$, then $(U,W) \smon{}{\star} (V,W')$.
\end{lemma}
\begin{proof}%[Proof of Lemma \ref{lem:m-safe}]
  Assume that $(U,W) \smon{m}{\star} (V,W')$.
  Then $(U,W) \smon{k}{\star} (V,W')$ for all $k \leq m$.
  Hence it remains to show that $(U,W) \smon{k}{\star} (V,W')$ for all $k > m$.
  
  Let $k > m$ and consider two state paths $\pi_U = u_1u_2 \dots u_k$
  and $\pi_W = w_1w_2 \dots w_k$ of $U$ and $W$, respectively, both of length $k$.
  By Lemma \ref{lem:cycles} there must exist $i < j \leq |S_U| \cdot |S_W| + 1$ such that
  $u_i = u_j$ and $w_i = w_j$.
  Since there exists a state path from $u_1$ to $u_i$,
  it must be possible to reach this state in less than $|S_U|$ steps,
  and likewise for $W$. Hence there must exist $l \leq \max\{|S_U|,|S_W|\}$,
  and state paths
  \[u_1 u_2' \dots u_l' u_i u_{i + 1} \dots u_j \quad \text{and} \quad w_1 w_2' \dots w_l' w_i w_{i + 1} \dots w_j\]
  of length
  $l + (j - i) \leq \max\{|S_U|,|S_W|\} + |S_U| \cdot |S_W| + 1 \leq m$.
  Hence we know that the conditions of Definition \ref{def:smon}
  are satisfied for $u_i \dots u_j$ and $w_i \dots w_j$.
  By removing the states $u_{i + 1} \dots u_j$ and $w_{i + 1} \dots w_j$
  from $\pi_U$ and $\pi_W$ we end up with two new state paths $\pi'_U$ and $\pi'_W$
  of length $k' = k - (j - i)$.
  We can keep doing this as long as $k' > m$,
  so at some point we must end up with state paths $\pi^*_U$ and $\pi^*_W$
  of length $k^* \leq m$,
  for which the conditions of Definition \ref{def:smon} are satisfied by assumption,
  and hence they are satisfied for all of $\pi_U$ and $\pi_W$.
  The same argument can be applied to two state paths $\pi_V$ and $\pi_{W'}$
  of $V$ and $W'$, so we conclude that $(U,W) \smon{}{\star} (V,W')$.
\end{proof}

We can now use the first-order theory of the reals to show
that strong monotonicity is a decidable property.

\begin{theorem}\label{thm:decidable}
  Consider the finite pointed SMDPs $(U,u_0)$, $(V,v_0)$, $(W,w_0)$, and $(W',w'_0)$.
  If for all state paths $\pi_U \in \spaths{m}{u_0}$,
  $\pi_V \in \spaths{m}{v_0}$, $\pi_W \in \spaths{m}{w_0}$, and $\pi_{W'} \in \spaths{m}{w'_0}$
  we have that $\{t \in \mathbb{R}_{\geq 0} \mid F_{\rho(\pi_U\sproj{i} \comp{\star} \pi_W\sproj{i})}(t) \geq F_{\rho_U(\pi_U\sproj{i})}(t)\}$
  and $\{t \in \mathbb{R}_{\geq 0} \mid F_{\rho_V(\pi_V\sproj{i})}(t) \geq F_{\rho(\pi_V\sproj{i} \comp{\star} \pi_{W'}\sproj{i})}(t)\}$
  are semialgebraic sets for all $1 \leq i \leq m$, then it is decidable whether $(U,W) \smon{}{\star} (V,W')$.
\end{theorem}
\begin{proof}%[Proof of Theorem \ref{thm:decidable}]
  Note first of all that since $L$ and $W'$ are finite,
  it is decidable whether $W'$ has a deterministic Markov kernel by looking at all the states.
  By Lemma \ref{lem:m-safe}, it suffices to check
  whether $(U,W) \smon{m}{\star} (V,W')$ where
  \[m = \max\{|S_U|\cdot|S_W|,|S_V|\cdot|S_{W'}|\} + \max\{|S_U|,|S_V|,|S_W|,|S_{W'}|\} + 1.\]
  This can be done by exploiting the decidability
  of the first-order theory of the reals in the following way.
  Since $L$ is finite and $U$, $V$, $W$, and $W'$ are all finite,
  there are finitely many state paths $\pi_U \in \spaths{m}{u_0}$,
  $\pi_V \in \spaths{m}{v_0}$, $\pi_W \in \spaths{m}{w_0}$, and $\pi_{W'} \in \spaths{m}{w'_0}$.
  Because of this, and since the sets
  \[\{t \in \mathbb{R}_{\geq 0} \mid F_{\rho(\pi_U\sproj{i} \comp{\star} \pi_W\sproj{i})}(t) \geq F_{\rho_U(\pi_U\sproj{i})}(t)\}\]
  and
  \[\{t \in \mathbb{R}_{\geq 0} \mid F_{\rho_V(\pi_V\sproj{i})}(t) \geq F_{\rho(\pi_V\sproj{i} \comp{\star} \pi_{W'}\sproj{i})}(t)\},\]
  which we need to check for the first condition,  
  were assumed to be semialgebraic,
  it is possible to express the conditions of Definition \ref{def:smon}
  in the first-order theory of the reals,
  using finitely many quantifiers and inequalities.
  Since the first-order theory of the reals is decidable,
  the truth value of the resulting formula is decidable.
\end{proof}

For uniform and exponential distributions with minimum or maximum composition,
the corresponding sets are all semialgebraic,
and the same is true for exponential distributions with product composition.
Theorem \ref{thm:decidable} can therefore be used for these types of composition.

Unfortunately, strong monotonicity is a very strict requirement.
In effect, it requires that there is only one possible action,
and hence rules out non-determinism.
However, strong monotonicity still makes sense as a requirement
on processes with no non-determinism,
since all our examples of timing anomalies in Section \ref{sec:anomalies}
are of this form.

\begin{proposition}\label{prop:singleton}
  If $(U,W) \smon{}{\star} (V,W')$, then $L$ is a singleton set or $u_0$ is a deadlock state.
\end{proposition}
\begin{proof}
  We prove the contrapositive.
  Suppose $|L| > 1$ and $u_0$ is not a deadlock state.
  Because $u_0$ is not a deadlock state,
  there must exist some state path $\pi_U$ such that
  $\pi_U\sproj{1} = u_0$ and $\tau_U(\pi_U\sproj{1},a)(\pi_U\sproj{2}) > 0$ for some $a \in L$.
  Since $|L| > 1$, we can find some $b \in L$ with $a \neq b$.
  Now construct schedulers given by $\sigma_{U,W}(s) = \delta_b$ and $\sigma_U(s) = \delta_a$
  for any state $s$.
  Then
  \[\tau^{\sigma_{U,W}}(\pi_U\sproj{1} \comp{\star} \pi_W\sproj{1}, a)(\tau_U\sproj{2} \comp{\star} \tau_W\sproj{2}) = 0\]
  but
  \[\tau_U^{\sigma_U}(\pi_U\sproj{1}, a)(\tau_U\sproj{2}) > 0,\]
  and hence the first condition of Definition \ref{def:smon}
  is violated.
\end{proof}

%% Conclusion
\section{Conclusion}
In this paper, we have investigated the notion of a process being faster than another process in the context of semi-Markov decision processes.
We have given a trace-based definition of a faster-than relation,
and shown that this definition is closely connected to convolutions of distributions.
The faster-than relation is unfortunately undecidable,
but we have shown how to approximate a time-bounded version of it.
By considering composition as being parametric in how the residence times of states are combined,
we have given examples showing that our faster-than relation gives rise to parallel timing anomalies
for many of the popular ways of composing rates.
We have therefore given sufficient conditions for how such parallel timing anomalies can be avoided,
and we have shown that these conditions are decidable.

The main challenge that we face when trying to construct algorithms
for the faster-than relation is that of schedulers,
and in particular the juxtaposition between the universal and existential
quantification over schedulers.
For example, we had to strengthen the existential quantifier
to a universal one in order to decide the conditions for avoiding parallel timing anomalies.
This is because we know that locally, for any scheduler $\sigma$,
there exists a scheduler $\sigma'$ which works.
However, it is not clear that all of these $\sigma'$
can be collected coherently into a single scheduler which works globally.
Solving this challenge would allow us to
decide the property of monotonicity instead of the too-strong property of strong monotonicity,
as well as prove decidability for so-called unambiguous processes.

The conditions we have given for avoiding timing anomalies
do not look at the context in isolation,
but depend also on the processes that are being swapped.
It would be preferable to have conditions on a context
that would guarantee the absence of parallel timing anomalies
no matter what processes are being swapped.

\defaultbib

  \cleardoublepage
  \setcounter{enumiii}{0}
  \setcounter{enumii}{0}
  \setcounter{enumiv}{0}
  \setcounter{enumi}{0}
  \setcounter{equation}{0}
  \setcounter{figure}{0}
  \setcounter{footnote}{0}
  \setcounter{mpfootnote}{0}
  \setcounter{paragraph}{0}
  \setcounter{parentequation}{0}
  \setcounter{part}{0}
  \setcounter{section}{0}
  \setcounter{subfigure}{0}
  \setcounter{subparagraph}{0}
  \setcounter{subsection}{0}
  \setcounter{subsubsection}{0}
  \setcounter{table}{0}
  \papertitlepage{%
  A Hemimetric Extension of Simulation for Semi-Markov Processes
}{paper:paperD}{%
  Mathias R. Pedersen, Giorgio Bacci, Kim G. Larsen, and Radu Mardare
}{%
  The paper has been published in the\\
  \textit{Proceedings of the 15th International Conference on Quantitative Evaluation of Systems}, pp.~339--355, 2018.
}{%
  \noindent\copyright\ 2018 Springer
  
  \noindent{\em The layout has been revised and the content extended.}
}

%% Abstract
\begin{abstract}
  Semi-Markov decision processes (SMDPs) are continuous-time Markov decision processes where 
  the residence-time on states is governed by generic distributions on the positive real line.

  In this paper we consider the problem of comparing two SMDPs
  with respect to their time-dependent behaviour.
  We propose a hemimetric between processes, which we call \emph{simulation distance},
  measuring the least acceleration factor by which a process
  needs to speed up its actions in order to 
  behave at least as fast as another process. We show that this distance can be computed in time 
  $\bigo(n^2 (f(l) + k) + m^2n)$,
  where $n$ is the number of states, $m$ the number of actions, $k$ the number of atomic propositions,
  and $f(l)$ the complexity of comparing the residence-time between states.
  The theoretical relevance and applicability of this distance is further argued by showing that
  (i) it is suitable for compositional reasoning with respect to CSP-like parallel composition and
  (ii) has a logical characterisation in terms of a simple Markovian logic.
\end{abstract}

%% Introduction
\section{Introduction}
Semi-Markov decision processes (SMDPs) are Markovian stochastic decision processes modelling 
the firing time of transitions via real-valued random variables describing the residence-time on states.
Semi-Markov decision processes provide a more permissive model than continuous-time Markov decision processes, 
since they allow as residence-time distributions any generic distribution on the positive real line,
rather than only exponential ones.
The generality offered by SMDPs has been found useful in modelling several real-case scenarios.
Successful examples include power plants~\cite{PST97} and power supply units~\cite{PTV04}, to name a few.

When considering such real-time stochastic processes,
non-functional requirements are important,
particularly requirements like response time and throughput,
which depend on the timing behaviour of the process.
We therefore wish to understand and be able to compare
the timing behaviour of different processes.

To cope with the need for comparing the timing behaviour of different systems,
in this paper we propose and study a quantitative extension of the simulation relation  
by Baier et al. \cite{baier2000},
called $\varepsilon$-simulation, which puts the focus on the timing aspect of processes.
The intuition is that a process $s_2$ \emph{$\varepsilon$-simulates}
another process $s_1$
if after accelerating the actions of $s_2$ by a factor $\varepsilon > 0$
it reacts to the inputs from the external environment as $s_1$ with at least the same speed.

This type of quantitative reasoning is not new in the literature,
and it dates back to the seminal work
of Jou and Smolka~\cite{jou1990,GJS90},
who proposed the concept of probabilistic $\varepsilon$-bisimulation.
This line of work has lead to much work on probabilistic bisimulation distances \cite{CBW12,DGJP04,FPP11}.
While our work is conceptually similar to the bisimulation distances,
it is technically very different.
This is because bisimulation distances are constructed from a coalgebraic view as fixed points of operators.
However, for the kind of timed systems that we are investigating,
the coalgebraic perspective is much less understood.
Moreover, since our distance generalises a preorder relation and not a congruence
as the other distances do, it is not symmetric,
which brings in new technical challenges

Following the work of Jou and Smolka,
our notion of $\varepsilon$-simulation naturally induces a distance between processes: 
For any pair of states $s_1$ and $s_2$, we define their \emph{simulation distance}
as the least acceleration factor needed by $s_2$ to speed up its actions 
in order to behave at least as fast as $s_1$. This definition does not provide a distance 
in the usual sense, but rather a \emph{multiplicative hemimetric},
i.e.\ an asymmetric notion of distance satisfying
a multiplicative version of the triangle inequality.
Such a notion is not new, as it is extensively applied in the context of differential privacy 
to measure information leakage of systems (see e.g.~\cite{ACMMPS14,CABP13}).

The theoretical relevance and applicability of the simulation distance
is argued by means of the following results,
which are the main technical contributions of the paper:
\begin{enumerate}
  \item We provide an algorithm for computing the simulation distance between arbitrary states
    of an SMDP running in time $\bigo(n^2 (f(l) + k) + m^2n)$, where $n$ is the 
    number of states, $m$ the number of actions, $k$ the number of 
    atomic propositions, and $f(l)$ the complexity of comparing the residence time distributions
    on states.
  \item We show that under some mild conditions on how 
    residence-time distributions are combined in the
    parallel composition of two states, CSP-like parallel 
    composition of SMDPs is non-expansive with respect to our hemimetric.
    This shows that the simulation distance is suitable for compositional reasoning.
  \item We provide a logical characterisation of the distance
    in terms of a simple Markovian logic,
    stating that the distance from $s_1$ to $s_2$ is less than or equal to $\varepsilon$ if and only if
    $s_2$ satisfies the $\varepsilon$-perturbation of any logical property that $s_1$ satisfies.
    Moreover, we prove that $\varepsilon$-simulation preserves the $\varepsilon$-perturbation
    of time-bounded reachability properties.
  \item We show that sets of formulas in our logic are closed in the topology induced by the distance.
    This means that approximate reasoning is sound in the limit:
    If a sequence of state converging to a limit all satisfy some logical property,
    then the limit will also satisfy this property.
\end{enumerate}

\subsubsection{Notation and Preliminaries.}
Let $\mathbb{N}$ denote the natural numbers,
$\mathbb{Q}_{\geq 0}$ the non-negative rational numbers,
$\mathbb{R}_{\geq 0}$ the non-negative real numbers,
and $\mathbb{R}_{> 0}$ the strictly positive ones.
We equip the real numbers with the usual Borel $\sigma$-algebra.
Given a set $X$, we will denote by $\dist(X)$
the set of all probability measures on $X$.
A probability measure $\mu \in \dist(X)$ is said to be \emph{finitely supported} if the set
$\{x \in X \mid \mu(x) > 0\}$ is finite.
If $\mu \in \mathcal{D}(\mathbb{R}_{\geq 0})$,
then the \emph{cumulative distribution function} (CDF)
will be denoted by $F_{\mu}$ and is given by $F_\mu(t) = \mu([0,t])$.
Any CDF $F$ is increasing, meaning that if $t \geq t'$, then $F(t) \geq F(t')$,
and also right-continuous,
meaning that for all $\varepsilon > 0$ there exists a $\delta > 0$
such that if $t < t' < t + \delta$, then $|F(t') - F(t)| < \varepsilon$.
For $x \in \mathbb{R}_{\geq 0}$, we will write $\delta_x$ for the Dirac measure at $x$,
which is defined by $\delta_x(E) = 1$ if $x \in E$ and $\delta_x(E) = 0$ otherwise.
For any $\theta \in \mathbb{R}_{> 0}$,
we will write $\Exp{\theta}$ for the CDF of an exponential distribution with rate $\theta$,
and for $a,b \in \mathbb{R}_{\geq 0}$ such that $a < b$,
we will write $\unif{a}{b}$ for the CDF of a uniform distribution.

We will use the convention that $\infty + x = \infty$ for $x \in \mathbb{R}_{\geq 0}$ and $\infty \cdot y = \infty$ for $y \in \mathbb{R}_{> 0}$.
A function $d \colon X \times X \rightarrow \mathbb{R}_{\geq 0} \cup \{\infty\}$ is called a
\emph{hemimetric} if it satisfies $d(x,x) = 0$ and the triangle inequality $d(x,z) \leq d(x,y) + d(y,z)$.
It is called a \emph{pseudometric} if it is also symmetric, i.e. $d(x,y) = d(y,x)$,
and it is called a \emph{metric} if it is symmetric and furthermore $d(x,y) = 0$ if and only if $x = y$.

%% Semi-Markov decision processes
\section{Semi-Markov Decision Processes}
In this section, we introduce semi-Markov decision processes,
which are continu\-ous-time reactive probabilistic systems.
A semi-Markov decision process has residence time on states governed 
by generic distributions on the positive real line and reacts to inputs from an external environment by making a probabilistic transition to a next state.

Hereafter, we consider a non-empty finite set of \emph{input actions} $A$,
and a non-empty, finite set of \emph{atomic propositions} $\ap$.
\begin{definition}\label{def:smdp}
  A \emph{semi-Markov decision process} (SMDP) is given by a tuple $M~=~(S, \tau, \rho, L)$ where
  \begin{itemize}
    \item $S$ is a non-empty, countable set of \emph{states},
    \item $\tau :  S \times A \rightarrow \mathcal{D}(S)$ is the \emph{transition function},
    \item $\rho : S \rightarrow \mathcal{D}(\mathbb{R}_{\geq 0})$ is the \emph{residence-time function}, and
    \item $L : S \rightarrow 2^{\ap}$ is the \emph{labelling function}. \qedhere
  \end{itemize}
\end{definition}

The operational behaviour of an SMDP is as follows. The SMDP at a given state $s \in S$, 
after receiving an input $a \in A$, goes to state $s' \in S$ within time $t$ 
with probability $\tau(s,a)(s') \cdot \rho(s)([0,t])$.
An SMDP is said to be \emph{finite} if it has a finite set of states,
and it is said to be \emph{finitely supported} if its transition function $\tau(s,a)$
is finitely supported for every $s$ and $a$.
For $s \in S$, we will write $F_s$ for the CDF of $\rho(s)$, i.e. $F_s(t) = \rho(s)([0,t])$.

Continuous-time Markov decision processes are a special case of SMDPs in which
all residence-time functions are exponentially distributed,
whereas discrete-time Markov decision processes are a special case of SMDPs where
the residence-time distribution in each state is the Dirac measure at $0$,
representing the fact that transitions are taken instantaneously.  

In defining simulation and bisimulation for SMDPs,
we will use ingredients from the definition of simulation
and bisimulation for Markov decision processes \cite{segala1995}
and simulation and bisimulation for continuous-time Markov chains \cite{BKHW05}.
However, since we are also generalising to arbitrary distributions on time
rather than just exponential distributions,
the condition on rates for exponential distributions must be replaced with
a more general condition on the distributions.
There is a rich literature on so-called stochastic orders \cite{shaked2007},
which impose an ordering on random variables.
We will consider here the most commonly used of these,
known as the \emph{usual stochastic order}.

\begin{definition}\label{def:simul}
  For an SMDP $M = (S, \tau, \rho, L)$,
  a relation $R \subseteq S \times S$ is a \emph{simulation}
  (resp.\ \emph{bisimulation}) on $M$ if for all $(s_1,s_2) \in R$ we have
  \begin{enumerate}
    \item $L(s_1) = L(s_2)$,
    \item $F_{s_2}(t) \geq F_{s_1}(t)$ (resp.\ $F_{s_2}(t) = F_{s_1}(t)$) for all 
    $t \in \mathbb{R}_{\geq 0}$, and
    \item for all $a \in A$ there exists a \emph{weight function} or \emph{coupling} $\Delta_a \colon S \times S \rightarrow [0,1]$ between $\tau(s_1,a)$ and $\tau(s_2,a)$ such that
    \begin{enumerate}
      \item $\Delta_a(s,s') > 0$ implies $(s,s') \in R$,
      \item $\tau(s_1,a)(s) = \sum_{s' \in S} \Delta_a(s,s')$ for all $s \in S$, and
      \item $\tau(s_2,a)(s') = \sum_{s \in S} \Delta_a(s,s')$ for all $s' \in S$.
    \end{enumerate}
  \end{enumerate}
  We say that $s_2$ \emph{simulates} (resp.\ is \emph{bisimilar to}) $s_1$, written
  $s_1~\simul~s_2$ (resp.\ $s_1 \sim s_2$), if there is a simulation (resp.\ bisimulation) relation
  containing $(s_1,s_2)$.
\end{definition}

It is easy to show that the \emph{similarity} relation $\simul$ is the largest simulation relation, 
and analogously that the \emph{bisimilarity} relation $\sim$ is the largest bisimulation relation.
The coupling ensures that the simulation relation is preserved by successor states.
Intuitively, $s_1$ simulates $s_2$ if the CDF of $\rho(s_2)$ is pointwise
greater than or equal to the CDF of $\rho(s_1)$, and the transition probability distribution 
of $s_1$ can be matched by the transition probability function $s_2$ by means of a coupling,
in such a way that if two successor states $s_1'$ and $s_2'$ have a non-zero coupling,
then $s_1'$ again simulates $s_2'$. For bisimulation, we instead require that the CDFs 
behave exactly the same in each step.

Given a set $C \subseteq S$ and a relation $R \subseteq S \times S$, let
\[R(C) = \{s' \in S \mid (s, s') \in R \text{ for some } s \in C\}\]
be the $R$-closure of $C$.
If $R$ is a preorder, $R(C)$ is the upward closure of $C$.

The following result, which is a trivial generalisation of \cite[Lemma 1]{SZ12},
gives a different but equivalent definition of simulation which is sometimes useful.

\begin{proposition}\label{prop:folk2}
  $R \subseteq S \times S$ is a simulation relation if and only if
  for any $(s_1,s_2) \in R$ we have
  \begin{enumerate}
    \item $L(s_1) = L(s_2)$,
    \item $F_{s_2}(t) \geq F_{s_1}(t)$ for all $t \in \mathbb{R}_{\geq 0}$, and
    \item $\tau(s_1,a)(C) \leq \tau(s_2,a)(R(C))$ for all $C \subseteq S$.
  \end{enumerate}
\end{proposition}

The following generalises \cite[Proposition 25(3)]{BKHW05} to the case of SMDPs.

\begin{proposition}\label{prop:folk3}
  ${\simul} \cap {\simul^{-1}} = {\sim}$.
\end{proposition}
\begin{proof}%[of Proposition \ref{prop:folk3}]
  First assume that $s_1 \sim s_2$.
  Then $L(s_1) = L(s_2)$.
  Also, $F_{s_2}(t) = F_{s_2}(t)$ for all $t$,
  so clearly $F_{s_2}(t) \geq F_{s_1}(t)$ and $F_{s_1}(t) \geq F_{s_2}(t)$ for all $t$.
  For any subset $C \subseteq S$ we have
  \[\tau(s_1,a)(C) \leq \tau(s_1,a)(\mathord{\sim}(C)) = \tau(s_2,a)(\mathord{\sim}(C))\]
  and
  \[\tau(s_2,a)(C) \leq \tau(s_2,a)(\mathord{\sim}(C)) = \tau(s_1,a)(\mathord{\sim}(C)) \enspace,\]
  so $s_1 \simul s_2$ and $s_1 \simul^{-1} s_2$.
  
  Now assume that $s_1 \simul s_2$ and $s_1 \simul^{-1} s_2$.
  Clearly $L(s_1) = L(s_2)$.
  Since $F_{s_2}(t) \geq F_{s_1}(t)$ and $F_{s_1}(t) \geq F_{s_2}(t)$ for all $t$,
  it follows that $F_{s_2}(t) = F_{s_1}(t)$ for all $t$.
  Now take an arbitrary subset $C \subseteq S$
  and let $B = \mathord{\simul \cap \simul^{-1}}(C)$, $C_1 = \mathord{\simul}(B)$,
  and $C_2 = C_1 \setminus B$. Then
  \[\tau(s_1,a)(C_1) = \tau(s_1,a)(\mathord{\simul}(C_1)) \leq \tau(s_2,a)(\mathord{\simul}(C_1)) = \tau(s_2,a)(C_1)\]
  and
  \[\tau(s_2,a)(C_1) = \tau(s_2,a)(\mathord{\simul}(C_1)) \leq \tau(s_1,a)(\mathord{\simul}(C_1)) = \tau(s_1,a)(C_1) \enspace,\]
  so $\tau(s_1,a)(C_1) = \tau(s_2,a)(C_1)$,
  and likewise we can show that $\tau(s_1,a)(C_2) = \tau(s_2,a)(C_2)$.
  Since we can write
  \[\tau(s_1,a)(C_1) = \tau(s_1,a)(B) + \tau(s_1,a)(C_2)\]
  and
  \[\tau(s_2,a)(C_1) = \tau(s_2,a)(B) + \tau(s_2,a)(C_2) \enspace,\]
  this together implies that $\tau(s_1,a)(B) = \tau(s_2,a)(B)$.
  Hence we conclude that $s_1 \sim s_2$.
\end{proof}

The above is analogous to a result stating that bisimulation and simulation equivalence
coincide for deterministic labelled transition systems \cite{baier2000}.
In our case, Proposition~\ref{prop:folk3}~holds because
reactive systems are inherently deterministic.

%% Functions
\section{Comparing the Speed of Residence-Time Distributions}\label{sec:func}
For comparing the random variables describing the residence time on states,
the similarity relation uses the usual stochastic order:
if $s_1 \simul s_2$ then, for all $t \in \mathbb{R}_{\geq 0}$,
$F_{s_1}(t) \leq F_{s_2}(t)$.
In words, if $s_2$ simulates $s_1$,
it is more likely that $s_2$ will take a transition before $s_1$,
that is, $s_2$ is stochastically faster than $s_1$ in reacting to an input. 

In this section, we propose a different way of comparing residence-time distributions. 
The idea is to get quantitative information about how much a distribution should 
be accelerated to become at least as fast as another distribution.

\begin{definition}\label{def:ft}
  Let $F$ and $G$ be CDFs and $\varepsilon \in \mathbb{R}_{> 0}$.
  We say that $F$ is \emph{$\varepsilon$-faster than} $G$,
  written $F \fft_\varepsilon G$,
  if for all $t$ we have $F(\varepsilon \cdot t) \geq G(t)$.
\end{definition}

Consider two states $s_1$ and $s_2$,
having residence time governed by the distributions $F_{s_1}$ and $F_{s_2}$, respectively,
and assume that $F_{s_1} \fft_\varepsilon F_{s_2}$ holds. 
If $0 < \varepsilon \leq 1$, then this means that
$s_1$ is stochastically faster than $s_2$, even if the residence time in
$s_1$ is slowed down by a factor $\varepsilon$.
If instead we have $\varepsilon > 1$,
then $s_1$ is stochastically slower than $s_2$,
but if we accelerate its residence-time distribution by a factor $\varepsilon$,
then it becomes stochastically faster than $s_2$. 

In the rest of the section we will argue that $\fft_\varepsilon$
is a good notion for gathering quantitative 
information about the speed of residence-time distributions on states.
We will do this by comparing the most common distributions used in the literature
for modelling residence time on states on stochastic systems:
Dirac distributions, exponential distributions, and uniform distributions.

The Dirac measure at zero is the fastest measure,
in the following sense.

\begin{proposition}\label{prop:dirac}
  Let $F$ be any CDF. The following holds for any $\varepsilon \in \mathbb{R}_{> 0}$.
  \begin{enumerate}
    \item $\dirac{0} \fft_{\varepsilon} F$.
    \item If $F \neq \dirac{0}$, then $F \not \fft_{\varepsilon} \dirac{0}$.
  \end{enumerate}
\end{proposition}
\begin{proof}%[of Proposition \ref{prop:dirac}]
  The first point is clear, since $\dirac{0}(t) = 1 \geq F(t)$ for any $t$.
  
  For the second point, note that $\dirac{0}$ is the only CDF such that
  \[\dirac{0}(0) = 1,\]
  and hence $F(\varepsilon \cdot 0) < \dirac{0}(0)$ for any $\varepsilon \geq 1$.
\end{proof}

For comparing exponential distributions,
it is simple to show that it is enough to accelerate by the ratio between the two rates.
The same is true for uniform distributions,
except we also need to consider whether the uniform distributions start at $0$,
since if a uniform distribution starts at $0$,
then we can only hope to make another uniform distribution faster than it
if this other uniform distribution also starts at $0$. 
To prove this, we make use of the following two lemmas.

\begin{lemma}\label{lem:exptransform}
  For $\varepsilon \in \mathbb{R}_{> 0}$ it holds that
  $\Exp{\theta}(\varepsilon \cdot t) = \Exp{\varepsilon \cdot \theta}(t)$.
\end{lemma}
\begin{proof}
  \[\Exp{\theta}(\varepsilon \cdot t) = 1 - e^{-\theta \cdot (\varepsilon \cdot t)} = 1 - e^{-(\theta \cdot \varepsilon) \cdot t} = \Exp{\varepsilon \cdot \theta}(t) . \qedhere\]
\end{proof}

\begin{lemma}\label{lem:uniftransform}
  For $\varepsilon \in \mathbb{R}_{> 0}$ it holds that
  $\unif{a}{b}(\varepsilon \cdot t) = \unif{\frac{a}{\varepsilon}}{\frac{b}{\varepsilon}}(t)$.
\end{lemma}
\begin{proof}
  \[\frac{\varepsilon \cdot t - a}{b - a} = 0 \implies \varepsilon \cdot t - a = 0 \implies t = \frac{a}{\varepsilon}\]
  and
  \[\frac{\varepsilon \cdot t - a}{b - a} = 1 \implies \varepsilon \cdot t - a = b - a \implies t = \frac{b}{\varepsilon} .\]
  Hence
  $\unif{a}{b}(\varepsilon \cdot t) = \unif{\frac{a}{\varepsilon}}{\frac{b}{\varepsilon}}(t)$.
\end{proof}

\begin{proposition}\label{prop:expandunif}
  \leavevmode
  \begin{enumerate}
    \item $\Exp{\theta_1} \fft_{\varepsilon} \Exp{\theta_2}$,
      where $\varepsilon = \frac{\theta_2}{\theta_1}$.
    \item If $c = 0$ and $a > 0$, then $\unif{a}{b} \not\fft_{\varepsilon} \unif{c}{d}$
      for any $\varepsilon \in \mathbb{R}_{> 0}$.
    \item If $c = 0$ and $a = 0$, then $\unif{a}{b} \fft_{\varepsilon} \unif{c}{d}$,
      where $\varepsilon = \frac{b}{d}$.
    \item If $c > 0$, then $\unif{a}{b} \fft_{\varepsilon} \unif{c}{d}$,
      where $\varepsilon = \max\left\{\frac{a}{c},\frac{b}{d}\right\}$.
  \end{enumerate}
  In all cases, the given $\varepsilon$ is the least such that
  the $\varepsilon$-faster than relation holds.
\end{proposition}
\begin{proof}%[of Proposition \ref{prop:expandunif}]
  \begin{enumerate}
    \item We see that
      \[\Exp{\theta_1}(\varepsilon \cdot t) = 1 - e^{-\theta_1 \varepsilon \cdot t} = 1 - e^{- \theta_2 t} = \Exp{\theta_2}(t)  .\]
      If $\varepsilon' < \varepsilon$, then take some $t > 0$ to get
      \[\Exp{\theta_1}(\varepsilon' \cdot t) = 1 - e^{-\theta_1 \varepsilon' \cdot t} < 1 - e^{-\theta_1 \varepsilon \cdot t} = \Exp{\theta_2}(t) ,\]
      and hence $\Exp{\theta_1} \not\fft_{\varepsilon'} \Exp{\theta_2}$.
    \item Let $c = 0$ and $a > 0$.
      Take an arbitrary $\varepsilon \in \mathbb{R}_{> 0}$
      and let $t = \frac{a}{\varepsilon} > 0$ in order to get
      $\unif{a}{b}(\varepsilon \cdot t) = \unif{a}{b}(a) = 0$.
      However, $\unif{c}{d}(t) > 0$ for any $t > 0$,
      so $\unif{a}{b}(\varepsilon \cdot t) < \unif{c}{d}(t)$.

    \item Let $c = 0$ and $a = 0$,
      and take $\varepsilon = \frac{b}{d}$.
      Then
      \[\unif{a}{b}(\varepsilon \cdot t) = \unif{a \cdot \frac{d}{b}}{d}(t) = \unif{0}{d}(t) = \unif{c}{d}(t) .\]
      To show that it is the least $\varepsilon$ such that the $\varepsilon$-faster-than
      relation holds, let $\varepsilon' < \frac{b}{d}$.
      Then
      \[\unif{a}{b}(\varepsilon' \cdot d) < \unif{a}{b}\left(\frac{b}{d} \cdot d\right) = 1 = \unif{c}{d}(d) .\]

    \item Now let $c > 0$ and $\varepsilon = \max\{\frac{a}{c},\frac{b}{d}\}$.
      If $\max\{\frac{a}{c},\frac{b}{d}\} = \frac{a}{c}$,
      then
      \[\unif{a}{b}(\varepsilon \cdot t) = \unif{c}{b \cdot \frac{c}{a}}(t).\]
      Since $\frac{a}{c} \geq \frac{b}{d}$ we get $\frac{c}{a} \leq \frac{d}{b}$,
      and hence
      \[\unif{c}{b \cdot \frac{c}{a}}(t) \geq \unif{c}{b \cdot \frac{d}{b}}(t) = \unif{c}{d}(t) .\]
      On the other hand, if $\max\{\frac{a}{c},\frac{b}{d}\} = \frac{b}{d}$,
      then we get $\frac{c}{a} \geq \frac{d}{b}$, and hence
      \[\unif{a}{b}(\varepsilon \cdot t) = \unif{a \cdot \frac{d}{b}}{d}(t) \geq \unif{c}{d}(t) .\]
      It remains to prove that this is the least $\varepsilon$ such that this relation holds.
      Let $\varepsilon' < \max\{\frac{a}{c},\frac{b}{d}\}$.
      If $\max\{\frac{a}{c},\frac{b}{d}\} = \frac{a}{c}$,
      then let $t = \frac{a}{\varepsilon'} > \frac{a}{\frac{a}{c}} = c$.
      Then $\unif{a}{b}(\varepsilon' \cdot t) = \unif{a}{b}(a) = 0$,
      but $\unif{c}{d}(t) > 0$ since $t > c$.
      On the other hand, if $\max\{\frac{a}{c},\frac{b}{d}\} = \frac{b}{d}$,
      then
      \[\unif{a}{b}(\varepsilon' \cdot d) < \unif{a}{b}\left(\frac{b}{d} \cdot d\right) = 1 = \unif{c}{d}(d) . \qedhere\]
  \end{enumerate}
\end{proof}

Moreover, an exponential distribution can never be made faster than a uniform distribution, 
since uniform distributions become $1$ eventually, but exponential distributions 
tend asymptotically to $1$ but never reach it.
Furthermore, whether or not a uniform distribution 
can be made faster than an exponential distribution depends on its value at $0$.

\begin{proposition}\label{prop:unifexp}
  \leavevmode
  \begin{enumerate}
    \item $\Exp{\theta} \not \fft_{\varepsilon} \unif{a}{b}$
      for all $\varepsilon \in \mathbb{R}_{> 0}$.
    \item If $a > 0$, then $\unif{a}{b} \not \fft_{\varepsilon} \Exp{\theta}$
      for all $\varepsilon \in \mathbb{R}_{> 0}$.
    \item If $a = 0$, then $\unif{a}{b} \fft_{\varepsilon} \Exp{\theta}$,
      where $\varepsilon = \theta \cdot b$.
      Furthermore, this is the least $\varepsilon$ such that the $\varepsilon$-faster-than relation holds.
  \end{enumerate}
\end{proposition}
\begin{proof}%[of Proposition \ref{prop:unifexp}]
  \begin{enumerate}
    \item We have $\unif{a}{b}(b) = 1$, but $\Exp{\theta}(t) < 1$ for all $t$,
      and hence $\Exp{\theta} \not \fft_{\varepsilon} \unif{a}{b}$
      for any $\varepsilon \in \mathbb{R}_{> 0}$.
    
    \item If $a > 0$, then let $\varepsilon \in \mathbb{R}_{> 0}$ be given,
      and let $t = \frac{a}{\varepsilon}$.
      Then $\unif{a}{b}(\varepsilon \cdot t) = \unif{a}{b}(a) = 0$,
      but $\Exp{\theta}(t) > 0$ since $t > 0$,
      and therefore $\unif{a}{b} \not \fft_{\varepsilon} \Exp{\theta}$.
    
    \item If $a = 0$, then let $\varepsilon = \theta \cdot b$.
      Clearly $\unif{a}{b}(\varepsilon \cdot 0) = 0$ and $\Exp{\theta}(0) = 0$.
      We see that $\unif{a}{b}(\varepsilon \cdot t) = \frac{\theta b t}{b}$
      and $\Exp{\theta}(t) = 1 - e^{- \theta t}$ have the same derivative from the right at $0$,
      namely $\theta$. Hence the slope of these two functions is the same in $0$,
      but since the slope of an exponential distribution is always decreasing,
      this means that $\unif{a}{b} \fft_{\varepsilon} \Exp{\theta}$.
      If $\varepsilon' < \theta \cdot b$,
      then the slope in $0$ of $\unif{a}{b}(\varepsilon' \cdot t)$
      must be less than that of $\Exp{\theta}$,
      so there will exist some $t > 0$ such that
      $\unif{a}{b}(\varepsilon' \cdot t) < \Exp{\theta}(t)$,
      and hence $\unif{a}{b} \not \fft_{\varepsilon'} \Exp{\theta}$. \qedhere
  \end{enumerate}
\end{proof}

The $\varepsilon$-faster-than relation enjoys a kind of monotonicity property,
which is simply a consequence of the fact that CDFs are increasing.

\begin{lemma}\label{lem:mono}
  Let $\varepsilon \leq \varepsilon'$ and assume that $F \fft_{\varepsilon} G$.
  Then $F \fft_{\varepsilon'} G$.
\end{lemma}
\begin{proof}%[of Lemma \ref{lem:mono}]
  $F \fft_{\varepsilon} G$ means that $F(\varepsilon \cdot t) \geq G(t)$ for all $t$.
  Since $F$ is non-decreasing and $\varepsilon \leq \varepsilon'$,
  this means that $F(\varepsilon' \cdot t) \geq F(\varepsilon \cdot t) \geq G(t)$ for all $t$,
  so $F \fft_{\varepsilon'} G$.
\end{proof}

The probability distribution of the sum of two independent random variables is the \emph{convolution}
of their individual distributions. The general formula for the convolution of two measures  
$\mu$ and $\nu$ on the real line is given by
\[(\mu * \nu)(E) = \int_0^\infty \nu(E - x) \;\mu(\wrt{x}). \]

Notably, the $\varepsilon$-faster-than relation is a congruence
with respect to convolution of measures.
\begin{proposition}\label{prop:conv}
  If $F_{\mu_1} \fft_{\varepsilon} F_{\mu_2}$ and $F_{\nu_1} \fft_\varepsilon F_{\nu_2}$,
  then $F_{(\mu_1 * \nu_1)} \fft_{\varepsilon} F_{(\mu_2 * \nu_2)}$.
\end{proposition}
\begin{proof}%[of Proposition \ref{prop:conv}]
  Define the transformation $T(x) = \varepsilon \cdot x$
  and let $\nu_1'([0,t]) = \nu_1\left(\left[0, \frac{t}{\varepsilon}\right]\right)$.
  Then we see that
  \begin{align*}
    \nu_1(T^{-1}([0,t]))
    &= \nu_1(\{x \mid x \cdot \varepsilon \in [0,t]\}) \\
    &= \nu_1\left(\left\{x \mid x \in \left[0, t/\varepsilon\right]\right\}\right) \\
    &= \nu_1\left(\left[0,\frac{t}{\varepsilon}\right]\right) \\
    &= \nu_1'([0,t]).
  \end{align*}
  Because $F_{\mu_1} \fft_\varepsilon F_{\mu_2}$
  we know that $\mu_1([0, \varepsilon \cdot t]) \geq \mu_2([0, t])$ for all $t$,
  and since $F_{\nu_1} \fft_\varepsilon F_{\nu_2}$,
  we know that $\nu_1([0, \varepsilon \cdot t) = \nu_1'([0,t]) \geq \nu_2([0,t])$
  for all $t$.
  We can therefore do following series of transformations \cite[Proposition 3.8]{panangaden2009}.
  \begin{align*}
    (\mu_1 * \nu_1)([0,\varepsilon \cdot t])
    &= \int_0^\infty \mu_1([0,\varepsilon \cdot t - x]) \; \nu_1(\wrt{x}) \\
    &= \int_0^\infty \mu_1([0,\varepsilon \cdot t - T(x)]) \; \nu_1'(\wrt{x}) \\
    &= \int_0^\infty \mu_1([0,\varepsilon (t - x)) \; \nu_1'(\wrt{x}) \\
    &\geq \int_0^\infty \mu_2([0, t - x]) \; \nu_1'(\wrt{x}) \\
    &\geq \int_0^\infty \mu_2([0, t - x]) \; \nu_2(\wrt{x}) \\
    &= (\mu_2 * \nu_2)([0, t]). \qedhere
  \end{align*}
\end{proof}

In Section~\ref{sec:linear} we will see that the above property
is essential for the preservation of reachability properties.
Intuitively, this is because convolution corresponds to
sequential composition of the residence-time behaviour.

There are other possible ways to compare the relative speed
of residence-time distributions quantitatively.
In the following we explore some alternative definitions
of the notion of the $\varepsilon$-faster-than relation,
and argue that none of them are preferable to the one given in Definition~\ref{def:ft}.
Given two CDFs $F$ and $G$, we consider the following three alternative definitions of
$F \fft_{\varepsilon} G$:
\begin{enumerate}
  \item \label{opt1}  for all $t$, $F(t) \cdot \varepsilon \geq G(t)$,
  \item \label{opt2}  for all $t$, $F(t) + \varepsilon \geq G(t)$, and
  \item \label{opt3}  for all $t$, $F(\varepsilon + t) \geq G(t)$.
\end{enumerate}

If $\fft_{\varepsilon}$ is defined as in \eqref{opt1}, then we see that $\unif{a}{b} \not\fft_\varepsilon \unif{c}{d}$, for any $\varepsilon \in \mathbb{R}_{> 0}$ whenever $c < a$. This is because 
$\unif{a}{b}(a) \cdot \varepsilon = 0 < \unif{c}{d}(a)$. Hence we lose the properties of 
Proposition \ref{prop:expandunif}.

If $\fft_{\varepsilon}$ is defined as in \eqref{opt2}, we trivially get that whenever 
$\varepsilon \geq 1$, $F \fft_\varepsilon G$, for any two CDFs $F$ and $G$.
Hence (2) is only interesting for $0 \leq \varepsilon < 1$.
However, even in this case we would still lose the properties of Proposition \ref{prop:expandunif}. 
Indeed, whenever $a \geq d$, $\unif{a}{b} \not\fft_\varepsilon \unif{c}{d}$, for any $0 \leq \varepsilon < 1$. This follows because $\unif{a}{b}(a) + \varepsilon = \varepsilon < 1 = \unif{c}{d}(a)$.

Lastly, if $\fft_{\varepsilon}$ is defined as in \eqref{opt3}, then it would not be a congruence
with respect to convolution of distributions, i.e., Proposition~\ref{prop:conv} would not hold. 
For a counterexample, take $F_{\mu_1} = \unif{2}{4}$, $F_{\mu_2} = \unif{1}{3}$,
$F_{\nu_1} = \unif{3}{4}$, and $F_{\nu_2} = \unif{2}{4}$.
Then $F_{\mu_1} \fft_1 F_{\mu_2}$ and $F_{\nu_1} \fft_1 F_{\nu_2}$,
but $F_{(\mu_1 * \mu_2)} \not\fft_1 F_{(\nu_1 * \nu_2)}$.

%% Simulation
\section{A Hemimetric for Semi-Markov Decision Processes}
In this section, we are going to extend the definition of simulation relation between SMDPs
to the quantitative setting.
We will see that this relation naturally induces a notion of distance between SMDPs,
describing the least acceleration factor required globally
on the residence-time distributions to make a given SMDP as fast as another one.

\begin{definition}\label{def:eps-sim}
  Let $\varepsilon \in \mathbb{R}_{> 0}$. For an SMDP $M = (S, \tau, \rho, L)$,
  a relation $R \subseteq S \times S$ is an \emph{$\varepsilon$-simulation relation}
  on $M$ if for all $(s_1,s_2) \in R$ we have
  that the first and third condition for simulation are satisfied,
  and $F_{s_2} \fft_{\varepsilon} F_{s_1}$.
  We say that \emph{$s_2$ $\varepsilon$-simulates $s_1$}, written $s_1 \simul_{\varepsilon} s_2$,
  if there is an $\varepsilon$-simulation relation $R$ such that $(s_1,s_2) \in R$.
\end{definition}

\begin{example}\label{ex:smdp}
  Let $A = \{a\}$ and consider the SMDP $M = (S, \tau, \rho, L)$ given by $S = \{s_1, s_2\}$,
  $\tau(s_1,a)(s_1) = 1 = \tau(s_2,a)(s_2)$,
  $F_{s_1} = \Exp{4}$, $F_{s_2} = \Exp{2}$, and $L(s_1) = L(s_2)$.
  By Proposition~\ref{prop:expandunif} we see that $s_1 \simul_{2} s_2$ and $s_2 \simul_{\frac{1}{2}} s_1$.
\end{example}

It is easy to show that the \emph{$\varepsilon$-similarity} relation $\simul_{\varepsilon}$ is the 
largest simulation relation, and with the previous section in mind, one immediately sees that
$\simul_{1}$ coincides with $\simul$. Moreover, the following holds.

\begin{proposition} \label{prop:lessthan1simul}
  For any $\varepsilon \leq 1$, if $s_1 \simul_{\varepsilon} s_2$, then $s_1 \simul s_2$. 
\end{proposition}
\begin{proof}%[of Proposition \ref{prop:lessthan1simul}]
  Let $R \subseteq S \times S$ be an $\varepsilon$-simulation relation
  such that $(s_1,s_2) \in R$.
  We will now argue that $R$ is also a simulation relation.
  The first condition is clear.
  For the second condition, we have
  \[F_{s_2}(t) \geq F_{s_2}(\varepsilon \cdot t) \geq F_{s_1}(t) .\]
  The third condition is satisfied because $R$ is an $\varepsilon$-simulation relation.
\end{proof}

If $\varepsilon > 1$, the above implication does not hold. For an easy counterexample
consider $s_1$ and $s_2$ from Example~\ref{ex:smdp} where $s_1 \simul_2 s_2$
but $s_1 \not\simul s_2$. 

For $\varepsilon > 1$, we can obtain a result similar to Proposition~\ref{prop:lessthan1simul} 
only if we ``accelerate'' the overall behaviour of $s_2$. Formally, for a given SMDP 
$M = (S, \tau, \rho, L)$, we define the SMDP 
$M_\varepsilon = (S_\varepsilon, \tau_\varepsilon, \rho_\varepsilon, L_\varepsilon)$ as follows:

\begin{align*}
  \begin{aligned}
    S_\varepsilon &= S \cup \{\pert{s}{\varepsilon} \mid  s \in S \} , \\
    L_\varepsilon(s) &= L(s) , \\
    L_\varepsilon(\pert{s}{\varepsilon}) &= L(s) , \\
    \rho_\varepsilon(s)([0, t]) &= \rho(s)([0, t])  , \\
    \rho_\varepsilon(\pert{s}{\varepsilon})([0, t]) &= \rho(s)([0, \varepsilon \cdot t])  ,
  \end{aligned}
  &&
  \begin{aligned}
    \tau_\varepsilon(s, a)(s') &= \tau(s,a)(s') , \\
    \tau_\varepsilon(s, a)(\pert{s'}{\varepsilon}) &= 0 , \\
    \tau_\varepsilon(\pert{s}{\varepsilon},a)(s') &= 0 , \\
    \tau_\varepsilon(\pert{s}{\varepsilon},a)(\pert{s'}{\varepsilon}) &= \tau(s,a)(s') .
  \end{aligned}
\end{align*} 

Intuitively, the states $s \in S$ in $M_\varepsilon$ are identical copies of those in $M$,
whereas the states $\pert{s}{\varepsilon}$ react to each input $a \in A$ functionally
identically to $s$ but faster, since the residence-time on the states
are all equally accelerated by a factor $\varepsilon$,
thus $\pert{s}{\varepsilon} \simul_{\varepsilon} s$. 
For this reason $\pert{s}{\varepsilon}$ is called the \emph{$\varepsilon$-acceleration of $s$}.

Given the definition of accelerated state, Proposition~\ref{prop:lessthan1simul} can be
generalised to arbitrary values of $\varepsilon \in \mathbb{R}_{> 0}$ in the following way.

\begin{proposition}\label{prop:simulpert}
  For any $\varepsilon \in \mathbb{R}_{> 0}$,
  $s_1 \simul_{\varepsilon} s_2$ if and only if $s_1 \simul \pert{s_2}{\varepsilon}$. 
\end{proposition}
\begin{proof}%[of Proposition \ref{prop:simulpert}]
  $(\implies)$ Let $R \subseteq S \times S$ be an $\varepsilon$-simulation relation
  with $(s_1, s_2) \in R$. Define
  \[R' = \{(s,\pert{s'}{\varepsilon} \in S_\varepsilon \times S_\varepsilon \mid (s,s') \in R\} ,\]
  and take an arbitrary $(s,\pert{s'}{\varepsilon}) \in R'$.
  The first condition of Definition \ref{def:eps-sim}
  is satisfied because $F_{\pert{s'}{\varepsilon}}(t) = F_{s'}(\varepsilon \cdot t) \geq F_s(t)$.
  
  For the second condition, we know that for any $a \in A$ there exists a coupling $\Delta_a$,
  and we now define
  \[\Delta_a'(s'',s''') =
  \begin{cases}
    0 & \text{if } s'' \notin S \text{ or } s''' \in S \\
    \Delta_a(s'',s''') & \text{otherwise.}
  \end{cases}\]
  Since
  \begin{align*}
    \Delta_a'(s'',\pert{s'''}{\varepsilon}) > 0
    &\implies \Delta_a(s'',s''') > 0 \\
    &\implies (s'',s''') \in R \\
    &\implies (s'', \pert{s'''}{\varepsilon}) \in R' ,
  \end{align*}
  condition (a) is also satisfied.
  For condition (b), first consider the case where $s'' \in S$.
  Then we get
  \begin{align*}
    \sum_{s''' \in S_\varepsilon} \Delta_a'(s'',s''')
    &= \sum_{\pert{s'''}{\varepsilon} \in S_\varepsilon} \Delta_a'(s'',\pert{s'''}{\varepsilon}) \\
    &= \sum_{s''' \in S} \Delta_a(s'',s''') \\
    &= \tau(s,a)(s'') \\
    &= \tau_\varepsilon(s,a)(s'') .
  \end{align*}
  For the case where $s'' \notin S$ we get
  \[\sum_{s''' \in S_\varepsilon} \Delta_a'(s'',s''') = 0\]
  and
  \[\tau_\varepsilon(s,a)(s'') = 0 .\]
  
  Likewise, for condition (c), first consider the case where $s''' \in S$.
  Then
  \[\sum_{s'' \in S_\varepsilon} \Delta_a'(s'',s''') = 0\]
  and
  \[\tau_\varepsilon(\pert{s'}{\varepsilon},a)(s''') = 0 .\]
  For the case where $s''' \notin S$, we get
  \begin{align*}
    \sum_{s'' \in S_\varepsilon} \Delta_a'(s'',s''')
    &= \sum_{s'' \in S} \Delta_a'(s'',s''') \\
    &= \sum_{s'' \in S} \Delta_a(s'',s''') \\
    &= \tau(s',a)(s''') \\
    &= \tau_\varepsilon(\pert{s'}{\varepsilon},a)(\pert{s'''}{\varepsilon}) .
  \end{align*}
  
  $(\impliedby)$ Let $R \subseteq S_\varepsilon \times S_\varepsilon$
  be a simulation relation with $(s_1, \pert{s_2}{\varepsilon}) \in R$
  and define
  \[R' = \{(s,s') \in S \times S \mid (s,\pert{s'}{\varepsilon}) \in R\} .\]
  For an arbitrary $(s,s') \in R'$ we get
  $F_{s'}(\varepsilon \cdot t) = F_{\pert{s'}{\varepsilon}} \geq F_{s_1}(t)$,
  thus satisfying the first condition.
  
  We know that for any $a \in A$ there exists a coupling $\Delta_a$,
  and we now define
  \[\Delta_a'(s'',s''') = \Delta_a(s'',\pert{s'''}{\varepsilon}) .\]
  This coupling satisfies condition (a) because
  \begin{align*}
    \Delta_a'(s'',s''') > 0
    &\implies \Delta_a(s'',\pert{s'''}{\varepsilon}) > 0 \\
    &\implies (s'',\pert{s'''}{\varepsilon}) \in R \\
    &\implies (s'',s''') \in R' .
  \end{align*}
  For condition (b), we see that
  \begin{align*}
    \sum_{s''' \in S} \Delta_a'(s'',s''')
    &= \sum_{\pert{s'''}{\varepsilon} \in S_\varepsilon} \Delta_a(s'',\pert{s'''}{\varepsilon}) \\
    &= \tau_\varepsilon(s,a)(s'') \\
    &= \tau(s,a)(s'') .
  \end{align*}
  Likewise, for condition (c) we have
  \begin{align*}
    \sum_{s'' \in S} \Delta_a'(s'',s''')
    &= \sum_{s'' \in S} \Delta_a(s'',\pert{s'''}{\varepsilon}) \\
    &= \tau_\varepsilon(\pert{s'}{\varepsilon},a)(\pert{s'''}{\varepsilon}) \\
    &= \tau(s',a)(s''') . \qedhere
  \end{align*}
\end{proof}

The relevance of the above statement is twofold:
it clarifies the relation between $\simul_{\varepsilon}$ and $\simul$,
and also provides a way to modify the behaviour 
of a state $s_2$ of an SMDP in order to simulate
a state $s_1$ whenever $s_1 \simul_{\varepsilon} s_2$ holds.

Having this characterisation of similarity in terms of acceleration of processes one can think 
about the following problem:
given two states, $s_1$ and $s_2$ such that $s_1 \not\simul s_2$, what 
is the least $\varepsilon \geq 1$ (if it exists) such that $s_1 \simul (s_2)_\varepsilon$ holds? 
We can answer this question by means of the following distance.

\begin{definition} \label{def:simuldist}
  The \emph{simulation distance} $\simdist \colon S \times S \to [1,\infty]$ 
  between two states $s_1$ and $s_2$ is given by
  \[\simdist(s_1,s_2) = \inf\{ \varepsilon \geq 1 \mid s_1 \simul_{\varepsilon} s_2\} . \qedhere\]
\end{definition}

As usual, if there is no $\varepsilon \geq 1$ such that $s_1 \simul_{\varepsilon} s_2$, then 
$\simdist(s_1,s_2) = \infty$, because $\inf \Emptyset = \infty$.
It is clear from the definition that $s_1 \simul s_2$ implies $\simdist(s_1,s_2) = 1$.
For finitely supported SMDPs, the converse is also true.
However, the proof of this makes use of the logical characterisation of the faster-than relation,
so we delay the proof until Section~\ref{sec:logic-d} where we discuss logical properties.

Note that the definition above does not give a distance in the usual sense,
for two reasons: $d$ is not symmetric and it does not satisfy the triangle inequality. 
One can show instead that $d$ satisfies a multiplicative version of the triangle inequality,
namely, that for all $s_1,s_2,s_3 \in S$, 
$\simdist(s_1,s_3) \leq \simdist(s_1,s_2) \cdot \simdist(s_2,s_3)$.
This is a direct consequence of the following properties of $\simul_\varepsilon$.
The first property states that $\simul_\varepsilon$ is monotonic with respect to 
increasing values of $\varepsilon$.

\begin{lemma}\label{lem:simul-mono}
  If $s_1 \simul_{\varepsilon} s_2$ and $\varepsilon \leq \varepsilon'$, then $s_1 \simul_{\varepsilon'} s_2$.
\end{lemma}
\begin{proof}%[of Lemma \ref{lem:simul-mono}]
  This follows from Lemma \ref{lem:mono}.
\end{proof}

The second property is a quantitative generalisation
of transitivity from which the multiplicative inequality
discussed above follows.

\begin{lemma}\label{lem:triangle}
  If $s_1 \simul_{\varepsilon} s_2$ and $s_2 \simul_{\varepsilon'} s_3$, then $s_1 \simul_{\varepsilon \cdot \varepsilon'} s_3$.
\end{lemma}
\begin{proof}%[of Lemma \ref{lem:triangle}]
  Since $s_1 \simul_\varepsilon s_2$ and $s_2 \simul_{\varepsilon'} s_3$,
  there exists an $\varepsilon$-simulation relation $R$ such that $(s_1,s_2) \in R$
  and an $\varepsilon'$-simulation relation $R'$ such that $(s_2,s_3) \in R'$.
  First construct a relation
  \[R'' = R \circ R' = \{(s_1',s_3') \in S \times S \mid \exists s_2' .( (s_1',s_2') \in R \text{ and } (s_2',s_3') \in R' )\}  .\]
  
  Now pick an arbitrary pair $(s_1',s_3') \in R''$.
  Clearly there exists $s_2'$ such that $(s_1', s_2') \in R$ and $(s_2',s_3') \in R'$.
  Hence $L(s_1') = L(s_2') = L(s_3')$ and $F_{s_2'}(\varepsilon \cdot t) \geq F_{s_1'}(t)$
  and $F_{s_3'}(\varepsilon' \cdot \varepsilon \cdot t) \geq F_{s_2'}(\varepsilon \cdot t)$,
  so $F_{s_3'}(\varepsilon' \cdot \varepsilon \cdot t) \geq F_{s_1'}(t)$,
  meaning $F_{s_3'} \fft_{\varepsilon \cdot \varepsilon'} F_{s_1'}$.
  Thus the first and second conditions are satisfied.
  
  Now let $a \in A$. There exists a coupling $\Delta_a$ between $\tau(s_1',a)$ and $\tau(s_2',a)$
  and another coupling $\Delta_a'$ between $\tau(s_2',a)$ and $\tau(s_3',a)$.
  Next we construct a coupling between $\tau(s_1',a)$ and $\tau(s_3',a)$ by
  \begin{equation}\label{eq:coupling}
    \Delta_a''(s,s'') = \sum_{s' \in \supp{\tau(s_2',a)}} \frac{\Delta_a(s,s') \cdot \Delta_a'(s',s'')}{\tau(s_2',a)(s')}  .
  \end{equation}
%  One can easily verify that this is a coupling.
  We first verify that this is a coupling.
  \begin{align*}
    \sum_{s'' \in S}\Delta_a''(s,s'') &= \sum_{s'' \in S}\sum_{s' \in \supp{\tau(s_2,a)}} \frac{\Delta_a (s,s') \cdot \Delta_a' (s',s'')}{\tau(s_2,a)(s')} \\
                                      &= \sum_{s' \in \supp{\tau(s_2,a)}} \frac{\Delta_a (s,s') \cdot \sum_{s'' \in S} \Delta_a' (s',s'')}{\tau(s_2,a)(s')} \\
                                      &= \sum_{s' \in \supp{\tau(s_2,a)}} \frac{\Delta_a (s,s') \cdot \tau(s_2,a)(s')}{\tau(s_2,a)(s')} \\
                                      &= \sum_{s' \in \supp{\tau(s_2,a)}} \Delta_a (s,s') \\
                                      &= \tau(s_1,a)(s) ,
  \end{align*}
  and likewise we can show that $\sum_{s \in S}\Delta_a''(s,s'') = \tau(s_3,a)(s'')$.
%  \begin{align*}
%    \sum_{s \in S}\Delta_a''(s,s'')
%                                    &= \sum_{s \in S}\sum_{s' \in \supp{\tau(s_2,a)}} \frac{\Delta_a (s,s') \cdot \Delta_a' (s',s'')}{\tau(s_2,a)(s')} \\
%                                    &= \sum_{s' \in \supp{\tau(s_2,a)}} \frac{\sum_{s \in S} \Delta_a (s,s') \cdot \Delta_a' (s',s'')}{\tau(s_2,a)(s')} \\
%                                    &= \sum_{s' \in \supp{\tau(s_2,a)}} \frac{\tau(s_2,a)(s') \cdot \Delta_a' (s',s'')}{\tau(s_2,a)(s')} \\
%                                    &= \sum_{s' \in \supp{\tau(s_2,a)}} \Delta_a' (s',s'') \\
%                                    &= \tau(s_3,a)(s'').
%  \end{align*}
  Now assume $\Delta_a''(s,s'') > 0$.
  By \eqref{eq:coupling}, this means that there must exist some
  \[s' \in \supp{\tau(s_2',a)} \quad \text{such that} \quad \Delta_a(s,s') > 0 \text{ and } \Delta_a'(s',s'') > 0.\]
  This implies that $(s,s') \in R$ and $(s',s'') \in R'$,
  so by the construction of $R''$ we get $(s,s'') \in R''$.
  
  Hence we have shown that $R''$ is an $\varepsilon \cdot \varepsilon'$-simulation relation.
  Since clearly $(s_1,s_3) \in R''$, it follows that $s_1 \simul_{\varepsilon \cdot \varepsilon'} s_3$.
\end{proof}

Typically, one still uses the term distance for such multiplicative distances,
because by applying the logarithm one does obtain a hemimetric.

\begin{theorem}\label{thm:metric}
  $\log \simdist$ is a hemimetric.
\end{theorem}
\begin{proof}%[of Theorem \ref{thm:metric}]
  Let $\lsimdist(s_1,s_2) = \log \simdist(s_1,s_2)$.
  Clearly, $\lsimdist(s_1,s_2) \geq 0$, and since $\simdist(s,s) = 1$, $\lsimdist(s,s) = \log(1) = 0$.
  Hence it only remains to verify the triangle inequality.
  
  If $\simdist(s_1,s_2) \cdot \simdist(s_2,s_3) = \infty$,
  then clearly $\simdist(s_1,s_3) \leq \simdist(s_1,s_2) \cdot \simdist(s_2,s_3)$.
  If $\simdist(s_1,s_2) \cdot \simdist(s_2,s_3) \neq \infty$,
  then the sets $\{\varepsilon \geq 1 \mid s_1 \simul_{\varepsilon} s_2\}$ and $\{\varepsilon' \geq 1 \mid s_2 \simul_{\varepsilon'} s_3\}$
  are both non-empty, so there must exist $\varepsilon, \varepsilon' \geq 1$
  such that $s_1 \simul_{\varepsilon} s_2$ and $s_2 \simul_{\varepsilon'} s_3$,
  so by Lemma \ref{lem:triangle} we have $s_1 \simul_{\varepsilon \cdot \varepsilon'} s_3$,
  and hence $\simdist(s_1,s_3) \neq \infty$.
  Taking the contrapositive of this,
  we get that $\simdist(s_1,s_3) = \infty$ implies that $\simdist(s_1,s_2) \cdot \simdist(s_2,s_3) = \infty$,
  and hence also $\simdist(s_1,s_3) \leq \simdist(s_1,s_2) \cdot \simdist(s_2,s_3)$.
  
  Now assume that $\simdist(s_1,s_3) \neq \infty$ and $\simdist(s_1,s_2) \cdot \simdist(s_2,s_3) \neq \infty$,
  and assume towards a contradiction that $\simdist(s_1,s_3) > \simdist(s_1,s_2) \cdot \simdist(s_2,s_3)$.
  Since $\simdist$ is defined as an infimum, there must exist $\varepsilon, \varepsilon' \geq 1$ such that
  $s_1 \simul_{\varepsilon} s_2$, $s_2 \simul_{\varepsilon'} s_3$, and
  \[\simdist(s_1,s_3) > \varepsilon \cdot \varepsilon' \geq \simdist(s_1,s_2) \cdot \simdist(s_2,s_3)  .\]
  However, by Lemma \ref{lem:triangle} we have $\varepsilon \cdot \varepsilon' \geq \simdist(s_1,s_3)$, a contradiction.
  Hence we get $\simdist(s_1,s_3) \leq \simdist(s_1,s_2) \cdot \simdist(s_2,s_3)$, and by taking logarithms,
  we get
  \[\lsimdist(s_1,s_3) \leq \lsimdist(s_1,s_2) + \lsimdist(s_2,s_3). \qedhere\]
\end{proof}

\begin{example}
  Consider again the SMDP from Example~\ref{ex:smdp}.
  We can now see that $\simdist(s_1,s_2) = 2$ and $\simdist(s_2,s_1) = \frac{1}{2}$.
  This also shows that our distance is not symmetric, and hence not a pseudometric.
\end{example}

\begin{remark}
  From a topological point of view, there is no real difference between 
  satisfying the standard triangle inequality or its multiplicative version.
  The difference essentially amounts to working either in the monoid $(\mathbb{R}_{\geq 0},+)$
  or the monoid $(\mathbb{R}_{\geq 1}, \cdot)$.
  However, these monoids are isomorphic
  via the bijection given by the logarithm and exponential functions. Since these 
  functions are also continuous, the isomorphism is actually a homeomorphism,
  so all topological properties are preserved.
\end{remark}

If one allowed $\varepsilon < 1$ in the distance,
then one would get strange results such as a constant sequence
which does not convergence to the constant element.
To see this, consider a sequence $\{s_n\}_{n \in \mathbb{N}}$,
where $s_n = s$ for any $n \in \mathbb{N}$.
Then $\simdist(s_n,s) = 1$ for any element of the sequence.
However, $\lim_{k \rightarrow \infty} s_k \ni s$
means that for any $\varepsilon > 0$,
there exists an $N \in \mathbb{N}$ such that $\simdist(s_n,s) < \varepsilon$
for all $n > N$.
However, this clearly does not hold for $0 < \varepsilon < 1$,
so $\lim_{k \rightarrow \infty} s_k \not\ni s$.

%% Algorithm
\section{Computing the Simulation Distance}\label{sec:compute}
In this section we provide an algorithm to compute
the simulation distance given in Definition~\ref{def:simuldist} for finite SMDPs.
The algorithm is shown to run in polynomial time for the distributions we have considered so far.

The following technical lemma will provide a sound basis for the correctness of the
algorithm. Given two CDFs $F$ and $G$, let
\begin{equation*}
  c(F,G)= \inf\{\varepsilon \geq 1 \mid F \fft_{\varepsilon} G\} 
\end{equation*}
denote the least acceleration factor needed by $F$ to be faster than $G$.
Given an SMDP $M$, we then define
\[\mathcal{C}(M) = \{c(F_{s'}, F_s) \mid s,s' \in S\}.\]

\begin{lemma}\label{lem:correctness}
  Let $M$ be a finite SMDP.
  If $\simdist(s_1,s_2) \neq \infty$, then
  \begin{itemize}
    \item $s_1 \simul_c s_2$, for some $c \in \mathcal{C}(M)$ and
    \item $\simdist(s_1,s_2) = \min\{c \in \mathcal{C}(M) \mid s_1 \simul_c s_2\}$.
  \end{itemize}
\end{lemma}
\begin{proof}%[of Lemma \ref{lem:correctness}]
  For the first claim, note that $\simdist(s_1,s_2) \neq \infty$
  implies that $s_1 \simul_{\varepsilon} s_2$ for some $\varepsilon \geq 1$.
  This is witnessed by some $\varepsilon$-simulation relation
  which we denote by $R$.
  Now let
  \[c^* = \max\{c \in \mathcal{C}(M) \mid c = c_{\rho(s'),\rho(s)} \text{ for some } (s,s') \in R \} .\]
  Then it is clear that $R$ is also a $c^*$-simulation relation,
  and hence $s_1 \simul_{c^*} s_2$.
  
  For the second claim, let
  \[c_* = \min\{c \in \mathcal{C}(M) \mid s_1 \simul_c s_2\}\]
  and
  \[X = \{\varepsilon \geq 1 \mid s_1 \simul_\varepsilon s_2\} .\]
  We first show that $c_*$ is a lower bound of $X$.
  If $s_1 \simul_\varepsilon s_2$,
  then by the previous argument we also have $s_1 \simul_{c^*} s_2$.
  Note that $\varepsilon \geq c^*$,
  because otherwise we would have had
  $F_{s'} \not \fft_{\varepsilon} F_s$ for some $(s,s') \in R$,
  contradicting the fact that $R$ is a $\varepsilon$-simulation relation.
  Hence
  \[\varepsilon \geq c^* \geq \min\{c \in \mathcal{C}(M) \mid s_1 \simul_c s_2\} = c_* ,\]
  so $c_*$ is a lower bound of $X$.
  Next we show that $c_*$ is the greatest lower bound of $X$.
  We know that $s_1 \simul_{c_*} s_2$, and hence $c_* \in X$,
  so if $\varepsilon > c_*$, then $\varepsilon$ can not be a lower bound of $X$.
  Hence $c_*$ is the greatest lower bound of $X$,
  and therefore we conclude that
  \[c_* = \min\{c \in \mathcal{C}(M) \mid s_1 \simul_c s_2\} = \inf X = \simdist(s_1,s_2) . \qedhere\]
\end{proof}

Lemma \ref{lem:correctness} provides a strategy for computing the simulation distance between any
two states $s_1$ and $s_2$ of a given SMDP $M$ as follows.
First, one constructs the set $\mathcal{C}(M)$.
If $s_1 \simul_c s_2$ does not hold for any $c \in \mathcal{C}(M)$,
then the distance must be infinite;
otherwise, it is the smallest $c \in \mathcal{C}(M)$ for which $s_1 \simul_c s_2$ holds.

In order for this strategy to work, we need two ingredients:
first, we should be able to compute the set $\mathcal{C}(M)$ and second,
for any $c \in \mathcal{C}(M)$, we need an algorithm for checking whether $s_1 \simul_c s_2$.

Recall that SMDPs allow for \emph{arbitrary} residence-time distributions in each state. 
Therefore, it is not guaranteed that for any SMDP $M$ the set $\mathcal{C}(M)$ can be computed.
With the following definition we identify the class of SMDPs for which this can be done.

\begin{definition}
  A class $\mathcal{C}$ of CDFs
  is \emph{effective} if for any $F,G \in \mathcal{C}$,
  $c(F,G)$ is computable.
  An SMDP $M$ is effective if $\{F_s \mid s \in S \}$ is an effective class.
\end{definition}

In particular, for any pair of states $s,s'$ of an effective SMDP $M$,
we can decide whether $F_{s'} \fft_{\varepsilon} F_s$ by simply checking whether
$\varepsilon \geq c(F_{s'},F_{s})$.
We will denote by $f(l)$ the complexity of computing $c(F_{s'},F_{s})$
for two arbitrary $s,s' \in S$ as a function of the length $l$
of the representation of the residence-time distributions of $M$.

Let $\due$ denote the class consisting of the Dirac distribution at $0$
as well as uniform and exponential distributions with rational parameters.
By Propositions~\ref{prop:dirac}--\ref{prop:unifexp}
we immediately see that $\due$ is an effective class,
and in fact it can be computed using only simple operations such as
multiplication, division, and taking maximum.
Hence $f(l)$ has constant complexity\footnote{As is standard, we consider numbers to be represented as floating points of bounded size in their binary representation.} whenever $M$
takes residence-time distributions from $\due$.

Next we consider how to decide $s_1 \simul_\varepsilon s_2$ for a given rational 
$\varepsilon \geq 1$.
A decision procedure can be obtained by adapting to our setting 
the algorithm by Baier et al.~\cite{baier2000} for deciding the simulation preorder 
between probabilistic labelled transition systems.
The algorithm from~\cite{baier2000} uses a partition refinement approach
to compute the largest simulation relation and runs in time $\bigo(mn^7 / \log n)$ for reactive systems,
where $m = |A|$ is the number of actions, and $n = |S|$ is the number of states.
Given $\varepsilon \geq 1$, we can proceed correspondingly to compute $\varepsilon$-similarity:
we start from the relation $R = S \times S$ and update it
by removing all the pairs $(s,s')$ of states
not satisfying the conditions of Definition~\ref{def:eps-sim}.
This process is repeated on the resulting updated relation until
no more pairs of states are removed.
The resulting relation is the largest $\varepsilon$-simulation.
Hence, checking $s_1 \simul_{\varepsilon} s_2$ corresponds to determining whether the pair 
$(s_1,s_2)$ is contained in the relation returned by the above algorithm.
The complexity can be improved to $\bigo(m^2n)$ by storing important information about
the previous iteration of the algorithm and use this in the current iteration \cite[Theorem 5.2.5]{zhang09}.

\begin{theorem}\label{thm:decidable-d}
  Let $M$ be a finite and effective SMDP.
  Given $s_1,s_2 \in S$ and $\varepsilon \geq 1$, deciding whether 
  $s_1 \simul_{\varepsilon} s_2$ can be done in time $\bigo(n^2(f(l) + k) + m^2n)$,
  where $k = |AP|$ is the number of atomic propositions.
\end{theorem}
\begin{proof}%[of Theorem \ref{thm:decidable-d}]
  The algorithm for deciding $s_1 \simul_{\varepsilon} s_2$
  is essentially the same as that for deciding untimed simulation.
  Since we have assumed effectiveness,
  when choosing whether to remove a pair $(s,s')$ from the current relation,
  we can check the conditions on Definition \ref{def:eps-sim}.
  However, we also need to check whether $L(s) = L(s')$.
  For this we assume that the set of atomic propositions $\ap$
  have an ordering $\ap = x_0, x_1, \dots$,
  and that $L(s)$ is represented as a binary array
  where the $i$th entry in the array is $1$ if $x_i \in L(s)$,
  and $0$ otherwise.
  Then checking whether $L(s) = L(s')$
  amounts to checking whether each array has the same entries,
  which can be done in time $k = |\ap|$.
  Testing for the existence of a coupling
  can be done in time $\bigo(m^2n)$ by using the algorithm from \cite{zhang09},
  Hence we get a time complexity of $\bigo(n^2(f(l) + k) + m^2n)$.
\end{proof}

The algorithm for computing the simulation distance
is given in Algorithm~\ref{alg:distance}.
The algorithm starts by ordering the elements of $\mathcal{C}(M)$ as 
$c_1, \dots, c_n$ while removing $\infty$ from the list.
Then it searches for the smallest $c_i$ such that $s_1 \simul_{c_i} s_2$ holds.
This is done by means of a bisection method.
If $s_1 \simul_{c_1} s_2$ holds, then $c_1$ is the smallest element
such that this holds, so we return it.
If $s_1 \simul_{c_n} s_2$ does not hold, then, by Lemma~\ref{lem:simul-mono},
$s_1 \simul_{c_i} s_2$ does not hold for any $1 \leq i \leq n$, so we return $\infty$.
If none of the above apply, at this point of the algorithm (line 4) we have that 
$s_1 \not\simul_{c_1} s_2$ and $s_1 \simul_{c_n} s_2$.

We use the variables $i$ and $j$, respectively, as the left and right endpoints of the bisection interval.
The bisection interval keeps track of those elements $c_n$ for which
we still do not know whether $s_1 \simul_{c_n} s_2$.
At the beginning, $i = 1$ and $j=n$. 
At line 7, $h = \left\lceil\frac{j-i}{2}\right\rceil$ is used as
the decrement factor for the length of the bisection interval at each step.
Since $h > 0$, the bisection interval decreases in size for each iteration.
If $s_1 \simul_{c_{j-h}} s_2$ holds, then
$j-h$ is the current smallest element in $\mathcal{C}(M)$ for which this holds,
hence $j-h$ will become the new right endpoint of the interval; otherwise $i+h$ is the new
left endpoint. The bisection method stops when the endpoints meet or cross each other,
at which point we know that $s_1 \not\simul_{c_n} s_2$ for all $n < j$
and $s_1 \simul_{c_n} s_2$ for all $n \geq j$, 
and hence we return $c_j$.

\begin{algorithm}[t]
  \SetAlgoLined
    Order the elements of $\mathcal{C}(M) \setminus \{\infty\}$
    such that $c_1 < c_2 < \dots < c_n$ \;
    \lIf{$s_1 \simul_{c_1} s_2$}{
      \Return $c_1$
    }
    \lElseIf{$s_1 \not\simul_{c_n} s_2$}{
      \Return $\infty$
    }
    \Else{ 
      $i \leftarrow 1, j \leftarrow n$ \;
      \While{$i < j$}{
        $h \leftarrow \left\lceil\frac{j-i}{2}\right\rceil$ \;
        \lIf{$s_1 \simul_{c_{j-h}} s_2$}{
          $j \leftarrow j - h$
        }
        \lElse{
          $i \leftarrow i + h$
        }
      }
      \Return $c_j$ \;
    }
  \caption{Computing the simulation distance between $s_1$ and $s_2$.}
  \label{alg:distance}
\end{algorithm}

Computing the set $\mathcal{C}(M)$ at line 1 has complexity $n^2 f(l)$,
and sorting it can be done in time $\bigo(n \log n)$ using mergesort.
By Theorem \ref{thm:decidable-d}, and since we have already computed $\mathcal{C}(M)$,
each of the $\varepsilon$-simulation checks
in lines $2$, $3$, and $8$ can be done in time $\bigo(n^2k + m^2n)$,
but the complexity $n^2k$ from comparing labels only needs to be computed once.
Since the bisection interval is halved each time, the while-loop is taken at most
$\log n$ times. We therefore get
an overall time complexity of $\bigo(n^2 (f(l) + k) + m^2n \cdot \log n)$.

\begin{theorem}\label{thm:compute}
  Let $M$ be a finite and effective SMDP. The simulation distance between any two states 
  can be computed in time $\bigo(n^2 (f(l) + k) + m^2n \cdot \log n)$.
\end{theorem}
\begin{proof}%[of Theorem \ref{thm:compute}]
  Consider the algorithm described in Algorithm \ref{alg:distance}.
  The correctness of the algorithm is given by Lemma \ref{lem:correctness}.
  We will now argue that the algorithm runs in time $\bigo(n^2 (f(l) + k) + m^2n \cdot \log n)$.
  The sorting in line $1$ of the algorithm can be done
  in time $\bigo(n \log n)$ using mergesort.
  The checks in line $2$ and $3$
  each has complexity $\bigo(n^2 (f(l) + k) + m^2n)$
  by Theorem \ref{thm:decidable-d}.
  However, we only need to compare labels and residence-time distributions
  once and then we can store the results for future iterations.
  Hence the complexity $\bigo(n^2 (f(l) + k))$
  is only incurred once, and not in each iteration.
  The while loop halves the number of elements left to check
  for each iteration, and hence it will loop at most $\log n$ times.
  Since in each iteration we make the check in line $8$,
  the complexity of the while loop becomes
  $\bigo(n^2 (f(l) + k) + m^2n \cdot \log n)$.
\end{proof}

%% Composition
\section{Compositional Properties}
One of the most successful principles of formal verification
is the notion of compositional reasoning,
in which a large system can be understood in terms of its smaller components \cite{CLM89}.
However, for this principle to work,
one must ensure that properties inferred about the components
carry over to the full, composite system.
For bisimulation, this means that one wants bisimulation
to be a congruence with respect to parallel composition,
and a precongruence in the case of simulation.
A natural generalisation of this is that of non-expansiveness,
which means that parallel composition does not increase (expand) the distance between states.
In this section we will prove that some natural notions of parallel composition 
on SMDPs are non-expansive with respect to the simulation distance.

First we define what it means to compose two SMDPs in parallel.
As argued in \cite{SV04}, the style of synchronous CSP is the one that is most suitable for 
SMDPs, so this is the one we will adopt here.

\begin{definition}
  A function $\star : \mathcal{D}(\mathbb{R}_{\geq 0}) \times \mathcal{D}(\mathbb{R}_{\geq 0}) \rightarrow \mathcal{D}(\mathbb{R}_{\geq 0})$
  is a \emph{residence-time composition function} if it is commutative.
\end{definition}

\begin{definition}
  Let $\star$ be a residence-time composition function.
  Then the \emph{$\star$-com\-po\-sition} of $M_1 = (S_1, \tau_1, \rho_1, L_1)$ and $M_2 = (S_2, \tau_2, \rho_2, L_2)$,
  denoted $M_1 \comp{\star} M_2 = (S, \tau, \rho, L)$, is given as follows, for
  arbitrary $s_1, s_1' \in S_1$, $s_2,s_2' \in S_2$, and $a \in A$.
  \begin{enumerate}
    \item $S = S_1 \times S_2$,
    \item $\tau((s_1,s_2),a)((s_1',s_2')) = \tau_1(s_1,a)(s_1') \cdot \tau_2(s_2,a)(s_2')$,
    \item $\rho((s_1,s_2)) = \star(\rho_1(s_1), \rho_2(s_2))$, and
    \item $L((s_1, s_2)) = L(s_1) \cup L(s_2)$. \qedhere
  \end{enumerate}
\end{definition}

Given a composite system $M_1 \comp{\star} M_2 = (S, \tau, \rho, L)$,
we write $s_1 \comp{\star} s_2$ to mean $(s_1,s_2) \in S$.
The residence-time composition function $\star$ allows us to accommodate many different ways of combining timing behaviour, including those found in the literature on process algebras.
We recall here some of these.
\begin{description}
  \item[Maximum composition:] $F_{\star(\mu,\nu)}(t) = \max(F_\mu(t), F_\nu(t))$.
\end{description}
For exponential distributions, $F_\mu = \Exp{\theta}$ and $F_\nu = \Exp{\theta'}$,
the following alternatives can be found.
\begin{description}
  \item[Product rate composition:] $F_{\star(\mu,\nu)} = \Exp{\theta \cdot \theta'}$.
  \item[Minimum rate composition:] $F_{\star(\mu,\nu)} = \Exp{\min\{\theta,\theta'\}}$.
  \item[Maximum rate composition:] $F_{\star(\mu,\nu)} = \Exp{\max\{\theta,\theta'\}}$. 
\end{description}

Maximum composition is used for interactive Markov chains \cite{hermanns2002},
product rate composition is used in SPA \cite{hermanns1998},
minimum rate composition is used in PEPA \cite{hillston2005},
and maximum rate composition is used in TIPP \cite{gotz1993}.

In order to have non-expansiveness for $\star$-composition of SMDPs, we will need to 
restrict to residence-time composition functions $\star$ that are monotonic.
\begin{definition}
  A residence-time composition function $\star$ is
  \emph{monotonic} if for all $\varepsilon \geq 1$ and 
  $\mu,\nu,\eta \in \mathcal{D}(\mathbb{R}_{\geq 0})$, it holds that
  \[F_\mu \fft_{\varepsilon} F_\nu \quad \text{implies} \quad F_{\star(\mu,\eta)} \fft_{\varepsilon} F_{\star(\nu,\eta)}. \qedhere\]
\end{definition}

Requiring monotonicity is not a significant restriction,
as many of the composition functions that are found in the literature are indeed monotonic.

\begin{lemma}\label{lem:monotonic}
  Maximum composition as well as product, minimum, and maximum rate composition
  are all monotonic.
\end{lemma}
\begin{proof}%[of Lemma \ref{lem:monotonic}]
  Let $\varepsilon \geq 1$ and
  assume that $F_\mu(\varepsilon \cdot t) \geq F_\nu(t)$ for all $t$.
  
  We first consider maximum composition.
  If $F_{\star(\mu,\eta)}(\varepsilon \cdot t) = F_\mu(\varepsilon \cdot t)$,
  then $F_\mu(\varepsilon \cdot t) \geq F_\eta(\varepsilon \cdot t) \geq F_\eta(t)$,
  so
  \[F_{\star(\mu,\eta)}(\varepsilon \cdot t) = F_\mu(\varepsilon \cdot t) \geq F_{\star(\nu,\eta)}(t) .\]
  On the other hand, consider the case where
  $F_{\star(\mu,\eta)}(\varepsilon \cdot t) = F_\eta(\varepsilon \cdot t)$.
  Then we know that
  $F_{\star(\mu,\eta)}(\varepsilon \cdot t) \geq F_\mu(\varepsilon \cdot t) \geq F_\nu(t)$.
  If it is the case that $F_{\star(\nu,\eta)}(t) = F_\nu(t)$, then
  \[F_{\star(\mu,\eta)}(\varepsilon \cdot t) \geq F_\nu(t) = F_{\star(\nu,\eta)}(t) .\]
  If $F_{\star(\nu,\eta)}(t) = F_\eta(t)$, then
  \[F_{\star(\mu,\eta)}(\varepsilon \cdot t) = F_\eta(\varepsilon \cdot t) \geq F_\eta(t) = F_{\star(\nu,\eta)}(t) .\]
  So we conclude that $F_{\star(\mu,\eta)}(t) \geq F_{\star(\nu,\eta)}(t')$.
  
  Next we consider the different rate compositions.
  Assume that $F_\mu = \Exp{\theta}$,
  $F_\nu = \Exp{\theta'}$, and $F_\eta = \Exp{\theta''}$.
  Since we have assumed $F_{\mu}(\varepsilon \cdot t) \geq F_\nu(t)$
  for all $t$, this implies by Lemma \ref{lem:exptransform} that
  $\varepsilon \cdot \theta \geq \theta'$.
  
  For product rate composition,
  note that $\varepsilon \cdot \theta \geq \theta'$
  implies $\varepsilon \cdot \theta \cdot \theta'' \geq \theta' \cdot \theta''$.
  Therefore
  \begin{align*}
    F_{\star(\mu,\eta)}(\varepsilon \cdot t) &= \Exp{\theta \cdot \theta''}(\varepsilon \cdot t) = \Exp{\varepsilon \cdot \theta \cdot \theta''}(t) \\
                                             &\geq \Exp{\theta' \cdot \theta''}(t) = F_{\star(\nu,\eta)}(t) .
  \end{align*}
  
  For minimum rate composition,
  we want to show that
  $\min\{\varepsilon \cdot \theta, \varepsilon \cdot \theta''\} \geq \min\{\theta',\theta''\}$.
  If $\min\{\varepsilon \cdot \theta, \varepsilon \cdot \theta''\} = \varepsilon \cdot \theta$,
  then 
  \[\min\{\varepsilon \cdot \theta, \varepsilon \cdot \theta''\} = \varepsilon \cdot \theta \geq \theta' \geq \min\{\theta',\theta''\} .\]
  Otherwise, if $\min\{\varepsilon \cdot \theta, \varepsilon \cdot \theta''\} = \varepsilon \cdot \theta''$, then
  \[\min\{\varepsilon \cdot \theta, \varepsilon \cdot \theta''\} = \varepsilon \cdot \theta'' \geq \theta'' \geq \min\{\theta',\theta''\} .\]
  Hence
  \[F_{\star(\mu,\eta)}(\varepsilon \cdot t) = \Exp{\min\{\varepsilon \cdot \theta, \varepsilon \cdot \theta''\}}(t) \geq \Exp{\min\{\theta',\theta''\}}(t) = F_{\star(\nu,\eta)}(t) .\]
  
  For maximum composition, we see that if $\max\{\theta',\theta''\} = \theta'$, then
  \[\max\{\varepsilon \cdot \theta, \varepsilon \cdot \theta''\} \geq \varepsilon \cdot \theta \geq \theta' = \max\{\theta',\theta''\} ,\]
  and if $\max\{\theta',\theta''\} = \theta''$, then
  \[\max\{\varepsilon \cdot \theta, \varepsilon \cdot \theta''\} \geq \varepsilon \cdot \theta'' \geq \theta'' = \max\{\theta',\theta''\} .\]
  Hence
  \[F_{\star(\mu,\eta)}(\varepsilon \cdot t) = \Exp{\max\{\varepsilon \cdot \theta, \varepsilon \cdot \theta''\}}(t) \geq \Exp{\max\{\theta',\theta''\}}(t) = F_{\star(\nu,\eta)}(t) .\]
\end{proof}

Now we can prove that the $\star$-composition of finite SMDPs
is indeed non-expansive with respect to the simulation distance,
provided that $\star$ is monotonic.

\begin{theorem}\label{thm:non-exp}
  For finite SMDPs and monotonic $\star$,
  \begin{equation*}
  \simdist(s_1,s_2) \leq \varepsilon 
  \quad \text{implies} \quad 
  \simdist(s_1 \comp{\star} s_3, s_2 \comp{\star} s_3) \leq \varepsilon.
  \end{equation*}
\end{theorem}
\begin{proof}%[of Theorem \ref{thm:non-exp}]
  Assume that $\simdist(s_1,s_2) \leq \varepsilon$.
  By Lemma \ref{lem:correctness},
  we have that $s_1 \simul_{\simdist(s_1,s_2)} s_2$,
  so by Lemma \ref{lem:simul-mono}
  we get that $s_1 \simul_{\varepsilon} s_2$.
  Hence, there exists a $\varepsilon$-simulation relation $R$ such that $(s_1,s_2) \in R$.
  Now construct
  \[R' = \{(s_1' \comp{\star} s_3, s_2' \comp{\star} s_3)       \mid (s_1',s_2') \in R \text{ and } s_3 \in S\}  ,\]
  and we want to show that $R'$ is a $\varepsilon$-simulation relation.
  
  Pick some $(s_1' \comp{\star} s_3, s_2' \comp{\star} s_3) \in R'$.
  Then we get
  \[L(s_1' \comp{\star} s_3) = L(s_1') \cup L(s_3) = L(s_2') \cup L(s_3) = L(s_2' \comp{\star} s_3) .\]
  Since $\star$ is monotonic, we immediately get
  \[\star(\rho(s_2'),\rho(s_3))([0, \varepsilon \cdot t]) \geq \star(\rho(s_1'),\rho(s_3))([0, t])\]
  for all $t$, so $F_{s_2' \comp{\star} s_3} \fft_{\varepsilon} F_{s_1' \comp{\star} s_3}$.
  Now let $a \in A$ be an arbitrary action and define
  \[\Delta_a'(s_1'' \comp{\star} s_3', s_2'' \comp{\star} s_3'') = \begin{cases}0 & \text{if } s_3' \neq s_3'' \\ \Delta_a(s_1'',s_2'') \cdot \tau(s_3,a)(s_3') & \text{otherwise.} \end{cases}\]
  If $\Delta_a'(s_1'' \comp{\star} s_3', s_2'' \comp{\star} s_3'') > 0$,
  then $s_3' = s_3''$ and also $\Delta_a(s_1'',s_2'') > 0$, so $(s_1'',s_2'') \in R$,
  and hence $(s_1'' \comp{\star} s_3', s_2'' \comp{\star} s_3'') \in R'$.
  Furthermore,
  \begin{align*}
    \sum_{s_2'' \comp{\star} s_3''} \Delta_a'(s_1'' \comp{\star} s_3', s_2'' \comp{\star} s_3'')
    &=  \sum_{s_2''} \Delta_a'(s_1'' \comp{\star} s_3', s_2'' \comp{\star} s_3') \\
    &= \sum_{s_2''} \Delta_a(s_1'',s_2'') \cdot \tau(s_3,a)(s_3') \\
    &= \tau(s_3,a)(s_3') \cdot \sum_{s_2''} \Delta_a(s_1'',s_2'') \\
    &= \tau(s_3,a)(s_3') \cdot \tau(s_1',a)(s_1'') \\
    &= \tau(s_1' \comp{\star} s_3,a)(s_1'' \comp{\star} s_3') ,
  \end{align*}
  and likewise we can show that
  \begin{align*}
    \sum_{s_1'' \comp{\star} s_3'} \Delta_a'(s_1'' \comp{\star} s_3', s_2'' \comp{\star} s_3'')
%                                                                                     &= \sum_{u'' \comp{\star} w_1} \Delta_a(u'',v'') \cdot \tau(w',a)(w_1) \cdot \tau(w',a)(w_2) \\
%                                                                                     &= \sum_{u''} \Delta_a(u'',v'') \cdot \tau(w',a)(w_2) \cdot \sum_{w_1} \tau(w',a)(w_1) \\
%                                                                                     &= \tau(v',a)(v'') \cdot \tau(w',a)(w_2) \\
                                                                                     &= \tau(s_2' \comp{\star} s_3,a)(s_2'' \comp{\star} s_3'') .
  \end{align*}
  
  We have thus shown that $R'$ is a $\varepsilon$-simulation relation,
  and hence
  \[s_1 \comp{\star} s_3 \simul_\varepsilon s_2 \comp{\star} s_3.\]
  Clearly, this implies that $\simdist(s_1 \comp{\star} s_3, s_2 \comp{\star} s_3) \leq \varepsilon$.
\end{proof}

We conclude this section by exploring the computational aspects of composition of SMDPs.
In particular, we would like to be able to also compute the distance between composite systems.

From Lemma~\ref{lem:correctness}, we know that computing the simulation distance amounts
to being able to compute the constants $c(F_s,F_{s'})$,
for each pair of states $s,s'$ of the SMDP. 
Hence we would like that, whenever two distributions $\mu$ and $\nu$ have effective CDFs then 
also their composition $\star(\mu,\nu)$ has an effective CDF.
By Proposition~\ref{prop:expandunif}, it is easy to see that this holds
for product, minimum, and maximum rate composition, 
since these compositions are still exponential distributions.

For maximum composition, the class $\due$ is unfortunately not closed under composition.
However, the following result holds.

\begin{proposition}\label{prop:compeffective}
  Let $\star$ be maximum composition.
  For any $\mu,\nu,\eta \in \due$,
  \begin{enumerate}
    \item $c(F_{\mu}, F_{\star(\nu,\eta)})$ is computable and
    \item $c(F_{\star(\mu,\eta)}, F_{\nu})$ is computable.
  \end{enumerate}
\end{proposition}
\begin{proof}%[of Proposition \ref{prop:compeffective}]
  \leavevmode
  \begin{enumerate}
    \item This follows from Lemmas \ref{lem:comp1}, \ref{lem:comp2}, \ref{lem:comp3}, \ref{lem:comp4}, and \ref{lem:comp5}.
    \item This follows from Lemmas \ref{lem:comp1}, \ref{lem:comp2}, \ref{lem:comp6}, \ref{lem:comp7}, and \ref{lem:comp8}. \qedhere
  \end{enumerate}
\end{proof}

The above results tells us that if
we are interested in computing the distance $\simdist(s_1, s_2 \comp{\star} s_3)$
or $\simdist(s_1 \comp{\star} s_2, s_3)$,
when $\star$ is maximum composition,
then we can indeed compute the constants $c$ that are needed
for Algorithm \ref{alg:distance} to work.

We now state and prove the lemmas necessary to prove Proposition \ref{prop:compeffective}.

\begin{lemma}\label{lem:comp1}
  Let $\star$ be maximum composition,
  and let $\mu_1$, $\mu_2$, $\nu_1$, and $\nu_2$ be measures.
  The following holds for any $\varepsilon \in \mathbb{R}_{> 0}$.
  \begin{enumerate}
    \item If one of $\mu_1$ and $\mu_2$ and one of $\nu_1$ and $\nu_2$
      is the Dirac measure at $0$,
      then $F_{\star(\mu_1,\mu_2)} \fft_{\varepsilon} F_{\star(\nu_1,\nu_2)}$.
    \item If one of $\mu_1$ and $\mu_2$ is the Dirac measure at $0$,
      but none of $\nu_1$ and $\nu_2$ are,
      then $F_{\star(\mu_1,\mu_2)} \fft_{\varepsilon} F_{\star(\nu_1,\nu_2)}$.
    \item If none of $\mu_1$ and $\mu_2$ are the Dirac measure at $0$,
      but one of $\nu_1$ and $\nu_2$ is,
      then $F_{\star(\mu_1,\mu_2)} \not\fft_{\varepsilon} F_{\star(\nu_1,\nu_2)}$.
  \end{enumerate}
\end{lemma}
\begin{proof}
  \begin{enumerate}
    \item We get $F_{\star(\mu_1,\mu_2)} = \dirac{0}$
      and $F_{\star(\nu_1,\nu_2)} = \dirac{0}$,
      so we can use Proposition \ref{prop:dirac}.
    \item We get $F_{\star(\mu_1,\mu_2)} = \dirac{0}$,
      so again we can use Proposition \ref{prop:dirac}.
    \item We get $F_{\star(\nu_1,\nu_2)} = \dirac{0}$,
      and $F_{\star(\mu_1,\mu_2)} \neq \dirac{0}$,
      so once more we can use Proposition \ref{prop:dirac}. \qedhere
  \end{enumerate}
\end{proof}

\begin{lemma}\label{lem:comp2}
  Let $\star$ be maximum composition,
  and let $F_{\mu_1} = \Exp{\theta_1}$, $F_{\mu_2} = \Exp{\theta_2}$,
  $F_{\nu_1} = \Exp{\lambda_1}$, and $F_{\nu_2} = \Exp{\lambda_2}$.
  Then $F_{\star(\mu_1,\mu_2)} \fft_{\varepsilon} F_{\star(\nu_1,\nu_2)}$
  where $\varepsilon = \frac{\max\{\lambda_1,\lambda_2\}}{\max\{\theta_1,\theta_2\}}$.
  Furthermore, this is the least $\varepsilon$
  such that the $\varepsilon$-faster-than relation holds.
\end{lemma}
\begin{proof}
  We have four cases to consider:
  \begin{enumerate}
    \item $\max\{\theta_1,\theta_2\} = \theta_1$
      and $\max\{\lambda_1,\lambda_2\} = \lambda_1$,
      in which case $F_{\star(\mu_1,\mu_2)} = F_{\mu_1}$
      and $F_{\star(\nu_1,\nu_2)} = F_{\nu_1}$.
    \item $\max\{\theta_1,\theta_2\} = \theta_2$
      and $\max\{\lambda_1,\lambda_2\} = \lambda_1$,
      in which case $F_{\star(\mu_1,\mu_2)} = F_{\mu_2}$
      and $F_{\star(\nu_1,\nu_2)} = F_{\nu_1}$.
    \item $\max\{\theta_1,\theta_2\} = \theta_1$
      and $\max\{\lambda_1,\lambda_2\} = \lambda_2$,
      in which case $F_{\star(\mu_1,\mu_2)} = F_{\mu_1}$
      and $F_{\star(\nu_1,\nu_2)} = F_{\nu_2}$.
    \item $\max\{\theta_1,\theta_2\} = \theta_2$
      and $\max\{\lambda_1,\lambda_2\} = \lambda_2$,
      in which case $F_{\star(\mu_1,\mu_2)} = F_{\mu_2}$
      and $F_{\star(\nu_1,\nu_2)} = F_{\nu_2}$.
  \end{enumerate}
  In all cases, the result then follows from Proposition \ref{prop:expandunif}.
\end{proof}

\begin{lemma}\label{lem:unifineq}
  Let $\star$ be maximum composition and let $F_\mu = \unif{a}{b}$ and $F_\nu = \unif{c}{d}$.
  If $a \leq c$ and $d \leq b$, then
  $F_{\star(\mu,\nu)}(t) \leq \unif{a}{d}(t)$ for all $t$.
\end{lemma}
\begin{proof}
  Note first that if $F_{\mu'} = \unif{a'}{b'}$ and $F_{\nu'} = \unif{c'}{d'}$
  with $a' \leq c'$ and $b' \leq d'$,
  then clearly $\unif{a'}{b'}(t) \geq \unif{c'}{d'}(t)$ for all $t$.
  
  Now, $a \leq a$ and $d \leq b$, so $\unif{a}{d}(t) \geq \unif{a}{b}(t)$ for all $t$.
  Likewise, $a \leq c$ and $d \leq d$, so $\unif{a}{d} \geq \unif{c}{d}(t)$ for all $t$.
  Hence
  \[\unif{a}{d}(t) \geq \max\{\unif{a}{b}(t),\unif{c}{d}(t)\} = F_{\star(\mu,\nu)} . \qedhere\]
\end{proof}

\begin{lemma}\label{lem:comp3}
  Let $\star$ be maximum composition,
  and let $F_\mu = \unif{a}{b}$, $F_{\nu_1} = \unif{c_1}{d_1}$,
  and $F_{\nu_2} = \unif{c_2}{d_2}$.
  \begin{enumerate}
    \item If $\min\{c_1,c_2\} = 0$ and $a > 0$,
      then $F_{\mu} \not\fft_{\varepsilon} F_{\star(\nu_1,\nu_2)}$
      for any $\varepsilon$.
    \item If $\min\{c_1,c_2\} = 0$ and $a = 0$,
      then $F_{\mu} \fft_{\varepsilon} F_{\star(\nu_1,\nu_2)}$
      where $\varepsilon = \frac{b}{\min\{d_1,d_2\}}$.
    \item If $\min\{c_1,c_2\} > 0$, then
      $F_{\mu} \fft_{\varepsilon} F_{\star(\nu_1,\nu_2)}$
      where
      \[\varepsilon = \max\left\{\frac{a}{\min\{c_1,c_2\}},\frac{b}{\min\{d_1,d_2\}}\right\} .\]
  \end{enumerate}
  In all cases, this is the least $\varepsilon$
  such that the $\varepsilon$-faster-than relation holds.
\end{lemma}
\begin{proof}
  \begin{enumerate}
    \item Take $t = \frac{a}{\varepsilon} > 0$ to get
      \[F_\mu(\varepsilon \cdot t) = \unif{a}{b}(\varepsilon \cdot t) = \unif{a}{b}(a) = 0 .\]
      However, since $\min\{c_1,c_2\} = 0$,
      at least one of $\unif{c_1}{d_1}(t) > 0$ and $\unif{c_2}{d_2}(t) > 0$
      must hold, and hence $F_{\star(\nu_1,\nu_2)}(t) > 0$.
    \item We get
      \begin{align*}
        F_{\mu}(\varepsilon \cdot t)
        &= \unif{a \cdot \frac{\min\{d_1,d_2\}}{b}}{\min\{d_1,d_2\}}(t) \\
        &= \unif{0}{\min\{d_1,d_2\}}(t) \\
        &\geq F_{\star(\nu_1,\nu_2)}(t) && \text{by Lemma \ref{lem:unifineq}} .
      \end{align*}
      If $\varepsilon' < \frac{b}{\min\{d_1,d_2\}}$, then
      \begin{align*}
        &\phantom{{}={}}F_\mu(\varepsilon' \cdot \min\{d_1,d_2\}) \\
        &= \unif{a}{b}(\varepsilon' \cdot \min\{d_1,d_2\}) \\
        &< \unif{a}{b}\left(\frac{b}{\min\{d_1,d_2\}} \cdot \min\{d_1,d_2\}\right) \\
        &= 1 \\
        &= \max\{\unif{c_1}{d_1}(\min\{d_1,d_2\}),\unif{c_2}{d_2}(\min\{d_1,d_2\})\} \\
        &= F_{\nu_1,\nu_2}(\min\{d_1,d_2\}) .
      \end{align*}
    \item We consider each case separately.
      \begin{description}
        \item[Case $c_1 \leq c_2$ and $d_1 \leq d_2$:]
          In this case we have $F_{\star(\nu_1,\nu_2)} = F_{\nu_1}$,
          so we can use Proposition \ref{prop:expandunif}.
        \item[Case $c_1 \leq c_2$ and $d_1 > d_2$:]
          In this case we get $\varepsilon = \max\left\{\frac{a}{c_1},\frac{b}{d_2}\right\}$.
          If $\varepsilon = \frac{a}{c_1}$,
          then $\frac{c_1}{a} \leq \frac{d_2}{b}$, so
          \begin{align*}
            F_{\mu}(\varepsilon \cdot t)
            &= \unif{c_1}{b \cdot \frac{d_2}{b}}(t) \\
            &\geq \unif{c_1}{b \cdot \frac{d_2}{b}}(t) \\
            &= \unif{c_1}{d_2}(t) \\
            &\geq F_{\star(\nu_1,\nu_2)}(t) && \text{by Lemma \ref{lem:unifineq}} .
          \end{align*}
          For any $\varepsilon' < \frac{a}{c_1}$,
          let $t = \frac{a}{\varepsilon'} > c_1$.
          Then $F_{\mu}(\varepsilon' \cdot t) = F_\mu(a) = 0$,
          but $F_{\star(\nu_1,\nu_2)}(t) > 0$
          since $c_1 \leq c_2$ and $t > c_1$.
          
          On the other hand, if $\varepsilon = \frac{b}{d_2}$,
          then $\frac{d_2}{b} \leq \frac{c_1}{a}$.
          This means that
          \begin{align*}
            F_{\mu}(\varepsilon \cdot t)
            &= \unif{a \cdot \frac{d_2}{b}}{d_2}(t) \\
            &\geq \unif{a \cdot \frac{c_1}{a}}{d_2}(t) \\
            &= \unif{c_1}{d_2}(t) \\
            &\geq F_{\star(\nu_1,\nu_2)}(t) && \text{by Lemma \ref{lem:unifineq}} .
          \end{align*}
          For any $\varepsilon' < \frac{b}{d_2}$
          we get
          \begin{align*}
            F_\mu(\varepsilon' \cdot d_2)
            &= \unif{a}{b}(\varepsilon' \cdot d_2) \\
            &< \unif{a}{b}(\frac{b}{d_2} \cdot d_2) \\
            &= 1 \\
            &= F_{\star(\nu_1,\nu_2)}(d_2)
          \end{align*}
          because $d_1 > d_2$.
        \item[Case $c_1 > c_2$ and $d_1 \leq d_2$:]
          In this case we get $\varepsilon = \max\left\{\frac{a}{c_2},\frac{b}{d_1}\right\}$.
          If $\varepsilon = \frac{a}{c_2}$,
          then $\frac{c_2}{a} \leq \frac{d_1}{b}$, and hence
          \begin{align*}
            F_{\mu}(\varepsilon \cdot t)
            &= \unif{c_2}{b \cdot \frac{c_2}{a}}(t) \\
            &\geq \unif{c_2}{b \cdot \frac{d_1}{b}}(t) \\
            &= \unif{c_2}{d_1}(t) \\
            &\geq F_{\star(\nu_1,\nu_2)}(t) && \text{by Lemma \ref{lem:unifineq}} .
          \end{align*}
          For any $\varepsilon' < \frac{a}{c_2}$,
          let $t = \frac{a}{\varepsilon'} > c_2$
          in order to get $F_\mu(\varepsilon' \cdot t) = \unif{a}{b}(a) = 0$,
          but $F_{\star(\nu_1,\nu_2)}(t) > 0$ since $c_1 > c_2$ and $t > c_2$.
          
          On the other hand, if $\varepsilon = \frac{b}{d_1}$,
          then $\frac{d_1}{b} \leq \frac{c_2}{a}$.
          Then we get
          \begin{align*}
            F_\mu(\varepsilon \cdot t)
            &= \unif{a \cdot \frac{d_1}{b}}{d_1}(t) \\
            &\geq \unif{c_2}{d_1}(t) \\
            &\geq F_{\star(\nu_1,\nu_2)}(t) && \text{by Lemma \ref{lem:unifineq}} .
          \end{align*}
          For any $\varepsilon' < \frac{b}{d_1}$ we get
          \begin{align*}
            F_{\mu}(\varepsilon' \cdot d_1)
            &= \unif{a}{b}(\varepsilon' \cdot d_1) \\
            &< \unif{a}{b}(\frac{b}{d_1} \cdot d_1) \\
            &= 1 \\
            &= F_{\star(\nu_1,\nu_2)}(d_1)
          \end{align*}
          since $d_1 \leq d_2$.
        \item[Case $c_1 > c_2$ and $d_1 > d_2$:]
          In this case we have $F_{\star(\nu_1,\nu_2)} = F_{\nu_2}$,
          so we can use Proposition \ref{prop:expandunif}. \qedhere
      \end{description}
  \end{enumerate}
\end{proof}

\begin{lemma}\label{lem:comp4}
  Let $\star$ be maximum composition,
  and let $F_{\mu} = \Exp{\theta}$, $F_{\nu_1} = \unif{a}{b}$, $F_{\nu_2} = \unif{c}{d}$.
  Then $F_{\mu} \not\fft_{\varepsilon} F_{\star(\nu_1,\nu_2)}$
  for any $\varepsilon$.
\end{lemma}
\begin{proof}
  $F_{\star(\nu_1,\nu_2)}(\min\{b,d\}) = 1$,
  but $F_\mu(t) < 1$ for all $t$,
  so
  \[F_\mu(\varepsilon \cdot \min\{b,d\}) < F_{\star(\nu_1,\nu_2)}(\min\{b,d\})\]
  for any $\varepsilon$.
\end{proof}

\begin{lemma}\label{lem:comp5}
  Let $\star$ be maximum composition,
  and let $\mu_1 = \Exp{\theta_1}$, $\mu_2 = \unif{a}{b}$,
  $\nu_1 = \Exp{\theta_2}$, and $\nu_2 = \unif{c}{d}$.
  \begin{enumerate}
    \item $F_{\mu_1} \not\fft_\varepsilon F_{\star(\nu_1,\nu_2)}$ for any $\varepsilon$.
    \item If $a > 0$, then $F_{\mu_2} \not\fft_\varepsilon F_{\star(\nu_1,\nu_2)}$
      for any $\varepsilon$.
    \item If $a = 0$ and $\frac{1}{d-c} \geq \theta_2$,
      then $F_{\mu_2} \fft_\varepsilon F_{\star(\nu_1,\nu_2)}$
      where $\varepsilon = \frac{b}{d}$.
    \item If $a = 0$ and $\frac{1}{d-c} < \theta_2$,
      then $F_{\mu_2} \fft_{\varepsilon} F_{\star(\nu_1,\nu_2)}$
      where $\varepsilon = \theta_2 \cdot b$.
  \end{enumerate}
  In all cases, this is the least $\varepsilon$
  such that the $\varepsilon$-faster-than relation holds.
\end{lemma}
\begin{proof}
  \begin{enumerate}
    \item $F_{\star(\nu_1,\nu_2)}(d) = 1$,
      but $F_{\mu_1}(t) < 1$ for all $t$,
      so $F_{\mu_1}(\varepsilon \cdot d) < F_{\star(\nu_1,\nu_2)}(d)$ for any $\varepsilon$.
    \item If $a > 0$, let $\varepsilon \in \mathbb{R}_{> 0}$ be given.
      Let $t = \frac{a}{\varepsilon} > 0$ to get
      $F_{\star(\nu_1,\nu_2)}(t) \geq \Exp{\theta_2}(t) > 0$
      since $t > 0$, but
      $F_{\mu_2}(\varepsilon \cdot t) = \unif{a}{b}(a) = 0$,
      and hence $F_{\mu_2}(\varepsilon \cdot t) < F_{\sat{\nu_1,\nu_2}}(t)$.
    \item If $a = 0$ and $\frac{1}{d-c} \geq \theta_2$,
      then the slope of $\unif{c}{d}$ is greater than that of $\Exp{\theta_2}$
      until $\unif{c}{d}$ hits $1$ and flattens out.
      Hence $F_{\star(\nu_1,\nu_2)} = \unif{c}{d}$,
      so we can use Proposition \ref{prop:expandunif}
      to get the result.
    \item If $a = 0$ and $\frac{1}{d-c} < \theta_2$,
      let $\varepsilon = \theta_2 \cdot b$.
      By Proposition \ref{prop:unifexp}
      we then get $F_{\mu_2}(\varepsilon \cdot t) \geq \Exp{\theta}(t)$
      for all $t$.
      Since the slope of $\unif{\frac{a}{\varepsilon}}{\frac{b}{\varepsilon}}$
      is $\theta_2$, it has greater slope than $\unif{c}{d}$,
      and hence also
      $F_{\mu_2}(\varepsilon \cdot t) = \unif{\frac{a}{\varepsilon}}{\frac{b}{\varepsilon}}(t) \geq \unif{c}{d}(t)$.
      We therefore get
      \[F_{\mu_2}(\varepsilon \cdot t) \geq \max\{\Exp{\theta_2}(t), \unif{c}{d}(t)\} = F_{\star(\nu_1,\nu_2)}(t).\]
      If $\varepsilon' < \theta_2 \cdot b$,
      then the slope in $0$ of
      $F_{\mu_2}(\varepsilon' \cdot t) = \unif{\frac{a}{\varepsilon'}}{\frac{b}{\varepsilon'}}(t)$
      must be less than that of $\Exp{\theta_2}$.
      Hence there exists some $t > 0$
      sufficiently close to $0$ such that
      $F_{\mu_2}(\varepsilon' \cdot t) < \Exp{\theta_2}(t) = F_{\star(\nu_1,\nu_2)}(t)$. \qedhere
  \end{enumerate}
\end{proof}

\begin{lemma}\label{lem:comp6}
  Let $\star$ be maximum composition,
  and let $F_{\mu_1} = \unif{a_1}{b_1}$, $F_{\mu_2} = \unif{a_2}{b_2}$,
  and $F_{\nu} = \unif{c}{d}$.
  \begin{enumerate}
    \item If $c = 0$ and $\min\{a_1,a_2\} > 0$,
      then $F_{\star(\mu_1,\mu_2)} \not\fft_{\varepsilon} F_\nu$
      for any $\varepsilon$.
    \item If $c = 0$ and $a_1 = a_2 = 0$,
      then $F_{\star(\mu_1,\mu_2)} \fft_{\varepsilon} F_\nu$
        where $\varepsilon = \frac{\min\{b_1,b_2\}}{d}$.
    \item If $c = 0$ and $\min\{a_1,a_2\} = 0$
      and either $a_1 < a_2$ and $b_1 \leq b_2$ or $a_1 > a_2$ and $b_1 > b_2$,
      then $F_{\star(\mu_1,\mu_2)} \fft_{\varepsilon} F_\nu$
      where $\varepsilon = \frac{\min\{b_1,b_2\}}{d}$.
    \item If $c = 0$ and $\min\{a_1,a_2\} = 0$
      and either $a_1 < a_2$ and $b_1 > b_2$ or $a_1 > a_2$ and $b_1 \leq b_2$,
      then $F_{\star(\mu_1,\mu_2)} \fft_{\varepsilon} F_\nu$
      where $\varepsilon = \frac{\max\{b_1,b_2\}}{d}$.
    \item If $c > 0$, $a_1 < a_2$, $b_1 > b_2$,
      $\frac{1}{c - \frac{a_1 \cdot c}{b_1}} \geq \frac{1}{d - c}$,
      and $\frac{1}{d - \frac{a_2 \cdot d}{b_2}} \leq \frac{1}{d - c}$,
      then $F_{\star(\mu_1,\mu_2)} \fft_{\varepsilon} F_\nu$
      where $\varepsilon = \min\left\{\frac{a_1}{c}, \frac{b_2}{d}\right\}$.
    \item If $c > 0$, $a_1 < a_2$, $b_1 > b_2$,
      $\frac{1}{c - \frac{a_1 \cdot c}{b_1}} < \frac{1}{d - c}$,
      and $\frac{1}{d - \frac{a_2 \cdot d}{b_2}} \leq \frac{1}{d - c}$,
      then $F_{\star(\mu_1,\mu_2)} \fft_{\varepsilon} F_\nu$
      where $\varepsilon = \frac{b_2}{d}$.
    \item If $c > 0$, $a_1 < a_2$, $b_1 > b_2$,
      $\frac{1}{c - \frac{a_1 \cdot c}{b_1}} \geq \frac{1}{d - c}$,
      and $\frac{1}{d - \frac{a_2 \cdot d}{b_2}} > \frac{1}{d - c}$,
      then $F_{\star(\mu_1,\mu_2)} \fft_{\varepsilon} F_\nu$
      where $\varepsilon = \frac{a_1}{c}$.
    \item If $c > 0$, $a_1 < a_2$, $b_1 > b_2$,
      $\frac{1}{c - \frac{a_1 \cdot c}{b_1}} < \frac{1}{d - c}$,
      and $\frac{1}{d - \frac{a_2 \cdot d}{b_2}} > \frac{1}{d - c}$,
      then $F_{\star(\mu_1,\mu_2)} \fft_{\varepsilon} F_\nu$
      where
      \[\varepsilon = \frac{(b_1 - a_1) \cdot k}{d \cdot k - c \cdot k - d \cdot a_1 + b_1 \cdot c}\]
      and $k = \frac{a_1 \cdot b_2 - a_2 \cdot b_1}{a_1 - b_1 - a_2 + b_2}$.
    \item If $c > 0$, $a_1 > a_2$, $b_1 < b_2$,
      $\frac{1}{c - \frac{a_2 \cdot c}{b_2}} \geq \frac{1}{d - c}$,
      and $\frac{1}{d - \frac{a_1 \cdot d}{b_1}} \leq \frac{1}{d - c}$,
      then $F_{\star(\mu_1,\mu_2)} \fft_{\varepsilon} F_\nu$
      where $\varepsilon = \min\left\{\frac{a_2}{c}, \frac{b_1}{d}\right\}$.
    \item If $c > 0$, $a_1 > a_2$, $b_1 < b_2$,
      $\frac{1}{c - \frac{a_2 \cdot c}{b_2}} < \frac{1}{d - c}$,
      and $\frac{1}{d - \frac{a_1 \cdot d}{b_1}} \leq \frac{1}{d - c}$,
      then $F_{\star(\mu_1,\mu_2)} \fft_{\varepsilon} F_\nu$
      where $\varepsilon = \frac{b_1}{d}$.
    \item If $c > 0$, $a_1 > a_2$, $b_1 < b_2$,
      $\frac{1}{c - \frac{a_2 \cdot c}{b_2}} \geq \frac{1}{d - c}$,
      and $\frac{1}{d - \frac{a_1 \cdot d}{b_1}} > \frac{1}{d - c}$,
      then $F_{\star(\mu_1,\mu_2)} \fft_{\varepsilon} F_\nu$
      where $\varepsilon = \frac{a_2}{c}$.
    \item If $c > 0$, $a_1 > a_2$, $b_1 < b_2$,
      $\frac{1}{c - \frac{a_2 \cdot c}{b_2}} < \frac{1}{d - c}$,
      and $\frac{1}{d - \frac{a_1 \cdot d}{b_1}} > \frac{1}{d - c}$,
      then $F_{\star(\mu_1,\mu_2)} \fft_{\varepsilon} F_\nu$
      where
      \[\varepsilon = \frac{(b_2 - a_2) \cdot k}{d \cdot k - c \cdot k - d \cdot a_2 + b_2 \cdot c}\]
      and $k = \frac{a_1 \cdot b_2 - a_2 \cdot b_1}{a_1 - b_1 - a_2 + b_2}$.
    \item Otherwise,
      $F_{\star(\mu_1,\mu_2)} \fft_{\varepsilon} F_\nu$
      where
      \[\varepsilon = \max\left\{\frac{\min\{a_1,a_2\}}{c},\frac{\min\{b_1,b_2\}}{d}\right\}.\]
  \end{enumerate}
  In all cases, this is the least $\varepsilon$
  such that the $\varepsilon$-faster-than relation holds.
\end{lemma}
\begin{proof}
  \begin{enumerate}
    \item Take an arbitrary $\varepsilon \in \mathbb{R}_{> 0}$
      and let $t = \frac{\min\{a_1,a_2\}}{\varepsilon} > 0$.
      Then $F_{\star(\mu_1,\mu_2)}(\varepsilon \cdot t) = F_{\star(\mu_1,\mu_2)}(\min\{a_1,a_2\}) = 0$,
      but $F_{\nu}(t) > 0$ since $c = 0$ and $t > 0$.
      Hence $F_{\star(\mu_1,\mu_2)}(\varepsilon \cdot t) < F_\nu(t)$.
    \item If $a_1 = a_2 = 0$,
      then $F_{\star(\mu_1,\mu_2)} = \unif{0}{\min\{b_1,b_2\}}$,
      so the result follows from Proposition \ref{prop:expandunif}.
    \item If $a_1 \leq a_2$ and $b_1 \leq b_2$,
      then $F_{\star(\mu_1,\mu_2)} = F_{\mu_1}$,
      and if $a_1 > a_2$ and $b_1 > b_2$,
      then $F_{\star(\mu_1,\mu_2)} = F_{\mu_2}$.
      In either case, we can then use Proposition \ref{prop:expandunif}
      to obtain the result.
    \item We consider here the case where $a_1 < a_2$ and $b_1 > b_2$.
      The case where $a_1 > a_2$ and $b_1 \leq b_2$ is symmetrical,
      noting that if $b_1 = b_2$, then $F_{\star(\mu_1,\mu_2)} = F_{\mu_1}$,
      in which case the result follows from Proposition \ref{prop:expandunif}.
      We have $c = \min\{a_1,a_2\} = a_1 = 0$
      and $\varepsilon = \frac{b_1}{d}$.
      Then
      \begin{align*}
        F_{\star(\mu_1,\mu_2)}(\varepsilon \cdot t)
        &\geq \unif{a_1}{b_1}(\varepsilon \cdot t) \\
        &= \unif{a_1 \cdot \frac{d}{b_1}}{b_1 \cdot \frac{d}{b_1}}(t) \\
        &= \unif{0}{d}(t) \\
        &= \unif{c}{d}(t) .
      \end{align*}
      To see that this is the least $\varepsilon$
      such that the $\varepsilon$-faster-than relation holds,
      first note that
      $\unif{\frac{a_1}{\varepsilon}}{\frac{b_1}{\varepsilon}}$ and
      $\unif{\frac{a_2}{\varepsilon}}{\frac{b_2}{\varepsilon}}$
      cross in the point
      \[t = \frac{a_1 \cdot b_2 - a_2 \cdot b_1}{\varepsilon \cdot (a_1 - b_1 - a_2 + b_2)}\]
      with $0 = a_1 < a_2 < t < b_2 < b_1$.
      From this it follows that
      \[1 > \unif{c}{d}(t) = \unif{a_1}{b_2}(\varepsilon \cdot t) = \unif{a_2}{b_2}(\varepsilon \cdot t) > 0 .\]
      Hence, if $\varepsilon' < \varepsilon$ we get
      \[\unif{c}{d}(t) = \unif{a_1}{b_1}(\varepsilon \cdot t) > \unif{a_1}{b_2}(\varepsilon' \cdot t)\]
      and
      \[\unif{c}{d}(t) = \unif{a_2}{b_2}(\varepsilon \cdot t) > \unif{a_2}{b_2}(\varepsilon' \cdot t) ,\]
      and therefore $F_{\star(\mu_1,\mu_2)}(\varepsilon' \cdot t) < \unif{c}{d}(t)$.
    \item $\frac{1}{c - \frac{a_1 \cdot c}{b_1}} \geq \frac{1}{d - c}$
      means that
      \[\unif{a_1}{b_2}\left(\frac{a_1}{c} \cdot t\right) = \unif{c}{b_1 \cdot \frac{c}{a_1}}(t)\]
      has greater slope than $\unif{c}{d}(t)$, so
      \[\unif{a_1}{b_1}\left(\frac{a_1}{c} \cdot t\right) \geq \unif{c}{d}(t) .\]
      Likewise, $\frac{1}{d - \frac{a_2 \cdot d}{b_2}} \leq \frac{1}{d - c}$
      means that
      \[\unif{a_2}{b_2}\left(\frac{b_2}{d} \cdot t\right) = \unif{a_2 \cdot \frac{d}{b_2}}{d}(t)\]
      has smaller slope than $\unif{c}{d}(t)$,
      and hence
      \[\unif{a_2}{b_2}(\frac{b_2}{d} \cdot t) \geq \unif{c}{d}(t) .\]
      We therefore conclude
      \[F_{\star(\mu_1,\mu_2)}\left(\min\left\{\frac{a_1}{c},\frac{b_2}{d}\right\} \cdot t\right) \geq \unif{c}{d}(t) .\]
      If $\varepsilon' < \varepsilon$,
      first assume that $\varepsilon = \frac{a_1}{c}$
      and let $t = \frac{a_1}{\varepsilon'} > c$.
      Then
      \[F_{\star(\mu_1,\mu_2)}(\varepsilon' \cdot t) = F_{\star(\mu_1,\mu_2)}(a_1) = 0,\]
      but $\unif{c}{d}(t) > 0$ since $t > c$.
      Now assume that $\varepsilon = \frac{b_2}{d}$.
      Then we get
      \[F_{\star(\mu_1,\mu_2)}(\varepsilon' \cdot d) < F_{\star(\mu_1,\mu_2))}(\varepsilon \cdot d) = F_{\star(\mu_1,\mu_2)}(b_2) = 1 = \unif{c}{d}(d) .\]
    \item This case is the same as case 5,
      only considering the part where $\varepsilon = \frac{b_2}{d}$.
    \item This case is the same as case 5,
      only considering the part where $\varepsilon = \frac{a_1}{c}$.
    \item $\varepsilon$ and $k$ are chosen such that
      $\frac{k}{\varepsilon}$ is the point in which
      $\unif{\frac{a_1}{\varepsilon}}{\frac{b_1}{\varepsilon}}$,
      $\unif{\frac{a_2}{\varepsilon}}{\frac{b_2}{\varepsilon}}$,
      and $\unif{c}{d}$ cross.
      Hence we get
      \[F_{\star(\mu_1,\mu_2)}\left(\varepsilon \cdot \frac{k}{\varepsilon}\right) = \unif{c}{d}\left(\frac{k}{\varepsilon}\right).\]
      We have
      \[F_{\star(\mu_1,\mu_2)}(\varepsilon \cdot t) = \unif{a_1}{b_1}(\varepsilon \cdot t) \geq \unif{c}{d}(t)\]
      for any $t \geq \frac{k}{\varepsilon}$ and
      \[F_{\star(\mu_1,\mu_2)}(\varepsilon \cdot t) = \unif{a_2}{b_2}(\varepsilon \cdot t) \geq \unif{c}{d}(t)\]
      for any $t \leq \frac{k}{\varepsilon}$.
      Hence we can conclude that $F_{\star(\mu_1,\mu_2)}(\varepsilon \cdot t) \geq \unif{c}{d}(t)$.
      If $\varepsilon' < \varepsilon$,
      then $F_{\star(\mu_1,\mu_2)}\left(\varepsilon' \cdot \frac{k}{\varepsilon}\right) < F_{\star(\mu_1,\mu_2)}\left(\varepsilon \cdot \frac{k}{\varepsilon}\right) = \unif{c}{d}\left(\frac{k}{\varepsilon}\right)$.
    \item Symmetric to case 5.
    \item Symmetric to case 6.
    \item Symmetric to case 7.
    \item Symmetric to case 8.
    \item In this case,
      we either get $F_{\star(\mu_1,\mu_2)} = F_{\mu_1}$
      or $F_{\star(\mu_1,\mu_2)} = F_{\mu_2}$,
      and the result is then obtained by applying Proposition \ref{prop:expandunif}. \qedhere
  \end{enumerate}
\end{proof}

\begin{lemma}\label{lem:comp7}
  Let $\star$ be maximum composition,
  and let $F_{\mu} = \Exp{\theta}$, $F_{\nu_1} = \unif{a}{b}$, $F_{\nu_2} = \unif{c}{d}$.
  \begin{enumerate}
    \item If $\min\{a,c\} > 0$, then
      $F_{\star(\nu_1,\nu_2)} \not\fft_{\varepsilon} F_\mu$ for any $\varepsilon$.
    \item If $a = 0$ and $c > 0$, then
      $F_{\star(\nu_1,\nu_2)} \fft_{\varepsilon} F_\mu$
      where $\varepsilon = \theta \cdot b$.
    \item If $a > 0$ and $c = 0$, then
      $F_{\star(\nu_1,\nu_2)} \fft_{\varepsilon} F_\mu$
      where $\varepsilon = \theta \cdot d$.
    \item If $a = 0$ and $c = 0$, then
      $F_{\star(\nu_1,\nu_2)} \fft_{\varepsilon} F_\mu$
      where $\varepsilon = \theta \cdot \min\{b,d\}$.
  \end{enumerate}
  In all cases, this is the least $\varepsilon$ such that
  the $\varepsilon$-faster-than relation holds.
\end{lemma}
\begin{proof}
  \begin{enumerate}
    \item If $\min\{a,c\} > 0$, let $\varepsilon$ be given,
      and let $t = \frac{\min\{a,c\}}{\varepsilon} > 0$. Then
      \[F_{\star(\nu_1,\nu_2)}(\varepsilon \cdot t) = F_{\star(\nu_1,\nu_2)}(\min\{a,c\}) = 0\]
      but $F_{\mu}(t) = \Exp{\theta}(t) > 0$ since $t > 0$.
    \item By Proposition \ref{prop:unifexp},
      we know that $\unif{a}{b} \fft_\varepsilon \Exp{\theta}$.
      Hence
      \[F_{\star(\nu_1,\nu_2)}(\varepsilon \cdot t) \geq \unif{a}{b}(\varepsilon \cdot t) \geq \Exp{\theta}(t).\]
      If $\varepsilon' < \theta \cdot b$,
      then there must exist some $t > 0$ sufficiently close to $0$
      such that
      \[F_{\star(\nu_1,\nu_2)}(\varepsilon' \cdot t) = \unif{a}{b}(\varepsilon' \cdot t) < \Exp{\theta}(t).\]
    \item Similar to the case where $a = 0$ and $c > 0$.
    \item If $a = 0$ and $c = 0$,
      then we get $F_{\star(\nu_1,\nu_2)} = \unif{a}{b}$ if $b \leq d$
      and $F_{\star(\nu_1,\nu_2)} = \unif{c}{d}$ if $b > d$.
      In either case,
      we can use Proposition \ref{prop:unifexp} to obtain the desired result. \qedhere
  \end{enumerate}
\end{proof}

\begin{lemma}\label{lem:comp8}
  Let $\star$ be maximum composition,
  and let $\mu_1 = \Exp{\theta_1}$, $\mu_2 = \unif{a}{b}$,
  $\nu_1 = \Exp{\theta_2}$, and $\nu_2 = \unif{c}{d}$.
  \begin{enumerate}
    \item If $a = 0$ and $\frac{1}{b-a} \geq \theta_1$,
      then $F_{\star(\mu_1,\mu_2)} \fft_{\varepsilon} F_{\nu_1}$
      where $\varepsilon = \theta_2 \cdot b$.
    \item If $a > 0$ or $\frac{1}{b-a} < \theta_1$,
      then $F_{\star(\mu_1,\mu_2)} \fft_\varepsilon F_{\nu_1}$
      where $\varepsilon = \frac{\theta_2}{\theta_1}$.
    \item If $\frac{1}{b - a} \geq \theta_1$ and $a = 0$,
      then $F_{\star(\mu_1,\mu_2)} \fft_\varepsilon F_{\nu_2}$
      where $\varepsilon = \frac{b}{d}$.
    \item Otherwise, $F_{\star(\mu_1,\mu_2)} \fft_\varepsilon F_{\nu_2}$ where
      \[\varepsilon = \max\left\{\frac{b}{d},\frac{(b-a) \cdot k}{d \cdot k - c \cdot k - d \cdot a + b \cdot c}\right\} ,\]
      $k = \frac{b \cdot \theta_1 + W(\theta_1 \cdot (a-b) \cdot e^{-b \cdot \theta_1})}{\theta_1}$,
      and $W$ is the Lambert $W$-function.
  \end{enumerate}
  In all cases, this is the least $\varepsilon$
  such that the $\varepsilon$-faster-than relation holds.
\end{lemma}
\begin{proof}
  \begin{enumerate}
    \item In this case, $F_{\star(\mu_1,\mu_2)} = F_{\mu_2}$,
      so we can use Proposition \ref{prop:unifexp} to obtain the result.
    \item We get
      \[F_{\star(\mu_1,\mu_2)}\left(\frac{\theta_2}{\theta_1} \cdot t\right) \geq \Exp{\theta_1}\left(\frac{\theta_2}{\theta_1} \cdot t\right) = \Exp{\theta_2}(t) = F_{\nu_1}(t) .\]
      To see that this is the least $\varepsilon$ such that the
      $\varepsilon$-faster-than relation holds,
      first note that because $a > 0$ or $\frac{1}{b-a} < \theta_1$,
      there must be some interval $[0,t]$
      where $F_{\star(\mu_1,\mu_2)}(t') = \Exp{\theta_1}(t')$
      for all $t' \in [0,t]$.
      If $\varepsilon' < \varepsilon$,
      then let $t' \in \left[0,\frac{t}{\varepsilon}\right]$,
      so that $\varepsilon \cdot t' \in [0,t]$,
      and also $\varepsilon' \cdot t' \in [0,t]$.
      Then we get
      \[F_{\star(\mu_1,\mu_2)}(\varepsilon' \cdot t') = \Exp{\theta_1}(\varepsilon' \cdot t') < \Exp{\theta_1}(\varepsilon \cdot t') = \Exp{\theta_2}(t').\]
    \item We get $F_{\star(\mu_1,\mu_2)} = F_{\mu_2}$,
      so we can use Proposition \ref{prop:expandunif}.
    \item In this case, $F_{\mu_1}$ and $F_{\mu_2}$
      will cross in some non-zero point.
      $k$ is chosen so that
      \[F_{\mu_1}(\varepsilon^* \cdot t^*) = F_{\mu_2}(\varepsilon^* \cdot t^*) = F_{\nu_2}(t^*)\]
      where
      \[\varepsilon^* = \frac{(b-a) \cdot k}{d \cdot k - c \cdot k - d \cdot a + b \cdot c}\]
      and
      \[t^* = \frac{k}{\varepsilon^*} .\]
      This also means that
      \[F_{\star(\mu_1,\mu_2)}(\varepsilon \cdot t) = \begin{cases} F_{\mu_1}(\varepsilon \cdot t) & \text{if } t \leq t^* \\ F_{\mu_2}(\varepsilon \cdot t) & \text{if } t \geq t^* .\end{cases}\]
      Now, if $\varepsilon = \frac{b}{d}$,
      then $F_{\mu_2}(\varepsilon \cdot d) = \unif{a \cdot \frac{d}{b}}{d}(d) = \unif{c}{d}(d)$,
      and $F_{\mu_2}(\varepsilon \cdot t^*) \geq F_{\mu_2}(\varepsilon^* \cdot t^*) = F_{\nu_2}(t^*)$.
      Hence $F_{\mu_2}(\varepsilon \cdot t) \geq F_{\nu_2}(t)$ for all $t \geq t^*$.
      For $t \leq t^*$ we get $F_{\mu_1}(\varepsilon \cdot t) \geq F_{\nu_2}(t)$,
      and hence $F_{\star(\mu_1,\mu_2)}(\varepsilon \cdot t) \geq F_{\nu_2}(t)$.
      If $\varepsilon' < \varepsilon$,
      then $F_{\star(\mu_1,\mu_2)}(\varepsilon' \cdot d) < F_{\star(\mu_1,\mu_2)}(\varepsilon \cdot d) = 1 = \unif{c}{d}(d)$.
      
      If $\varepsilon = \varepsilon^*$,
      then $F_{\mu_1}(\varepsilon \cdot t^*) = F_{\mu_2}(\varepsilon \cdot t^*) = F_{\nu_2}(t^*)$.
      Since
      \[F_{\mu_2}(\varepsilon \cdot d) \geq F_{\mu_2}\left(\frac{b}{d} \cdot d\right) = 1 = \unif{c}{d}(d) ,\]
      we get $F_{\mu_2}(\varepsilon \cdot t) \geq F_{\nu_2}(t)$
      for all $t \geq t^*$.
      For $t \leq t^*$ we get $F_{\mu_1}(\varepsilon \cdot t) \geq F_{\nu_2}(t)$,
      and hence we conclude
      $F_{\star(\mu_1,\mu_2)}(\varepsilon \cdot t) \geq F_{\nu_2}(t)$.
      If $\varepsilon' < \varepsilon$,
      then $F_{\star(\mu_1,\mu_2)}(\varepsilon' \cdot t^*) < F_{\star(\mu_1,\mu_2)}(\varepsilon \cdot t^*) = F_{\nu_2}(t^*)$. \qedhere
  \end{enumerate}
\end{proof}

%% Logic
\section{Logical Properties of the Simulation Distance}\label{sec:logic-d}
If the distance between two processes is small,
then we would also expect that they satisfy almost the same properties.
In order to make this idea precise,
in this section we introduce and study a slight extension of Markovian logic~\cite{KMP13},
which we will call \emph{timed Markovian logic} ($\mlp$).
The syntax of $\mlp$ is given by the following grammar,
where $\alpha \in \ap$, $p \in \mathbb{Q}_{\geq 0} \cap [0,1]$, $t \in \mathbb{Q}_{\geq 0}$, and $a \in A$.

\[\mlp: \quad \varphi ::= \alpha \mid \neg \alpha \mid \ell_p t \mid m_p t \mid L_p^a \varphi \mid M_p^a \varphi \mid \varphi \land \varphi' \mid \varphi \lor \varphi'\]

The semantics of $\mlp$ is given by
\[\begin{array}{l l l l l l l}
  s \models \alpha & \; \text{iff} \; & \alpha \in L(s) & \quad & s \models \ell_p t & \; \text{iff} \; & F_s(t) \geq p \\
  s \models \neg \alpha & \; \text{iff} \; & \alpha \notin L(s) & \quad & s \models m_p t & \; \text{iff} \; & F_s(t) \leq p \\
  s \models \varphi \land \varphi' & \; \text{iff} \; & s \models \varphi \text{ and } s \models \varphi' & \quad & s \models L_p^a \varphi           & \; \text{iff} \; & \tau(s,a)(\sat{\varphi}) \geq p \\
  s \models \varphi \lor \varphi' & \; \text{iff} \; & s \models \varphi \text{ or } s \models \varphi' & \quad & s \models M_p^a \varphi           & \; \text{iff} \; & \tau(s,a)(\sat{\varphi}) \leq p \\
\end{array}\]
where $\sat{\varphi} = \{s \in S \mid s \models \varphi\}$
is the set of states satisfying $\varphi$.

We also isolate the following two fragments of $\mlp$.

\[\mlpgeq: \quad \varphi ::= \alpha \mid \neg \alpha \mid \ell_p t \mid L_p^a \varphi \mid \varphi \land \varphi' \mid \varphi \lor \varphi'\]
\[\mlpleq: \quad \varphi ::= \alpha \mid \neg \alpha \mid m_p t \mid M_p^a \varphi \mid \varphi \land \varphi' \mid \varphi \lor \varphi'\]

Intuitively, the modal formula $L_p^a \varphi$ says
that the probability of taking an $a$-transition to where $\varphi$ holds is \emph{at least} $p$,
and $M_p^a \varphi$ says the probability is \emph{at most} $p$.
$\ell_p t$ and $m_p t$ are similar in spirit,
but talk about the probability of firing a transition instead.
Thus, $\ell_p t$ says that the probability
of firing a transition before time $t$ is \emph{at least} $p$,
whereas $m_p t$ says that the probability is \emph{at most} $p$.

For any $\varphi \in \mlp$ and $\varepsilon \geq 1$
we denote the \emph{$\varepsilon$-perturbation} of $\varphi$ by $\pert{\varphi}{\varepsilon}$
and define it inductively as
\[\begin{array}{l l l}
  \pert{\alpha}{\varepsilon} = \alpha & \; & \pert{\ell_p t}{\varepsilon} = \ell_p \varepsilon \cdot t \\
  \pert{\neg \alpha}{\varepsilon} = \neg \alpha & \; & \pert{m_p t}{\varepsilon} = m_p \varepsilon \cdot t \\
  \pert{\varphi \land \varphi'}{\varepsilon}  = \pert{\varphi}{\varepsilon} \land \pert{\varphi'}{\varepsilon} & \; & \pert{L_p^a \varphi}{\varepsilon} = L_p^a \pert{\varphi}{\varepsilon} \\
  \pert{\varphi \lor \varphi'}{\varepsilon} = \pert{\varphi}{\varepsilon} \lor \pert{\varphi'}{\varepsilon} & \; & \pert{M_p^a \varphi}{\varepsilon} = M_p^a \pert{\varphi}{\varepsilon}. \\
\end{array}\]

By making use of the alternative characterisation
for simulation given in Proposition \ref{prop:folk2}
and drawing upon ideas from \cite{desharnais99},
we can now prove the following logical characterisation of the $\varepsilon$-simulation relation.

\begin{theorem}\label{thm:logic-d}
  Let $\varepsilon \in \mathbb{Q}_{\geq 0}$ with $\varepsilon \geq 1$.
  Then the following holds.
  \begin{itemize}
    \item $s_1 \simul_\varepsilon s_2 \quad$ if and only if $\quad \forall \varphi \in \mlpgeq. s_1 \models \varphi \implies s_2 \models \pert{\varphi}{\varepsilon}$.
    \item $s_1 \simul_\varepsilon s_2 \quad$ if and only if $\quad \forall \varphi \in \mlpleq. s_2 \models \pert{\varphi}{\varepsilon} \implies s_1 \models \varphi$.
  \end{itemize}
\end{theorem}
\begin{proof}%[of Theorem \ref{thm:logic-d}]
  $\bullet$ We first prove the first item.
  
  ($\implies$)
  We proceed by induction on $\varphi$.
  The cases of conjunction and disjunction are standard.
  
  Case $\varphi = \alpha$: $s_1 \models \alpha$ means that $\alpha \in L(s_1)$.
    Since $L(s_1) = L(s_2)$ we then get $s_2 \models \alpha$.
    
  Case $\varphi = \neg \alpha$: $s_1 \models \neg \alpha$ means that $\alpha \notin L(s_1)$.
    Since $L(s_1) = L(s_2)$ we then get $s_2 \models \neg \alpha$.
  
  Case $\varphi = \ell_p t$: $s_1 \models \ell_p t$ means that $F_{s_1}(t) \geq p$,
    and since $F_{s_2}(\varepsilon \cdot t) \geq F_{s_1}(t)$, we get $s_2 \models \pert{\varphi}{\varepsilon}$.
  
  Case $\varphi = L_p^a \varphi'$: $s_1 \models L_p^a \varphi'$
    means that $\tau(s_1,a)(\sat{\varphi'}) \geq p$.
    There exists a coupling $\Delta_a(s,s')$ such that
    \begin{align*}
      \tau(s_1,a)(\sat{\varphi'}) &= \sum_{s \in \sat{\varphi'}} \tau(s_1,a)(s) \\
                                  &= \sum_{s \in \sat{\varphi'}} \sum_{s' \in S} \Delta_a(s,s') \\
                                  &= \sum_{s \in \sat{\varphi'}} \sum_{s' \in \sat{\pert{\varphi'}{\varepsilon}}} \Delta_a(s,s') && \text{(ind. hyp.)} \\
                                  &\leq \sum_{s' \in \sat{\pert{\varphi'}{\varepsilon}}} \tau(s_2,a)(s') \\
                                  &= \tau(s_2,a)(\sat{\pert{\varphi'}{\varepsilon}}) ,
    \end{align*}
    and hence $s_2 \models \pert{\varphi}{\varepsilon}$.
  
  ($\impliedby$)
  We construct the relation
  \[\mathcal{R} = \{(s,s') \in S \times S \mid \forall \varphi \in \mlpgeq. s \models \varphi \implies s' \models \varphi\}\]
  and we must show that it is a $\varepsilon$-simulation relation.
  Let $(s_1',s_2') \in \mathcal{R}$ be arbitrary.
  We will first show that $L(s_1') = L(s_2')$.
  If $\alpha \in L(s_1')$, then $s_1' \models \alpha$,
  and hence $s_2' \models \alpha$, which means that $\alpha \in L(s_2')$.
  If $\alpha \notin L(s_1')$, then $s_1' \models \neg \alpha$,
  implying that $s_2' \models \neg \alpha$, so $\alpha \notin L(s_2')$.
  Therefore $L(s_1') = L(s_2')$.
  
  Next we will show that $F_{s_1'}(t) \leq F_{s_2'}(\varepsilon \cdot t)$
  for all $t \in \mathbb{R}_{\geq 0}$.
  Assume towards a contradiction that
  $F_{s_1'}(t) > F_{s_2'}(\varepsilon \cdot t)$ for some $t \in \mathbb{Q}_{\geq 0}$.
  Then there exists $q \in \mathbb{Q}_{\geq 0}$ such that
  $F_{s_1'}(t) > q > F_{s_2'}(\varepsilon \cdot t)$.
  But then $s_1' \models \ell_q t$ whereas $s_2' \not \models \ell_q \varepsilon \cdot t$,
  which contradicts how $\mathcal{R}$ was constructed.
  Hence $F_{s_1'}(t) \leq F_{s_2'}(\varepsilon \cdot t)$ for all $t \in \mathbb{Q}_{\geq 0}$.
  Now assume towards a contradiction that $F_{s_1'}(t) > F_{s_2'}(\varepsilon \cdot t)$
  for some $t \in \mathbb{R}_{\geq 0}$
  and let $\varepsilon' = F_{s_1'}(t) - F_{s_2'}(\varepsilon \cdot t) > 0$.
  By right-continuity, there exists $\delta > 0$ such that
  $\varepsilon \cdot t < c < \varepsilon \cdot t + \delta$
  implies $|F_{s_2'}(c) - F_{s_2'}(\varepsilon \cdot t)| < \varepsilon'$.
  Now pick some $q \in \mathbb{Q}_{\geq 0}$
  such that $\varepsilon \cdot t < q < \varepsilon \cdot t + \delta$.
  Then
  \[F_{s_2'}(\varepsilon \cdot t) \leq F_{s_2'}(q) < F_{s_1'}(t) \leq F_{s_1'}(q) ,\]
  which is a contradiction.
  Hence we conclude that $F_{s_1'}(t) \leq F_{s_2'}(\varepsilon \cdot t)$
  for all $t \in \mathbb{R}_{\geq 0}$.
  
  Finally we will show that $\tau(s_1',a)(C) \leq \tau(s_2',a)(R(C))$ for all $a$ and $C \subseteq S$.
  Pick an arbitrary $a$ and $C \subseteq S$.
  By construction of $\mathcal{R}$, we know that
  \[\tau(s_1',a)(\sat{\varphi}) \geq p \quad \text{implies} \quad \tau(s_2',a)(\sat{\pert{\varphi}{\varepsilon}}) \geq p \quad \text{for any } p.\]
  This implies that
  \begin{equation}\label{eq:logic}
    \tau(s_1',a)(\sat{\varphi}) \leq \tau(s_2',a)(\sat{\pert{\varphi}{\varepsilon}})
  \end{equation}
  for all $\varphi \in \mlpgeq$.
  The strategy will now be to construct a formula $\varphi$
  to exploit the inequality in Equation \eqref{eq:logic}.
  To do this, we introduce the following notation.
  For a state $s \in S$, let
  \[\tas{s} = \{\varphi \in \mlpgeq \mid s \models \varphi\} \text{ and } \tas{s}^\varepsilon = \{\varphi \in \mlpgeq \mid s \models \pert{\varphi}{\varepsilon}\} .\]
  Given a formula $\varphi \in \mlpgeq$,
  we let $\mathbb{Q}(\varphi)$ be the set of values $t \in \mathbb{Q}_{\geq 0}$
  and $p \in \mathbb{Q}_{\geq 0} \cap [0,1]$ that are used in $\varphi$.
  Furthermore, we define the depth $\depth(\varphi)$ as
  \[\depth(\varphi) =
    \begin{cases}
      0                                               & \text{if } \varphi = \ell_p t \text{ or } \varphi = m_p t , \\
      1 + \depth(\varphi')                            & \text{if } \parbox[t]{.3\textwidth}{$\varphi = L_p^a \varphi'$, $\varphi = M_p^a \varphi'$, or $\varphi = \neg \varphi'$,} \\
      1 + \max\{\depth(\varphi_1),\depth(\varphi_2)\} & \text{if } \varphi = \varphi_1 \land \varphi_2 .
    \end{cases}\]
  Finally, we let
  \[I_k = \{q \in \mathbb{Q}_{\geq 0} \mid q = l \cdot \frac{1}{j} \text{ for some } l \in \mathbb{N}_0 \text{ and } j \in \mathbb{N} \text{ where } l \leq k \text{ and } j \leq k\} .\]
  Then we can define
  \[F_k = \{\varphi \in \mlpgeq \mid \depth(\varphi) \leq k \text{ and } \mathbb{Q} \subseteq I_k\}\]
  as a finite fragment of $\mlpgeq$ and
  \[\tas{s}_k = \tas{s} \cap F_k = \{\varphi \in F_k \mid s \models \varphi\}\]
  as the restriction of $\tas{s}$ to $F_k$.
  Intuitively, $F_k$ is a better and better approximation of all formulas of $\mlpgeq$
  as $k$ increases.
  Formally, this means that $\bigcup_{k \in \mathbb{N}} F_k = \mlpgeq$,
  and hence any formula in $\mlpgeq$ will be in one of the $F_k$ for some $k$.
  Now we can construct a formula that describes the set $R(C)$.
  Note that by construction of $\mathcal{R}$ we have
  \[\mathcal{R}(C) = \{s' \in S \mid \exists s \in C. (s,s') \in \mathcal{R}\} = \{s' \in S \mid \exists s \in C. \tas{s} \subseteq \tas{s'}^\varepsilon\}\]
  and we also have
  \[\bigcup_{s \in C} \bigcap_{\varphi \in \tas{s}} \sat{\pert{\varphi}{\varepsilon}} = \bigcup_{s \in C}\{s' \in S \mid \tas{s} \subseteq \tas{s'}^\varepsilon\} = \{s' \in S \mid \exists s \in C. \tas{s} \subseteq \tas{s'}^\varepsilon\} ,\]
  so $\mathcal{R}(C) = \bigcup_{s \in C}\bigcap_{\varphi \in \tas{s}} \sat{\pert{\varphi}{\varepsilon}}$.
  
  We consider the case where $C$ is finite and the case where $C$ is infinite separately.
  Assume first that $C$ is finite.
  Now let
  \[\chi^C_k = \bigvee_{s \in C}\bigwedge_{\varphi \in \tas{s}_k} \varphi, \quad \pert{\chi^C_k}{\varepsilon} = \bigvee_{s \in C}\bigwedge_{\varphi \in \tas{s}_k} \pert{\varphi}{\varepsilon}, \quad \text{and}\]
  \[\chi^C = \bigvee_{s \in C}\bigwedge_{\varphi \in \tas{s}} \varphi.\]
  Because $C$ is finite, $\chi^C_k$, $\pert{\chi^C_k}{\varepsilon}$, and $\chi^C$
  consist of finitely many disjunctions and conjunctions and are hence formulas.
  These formulas will be the ones we use in Equation \eqref{eq:logic}.
  Note that $\sat{\chi^C} \subseteq \sat{\chi^C_k}$ for any $k$.
  Now let
  \[C_k = \sat{\chi^C_k} \quad \text{and} \quad C^\varepsilon_k = \sat{\pert{\chi^C_k}{\varepsilon}}.\]
  Then we get decreasing chains
  \[C_1 \supseteq C_2 \supseteq \dots \quad \text{and} \quad C^\varepsilon_1 \supseteq C^\varepsilon_2 \supseteq \dots\]
  of finite sets, and we will now prove that $\bigcap_{k \in \mathbb{N}} C^\varepsilon_k = \mathcal{R}(C)$.
  If $s' \in \mathcal{R}(C)$,
  then there exists $s \in C$ such that $\tas{s} \subseteq \tas{s'}^\varepsilon$,
  and hence $s' \models \bigwedge_{\varphi \in \tas{s}_k} \pert{\varphi}{\varepsilon}$
  for all $k$, so $s' \in \bigcap_{k \in \mathbb{N}} C^\varepsilon_k$.
  If $s' \notin \mathcal{R}(C)$,
  then for all $s \in C$ there exists $\varphi_s \in \mlpgeq$
  such that $s \models \varphi_s$ but $s' \not \models \pert{\varphi_s}{\varepsilon}$.
  Since $C$ is finite,
  we can fix $k'$ such that $\varphi_s \in F_{k'}$ for all $s \in C$.
  Then $s' \notin C^\varepsilon_{k'}$ because
  $s' \not \models \bigvee_{s \in C}\bigwedge_{\varphi \in \tas{s}_{k'}} \pert{\varphi}{\varepsilon}$
  since $\varphi_s \in \tas{s}_{k'}$.
  Therefore $s' \notin \bigcap_{k \in \mathbb{N}} C^\varepsilon_k$,
  so we conclude that $\mathcal{R}(C) = \bigcap_{k \in \mathbb{N}} C^\varepsilon_k$.
  
  Now, by Equation \eqref{eq:logic}, we get
  \begin{align*}
    &\phantom{{}\implies{}} \tau(s_1',a)\left(\sat{\chi^C_k}\right) \leq \tau(s_2',a)\left(\sat{\pert{\chi^C_k}{\varepsilon}}\right) \\
    &\implies \tau(s_1',a)(C_k) \leq \tau(s_2',a)(C^\varepsilon_k) && \text{for all } k \\
    &\implies \tau(s_1',a)(C_{k'}) \leq \lim_{k \rightarrow \infty} \tau(s_2',a)(C^\varepsilon_k) && \text{for a fixed } k' \\
    &\implies \tau(s_1',a)(C_{k'}) \leq \tau(s_2',a)\left(\bigcap_{k \in \mathbb{N}} C^\varepsilon_k\right) && \text{(cont. of measures)} \\
    &\implies \tau(s_1',a)(\sat{\chi^C}) \leq \tau(s_2',a)(\mathcal{R}(C)) && \sat{\chi^C} \subseteq C_{k'} \\
    &\implies \tau(s_1',a)(C) \leq \tau(s_2',a)(\mathcal{R}(C)) && C \subseteq \sat{\chi^C}.
  \end{align*}
  
  Next assume that $C$ is countably infinite and let $\{C_k\}_{k \in \mathbb{N}}$
  be an increasing sequence of finite sets such that $\bigcup_{k \in \mathbb{N}} C_k = C$.
  Since every $C_k$ is finite, we get
  \[\tau(s_1',a)(C_k) \leq \tau(s_2',a)(\mathcal{R}(C_k))\]
  from what we just proved for the finite case.
  By continuity of measures, this implies
  \[\tau(s_1',a)(C) = \tau(s_1',a)\left(\bigcup_{k \in \mathbb{N}} C_k\right) \leq \tau(s_2',a)\left(\bigcup_{k \in \mathbb{N}} \mathcal{R}(C_k)\right) = \tau(s_2',a)(\mathcal{R}(C)).\]

  $\bullet$ We now prove the second item.
  
  ($\implies$)
  For this we first prove that
  \[s_1 \simul s_2 \quad \text{implies} \quad \forall \varphi \in \mlpleq. s_2 \models \varphi \implies s_1 \models \varphi\]
  by induction on $\varphi$.
  We only consider here the cases of $\varphi = m_p t$ and $\varphi = M^a_p \varphi'$,
  since the other cases are as in the first item.
  
  Case $\varphi = m_p t$: $s_2 \models m_p t$ means that $F_{s_2}(t) \leq p$.
    Since $s_1 \simul s_2$, we know that $F_{s_1}(t) \leq F_{s_2}(t)$,
    so $F_{s_2}(t) \leq p$, implying $s_1 \models m_p t$.
  
  Case $\varphi = M^a_p \varphi'$: $s_2 \models M^a_p \varphi'$ means that $\tau(s_2,a)(\sat{\varphi'}) \leq p$.
    By induction hypothesis, we know that $\sat{\varphi'}$ is $\simul$-closed,
    and hence we get $\tau(s_1,a)(\sat{\varphi'}) \leq \tau(s_2,a)(\sat{\varphi'})$,
    so $s_1 \models M^a_p \varphi'$.
  
  We now prove the claim that
  \[s_1 \simul_\varepsilon s_2 \quad \text{implies} \quad \forall \varphi \in \mlpleq. s_2 \models \pert{\varphi}{\varepsilon} \implies s_1 \models \varphi\]
  by induction on $\varphi$.
  The only case left to consider is the case where $\varphi = M^a_p \varphi'$,
  since the remaining cases are as before
  
  Case $\varphi = M^a_p \varphi'$:
    We have $\tau(s_2,a)(\sat{\pert{\varphi'}{\varepsilon}}) \leq p$.
    Now, $\tau(s_2,a)(\sat{\pert{\varphi'}{\varepsilon}}) = \tau(\pert{s_2}{\varepsilon},a)(\sat{\varphi'})$,
    and since we now know that $\sat{\varphi'}$ is $\simul$-closed, we get
    \[p \geq \tau(\pert{s_2}{\varepsilon},a)(\sat{\varphi'}) \geq \tau(s_1,a)(\sat{\varphi'}),\]
    meaning that $s_1 \models M^a_r \varphi'$.
  
  ($\impliedby$) Same as the first item.
\end{proof}

As a special case of Theorem \ref{thm:logic-d},
we have also shown that $\mlpgeq$ and $\mlpleq$ characterise simulation for SMDPs.
Conceptually, Theorem \ref{thm:logic-d} says
that if $s_1$ $\varepsilon$-simulates $s_2$,
then $s_2$ satisfies the $\varepsilon$-perturbation of any property that $s_2$ satisfies
for the $\mlpgeq$ fragment of $\mlp$,
and vice versa for the $\mlpleq$ fragment.

By Lemma \ref{lem:correctness} and Theorem \ref{thm:logic-d},
we get the following corollary,
connecting our simulation distance with the properties
expressible in the logic $\mlp$.

\begin{corollary}
  Let $\varepsilon \in \mathbb{Q}_{\geq 0}$ with $\varepsilon \geq 1$.
  For finite SMDPs the following holds.
  \begin{itemize}
    \item $\simdist(s_1,s_2) \leq \varepsilon \quad$ if and only if $\quad \forall \varphi \in \mlpgeq. s_1 \models \varphi \implies s_2 \models \pert{\varphi}{\varepsilon}$.
    \item $\simdist(s_1,s_2) \leq \varepsilon \quad$ if and only if $\quad \forall \varphi \in \mlpleq. s_2 \models \pert{\varphi}{\varepsilon} \implies s_1 \models \varphi$.
  \end{itemize}
\end{corollary}

By Proposition \ref{prop:folk3},
we also get a logical characterisation of bisimulation for SMDPs in terms of $\mlp$,
which is simpler than the one given in \cite{neuhausser2007,SZG14}.

\begin{theorem}\label{thm:logic2}
  \[s_1 \sim s_2 \quad \text{if and only if} \quad \forall \varphi \in \mlp. \; s_1 \models \varphi \iff s_2 \models \varphi .\]
\end{theorem}
\begin{proof}%[of Theorem \ref{thm:logic2}]
  ($\implies$)
  We first prove that $s_1 \models \varphi$ implies $s_2 \models \varphi$ for all $\varphi \in \mlp$.
  The proof proceeds by induction on $\varphi$.
  The cases of disjunction and conjunction are standard,
  and the cases of $\varphi = \alpha$ and $\varphi = \neg \alpha$
  are the same as in the proof of Theorem \ref{thm:logic}.
  
  Case $\varphi = \ell_p t$: $s_1 \models \ell_p t$ means $F_{s_1}(t) \geq p$,
  and since $F_{s_1}(t) = F_{s_2}(t)$, we get $s_2 \models \ell_p t$.
  
  Case $\varphi = m_p t$: Same argument as $\ell_p t$.
  
  Case $\varphi = L_p^a \varphi'$: $s_1 \models L_p^a \varphi'$
  means $\tau(s_1,a)(\sat{\varphi'}) \geq p$.
  We know that there exists a coupling $\Delta_a$ such that
  \begin{align*}
    \tau(s_1,a)(\sat{\varphi'}) &= \sum_{s \in \sat{\varphi'}} \tau(s_1,a)(s) \\
                                &= \sum_{s \in \sat{\varphi'}} \sum_{s' \in S} \Delta_a(s,s') \\
                                &= \sum_{s \in \sat{\varphi'}} \sum_{s' \in \sat{\varphi'}} \Delta_a(s,s') && \text{(ind. hyp.)} \\
                                &\leq \sum_{s' \in \sat{\varphi'}} \tau(s_2,a)(s') \\
                                &= \tau(s_2,a)(\sat{\varphi'}) ,
  \end{align*}
  so $s_2 \models L_p^a \varphi'$.
  
  Case $\varphi = M_p^a \varphi'$: Same argument as $L_p^a \varphi'$.

  Next we prove that $s_2 \models \varphi$
  implies $s_1 \models \varphi$ for all $\varphi \in \mlp$,
  again by induction on $\varphi$.
  All cases except $\varphi = L_p^a \varphi'$ and $\varphi = M_p^a \varphi'$ are as before.
  
  Case $\varphi = L_p^a \varphi'$:
    We have $\tau(s_2,a)(\sat{\varphi'}) \geq p$.
    Since $s_1 \sim s_2$, in particular we have $s_2 \simul s_1$,
    so by Theorem~\ref{thm:logic-d} and Proposition~\ref{prop:folk2} we get
    \[\tau(s_2,a)(\sat{\varphi'}) \leq \tau(s_1,a)(\sat{\varphi'}),\]
    and hence $s_1 \models L_p^a \varphi'$.
    
  Case $\varphi = M_p^a \varphi'$: Same argument as $L_p^a \varphi'$.
  
  ($\impliedby$)
  We have assumed that $\forall \varphi \in \mlp. s_1 \models \varphi \iff s_2 \models \varphi$,
  and hence we also get
  $\forall \varphi \in \mlpgeq. s_1 \models \varphi \implies s_2 \models \varphi$ and
  $\forall \varphi \in \mlpleq. s_2 \models \varphi \implies s_1 \models \varphi$.
  By Theorem \ref{thm:logic} we then get $s_1 \simul s_2$ and $s_2 \simul s_1$,
  and hence Proposition \ref{prop:folk3} implies $s_1 \sim s_2$.
\end{proof}

With the logical characterisation of $\varepsilon$-simulation in hand,
we can now prove the promised result that the kernel of the simulation distance is simulation.
For this, we first need the following technical lemma.
Given a weight function $\Delta_a$, we let
\[\supp{\Delta_a} = \{(s,s') \in S \times S \mid \Delta_a(s,s') > 0\}.\]

\begin{lemma}\label{lem:finite-support}
  Assume that $M = (S, \tau, \rho, L)$ is finitely supported
  and consider $s_1,s_2 \in S$.
  If $s_1 \simul_\varepsilon s_2$ for all $\varepsilon > 1$ then
  for all $\varepsilon > 1$ and $a \in L$ there exists a $\varepsilon$-simulation relation $R_\varepsilon$
  and a weight function $\Delta_a$ such that for any $\varepsilon > \varepsilon' > 1$
  there exists a $\varepsilon'$-simulation relation $R_{\varepsilon'}$
  with $\supp{\Delta_a} \subseteq R_{\varepsilon'}$.
\end{lemma}
\begin{proof}
  Let $\varepsilon > 1$ and $a \in L$. Because we know that $s_1 \simul_\varepsilon s_2$,
  there exists a $\varepsilon$-simulation relation $R_\varepsilon$ and a weight function $\Delta_a$ with $\supp{\Delta_a} \subseteq R_\varepsilon$.
  Now let $\varepsilon > \varepsilon_1 > 1$.
  Because $s_1 \simul_{\varepsilon_1} s_2$,
  we know that there exists a $\varepsilon_1$-simulation relation $R_{\varepsilon_1}$ and a weight function $\Delta_a^1$ such that $\supp{\Delta_a^1} \subseteq R_{\varepsilon_1}$.
  
  If $\supp{\Delta_a} = \supp{\Delta_a^1}$, we are done.
  If not, it may be the case that for any $\varepsilon_1 > \varepsilon_2 > 1$ there exists a $\varepsilon_2$-simulation relation $R_{\varepsilon_2}$
  and a weight function $\Delta_a^2$ such that $\supp{\Delta_a^2} \subseteq R_{\varepsilon_2}$.
  This would imply by monotonicity that for all $\varepsilon_1 > 1$ there exists a $\varepsilon_1$-simulation relation and a weight function $\Delta_a^1$
  such that for any $\varepsilon_1 > \varepsilon_2 > 1$ there exists a $\varepsilon_2$-simulation relation $R_{\varepsilon_2}$ such that $\supp{\Delta_a^1} \subseteq R_{\varepsilon_2}$,
  in which case we are also done.
  
  If this is not the case, then there must exist some $\varepsilon_1 > \varepsilon_2 > 1$ such that $\supp{\Delta_a^1} \not\subseteq R_{\varepsilon_2}$
  for any $\varepsilon_2$-simulation relations $R_{\varepsilon_2}$.
  However, we know that $s_1 \simul_{\varepsilon_2} s_2$,
  so there exists a $\varepsilon_2$-simulation relation $R_{\varepsilon_2}$ and a weight function $\Delta_a^2$ such that $\supp{\Delta_a^2} \subseteq R_{\varepsilon_2}$.
  Note that we must have $\supp{\Delta_a^1} \neq \supp{\Delta_a^2}$
  because we have $\supp{\Delta_a^1} \not\subseteq R_{\varepsilon_2}$ but $\supp{\Delta_a^2} \subseteq R_{\varepsilon_2}$.
  If $\supp{\Delta_a} = \supp{\Delta_a^2}$, we are done.
  If not, it may be the case that for any $\varepsilon_2 > \varepsilon_3 > 1$
  there exists a $\varepsilon_3$-simulation relation $R_{\varepsilon_3}$ with $\supp{\Delta_a^2} \subseteq R_{\varepsilon_3}$.
  This would by monotonicity again imply that we are done.
  
  If this is not the case, then there must exist some $\varepsilon_2 > \varepsilon_3 > 1$
  such that $\supp{\Delta_a^2} \not\subseteq R_{\varepsilon_3}$ for all $\varepsilon_3$-simulation relations $R_{\varepsilon_3}$.
  However, we know that $s_1 \simul_{\varepsilon_3} s_2$,
  so there exists a $\varepsilon_3$-simulation relation $R_{\varepsilon_3}$ and a weight function $\Delta_a^3$ such that $\supp{\Delta_a^3} \subseteq R_{\varepsilon_3}$.
  Note that $\supp{\Delta_a^2} \neq \supp{\Delta_a^3}$ and also $\supp{\Delta_a^1} \neq \supp{\Delta_a^3}$
  since $R_{\varepsilon_3}$ is also a $\varepsilon_2$-simulation relation.
  If $\supp{\Delta_a} = \supp{\Delta_a^3}$, we are done.
  If not, it may be the case that for any $\varepsilon_3 > \varepsilon_4 > 1$
  there exists a $\varepsilon_4$-simulation relation $R_{\varepsilon_4}$ with $\supp{\Delta_a^3} \subseteq R_{\varepsilon_4}$
  which, by monotonicity, would imply that we are done.
  
  Continuing in this way, we get a sequence
  \[\supp{\Delta_a^1}, \supp{\Delta_a^2}, \supp{\Delta_a^3}, \dots,\]
  all pairwise different from each other.
  However, since
  \[\Delta_a(s,s') > 0 \quad \text{implies} \quad \tau(s_1,a)(s) > 0 \text{ and } \tau(s_2,a)(s') > 0,\]
  $\supp{\Delta_a}$ seen as a function of $\Delta_a$ can only take on finitely many values.
  Hence the process must eventually stop and we find witnesses for the statement of the lemma.
\end{proof}

\begin{theorem}
  \[s_1 \simul s_2 \quad \text{implies} \quad \simdist(s_1,s_2) = 1.\]
  For finitely supported SMDPs, it also holds that
  \[\simdist(s_1,s_2) = 1 \quad \text{implies} \quad s_1 \simul s_2.\]
\end{theorem}
\begin{proof}
  The first point is immediate: If $s_1 \simul s_2$, this means that $s_1 \simul_1 s_2$, so $\simdist(s_1,s_2) = 1$.
  
  For the second point, assume that $\simdist(s_1,s_2) = 1$.
  This means that either $s_1 \simul_1 s_2$, in which case we are done,
  or $s_1 \simul_\varepsilon s_2$ for all $\varepsilon > 1$,
  in which case we wish to prove that this implies that
  \begin{equation}\label{eq:logic-implication}
    s_1 \models \varphi \implies s_2 \models \varphi \quad \text{for all } \varphi \in \mlpgeq.
  \end{equation}
  This would imply, by Theorem~\ref{thm:logic-d},
  that $s_1 \simul s_2$, and we are done.
  Hence we now prove the claim in \eqref{eq:logic-implication} by induction on $\varphi$.
  
  ($\varphi = \alpha$ or $\varphi = \neg \alpha$):
  Choose some $\varepsilon > 1$.
  Then there exists a $\varepsilon$-simulation relation $R$
  such that $s_1 R s_2$.
  This implies that $L(s_1) = L(s_2)$,
  so if $s_1 \models \varphi$,
  then also $s_2 \models \varphi$.
  
  ($\varphi = \varphi_1 \lor \varphi_2$):
  If $s_1 \models \varphi_1 \lor \varphi_2$,
  then $s_1 \models \varphi_1$ or $s_1 \models \varphi_2$.
  By induction hypothesis, this implies that $s_2 \models \varphi_1$ or $s_2 \models \varphi_2$,
  so $s_2 \models \varphi_1 \lor \varphi_2$.
  
  ($\varphi = \varphi_1 \land \varphi_2$):
  Similar to the case $\varphi = \varphi_1$.
  
  ($\varphi = \ell_p t$):
  Assume $s_1 \models \ell_p t$,
  which means that $F_{s_1}(t) \geq p$.
  Assume towards a contradiction that $F_{s_2}(t) < p$,
  and let $\varepsilon' = p - F_{s_2}(t) > 0$.
  Then there exists $\delta > 0$ such that for any $t < x < t + \delta$
  we have $F_{s_2}(x) - F_{s_2}(t) < \varepsilon'$.
  If $t = 0$ then for any $\varepsilon > 0$
  we have
  \[p > F_{s_2}(t) = F_{s_2}(0) = F_{s_2}(\varepsilon \cdot t) \geq F_{s_1}(t) \geq p,\]
  which is a contradiction.
  If $t > 0$, then choose an $\varepsilon > 0$ such that
  $1 < \varepsilon < \frac{t + \delta}{t}$,
  meaning that $t < \varepsilon \cdot t < t + \delta$.
  By right-continuity, this implies that $F_{s_2}(\varepsilon \cdot t) - F_{s_2}(t) < \varepsilon'$.
  Hence we get
  \[p > F_{s_2}(t) \geq F_{s_2}(\varepsilon \cdot t) \geq F_{s_1}(t) \geq p,\]
  which is also a contradiction.
  
  ($\varphi = L_p^a \varphi'$):
  Assume $s_1 \models L_p^a \varphi'$, meaning that $\tau(s_1,a)(\sat{\varphi'}) \geq p$,
  and choose some $\varepsilon > 1$.
  By Lemma~\ref{lem:finite-support},
  there exists a $\varepsilon$-simulation relation $R_\varepsilon$
  and a coupling $\Delta_a$ with $\supp{\Delta_a} \subseteq R_\varepsilon$ such that
  for any $\varepsilon > \varepsilon' > 1$
  there exists a $\varepsilon'$-simulation relation $R_{\varepsilon'}$ with $\supp{\Delta_a} \subseteq R_{\varepsilon}'$.
  We then get
  \begin{align*}
    p &\leq \tau(s_1,a)(\sat{\varphi'}) \\
      &= \sum_{s \in \sat{\varphi'}} \tau(s_1,a)(s) \\
      &= \sum_{s \in \sat{\varphi'}} \sum_{s' \in S} \Delta_a(s,s') \\
      &= \sum_{(s,s') \in (\sat{\varphi'} \times S) \cap \supp{\Delta_a}} \Delta_a(s,s').
  \end{align*}
  
  Now, we know that for any $\varepsilon > \varepsilon' > 1$
  there exists a $\varepsilon'$-simulation relation $R_{\varepsilon'}$ such that $\supp{\Delta_a} \subseteq R_{\varepsilon'}$.
  This means that for any $(s,s') \in (\sat{\varphi'} \times S) \cap \supp{\Delta_a}$
  we have $s \simul_{\varepsilon'} s'$.
  By monotonicity, we therefore get that $s \simul_{\varepsilon'} s'$ for any $\varepsilon' > 1$.
  The induction hypothesis then gives
  \begin{align*}
    &\phantom{{}={}} \sum_{(s,s') \in (\sat{\varphi'} \times S) \cap \supp{\Delta_a}} \Delta_a(s,s') \\
    &= \sum_{(s,s') \in (\sat{\varphi'} \times \sat{\varphi'}) \cap \supp{\Delta_a}} \Delta_a(s,s') \\
    &= \sum_{s \in \sat{\varphi'}} \sum_{s' \in \sat{\varphi'}} \Delta_a(s,s') \\
    &\leq \sum_{s \in S} \sum_{s' \in \sat{\varphi'}} \Delta_a(s,s') \\
    &= \sum_{s' \in \sat{\varphi'}} \tau(s_2,a)(s') \\
    &= \tau(s_2,a)(\sat{\varphi'}),
  \end{align*}
  which implies that $s_2 \models L_p^a \varphi'$.
\end{proof}

\subsection{Reachability Properties}\label{sec:linear}
We will now argue that the simulation distance
behaves nicely also with respect to linear-time properties,
by proving preservation of reachability properties up to perturbations.

The probability of reaching a given set of states in an SMDP depends on the
choice of actions in each state. 
The non-determinism introduced by this choice is typically
resolved by means of \emph{schedulers}. 
Here we consider probabilistic schedulers $\sigma$ of type $S^* \rightarrow \dist(A)$,
telling us what the probability is of selecting an action $a \in A$
depending on the history of the states visited so far.

Given a SMDP $M = (S, \tau, \rho, L)$,
a \emph{path} in $M$ is a sequence
\[\pi = (s_1, t_1), (s_2,t_2), \dots ,\]
where $s_i \in S$ and $t_i \in \mathbb{R}_{\geq 0}$.
Intuitively, a path $\pi$ denotes an execution of the SMDP,
where $s_i$ denotes the $i$th state visited,
and $t_i$ denotes the time spent in $s_i$.
We denote by $\paths(M)$ the set of all paths in $M$,
and we let $\pi\sproj{i} = s_i$ and $\pi\tproj{i} = t_i$.

Let $X \subseteq S$. Then
\[\eventually{t} X = \{\pi \in \Pi(M) \mid \exists i \in \mathbb{N}. \pi\sproj{i} \in X \text{ and } \sum_{j = 1}^{i - 1} \pi\tproj{j} \leq t\}\]
is the set of paths that eventually reach a state in $X$
and does so within time $t$.

Given a scheduler $\sigma$, we define a probability
\begin{align*}
  \prob^\sigma_{s}(S_1) &= \sum_{a \in A}\sum_{s' \in S} \tau^\sigma(s,a)(s') \cdot \rho(s) \\
  \prob^{\sigma}_{s}(S_1, S_2, \dots, S_n) &= \sum_{a \in A}\sum_{s' \in S} \tau^\sigma(s,a)(s') \cdot (\rho(s) * \prob^\sigma_{s'}(S_2, \dots, S_n))
\end{align*}
through the usual cylinder construction.
Then $\prob^\sigma_s(S_1, \dots, S_n)(t)$
is the probability, starting from $s$ and under the scheduler $\sigma$,
to first visit a state in $S_1$,
then a state in $S_2$, and so on, until a state in $S_n$ is reached,
and the total time elapsed is at most $t$.

\begin{lemma}\label{lem:cylinders}
  Let $\beta$ be a Boolean combination of atomic propositions.
  If $s_1 \simul_\varepsilon s_2$,
  then for any scheduler $\sigma$ there exists a scheduler $\sigma'$
  such that
  \[\prob^\sigma_{s_1}(\underbrace{\sat{\beta}^c, \dots, \sat{\beta}^c}_{n - 1 \text{ times}}, \sat{\beta})(t) \leq \prob^{\sigma'}_{s_2}(\underbrace{\sat{\beta}^c, \dots, \sat{\beta}^c}_{n-1 \text{ times}}, \sat{\beta})(\varepsilon \cdot t)\]
  for all $n \in \mathbb{N}$ and $t \in \mathbb{R}_{\geq 0}$.
\end{lemma}
\begin{proof}
  Let $R$ be a $\varepsilon$-simulation relation
  witnessing that $s_1 \simul_\varepsilon s_2$.
  
  \textbf{Case $n = 1$:}
  For each $a \in A$ there exists a coupling $\Delta_a$ such that
  \begin{align*} 
    \prob^\sigma_{s_1}(\sat{\beta})(t)
    &= \sum_{a \in A}\sum_{s \in \sat{\beta}} \tau(s_1,a)(s) \cdot \sigma(s_1)(a) \cdot \rho(s_1)(t) \\
    &= \sum_{a \in A}\sum_{s \in \sat{\beta}}\sum_{s' \in S} \Delta_a(s,s') \cdot \sigma(s_1)(a) \cdot \rho(s_1)(t) .
  \end{align*}
  If $s \in \sat{\beta}$ and $s' \notin \sat{\beta}$,
  then $s \not\simul_{\varepsilon} s'$,
  and hence $(s,s') \notin R$, so $\Delta_a(s,s') = 0$.
  We therefore get
  \begin{align*}
    \prob^\sigma_{s_1}(\sat{\beta})(t)
    &= \sum_{a \in A}\sum_{s \in \sat{\beta}}\sum_{s' \in \sat{\beta}} \Delta_a(s,s') \cdot \sigma(s_1)(a) \cdot \rho(s_1)(t) \\
    &= \sum_{a \in A}\sum_{s' \in \sat{\beta}}\sum_{s \in \sat{\beta}} \Delta_a(s,s') \cdot \sigma(s_1)(a) \cdot \rho(s_1)(t) \\
    &\leq \sum_{a \in A}\sum_{s' \in \sat{\beta}}\sum_{s \in S} \Delta_a(s,s') \cdot \sigma(s_1)(a) \cdot \rho(s_1)(t) \\
    &= \sum_{a \in A}\sum_{s' \in \sat{\beta}} \tau(s_2,a)(s') \cdot \sigma(s_1)(a) \cdot \rho(s_1)(t) .
  \end{align*}
  Now we define $\sigma'(s_2)(a) = \sigma(s_1)(a)$
  and observe that $\rho(s_1)(t) \leq \rho(s_2)(\varepsilon \cdot t)$
  since we have assumed $s_1 \simul_\varepsilon s_2$.
  Hence we get
  \begin{align*}
    \prob^\sigma_{s_1}(\sat{\beta})(t) 
    &\leq \sum_{a \in A}\sum_{s' \in \sat{\beta}} \tau(s_2,a)(s') \cdot \sigma'(s_2)(a) \cdot \rho(s_2)(\varepsilon \cdot t) \\
    &= \prob^{\sigma'}_{s_2}(\sat{\beta})(\varepsilon \cdot t) .
  \end{align*}
  
  \textbf{Case $n > 1$:}
  For any $a \in A$ we again get a coupling $\Delta_a$ such that
  \begin{align*}
    &\phantom{{}={}}\prob^\sigma_{s_1}(\underbrace{\sat{\beta}^c, \dots, \sat{\beta}^c}_{n-1 \text{ times}}, \sat{\beta})(t) \\
    &= \sum_{a \in A}\sum_{s \in \sat{\beta}^c} \tau(s_1,a)(s) \cdot \sigma(s_1)(a) \cdot (\rho(s_1) * \prob^\sigma_s(\underbrace{\sat{\beta}^c, \dots, \sat{\beta}^c}_{n - 2 \text{ times}}, \sat{\beta}))(t) \\
    &= \sum_{a \in A}\sum_{s \in \sat{\beta}^c}\sum_{s' \in S} \Delta_a(s,s') \cdot \sigma(s_1)(a) \cdot (\rho(s_1) * \prob^\sigma_s(\underbrace{\sat{\beta}^c, \dots, \sat{\beta}^c}_{n-2 \text{ times}}, \sat{\beta}))(t) \\
    &= \sum_{a \in A}\sum_{s \in \sat{\beta}^c}\sum_{s' \in \sat{\beta}^c} \Delta_a(s,s') \cdot \sigma(s_1)(a) \cdot (\rho(s_1) * \prob^\sigma_s(\underbrace{\sat{\beta}^c, \dots, \sat{\beta}^c}_{n-2 \text{ times}}, \sat{\beta}))(t) .
  \end{align*}
  Since $s \not\simul_\varepsilon s'$ implies $\Delta_a(s,s') = 0$,
  any term where $s \not\simul_\varepsilon s'$ contributes nothing to the sum.
  Hence we may assume that $s \simul_\varepsilon s'$.
  By induction hypothesis, we then get that for any $s'$ there exists $\sigma'_{s'}$ such that
  \[\prob^\sigma_s(\underbrace{\sat{\beta}^c, \dots, \sat{\beta}^c}_{n-2 \text{ times}}, \sat{\beta})(t) \leq \prob^{\sigma'_{s'}}_{s'}(\underbrace{\sat{\beta}^c, \dots, \sat{\beta}^c}_{n-2 \text{ times}}, \sat{\beta})(\varepsilon \cdot t) .\]
  Now let $w \in S^*$ and define
  \[\sigma''(s' w)(a) = \sigma'_{s'}(w)(a) \quad \text{and} \sigma''(s_2)(a) = \sigma(s_1)(a).\]
  By Proposition \ref{prop:conv} we get
  \[(\rho(s_1) * \prob^{\sigma}_s(\underbrace{\sat{\beta}^c, \dots, \sat{\beta}^c}_{n-2 \text{ times}}, \sat{\beta}))(t) \leq (\rho(s_2) * \prob^{\sigma'_{s'}}_{s'}(\underbrace{\sat{\beta}^c, \dots, \sat{\beta}^c}_{n-2 \text{ times}}, \sat{\beta}))(\varepsilon \cdot t) .\]
  Hence we get
  \begin{align*}
    &\phantom{{}={}} \prob^\sigma_{s_1}(\underbrace{\sat{\beta}^c, \dots, \sat{\beta}^c}_{n-1 \text{ times}}, \sat{\beta})(t) \\
    &\leq \sum_{a \in A}\sum_{s' \in \sat{\beta}^c} \tau(s_2,a)(s') \cdot \sigma''(s_2)(a) \cdot (\rho(s_2) * \prob^{\sigma''}_{s'}(\underbrace{\sat{\beta}^c, \dots, \sat{\beta}^c}_{n-2 \text{ times}}, \sat{\beta}))(\varepsilon \cdot t) \\
    &= \prob^{\sigma''}_{s_2}(\underbrace{\sat{\beta}^c, \dots, \sat{\beta}^c}_{n-1 \text{ times}}, \sat{\beta})(\varepsilon \cdot t) . \qedhere
  \end{align*}
\end{proof}

Given our notion of $\varepsilon$-simulation, we can prove the following result.

\begin{theorem}\label{thm:linear}
  Let $\beta$ be a Boolean combination of atomic propositions.
  If we have $s_1 \simul_\varepsilon s_2$, then
  for any scheduler $\sigma$ there exists a scheduler $\sigma'$
  such that
  \begin{align*}
    \prob^\sigma_{s_1}(\eventually{t} \sat{\beta}) \leq 
    \prob^{\sigma'}_{s_2}(\eventually{\varepsilon \cdot t} \sat{\beta}) 
    &&
    (\text{or equivalently, } \prob^\sigma_{s_1}(\neg \eventually{t} \sat{\beta}) \geq 
    \prob^{\sigma'}_{s_2}(\neg \eventually{\varepsilon \cdot t} \sat{\beta}) ).
  \end{align*}
\end{theorem}
\begin{proof}%[of Theorem \ref{thm:linear}]
  First note that for any $s$ and $\sigma$,
  we have
  \[\prob^\sigma_s(\eventually{t} \sat{\beta}) = \sum_{n \in \mathbb{N}} \prob^\sigma_s(\sat{\beta}^{t}_n) ,\]
  where
  \[\sat{\beta}^t_n = \{\pi \in \paths(M) \mid \pi\sproj{n} \in \sat{\beta}, \forall k < n. \pi\sproj{k} \notin \sat{\beta}, \text{ and } \sum_{j = 1}^{n-1} \pi\tproj{j} \leq t\} .\]
  We will now argue that for any $\sigma$ there exists $\sigma'$ such that
  \[\prob^\sigma_{s_1}(\sat{\beta}^t_n) \leq \prob^{\sigma'}_{s_2}(\sat{\beta}^{\varepsilon \cdot t}_n)\]
  for any $n \in \mathbb{N}$ and $t \in \mathbb{R}_{\geq 0}$.
  
  \textbf{Case $n = 1$:}
  In this case we have $\prob^\sigma_{s_1}(\sat{\beta}^t_n) = \mathbbm{1}_{\sat{\beta}}(s_1)$
  and $\prob^{\sigma'}_{s_2}(\sat{\beta}^{\varepsilon \cdot t}_n) = \mathbbm{1}_{\sat{\beta}}(s_2)$.
  Since $s_1 \simul_\varepsilon s_2$, we get $s_1 \in \sat{\beta}$ if and only if $s_2 \in \sat{\beta}$,
  and hence $\prob^\sigma_{s_1}(\sat{\beta}^t_n) = \prob^{\sigma'}_{s_2}(\sat{\beta}^{\varepsilon \cdot t}_n)$
  for any $\sigma$ and $\sigma'$.
  
  \textbf{Case $n > 1$:}
  In this case we have
  \[\prob^\sigma_{s_1}(\sat{\beta}^t_n) = \mathbbm{1}_{\sat{\beta}^c}(s_1) \cdot \prob^\sigma_{s_1}(\underbrace{\sat{\beta}^c, \dots, \sat{\beta}^c}_{n-1 \text{ times}}, \sat{\beta})(t)\]
  and
  \[\prob^{\sigma'}_{s_2}(\sat{\beta}^{\varepsilon \cdot t}_n) = \mathbbm{1}_{\sat{\beta}^c}(s_2) \cdot \prob^{\sigma'}_{s_2}(\underbrace{\sat{\beta}^c, \dots, \sat{\beta}^c}_{n-1 \text{ times}}, \sat{\beta})(\varepsilon \cdot t). \]
  Since $s_1 \simul_\varepsilon s_2$,
  we have $\mathbbm{1}_{\sat{\beta}^c}(s_1) = \mathbbm{1}_{\sat{\beta}^c}(s_2)$.
  The result then follows from Lemma~\ref{lem:cylinders}.
\end{proof}

Note that the above result might find useful applications for speeding up the computation 
time required by model checking tools to disprove certain types of reachability properties.
For example, consider the atomic proposition $\mathsf{bad}$,
identifying all the states considered ``not safe'' in the SMDP.
Usually, given a process $s$, one wants to verify that,
under all possible schedulers $\sigma$,
the probability $\prob^\sigma_{s}(\neg \eventually{t} \sat{\mathsf{bad}})$
is above a certain threshold value $\delta \leq 1$,
meaning that the SMDP is unlikely to end up
in an unsafe configuration within a time horizon bounded by $t$. 
Then, to disprove this property one only needs to provide a scheduler 
$\sigma'$ and a process $s'$ such that $s' \simul_\varepsilon s$
and $\prob^{\sigma'}_{s'}(\neg \eventually{\frac{t}{\varepsilon}} \sat{\mathsf{bad}}) < \delta$.
Indeed, given that $s' \simul_\varepsilon s$,  by Theorem~\ref{thm:linear}
\[
  \prob^{\sigma'}_{s'}\left(\neg \eventually{\frac{t}{\varepsilon}} \sat{\mathsf{bad}}\right) < \delta
  \stackrel{Th. \ref{thm:linear}}{\implies}
  \exists \sigma. \; \prob^\sigma_{s}(\neg \eventually{t} \sat{\mathsf{bad}}) < \delta  .
\]
Since $s$ simulates $s'$, $s'$ can be thought of as a simplified abstraction of $s$,
which is usually a smaller process.
Hence, finding a scheduler $\sigma'$ for $s'$ which gives a counterexample
may be much simpler than finding one for $s$.
Moreover, the above technique is robust to perturbations of $\varepsilon$.

%% Topology
\section{The Topology of TML}
A common practice in science and engineering is that of approximating and refining models.
We would therefore like to ensure that whenever we make better and better approximations
of a model, whatever property holds for the approximations should also hold
for the model that is approximated.
In our terms, this means that we would like the sets
satisfying formulas in $\mlp$ to be \emph{closed},
because then we would know that whenever a sequence of states $\{s_n\}$
which converges to some state $s$ satisfies some property of $\mlp$,
then $s$ also satisfies that property.

Since the concept of open and closed sets are topological concepts,
we introduce the topology generated by our distance.
Note that the concept of closed set and sequentially closed set
need not coincide in arbitrary topological spaces.
However, they do coincide for hemimetric spaces,
since these are first-countable \cite{goubault2013}.
Moreover, hemimetric spaces are not in general Hausdorff,
so limits need not be unique.

Because the distance is non-symmetric,
we can in fact generate two different topologies.
For $r > 1$, the open balls of the form
\[\mathcal{B}_r^L(s) = \{s' \mid \simdist(s,s') < r\}\]
generate the \emph{left-centered topology} and open balls of the form
\[\mathcal{B}_r^R(s) = \{s' \mid \simdist(s',s) < r\}\]
generate the \emph{right-centered topology}.
These two topologies behave differently, as we will now show.

\begin{lemma}\label{lem:right}
  The following holds in the right-centered topology.
  \begin{enumerate}
    \item $\sat{\ell_p t}$ is closed.
    \item If $p = 0$, then $\sat{\ell_p t}$ is open.
    \item If $p > 0$, then $\sat{\ell_p t}$ is not open.
    \item If $p = 1$, then $\sat{m_p t}$ is closed.
    \item If $p < 1$, then $\sat{m_p t}$ is not closed.
  \end{enumerate}
\end{lemma}
\begin{proof}
  \begin{enumerate}
    \item Let $\{s_k\}$ be a sequence of states such that $s_k \in \sat{\ell_p t}$ for all $k$
      and $\lim_{k \rightarrow \infty} s_k \ni s$. We must show that $s \in \sat{\ell_p t}$.
      Assume towards a contradiction that $s \notin \sat{\ell_p t}$
      and let $\varepsilon > 1$.
      First note that if $t = 0$, then $F_{s_k}(0) \geq p$ for all $k$.
      Since $\lim_{k \rightarrow \infty} s_k \ni s$,
      there must exist some $n$ such that $\simdist(s_n, s) < \varepsilon$.
      But then
      \[p \leq F_{s_n}(0) \leq F_s(\varepsilon \cdot 0) = F_s(0) < p,\]
      which is a contradiction.
      
      We can therefore now assume that $t > 0$.
      Let $p - F_s(t) = \varepsilon > 0$.
      By right-continuity, there exists a $\delta > 0$ such that
      $x < c < x + \delta$ implies $|F_s(x) - F_s(c)| < \varepsilon$.
      Now choose $\varepsilon'$ such that $1 < \varepsilon' < \frac{t + \delta}{t}$,
      which means that $t < \varepsilon' \cdot t < t + \delta$.
      Then we get
      \[|F_s(t) - F_s(\varepsilon' \cdot t)| < \varepsilon\]
      which implies $F_s(t) \leq F_s(\varepsilon' \cdot t) < p$.
      However, since $\lim_{k \rightarrow \infty} s_k \ni s$,
      we know that there must exist some $n$ such that
      \[p \leq F_{s_n}(t) \leq F_s(\varepsilon' \cdot t) < p,\]
      which is a contradiction.
  
    \item If $p = 0$, then $F_s(t) \geq p$ for any $s$,
      so $s \in \sat{\ell_p t}$ for any $s$.
      
    \item If $p > 0$, let $F_s = \unif{a}{t}$ and $F_{s'} = \unif{t}{b}$
      for some $a$ and $b$ (if $t = 0$, $F_s = \delta_0$ instead).
      Then $F_s(t) = 1 \geq p$, so $s \in \sat{\ell_p t}$,
      but $F_{s'}(t) = 0 < p$, and hence $s' \notin \sat{\ell_p t}$.
      However, for any $r > 1$, we must have $s' \in \mathcal{B}^R_r(u)$,
      since $F_s(t) \geq F_{s'}(t)$ for any $t$.
      Hence $\mathcal{B}^R_r(s) \not\subseteq \sat{\ell_p t}$ for any $r$.
  
    \item If $p = 1$, then $F_s(t) \leq p$ for any $s$,
      so $s \in \sat{m_p t}$ for any $s$.
  
    \item If $p < 1$, let $s$ be a state with $F_s = \unif{a}{b}$
      and let $\{s_k\}$ be a sequence of states such that
      $F_{s_k} = \unif{b}{c}$ for any $k$.
      Then $\lim_{k \rightarrow \infty} s_k \ni s$
      and $F_{s_k}(b) = 0 \leq p$, so $s_k \in \sat{m_p t}$.
      However, $p < 1 = F_s(b)$, so $u \notin \sat{m_p t}$. \qedhere
  \end{enumerate}
\end{proof}

\begin{lemma}\label{lem:left}
  The following holds in the left-centered topology.
  \begin{enumerate}
    \item If $p = 1$, then $\sat{m_p t}$ is open.
    \item If $p < 1$, then $\sat{m_p t}$ is not open.
    \item If $p = 0$, $\sat{\ell_p t}$ is closed.
    \item If $p > 0$, $\sat{\ell_p t}$ is not closed.
  \end{enumerate}
\end{lemma}
\begin{proof}
  \begin{enumerate}
    \item If $p = 1$, then $F_s(t) \leq p$ for any $s$.
    
    \item If $p < 1$, then let $F_{s'} = \unif{a}{t}$
      and $F_s = \unif{t}{b}$ for some $a$ and $b$
      (if $t = 0$, let $F_{s'} = \delta_0$ instead).
      Then $F_s(t) = 0 \leq p$ and hence $s \in \sat{m_p t}$,
      but $F_{s'}(t) = 1 > p$, so $s' \notin \sat{m_p t}$.
      However, $s' \in \mathcal{B}^L_r(u)$ for any $r > 1$
      because $F_{s'}(t) \geq F_s(t)$ for all $t$.
      Hence $\mathcal{B}^L_r(s) \not\subseteq \sat{m_p t}$ for any $r$. 
    
    \item If $p = 0$, then $F_s(t) \geq p$ for any $s$.
    
    \item If $p > 0$, let $s$ be a state such that $F_s = \unif{b}{c}$
      and let $\{s_k\}$ be a sequence of states such that $F_{s_k} \sim \unif{a}{b}$ for any $k$.
      Then $\lim_{k \rightarrow \infty} s_k \ni s$
      and $F_{s_k}(b) = 1 \geq p$, and hence $s_k \in \sat{\ell_p t}$ for any $k$.
      However, $p > 0 = F_s(b)$, so $s \notin \sat{\ell_p t}$. \qedhere
  \end{enumerate}
\end{proof}

The case of $\sat{m_p t}$ is missing in Lemma \ref{lem:left}.
We have not been able to determine whether $\sat{m_p t}$ is closed in the left-centered topology,
but we strongly suspect that this is the case.
Hence we have the following conjecture.

\begin{conjecture}\label{conj:closed}
  In the left-centered topology, $\sat{m_p t}$ is closed.
\end{conjecture}

We can now show that approximate reasoning in $\mlpgeq$ is sound
with respect to the right-centered topology,
in the sense that if we have a sequence of states $s_1, s_2, \dots$
that approximate some state $s$ better and better,
then if a property holds for all $s_1, s_2, \dots$,
it will also hold for $s$.

\begin{theorem}\label{thm:closed}
  For any $\varphi \in \mlpgeq$, $\sat{\varphi}$ is closed in the right-centered topology.
\end{theorem}
\begin{proof}
  Let $\varphi \in \mlpgeq$.
  We prove by induction on $\varphi$ that
  $\sat{\varphi}$ is closed in the right-centered topology.
  
  Case $\varphi = \ell_p t$: This follows by Lemma \ref{lem:right}.
  
  Case $\varphi = L_p^a \varphi'$: Let $\{s_k\}$ be a sequence
  of states such that $\lim_{k \rightarrow \infty} s_k \ni s$
  and $s_k \models L_p^a \varphi'$.
  Let $\varepsilon > 1$. Then there exists $k'$ such that $s_{k'} \simul_\varepsilon s$,
  and by assumption, $\tau(s_{k'},a)(\sat{\varphi'}) \geq p$.
  By Theorem \ref{thm:logic-d}, this implies $\tau(s,a)(\sat{\pert{\varphi'}{\varepsilon}}) \geq p$.
  Since this holds for any $\varepsilon > 1$ and $\lim_{\varepsilon \rightarrow 1} \pert{\varphi'}{\varepsilon} = \varphi'$,
  we then get
  \[\tau(s,a)(\sat{\varphi'}) \geq p.\]
  
  Case $\varphi = \varphi_1 \land \varphi_2$: Since a finite intersection of closed sets
  is again closed, we get that
  $\sat{\varphi_1 \land \varphi_2} = \sat{\varphi_1} \cap \sat{\varphi_2}$
  is closed by the induction hypothesis.
  
  Case $\varphi = \varphi_1 \lor \varphi_2$: Since a finite union of closed sets is again closed,
  we get that
  $\sat{\varphi_1 \lor \varphi_2} = \sat{\varphi_1} \cup \sat{\varphi_2}$
  is closed by the induction hypothesis.
\end{proof}

Furthermore, if we assume Conjecture~\ref{conj:closed},
then we can use a symmetric argument as in the proof of Theorem~\ref{thm:closed}
to show that approximate reasoning in $\mlpleq$ is sound with respect to the left-centered topology.

\begin{conjecture}
  For any $\varphi \in \mlpleq$, $\sat{\varphi}$ is closed in the left-centered topology.
\end{conjecture}

%% Conclusion
\section{Conclusion and Open Problems}
We have proposed a quantitative extension of the notion of simulation relation on SMDPs,
called $\varepsilon$-simulation, comparing the relative speed of 
different processes. This quantitative notion of simulation relation induces a multiplicative 
hemimetric, which we call simulation distance, measuring the least acceleration factor 
needed by a process to speed up its actions in order to behave at least as fast as 
another process.

We have given an efficient algorithm to compute the simulation distance
and identified a class of distributions for which the algorithm works on 
finite SMDPs.
Furthermore, we have shown that, under mild conditions on the composition of 
residence-time distributions on states, a generalised version of CSP-like parallel composition
on SMDPs is non-expansive with respect to this distance, showing that our distance
is suitable for compositional reasoning.
We have also shown the connection between our distance
and properties expressible in a timed extension of Markovian logic.
Namely, we have shown 
that if the simulation distance between $s_1$ and $s_2$ is at most $\varepsilon$,
then $s_1$ satisfies the $\varepsilon$-perturbation of any property that $s_2$ satisfies.
This result also gives a novel logical characterisation of simulation and bisimulation
for semi-Markov decision processes.
Lastly we have investigated the topological properties of our distance for this logic,
and shown that approximate reasoning is sound in the limit.

Instead of using the usual stochastic order to relate the timing behaviour of states as we have done,
one could also consider many other kinds of stochastic orders,
for example ones that compare the expected value of the distributions.
This may be more natural for applications where one wants
to consider an exponential distribution with a high enough rate to be faster than a uniform distribution.

We have shown that the timing distributions that are obtained
when composing systems are compatible with the algorithm for computing the distance
only in the case when composing systems either on the left or on the right.
A more general result showing that this also
happens when composing on both sides an arbitrary number of components
seems difficult. Nonetheless, we are confident that such a result can be obtained
for any concrete case involving common types of distributions
used in the literature.

\defaultbib

\end{document}